\documentclass[11pt, oneside, a4paper, oldfontcommands]{memoir}

\usepackage[utf8]{inputenc}
\usepackage[T1]{fontenc}
\usepackage{lmodern}
\usepackage[a4paper,width=152mm,top=35mm,bottom=35mm,bindingoffset=13mm]{geometry}
\usepackage{url}
\usepackage{titlesec}
\titleformat*{\section}{\Large\bfseries\boldmath}
\titleformat*{\subsection}{\large\bfseries\boldmath}
\titleformat*{\subsubsection}{\large\itshape}
\usepackage{graphicx}
\usepackage[colorlinks=true,linkcolor=blue,citecolor=blue]{hyperref}
\usepackage[dvipsnames]{xcolor}
\usepackage{rotating}
\usepackage{amssymb}
\usepackage[normalem]{ulem}
\usepackage{pdflscape}
\usepackage{mathrsfs,amsmath}
\usepackage{units}
\usepackage{siunitx}
\usepackage{acronym}
\usepackage{booktabs}
\usepackage[UKenglish]{babel}
\usepackage{bibentry}
\usepackage{subcaption}
\usepackage{natbib}
\usepackage{datetime}
\newdateformat{monthyeardate}{%
  \monthname[\THEMONTH] \THEYEAR}

\usepackage{tikz}
\usetikzlibrary{positioning}
\usepackage{xspace}
\usetikzlibrary{backgrounds}
\usetikzlibrary{arrows}
\usetikzlibrary{calc}
\setcitestyle{authoryear,open={(},close={)}}
\usepackage{perpage}
\MakePerPage{footnote}

\newsubfloat{figure}

\newcommand{\thesisTitle}{A general framework for unbiased tests of gravity using galaxy clusters}

\newcommand{\thesisName}{Myles Arthur Mitchell}

\newcommand{\thesisSupervisor}{Prof.~Baojiu Li, Dr.~Christian Arnold and Prof.~Carlton M.~Baugh}

\makechapterstyle{mymadsen}{%
  \chapterstyle{default}

  \setlength{\chapindent}{0.0in}
  \setlength\midchapskip{1ex}
  
}

\chapterstyle{mymadsen}

\DoubleSpacing

\setsecnumdepth{subsection}

\nobibliography*

\makepagestyle{myruled}
\makeheadrule {myruled}{\textwidth}{\normalrulethickness}
\makefootrule {myruled}{\textwidth}{\normalrulethickness}{\footruleskip}
\makeevenhead {myruled}{}{\small\itshape\leftmark}{}
\makeoddhead {myruled}{}{\small\itshape\rightmark}{}
\makeevenfoot {myruled}{\small \thepage}{}{}
\makeoddfoot {myruled}{}{}{\small \thepage}

\makepsmarks {myruled}{
\nouppercaseheads
\createmark {chapter} {both} {shownumber}{}{. \ }
\createmark {section} {both}{shownumber}{} {. \ }
\createmark {subsection} {both}{shownumber}{} {. \ }
\createmark {subsubsection}{both}{shownumber}{} {. \ }
\createplainmark {toc} {both} {\contentsname}
\createplainmark {lof} {both} {\listfigurename}
\createplainmark {lot} {both} {\listtablename}
\createplainmark {bib} {both} {\bibname}
\createplainmark {index} {both} {\indexname}
\createplainmark {glossary} {both} {\glossaryname}
}

\makepagestyle{plain}
\makefootrule {plain}{\textwidth}{\normalrulethickness}{\footruleskip}
\makeevenfoot {plain}{\small \thepage}{}{}
\makeoddfoot {plain}{}{}{\small \thepage}

\makepsmarks {plain}{
\nouppercaseheads
\createmark {chapter} {both} {shownumber}{}{. \ }
\createmark {section} {both}{shownumber}{} {. \ }
\createmark {subsection} {both}{shownumber}{} {. \ }
\createmark {subsubsection}{both}{shownumber}{} {. \ }
\createplainmark {toc} {both} {\contentsname}
\createplainmark {lof} {both} {\listfigurename}
\createplainmark {lot} {both} {\listtablename}
\createplainmark {bib} {both} {\bibname}
\createplainmark {index} {both} {\indexname}
\createplainmark {glossary} {both} {\glossaryname}
}

\setsecnumdepth{subsubsection}
\begin{document}
\frontmatter
\pagenumbering{roman}           
\pagestyle{empty}               

\begin{titlingpage}
    \begin{center}
        \vspace*{1cm}

        \huge
        \textbf{\thesisTitle{}}

        \Large
        \vspace{0.5cm}

        \vspace{1.5cm}

        \textbf{\thesisName}

        \vfill

        \large
        A thesis presented for the degree of\\
        Doctor of Philosophy

        \vspace{0.8cm}

        \includegraphics[width=0.40\textwidth]{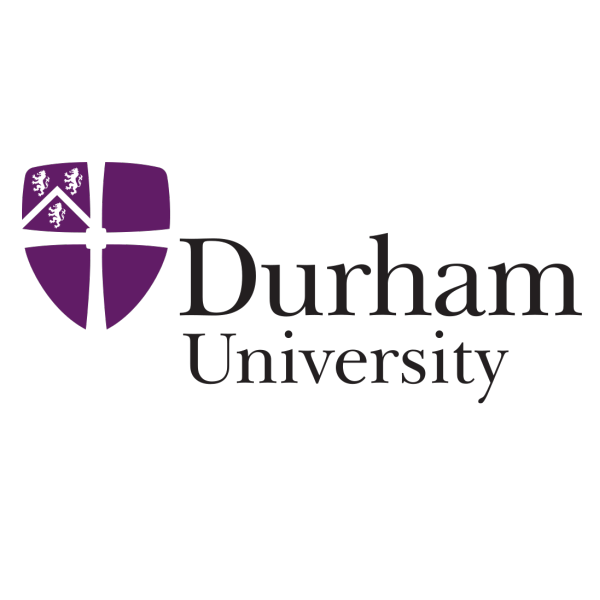}

        Institute for Computational Cosmology\\
        Department of Physics\\
        Durham University\\
        United Kingdom\\
        \monthyeardate\today

    \end{center}
\end{titlingpage}
\cleardoublepage

\pagestyle{plain}               
%
\hfill
\begin{center}
\textbf{{\Large \thesisTitle}}\\
\textbf{{\small \thesisName}}\\
\end{center}

\begin{abstract}
We present a Markov chain Monte Carlo pipeline which can be used for unbiased large-scale tests of gravity using galaxy cluster observations. The pipeline, which currently uses cluster number counts to constrain the present-day background scalar field $f_{R0}$ of Hu-Sawicki $f(R)$ gravity, fully accounts for the effects of the fifth force on cluster properties including the dynamical mass, the halo concentration and the observable-mass scaling relations. This is achieved using general models which have been calibrated over a wide and continuous mass range ($10^{11}M_{\odot}\lesssim M\lesssim10^{15}M_{\odot}$) using a large suite of cosmological simulations, including the first to simultaneously incorporate both screened modified gravity and full baryonic physics. We show, using mock cluster catalogues, that an incomplete treatment of the observable-mass scaling relations in $f(R)$ gravity, which do not necessarily follow the usual power-law behaviour, can lead to unbiased and imprecise constraints. It is therefore essential to fully account for these effects in future cosmological tests of gravity that will make use of vast cluster catalogues from ongoing and upcoming galaxy surveys. Our constraint framework can be easily extended to other gravity models; to demonstrate this, we have carried out a similar modelling of cluster properties in the normal-branch Dvali-Gabadadze-Porrati model (nDGP), which features a very different screening mechanism. Using our full-physics simulations, we also study the angular power spectra of the thermal and kinetic Sunyaev-Zel'dovich effects in $f(R)$ gravity and nDGP, and demonstrate the potential for precise constraints of gravity using data from upcoming CMB experiments. Finally, we present a retuned baryonic physics model, based on the IllustrisTNG model, which can be used for full-physics simulations within large cosmological volumes. This can be used to study the properties of galaxy groups and clusters in screened modified gravity over the mass range $10^{13}M_{\odot}\leq M\lesssim10^{15}M_{\odot}$.

\end{abstract}

\vfill
    {\small Supervisors: \thesisSupervisor}
\cleardoublepage
%
%

\chapter{Declaration}
\label{sec:declaration}
\begin{OnehalfSpacing}
The research described in this thesis was carried out between 2017 and 2021 while the author was a research student under the supervision of Prof.~Baojiu Li, Dr.~Christian Arnold and Prof.~Carlton M.~Baugh at the Institute for Computational Cosmology in the Department of Physics, Durham University. No part of this work has been submitted for any other degree or qualification at Durham University or any other university. 

\bigskip

The contents of this thesis have appeared in the following papers:
\begin{itemize}
    \item \textbf{Mitchell M.~A.}, He J.-h., Arnold C., Li B., 2018, A general framework to test gravity using galaxy clusters I: Modelling the dynamical mass of haloes in $f(R)$ gravity, \textit{MNRAS}, \textbf{477}, 1133 \textbf{(Chapter \ref{chapter:mdyn})}
    \item \textbf{Mitchell M.~A.}, Arnold C., He J.-h., Li B., 2019, A general framework to test gravity using galaxy clusters II: A universal model for the halo concentration in $f(R)$ gravity, \textit{MNRAS}, \textbf{487}, 1410 \textbf{(Chapter \ref{chapter:concentration})}
    \item \textbf{Mitchell M.~A.}, Arnold C., Hern\'andez-Aguayo C., Li B., 2021, The impact of modified gravity on the Sunyaev-Zel'dovich effect, \textit{MNRAS}, \textbf{501}, 4565 \textbf{(Chapter \ref{chapter:sz_power_spectrum})}
    \item \textbf{Mitchell M.~A.}, Arnold C., Li B., 2021, A general framework to test gravity using galaxy clusters III: Observable-mass scaling relations in $f(R)$ gravity, \textit{MNRAS}, \textbf{502}, 6101 \textbf{(Chapter \ref{chapter:scaling_relations}, sections 2 to 5)}
    \item \textbf{Mitchell M.~A.}, Hern\'andez-Aguayo C., Arnold C., Li B., 2021, A general framework to test gravity using galaxy clusters IV: Cluster and halo properties in DGP gravity, \textit{accepted by MNRAS}, arXiv:2106.13815 [astro-ph.CO] \textbf{(Chapter \ref{chapter:DGP_clusters}, sections 2 to 4)}
    \item \textbf{Mitchell M.~A.}, Arnold C., Li B., 2021, A general framework to test gravity using galaxy clusters V: A self-consistent pipeline for unbiased constraints of $f(R)$ gravity, \textit{accepted by MNRAS}, arXiv:2107.14224 [astro-ph.CO] \textbf{(Chapter \ref{chapter:constraint_pipeline}, sections 2 to 6)}
    \item \textbf{Mitchell M.~A.}, Arnold C., Li B., 2021, A general framework to test gravity using galaxy clusters VI: Realistic galaxy formation simulations to study clusters in modified gravity, \textit{submitted to MNRAS}, arXiv:2109.01147 [astro-ph.CO] \textbf{(Chapter \ref{chapter:baryonic_fine_tuning})}
\end{itemize}


The results presented in this thesis rely on cosmological simulations which have either been run by the author (the simulations presented in Chapter \ref{chapter:baryonic_fine_tuning}) or by collaborators (the simulations described in Chapters \ref{chapter:mdyn}-\ref{chapter:sz_power_spectrum}).

%

%
%
%

\vfill
    \textbf{Copyright \textcopyright~2021 by \thesisName.}

    \emph{The copyright of this thesis rests with the author. No quotation from it should be published without the author's prior written consent and information derived from it should be acknowledged}.

\end{OnehalfSpacing}

\cleardoublepage
%
%
\chapter{Acknowledgements}
\label{sec:acknowledgement}

\begin{OnehalfSpacing}
First of all, I am grateful to my supervisory team for giving me the opportunity to be involved in such fascinating and exciting projects during my PhD. I would like to thank Baojiu and Christian for always being so supportive and enthusiastic about my research- it has been a pleasure working with both of you. I also thank Carlton for getting me involved in the CDT programme, which I have thoroughly enjoyed being a part of.\bigskip

I am also grateful to my collaborators: Jianhua He for his help and guidance during the early stages of my project, and C\'esar Hern\'andez-Aguayo for letting me use his simulations. Thank you both for contributing to the work presented in this thesis. I also thank Matteo Cataneo, Weiguang Cui, Lucas Lombriser and Ian McCarthy for their helpful comments and discussions. In particular, I thank the COSMA support team for always being on hand to help.\bigskip

Last but not least, I am grateful to all of the friends that I have made throughout my time at Durham, particularly my office mates and everyone involved in the CDT programme. A special mention goes to Giorgio- I will miss our afternoon walks and lunchtime banter.\bigskip

I was supported by a studentship with the Durham Centre for Doctoral Training in Data Intensive Science, funded by the UK Science and Technology Facilities Council (STFC, ST/P006744/1) and Durham University. This thesis used the DiRAC@Durham facility managed by the Institute for Computational Cosmology on behalf of the STFC DiRAC HPC Facility (\url{www.dirac.ac.uk}). The equipment was funded by BEIS capital funding via STFC capital grants ST/K00042X/1, ST/P002293/1, ST/R002371/1 and ST/S002502/1, Durham University and STFC operations grant ST/R000832/1. DiRAC is part of the National e-Infrastructure.

\end{OnehalfSpacing}
\cleardoublepage
\setcounter{tocdepth}{3}        
{
\hypersetup{hidelinks}
\tableofcontents*                
}
\cleardoublepage
{
\hypersetup{hidelinks}
\listoffigures
}
\cleardoublepage
{
\hypersetup{hidelinks}
\listoftables
}
\cleardoublepage

\newpage

\mainmatter

\pagestyle{myruled}

\setlength{\parindent}{5ex}

\setcounter{equation}{0}
\graphicspath{{./gfx/}}

\chapter{Introduction}
\label{chapter:intro}

\section{\texorpdfstring{$\Lambda$}{L}CDM: the standard model of cosmology}

Cosmology is the study of the origin and evolution of the Universe. The primary goal is to understand what happened between the `Big Bang' ($\sim14$ billion years ago) and the present-day and to understand the fundamental physics driving this evolution; how and why quantum fluctuations evolved into density perturbations and eventually into galaxies like the Milky Way in which we were formed. The best way to answer these questions is to develop theoretical models whose predictions can be compared with astronomical observations. 

\sloppy The current standard model of cosmology is the `$\Lambda$-cold-dark-matter' ($\Lambda$CDM) paradigm. This is the simplest model whose predictions can match a number of key observations, including the temperature fluctuations in the cosmic microwave background (CMB) \citep[e.g.,][]{WMAP:2003elm}, the late-time accelerated expansion of the cosmos \citep[e.g.,][]{SupernovaSearchTeam:1998fmf}, and the distribution of galaxies on large scales \citep[e.g.,][]{SDSS:2005xqv}. The model makes the following assumptions: gravity, which is the dominant force on cosmological scales, obeys Einstein's General Relativity (GR); matter is primarily made up of a `cold dark matter' component, which is assumed to consist of massive particles which are yet to be detected experimentally; a smaller proportion ($\sim15\%$) of matter is made up of `visible' matter, including gas and stars (we will refer to this as `baryonic' matter); and the accelerated expansion is brought about by assigning a positive value to the cosmological constant $\Lambda$ which appears in the framework of GR. 

Throughout this chapter, we will use the unit convention $c=1$ for the speed of light in a vacuum. Greek indices can take values 0, 1, 2 and 3, and, unless otherwise stated, a subscript $_0$ denotes the present-day value of a quantity.

\subsection{Cosmic expansion in \texorpdfstring{$\Lambda$}{L}CDM}

The size of the Universe at a given time $t$ can be parameterised using the cosmic scale factor $a$, which is normalised such that $a(t=t_0)\equiv a_0=1$ at the present-day. Because the Universe is expanding, the light from distant galaxies is observed to be redshifted. The redshift $z$ is related to the scale factor by the relation $1+z=1/a$, and both $z$ and $a$ can be used as alternative coordinates of time. 

The expansion of space can be determined using the Einstein field equations which govern GR:
\begin{equation}
    G_{\alpha\beta} + \Lambda g_{\alpha\beta} = 8\pi GT_{\alpha\beta},
    \label{eq:einstein_eqn}
\end{equation}
where $G_{\alpha\beta}$ is the Einstein tensor, $\Lambda$ is the cosmological constant, $g_{\alpha\beta}$ is the metric tensor, $G$ is Newton's gravitational constant and $T_{\alpha\beta}$ is the stress-energy tensor. This tensor equation describes how the curvature of spacetime (encapsulated in $G_{\alpha\beta}$) is related to the distribution of energy and momentum (encapsulated in $T_{\alpha\beta}$). By using the Friedmann-Lema\^itre-Robertson-Walker metric for a homogeneous and isotropic expanding universe, the Friedmann equation can be derived from the 00 component of Eq.~(\ref{eq:einstein_eqn}):
\begin{equation}
    \left(\frac{\dot{a}}{a}\right)^2\equiv H^2=\frac{8\pi G}{3}\left(\rho_{\rm M}+\rho_{\rm r}\right) + \frac{K}{a^2} + \frac{\Lambda}{3},
    \label{eq:friedmann}
\end{equation}
where $H$ is the Hubble parameter. This equation can determine the cosmic expansion rate (where $\dot{a}$ is the time-derivative of $a$) as a function of the background densities of matter $\rho_{\rm M}$ and radiation $\rho_{\rm r}$, the curvature $K$ and the cosmological constant. It can be shown that a positive cosmological constant is required for an accelerating expansion: physically, this can be thought of as a vacuum energy with constant density $\rho_{\rm v}=\Lambda/(8\pi G)$.

Observations \citep[e.g.,][]{Planck:2018vyg} indicate that the Universe is extremely close to being spatially flat ($K\approx0$). Therefore, the second term in the right-hand side of Eq.~(\ref{eq:friedmann}) can be neglected and the sum of the energy densities $\rho_{\rm M}$, $\rho_{\rm r}$ and $\rho_{\rm v}$ is equal to the critical density for a spatially flat universe:
\begin{equation}
    \rho_{\rm crit} = \frac{3H^2}{8\pi G}.
    \label{eq:rho_crit}
\end{equation}
In the $\Lambda$CDM framework, the energy densities are often reparameterised in terms of the dimensionless ratios of the present-day densities with respect to the critical density: $\Omega_{\rm M}=\rho_{\rm M,0}/\rho_{\rm crit,0}$, $\Omega_{\rm r}=\rho_{\rm r,0}/\rho_{\rm crit,0}$ and $\Omega_{\Lambda}=\rho_{\rm v}/\rho_{\rm crit,0}$. The Friedmann equation can be rewritten in terms of these parameters:
\begin{equation}
    \frac{H(z)}{H_0} \equiv E(z) = \sqrt{\Omega_{\rm M}(1+z)^3 + \Omega_{\rm r}(1+z)^4 + \Omega_{\Lambda}},
    \label{eq:hubble_evolution}
\end{equation}
where $H_0$ is the present-day value of the Hubble parameter (the `Hubble constant'), $E(z)$ is the time evolution of the Hubble parameter, and we have used the time-dependencies $\rho_{\rm M}(z)\propto(1+z)^3$ and $\rho_{\rm r}(z)\propto(1+z)^4$ to derive the right-hand side. For a given set of values of the parameters $H_0$, $\Omega_{\rm M}$, $\Omega_{\rm r}$ and $\Omega_{\Lambda}$, the expansion history of the Universe can be inferred using Eq.~(\ref{eq:hubble_evolution}). Observations show that the radiation component $\Omega_{\rm r}$ ($\approx10^{-4}$) is very small compared to $\Omega_{\rm M}$ ($\approx0.3$) and $\Omega_{\Lambda}$ ($\approx0.7$) \citep[e.g.,][]{Planck:2018vyg}. Our work is primarily focused on late times (low $z$), where the radiation contribution in Eq.~(\ref{eq:hubble_evolution}) is negligible; we will therefore assume $\Omega_{\rm r}=0$ for the remainder of this thesis.

\subsection{Strengths and weakness of \texorpdfstring{$\Lambda$}{L}CDM}

One of the most notable successes of $\Lambda$CDM is in accurately predicting the statistical properties of the CMB temperature fluctuations. This is achieved using linear perturbation theory, which uses fluid dynamics to predict the evolution of primordial perturbations in the matter density field (which are seeded by quantum fluctuations) from inflation to the epoch of recombination. One of the key differential equations in perturbation theory, which will appear later in this thesis, is the Poisson equation:
\begin{equation}
    \nabla^2\Phi = 4\pi G\delta\rho_{\rm M},
    \label{eq:poisson_eqn}
\end{equation}
where $\nabla^2$ is a (second-order) space derivative rather than a spacetime derivative. This is used to relate the Newtonian gravitational potential $\Phi$ to the matter density perturbations $\delta\rho_{\rm M}$ of the cosmic fluid. By comparing observations of the CMB with the predictions from linear perturbation theory, precise constraints can be made of $\Lambda$CDM parameters including $\Omega_{\rm M}$, $\Omega_{\Lambda}$ and $H_0$ \citep[e.g.,][]{Planck:2013pxb}.

At later times, the evolution of the density fluctuations (which start to undergo gravitational collapse) becomes nonlinear and extremely difficult to solve analytically. Instead, computer simulations can be used to evolve the density field from some early time (based on initial conditions that are consistent with CMB observations) to the present-day \citep[e.g.,][]{2009MNRAS.398.1150B}. This numerical approach is able to reproduce the large-scale distribution of galaxies at the present-day (the `cosmic web'). This close match to galaxy observations is another success of $\Lambda$CDM, and comparing predictions and observations of galaxy statistics (for example, the abundance of galaxy clusters) offers another means of constraining the model parameters \citep[e.g.,][]{2012MNRAS.427.3435A,Planck_SZ_cluster}.

Arguably the greatest success of $\Lambda$CDM is that it requires only six independent parameters to accurately reproduce these observations and other phenomena such as the late-time accelerated expansion. This simplicity is why it has become the standard working model of cosmology.

Despite these successes, $\Lambda$CDM also has a number of weaknesses. For example: dark matter particles have still not been detected by particle physics experiments; numerical simulations predict too many dwarf-size galaxies and Milky Way satellites \citep[e.g.,][]{Weinberg:2013aya}; and estimates of the vacuum energy density from quantum electrodynamics are many orders of magnitude larger than the value derived from cosmological constraints of $\Lambda$ \citep[e.g.,][]{Martin:2012bt}. In this thesis, we will be addressing the latter, namely the possibility that a different mechanism is driving the accelerated cosmic expansion. One possibility is that the acceleration is caused by some exotic matter species that has not yet been detected; however, we will be focusing on another idea that the acceleration arises due to departures from GR on large scales.

\section{Modifications to General Relativity}

A wide range of modified gravity (MG) models have been used to explain the accelerated cosmic expansion \citep[see, e.g.,][]{Koyama:2015vza}. We will be focusing on two popular models --- $f(R)$ gravity \citep[e.g.,][]{Sotiriou:2008rp,DeFelice:2010aj} and the Dvali-Gabadadze-Porrati (DGP) model \citep{DVALI2000208} --- which both feature an extra scalar field that mediates a force between matter particles. Since there are already four observed fundamental interactions, this is called the `fifth force'. When this force is able to act, the total strength of gravity is enhanced; this can speed up the formation of large-scale structure and create observational signatures, such as an enhanced abundance of galaxy clusters, which can be used to probe gravity on large scales. However, tests of gravity within our Solar System have already verified GR to a remarkably high precision \citep[e.g.,][]{Will:2014kxa}, ruling out a strengthened gravitational force within our local environment; to get around this, both models feature `screening' mechanisms which suppress the fifth force in particular regimes, including sufficiently high-density regions or regions where the second derivatives of the scalar field are large, enabling the models to evade Solar System tests.

The following sections will provide some background on these models, including details of the underlying theory, the screening mechanisms and the strength of the fifth force.

\subsection{\texorpdfstring{$f(R)$}{f(R)} gravity}
\label{sec:intro:fR}

The $f(R)$ gravity model is an extension of GR which is constructed by adding a nonlinear function of the Ricci scalar curvature, $f(R)$, to the $R$ term in the Einstein-Hilbert action:
\begin{equation}
    S=\int {\rm d}^4x\sqrt{-g}\left[\frac{R+f(R)}{16\pi G}+\mathcal{L}_{\rm M}\right],
\label{eq:fR_action}
\end{equation}
where $g$ is the determinant of the metric and $\mathcal{L}_{\rm M}$ is the Lagrangian density for matter fields (as mentioned above, we will mainly focus on late-time behaviour, where matter is non-relativistic). By setting the variation of the action with respect to the metric to zero, we obtain the modified Einstein field equations, which now contain a new tensor $X_{\alpha \beta}$:
\begin{equation}
    G_{\alpha \beta} + X_{\alpha \beta} = 8\pi GT_{\alpha \beta},
\label{eq:fr_field_equations}
\end{equation}
where:
\begin{equation}
    X_{\alpha \beta} = f_RR_{\alpha \beta} - \left(\frac{f}{2}-\Box f_R\right)g_{\alpha \beta} - \nabla_{\alpha}\nabla_{\beta}f_R.
\label{eq:GR_modification}
\end{equation}
The tensor $R_{\alpha \beta}$ represents the Ricci curvature tensor, $\nabla_{\alpha}$ is the covariant derivative associated with the metric, and $\Box$ is the d'Alembert operator. The derivative $f_R\equiv{\rm d}f(R)/{\rm d}R$ represents the extra dynamic degree of freedom in this model, and can be treated as a scalar field whose dynamics is governed by the following equation of motion:
\begin{equation}
    \Box f_R = \frac{1}{3}(R - f_RR + 2f + 8\pi G\rho_{\rm M}),
\end{equation}
which is derived from the trace of Eq.~(\ref{eq:fr_field_equations}). The scalar field mediates a fifth force whose physical range is set by the Compton wavelength:
\begin{equation}
    \lambda_{\rm C} = a^{-1}\left(3\frac{{\rm d}f_R}{{\rm d}R}\right)^{\frac{1}{2}}.
\label{eq:compton_wavelength}
\end{equation}
The fifth force is an attractive force which is felt by massive particles. In sufficiently low-density environments (for example, cosmic voids and the outer regions of galaxy clusters), it enhances the strength of gravity by up to a factor of $1/3$. However, in high-density regions, it is suppressed and gravity behaves according to GR. This is caused by the chameleon screening mechanism \citep[e.g.,][]{Khoury:2003aq,Khoury:2003rn,Mota:2006fz}, which was included in the model to ensure consistency with Solar System tests. The screening is brought about by an environment-dependent effective mass of the scalar field which becomes very heavy in dense regions so that the fifth force becomes very short-ranged and undetectable. Consequently, the fifth force can only act in regions where the gravitational potential well is not too deep.

Structure formation in $f(R)$ gravity is governed by the modified Poisson equation, which, in the weak-field and quasi-static limits, is given by \citep[e.g.,][]{Li:2011vk}:
\begin{equation}
    \nabla^2\Phi = \frac{16\pi G}{3}\delta\rho_{\rm M} - \frac{1}{6}\delta R,
    \label{eq:fr_poisson_eqn}
\end{equation}
where $\delta R$ represents the perturbation of the Ricci scalar curvature and $\Phi$ now represents the modified gravitational potential. The scalar field is related to the curvature and density perturbations as follows:
\begin{equation}
    \nabla^2f_R = \frac{1}{3}(\delta R - 8\pi G\delta\rho_{\rm M}).
\end{equation}

\begin{figure}
\centering
\includegraphics[width = 0.9\columnwidth]{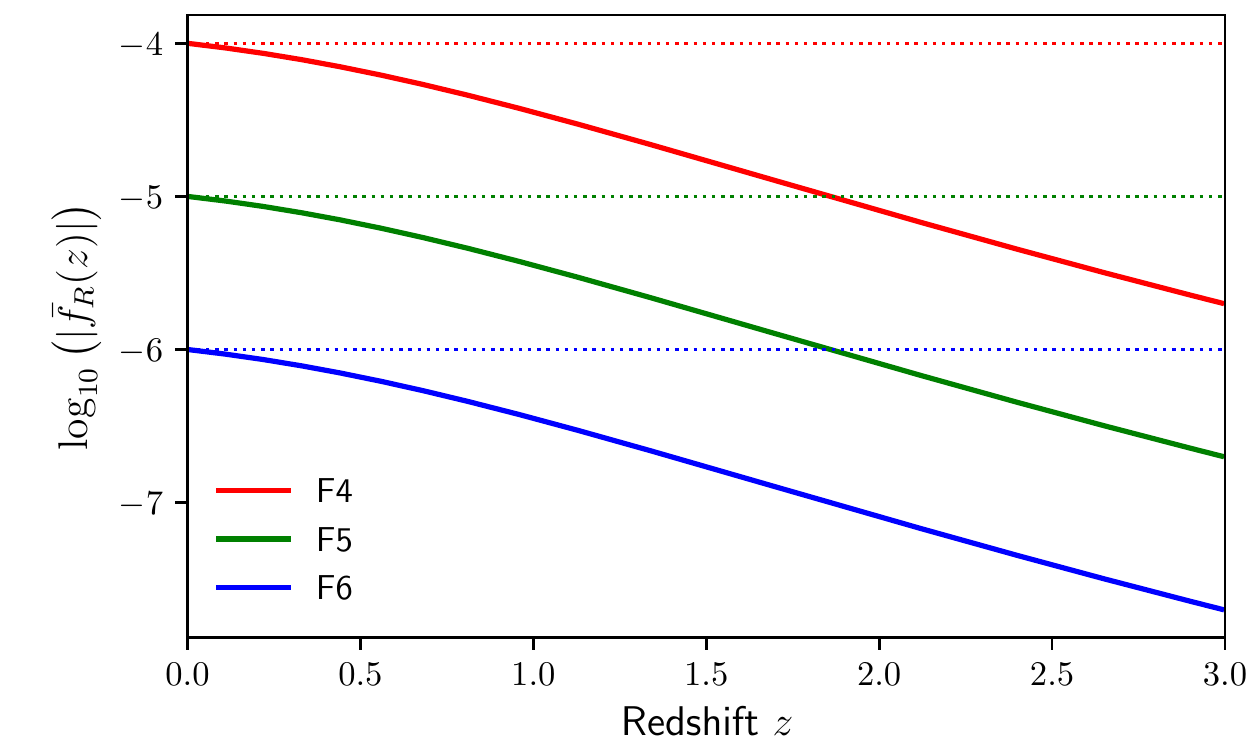}
\caption[Absolute value of the background scalar field in HS $f(R)$ gravity as a function of redshift.]{Absolute value of the background scalar field in HS $f(R)$ gravity plotted as a function of redshift for the F4, F5 and F6 models (from top to bottom), assuming the Hu-Sawicki $f(R)$ model with parameters $n=1$ and $|f_{R0}|=10^{-4}, 10^{-5}, 10^{-6}$ respectively. Cosmological parameters $\Omega_{\Lambda}=0.719$ and $\Omega_{\rm M}=0.281$ are used.}
\label{fig:fR_z_dependence}
\end{figure}

This work will focus on the Hu-Sawicki (HS) model \citep{Hu:2007nk} of $f(R)$ gravity, which uses the following prescription for $f(R)$:
\begin{equation}
    f(R) = -m^2\frac{c_1\left(-R/m^2\right)^n}{c_2\left(-R/m^2\right)^n+1},
\label{eq:hu_sawicki}
\end{equation}
where $m^2\equiv8\pi G\rho_{\rm M,0}/3=H_0^2\Omega_{\rm M}$. The theory has three model parameters: $c_1$, $c_2$ and $n$. By choosing $c_1/c_2=6\Omega_{\Lambda}/\Omega_{\rm M}$ and assuming that the curvature obeys the inequality $-R\gg m^2$, it can be shown that the function $f(R)$ behaves as a cosmological constant in background cosmology \citep{Hu:2007nk}, giving rise to the observed late-time accelerated expansion. Under these assumptions, and by choosing $n=1$ (which is commonly used in $f(R)$ literature), it follows that:
\begin{equation}
    f_R \simeq -\frac{c_1}{c_2^2}\left(\frac{m^2}{-R}\right)^2.
\label{eq:fR}
\end{equation}
Therefore, the background value of the scalar field, $\bar{f}_R$, can be expressed as:
\begin{equation}
    \bar{f}_R(a) = \bar{f}_{R0}\left(\frac{\bar{R}_0}{\bar{R}(a)}\right)^2,
\label{eq:fR_background}
\end{equation}
where, assuming that the background expansion history is almost indistinguishable from that of $\Lambda$CDM, the background curvature $\bar{R}$ is given by:
\begin{equation}
    \bar{R} \simeq -3m^2\left[a^{-3} + 4\frac{\Omega_{\Lambda}}{\Omega_{\rm M}}\right].
\label{eq:ricci_scalar}
\end{equation}
We note that $-\bar{R}\gg m^2$ is a good approximation for a realistic choice of cosmological parameters. Therefore, using realistic approximations, we are able to work with just a single parameter: $\bar{f}_{R0}$, the present-day background scalar field (the over-bar of $\bar{f}_{R0}$ will be omitted for the remainder of this thesis). The amplitude $|f_{R0}|$ represents the highest value of the scalar field in cosmic history: see Fig.~\ref{fig:fR_z_dependence} for the time evolution between redshifts 0 and 3. A higher value of $|f_{R0}|$ corresponds to a stronger modification of GR, allowing regions of higher density to be unscreened at a given time. As a convention, we will refer to models of HS $f(R)$ gravity with $|f_{R0}|=10^{-6}$, $|f_{R0}|=10^{-5.5}$, $|f_{R0}|=10^{-5}$, ... as F6, F5.5, F5, etc.

\subsection{Dvali-Gabadadze-Porrati model}
\label{sec:intro:nDGP}

The DGP model consists of two branches: a `self-accelerating' branch (sDGP) and a `normal' branch (nDGP). The former is able to reproduce the late-time cosmic acceleration without requiring any dark energy; however, it is also prone to ghost instabilities which are absent in the latter \citep[e.g.,][]{Koyama:2007za}. Because of this, nDGP is the preferred model of gravity. One caveat of this model is that it requires an extra dark energy component in order to be viable; nevertheless, it is still a useful toy model that can be used to search for deviations from GR.

\sloppy In nDGP, the Universe is modelled as a 4-dimensional brane embedded in a 5-dimensional bulk spacetime. The gravitational action of the model consists of two terms:
\begin{equation}
    S=\int_{\rm brane} {\rm d}^4x\sqrt{-g}\left(\frac{R}{16\pi G}\right) + \int {\rm d}^5x\sqrt{-g^{(5)}}\left(\frac{R^{(5)}}{16\pi G^{(5)}}\right),
\label{eq:DGP_action}
\end{equation}
where the first term is the usual Einstein-Hilbert action of GR and the second term represents the contribution from the 5D bulk (where $g^{(5)}$, $R^{(5)}$ and $G^{(5)}$ are analogous to $g$, $R$ and $G$). A characteristic length scale, known as the cross-over scale $r_{\rm c}$, can be defined: 
\begin{equation}
    r_{\rm c} = \frac{1}{2}\frac{G^{(5)}}{G}.
\label{eq:DGP_crossover}
\end{equation}
The second term of Eq.~(\ref{eq:DGP_action}) will dominate on scales larger than the cross-over scale, where gravity transitions to 5D. Taking the variation of Eq.~(\ref{eq:DGP_action}) and setting this to zero leads to the modified Friedmann equation:
\begin{equation}
    \frac{H(a)}{H_0} = \sqrt{\Omega_{\rm M}a^{-3} + \Omega_{\rm DE}(a) + \Omega_{\rm rc}} - \sqrt{\Omega_{\rm rc}},
\label{eq:DGP_friedmann}
\end{equation}
where $\Omega_{\rm DE}(a)$ is the dimensionless density parameter for the additional dark energy component mentioned above, and $\Omega_{\rm rc}$ is given by:
\begin{equation}
    \Omega_{\rm rc} \equiv \frac{1}{4H_0^2r_{\rm c}^2}.
\label{eq:DGP_omega}
\end{equation}
We will assume that $\Omega_{\rm DE}(a)$ takes a form that makes $H(a)$ identical to a $\Lambda$CDM expansion history, and that the dark energy component has negligible clustering on the sub-horizon scales that are the focus of this work. The quantity $H_0r_{\rm c}$ is often used to quantify deviations from GR, with a larger value representing a smaller departure from GR. As a convention, we will refer to models with $H_0r_{\rm c}=5$, $2$, $1$ and $0.5$ as N5, N2, N1, and N0.5, respectively. 

The modified Poisson equation in nDGP, in the weak-field and quasi-static limits, is given by \citep{PhysRevD.75.084040}:
\begin{equation}
    \nabla^2\Phi = 4\pi Ga^2\delta\rho_{\rm M} + \frac{1}{2}\nabla^2\varphi,
\label{eq:DGP_pot}
\end{equation}
where $\varphi$ is the extra scalar field of the model. This satisfies the following dynamical equation of motion:
\begin{equation}
    \nabla^2\varphi + \frac{r_{\rm c}^2}{3\beta a^2}\left[(\nabla^2\varphi)^2 - (\nabla_i\nabla_j\varphi)(\nabla^i\nabla^j\varphi)\right] = \frac{8\pi Ga^2}{3\beta}\delta\rho_{\rm M},
\label{eq:DGP_scalar_field}
\end{equation}
where the function $\beta$ is given by:
\begin{equation}
    \beta(a) = 1 + 2Hr_{\rm c}\left(1 + \frac{\dot{H}}{3H^2}\right) = 1 + \frac{\Omega_{\rm M}a^{-3} + 2\Omega_{\Lambda}}{2\sqrt{\Omega_{\rm rc}\left(\Omega_{\rm M}a^{-3} + \Omega_{\Lambda}\right)}}.
\label{eq:DGP_beta}
\end{equation}
The fifth force can act on sufficiently large scales, where the nonlinear terms in the square bracket of Eq.~(\ref{eq:DGP_scalar_field}) can be ignored. This enhances the strength of gravity by a factor of $[1 + 1/(3\beta)]$, which is larger at later times with present-day enhancements of approximately $1.04$ for N5, $1.08$ for N2, $1.12$ for N1 and $1.18$ for N0.5. On small scales, the nonlinearity of the scalar field causes the fifth force to be suppressed via the Vainshtein screening mechanism \citep{VAINSHTEIN1972393}. This efficiently suppresses the fifth force in regimes including the Solar System, galaxies and galaxy clusters.

\section{Testing gravity with large-scale structure}
\label{sec:intro:constraints}

The fifth forces in $f(R)$ gravity and nDGP leave numerous observational signatures in large-scale cosmic structure. For example, the sped-up structure formation that results from a strengthened gravity can lead to a greater number of galaxy clusters being formed by the present-day. Galaxy clusters are the largest objects to have been observed in the Universe: they are gravitationally-bound groups of thousands of galaxies that are found within vast dark matter haloes whose mass ranges from $\sim10^{14}M_{\odot}$ to $\sim10^{16}M_{\odot}$. Galaxy clusters are therefore thought to trace the highest peaks of the primordial density field, and are therefore highly sensitive to the values of cosmological parameters that affect cosmic structure formation, including $f_{R0}$ and $\Omega_{\rm rc}$ which control the effect of the fifth forces in $f(R)$ gravity and nDGP, respectively. Clusters can also be detected using a number of means, including the X-ray emission from the hot intra-cluster gas and the Sunyaev-Zel'dovich (SZ) effect, which is caused by inverse Compton scattering of CMB photons off high-energy electrons in the intra-cluster gas. By comparing observational data with theoretical predictions, statistics including cluster number counts can be used to probe the strength of gravity on the largest scales.


The following sections are laid out as follows: Sec.~\ref{sec:introduction:previous_work} will summarise previous tests of $f(R)$ gravity and nDGP using large scale structure; Sec.~\ref{sec:introduction:observations} will outline ongoing and future galaxy cluster surveys and discuss some key considerations for future gravity tests; Sec.~\ref{sec:introduction:simulations} will provide an outline of $N$-body and hydrodynamical numerical simulations, which provide the best means of modelling the effects of a strengthened gravity on the properties of galaxy clusters; finally, Sec.~\ref{sec:introduction:mcmc} will provide an introduction to Markov chain Monte Carlo (MCMC) sampling, which is a popular method for inferring parameter constraints using observations and theoretical predictions.

\subsection{Previous work}
\label{sec:introduction:previous_work}

Previous works have tested $f(R)$ gravity using cluster number counts \citep[e.g.,][]{PhysRevD.92.044009,Liu:2016xes,Peirone:2016wca} and the clustering of clusters \citep{Arnalte-Mur:2016alq}, which are both enhanced by the sped-up structure formation. Meanwhile, the effects of the fifth force on galaxy velocities have been tested using redshift-space distortions \citep[e.g.,][]{Bose:2017dtl,2018NatAs...2..967H,Hernandez-Aguayo:2018oxg}. The temperature of the intra-cluster gas, which is closely correlated with the total gravitational potential, is also enhanced by the fifth force; the model can therefore be tested using the temperature-mass relation \citep[see, e.g.,][]{Hammami:2016npf,DelPopolo:2019oxn}. A raised temperature will produce greater X-ray emission from high-energy electrons and a higher level of inverse Compton scattering of CMB photons. The $f(R)$ model can therefore be tested using the cluster SZ profile \citep{deMartino:2016xso}, the cluster gas mass fraction probed through the gas temperature and X-ray luminosity \citep[e.g.,][]{Li:2015rva}, and by comparing weak lensing measurements of clusters (which are unaffected by the fifth force for realistic models) with X-ray and SZ observations \citep[e.g.,][]{Terukina:2013eqa,Wilcox:2015kna}. The model has also been tested using weak lensing by cosmic voids \citep{Cautun:2017tkc}, which, owing to their extremely low density, are largely unaffected by the chameleon screening.

The Vainshtein mechanism of nDGP is more efficient than the chameleon mechanism of $f(R)$ gravity at screening out the fifth force within galaxy clusters. However, the nDGP fifth force can still speed up the formation of clusters, and in recent years the model has been tested using cluster number counts \citep[e.g.,][]{2009PhRvD..80l3003S,vonBraun-Bates:2018lxq} and redshift-space distortions \citep[e.g.,][]{PhysRevD.94.084022,Hernandez-Aguayo:2018oxg}. Again, cosmic voids are largely unaffected by the screening and can be used to test the model \citep[e.g.,][]{Falck:2017rvl,10.1093/mnras/stz022}. Other models which feature Vainshtein screening have also been tested by, for example, comparing weak lensing data with SZ and X-ray cluster observations \citep{Terukina:2015jua}.

\subsection{Current and future observations}
\label{sec:introduction:observations}

It is currently an exciting time for cluster cosmology: various ongoing and upcoming astronomical surveys are generating vast cluster catalogues using all means of detection, including the clustering of galaxies in optical/infrared surveys \citep[e.g.,][]{ukidss,lsst,euclid,desi}, distortions of the CMB by the SZ effect \citep[e.g.,][]{act,Planck_SZ_cluster,Abazajian:2016yjj,Ade:2018sbj}, and X-ray emission from the hot intra-cluster gas \citep[][]{chandra,xmm-newton,erosita}. These will be many times larger than previous catalogues, and have the potential to significantly advance our understanding of gravity at the largest scales and the mechanisms driving the accelerated cosmic expansion. 

Current X-ray surveys include XMM-Newton \citep{xmm-newton} and Chandra \citep{chandra}, which together provide coverage of $\sim$1000 clusters. The ongoing X-ray survey eROSITA \citep{erosita} will detect $\sim$125,000 objects with mass $M>10^{13}h^{-1}M_{\odot}$ over redshifts $z<1$, most of which will be galaxy groups. The cluster number count data from eROSITA is expected to produce very tight ($\sim1\%$) constraints on $\Omega_{\rm M}$ and the root-mean-squared linear matter density fluctuation $\sigma_8$. Another planned X-ray survey, Athena \citep{2017AN....338..153B}, will launch in the early 2030s and detect clusters with mass $M>5\times10^{13}M_{\odot}$, including distant objects at $z>2$.

Current CMB experiments include Planck \citep[e.g.,][]{Planck_SZ_cluster}, the South Pole Telescope \citep[SPT, e.g.,][]{Bocquet:2018ukq} and the Atacama Cosmology Telescope \citep[ACT, e.g.,][]{act}. The Planck survey has detected over $400$ clusters spanning $2\times10^{14}M_{\odot}$ to $10^{15}M_{\odot}$ over redshifts $z<1$. The Advanced ACTPol (AdvACT) receiver has detected over 4000 clusters since it launched in 2016 \citep{ACT:2020lcv}. The upcoming SPT-3G survey \citep{SPT-3G:2014dbx} will extend the work of the existing SPT-SZ survey and detect $\sim$5000 clusters with mass $M\gtrsim10^{14}M_{\odot}$ by 2023. Further upcoming ground-based experiments include the Simons Observatory \citep[2022,][]{Ade:2018sbj}, which will produce a legacy catalogue of $\sim$16,000 clusters, and CMB-S4 \citep[2029,][]{Abazajian:2016yjj}, which is expected to detect $\sim$140,000 clusters with mass exceeding $6-8\times10^{13}M_{\odot}$.

Existing optical/infrared surveys include the ongoing Dark Energy Survey \citep[DES,][]{DES:2016jjg}, which will detect $\sim10^5$ clusters spanning $z\lesssim1$, and the Dark Energy Spectroscopic Instrument \citep[DESI,][]{desi}. Upcoming surveys include the space-based survey Euclid \citep[e.g.,][]{Euclid:2019bue}, and the ground-based Vera Rubin Observatory \citep{lsst} and 4MOST \citep[e.g.,][]{4MOST:2019qkp}. Together, these surveys will be capable of detecting 100,000s of clusters out to high redshifts and an unprecedented number of galaxy groups. For example, Euclid (starting 2022) will detect $>10^5$ clusters out to redshift $z\sim2$ and down to mass $10^{13.5}M_{\odot}$, while the Vera Rubin Observatory (starting 2022-2023) expects to obtain a nearly complete sample out to $z\sim1.2$, with $\sim10^5$ massive clusters and $\sim10^6$ galaxy groups. Meanwhile, 4MOST will provide spectroscopic data for $\sim$40,000 groups and clusters detected by eROSITA.

Synergies between these surveys will be vital for precise constraints. For example, a major challenge in cluster cosmology is precise calibration of the cluster mass. In the most recent cluster abundance constraints from Planck \citep{Planck_SZ_cluster}, it was noted that the cluster mass calibration, estimated to have a precision of $10-15\%$, was the single largest source of uncertainty. Recent works have combined weak lensing data, which can be used to estimate the absolute mass, with SZ cluster data in order to reduce systematics related to mass calibration \citep[e.g.,][]{Bocquet:2018ukq}. The weak lensing mass calibration used by the upcoming optical/infrared surveys listed above is expected to achieve $1-2\%$ precision, so crossovers with SZ and X-ray selected catalogues will be very helpful. For example, it is planned that the DES cluster sample will overlap with 1000s of SZ and X-ray selected clusters. The upcoming CMB surveys described above will also use CMB lensing as a novel mass calibration technique, which will help to further reduce these systematics.

In order to make the best use of this wealth of observational data for MG tests, it is now vital to prepare accurate and robust theoretical predictions which can be safely combined with the observations to make precise and unbiased constraints. As mentioned in Sec.~\ref{sec:introduction:previous_work}, the fifth force can affect the total gravitational potential and the thermal properties of clusters, and it can also affect the density profile. If these effects are not fully accounted for in tests that use, for example, cluster number counts, the measurements of the cluster mass could be biased. 

For example, previous works \citep[e.g.,][]{He:2015mva} have shown that the $f(R)$ fifth force can alter the amplitude of the intra-cluster gas temperature and related observables, including the integrated SZ flux, by up to $33\%$ (i.e., the maximum enhancement of the total gravitational potential). For the case of the integrated SZ flux, which according to \citet{Planck_SZ_cluster} varies with the cluster mass as $\sim M^{1.79}$, failing to account for the effect of the fifth force could lead to an estimate of the mass that is biased by up to $\sim15\%$: this is similar to the $10-15\%$ precision of the mass calibration used by Planck. Therefore, as weak lensing and novel CMB lensing mass calibrations continue to approach the $1-2\%$ level targeted by upcoming surveys, it will be vital to conduct a full modelling of the effects of the fifth force on cluster properties in order to avoid biased constraints in future tests.

Another exciting prospect is the inclusion of data from the galaxy group regime ($10^{13}M_{\odot}\lesssim M\lesssim10^{14}M_{\odot}$) in cluster catalogues for the first time \citep[e.g.,][]{2021Univ....7..139L}. This has important implications for gravity tests. For example, the screening of the fifth force in $f(R)$ gravity is typically less efficient for lower-mass objects, making departures from GR more prevalent in low-mass clusters and groups than in massive clusters. The detections of galaxy groups down to $10^{13}M_{\odot}$ will consequently make it possible to rule out weaker models of $f(R)$ gravity.

\subsection{$N$-body and hydrodynamical simulations}
\label{sec:introduction:simulations}

The best way to model and understand cluster properties is to use numerical cosmological simulations which can accurately reproduce the formation of large-scale structure. The simplest examples are $N$-body simulations which assume that the matter content of the Universe is entirely in the form of collisionless dark matter. These consist of massive `particles' which only interact via gravity, where the gravitational potential is evaluated at each timestep using the Poisson equation, which for GR is given by Eq.~(\ref{eq:poisson_eqn}). The simulation is started at some early time using initial conditions which are consistent with, for example, cosmological constraints from CMB temperature fluctuations. Over the course of the simulation, the dark matter particles will attract and cluster together to form gravitationally-bound structures called `haloes' and `subhaloes': these are the theoretical counterparts of galaxy groups and clusters and galaxies, respectively.

Dark-matter-only (DMO) simulations have been widely used in past works: for example, they can successfully reproduce the web-like structure of the Universe on large scales \citep[e.g.,][]{2009MNRAS.398.1150B}, and they can be used for accurate theoretical predictions of the cluster abundance \citep[e.g.,][]{Tinker:2008ff}, which is quantified using the halo mass function (HMF). However, a notable limitation of these simulations is the absence of baryonic matter, which behaves very differently to collisionless dark matter. This can be incorporated in the form of gas, which, in addition to interacting via gravity, must also obey the hydrodynamical fluid equations.

There are a number of ways to include gas. For example, some simulation codes use smoothed particle hydrodynamics (SPH), where the continuum of the gas fluid is approximated using particles with a local resolution that automatically follows the mass flow \citep[e.g.,][]{Springel:2005mi}. Another example is the \textsc{arepo} code \citep{2010MNRAS.401..791S}, which has been used to run all of the hydrodynamical simulations presented in this thesis. This tracks the gas using a moving, unstructured Voronoi mesh, which is made up of gas cells that adaptively refine (split) and derefine (merge) such that the mass of any cell does not differ by more than a factor of two from the mean. The \textsc{arepo} code can also account for the presence of a magnetic field which dynamically couples to the gas through magnetic pressure. This is achieved by solving the ideal magneto-hydrodynamics equations \citep[see][]{Pakmor2011, Pakmor2013}.

In order to develop an accurate model for galaxy formation, it is necessary to account for additional effects such as radiative gas cooling, star formation, and stellar and black hole feedback. Significant progress has been made in recent years in the development of sub-resolution models for these additional baryonic processes \citep[e.g.,][]{Schaye:2014tpa,2017MNRAS.465.3291W,Pillepich:2017jle}. Throughout this thesis, we will refer to these as `full-physics' models, since they provide the most complete treatments of the underlying baryonic physics that are currently available.

For example, the IllustrisTNG model \citep{2017MNRAS.465.3291W,Pillepich:2017jle}, which is implemented in \textsc{arepo}, employs a sub-resolution scheme for star formation which is based on the \citet{Springel:2002uv} model: for gas cells which exceed a particular threshold density, a fraction of the gas mass is converted to mass in star particles at each simulation timestep according to the Kennicutt-Schmidt law. A portion of the gas mass is also converted into wind particles which are launched in random directions: these represent supernova-driven galactic winds. Having travelled outside their local dense interstellar medium, these wind particles will recouple to the gas, transferring their thermal energy and metal content. Gas also undergoes radiative cooling, which is modulated by a time-dependent ultraviolet background. The model also accounts for supermassive black holes, which are seeded at the centre of haloes which exceed a particular mass threshold. These grow through a combination of black hole mergers and Eddington-limited Bondi gas accretion. In high accretion states, a thermal feedback model is employed which heats up the surrounding gas, while in low accretion states a kinetic feedback model is employed which produces black hole-driven winds \citep[see][]{2017MNRAS.465.3291W, Vogelsberger2013}.

Significant advances have been made in recent years in the development of numerical simulations of screened modified gravity \citep[see, e.g.,][]{Winther:2015wla}. For example, the MG solver implemented by \textsc{arepo} can be used to run $N$-body and hydrodynamical simulations of $f(R)$ gravity and nDGP, where structure formation is now governed by the modified Poisson equations, which are given by Eqs.~(\ref{eq:fr_poisson_eqn}) and (\ref{eq:DGP_pot}), respectively. Both models feature a highly nonlinear scalar field which is calculated on an adaptively refining mesh, ensuring accurate calculations in high-density regions. The fifth force is then computed on this mesh and interpolated to the simulation particles and gas cells. This calculation is carried out less frequently in high-density regions where the fifth force is screened out, making the code highly efficient. For the first time, it is now possible to incorporate full-physics baryonic models into these MG simulations \citep[e.g.,][]{Arnold:2019vpg,Hernandez-Aguayo:2020kgq}, and this will be invaluable in helping to fully understand the effects of a strengthened gravitational force on the properties of clusters.

As a convention throughout this thesis, when studying $N$-body and hydrodynamical simulations, we will define the halo mass as the total mass contained within the sphere that is centred on the gravitational potential minimum of the halo and encloses an average density of $\Delta$ times the critical density of the Universe at the halo redshift. We will typically consider overdensities $\Delta=200$, $500$ or $300\Omega_{\rm M}(z)$, where the corresponding halo mass (radius) is labelled $M_{\rm 200c}$ ($R_{\rm 200c}$), $M_{\rm 500c}$ ($R_{\rm 500c}$) and $M_{\rm 300m}$ ($R_{\rm 300m}$), respectively (we will often neglect the `c' in the subscripts). We note that for $\Delta=300\Omega_{\rm M}(z)$, the average enclosed density is equal to 300 times the mean matter density of the Universe.

\subsection{Markov chain Monte Carlo sampling}
\label{sec:introduction:mcmc}

Cosmological parameter constraints can be obtained by combining observations with simulation-based theoretical predictions using MCMC sampling. This is an iterative random sampling technique based on Bayes' theorem, which states that the conditional probability of the parameter values $\theta\equiv\{\theta_i\}$, given the observed data $D$, is given by:
\begin{equation}
    P(\theta|D) = \frac{P(D|\theta)P(\theta)}{P(D)} \propto P(D|\theta)P(\theta),
\end{equation}
where $P(D|\theta)$ is the conditional probability of the data given the parameter values (the `likelihood' $\mathcal{L}$), $P(\theta)$ is the \textit{a priori} probability of the parameter values (the `prior') and $P(D)$ is the probability of the observed data (the latter is a constant, since the data is fixed during the sampling). Because sums are easier to work with than products, it is common to use logarithms:
\begin{equation}
    \ln P(\theta|D) = \ln \mathcal{L} + \ln P(\theta) + {\rm const}
\end{equation}
The sampling works as follows: starting with an initial point in the parameter space, a new point is generated by perturbing the initial point by a small, random amount. This new point is either accepted or rejected based on the effect of the parameter change on the \textit{a posteriori} probability $P(\theta|D)$ (the `posterior'). An example is the Metropolis-Hastings algorithm: this accepts the new point if the new posterior probability is higher, and it may still accept it if the new posterior is lower by a small amount. This process is repeated such that a chain of points is generated which eventually converges on a small region in the parameter space containing the maximum of the posterior. Typically, multiple chains (`walkers') are used, which are generated using different initial points, to ensure that the global maximum of the posterior distribution is successfully located.

This process requires two ingredients: the likelihood and the prior. For the case of cluster number counts, the Poisson likelihood can be used, which gives the probability of counting the observed number of clusters given the parameter-dependent theoretical prediction. For the parameter priors, either a flat (constant) prior is assumed (i.e., no prior knowledge of the parameters is included) or a Gaussian prior can be used which is based on a previous constraint using different observables. For example, the Planck 2015 cluster counts analysis \citep{Planck_SZ_cluster} adopted Gaussian priors, which are based on previous constraints from CMB temperature fluctuations, Big Bang nucleosynthesis and baryon acoustic oscillations, for parameters including $H_0$ and the dimensionless baryonic density parameter $\Omega_{\rm b}$, while flat priors can be used for $\Omega_{\rm M}$ and $\sigma_8$ since these are well-constrained by cluster number counts. The prior is then given by:
\begin{equation}
    \ln P(\theta) = -\sum_i\frac{(\theta_i - \mu_i)^2}{2\sigma_i^2} + {\rm const},
\end{equation}
where $\mu_i$ and $\sigma_i$ are the mean and standard deviation inferred by the previous constraints of parameter $i$, and the sum only carries over parameters with a Gaussian prior.

\section{Thesis overview}
\label{sec:intro:overview}

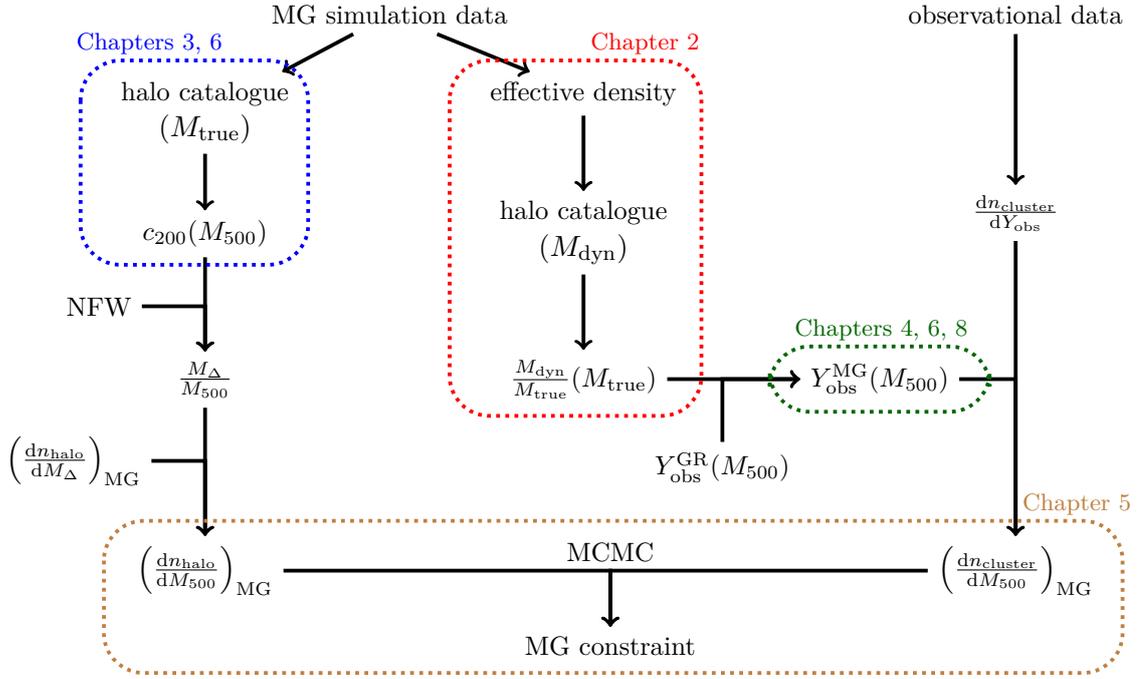
\begin{figure}
\centering
\begin{tikzpicture}
\tikzstyle{myarrow}=[line width=0.5mm,draw=black,-triangle 45,postaction={draw, line width=0.5mm, shorten >=4mm, -}]

\node    (simulations)    {\small MG simulation data};
\node    (cat_true)    [below left = 0.5cm and -0.5cm of simulations]   [align=center]{\small halo catalogue \\ ($M_{\rm true}$)};
\node    (c_m)    [below = 0.75cm of cat_true] [align=center]   {\small $c_{200}(M_{500})$};
\node   (m300_m500)    [below = 1.25cm of c_m] [align=center]    {\small $\frac{M_{\Delta}}{M_{500}}$};
\node    (nfw)    [below left = 0.65cm and 1.0cm of c_m]  [anchor=west]   {\small NFW};
\node    (hmf_m300)    [below left = 2.7cm and 1.8cm of c_m]  [anchor=west]   {\small $\left( \frac{{\rm d} n_{\rm halo}}{{\rm d}M_{\Delta}} \right)_{\rm MG}$};
\node    (hmf_th)    [below = 1.7cm of m300_m500]  [align=center]   {\small $\left( \frac{{\rm d} n_{\rm halo}}{{\rm d} M_{500}} \right)_{\rm MG}$};
\node    (rho_eff)    [below right = 0.5cm and -0.5cm of simulations]   [align=center]  {\small effective density};
\node    (cat_mdyn)    [below = 1.0cm of rho_eff]    [align=center] [align=center]{\small halo catalogue \\ ($M_{\rm dyn}$)};
\node    (mdyn_mtrue)    [below = 1.0cm of cat_mdyn]  [align=center]   {\small $\frac{M_{\rm dyn}}{M_{\rm true}}(M_{\rm true})$};
\node    (observations)    [right = 5cm of simulations]    [align=center] {\small observational data};
\node    (n_Y)    [below = 2.0cm of observations]   [align=center]  {\small $\frac{{\rm d} n_{\rm cluster}}{{\rm d}Y_{\rm{obs}}}$};
\node    (hmf_obs)  at (hmf_th -| n_Y)  [ align=center]   {\small $\left(\frac{{\rm d} n_{\rm cluster}}{{\rm d} M_{\rm 500}}\right)_{\rm MG}$};
\node    (scaling_relation)    [right = 1.75cm of mdyn_mtrue]    [align=center]{\small $Y_{\rm{obs}}^{\rm MG}(M_{500})$};
\node    (scaling_relation_lcdm)    [below left= 0.5cm and -0cm of scaling_relation]    [align=center]{\small $Y_{\rm{obs}}^{\rm GR}(M_{500})$};

\node    (mcmc)    at ($(hmf_th)!0.5!(hmf_obs)+(0.0,0.25)$)   [align=center]  {\small MCMC};
\node    (constraint)  at ($(hmf_th)!0.5!(hmf_obs)+(0.0,-1.0)$) [align=center]  {\small MG constraint};

\draw[->, line width=0.5mm] (simulations) -- (cat_true);
\draw[->, line width=0.5mm] (cat_true) -- (c_m);
\draw[->, line width=0.5mm] (c_m) -- (m300_m500);
\draw[->, line width=0.5mm, to path={-| (\tikztotarget)}] (nfw) edge (m300_m500);
\draw[->, line width=0.5mm] (m300_m500) -- (hmf_th);
\draw[->, line width=0.5mm, to path={-| (\tikztotarget)}] (hmf_m300) edge (hmf_th);
\draw[->, line width=0.5mm] (simulations) -- (rho_eff);
\draw[->, line width=0.5mm] (rho_eff) -- (cat_mdyn);
\draw[->, line width=0.5mm] (cat_mdyn) -- (mdyn_mtrue);

\draw[->, line width=0.5mm] (observations) -- (n_Y);
\draw[->, line width=0.5mm] (n_Y) -- (hmf_obs);
\draw[->, line width=0.5mm, to path={-| (\tikztotarget)}] (scaling_relation) edge (hmf_obs);
\draw[->, line width=0.5mm] (mdyn_mtrue) -- (scaling_relation);
\draw[->, line width=0.5mm, to path={|- (\tikztotarget)}] (scaling_relation_lcdm) edge (scaling_relation);

\draw[->, line width=0.5mm, to path={-| (\tikztotarget)}] (hmf_th) edge (constraint);
\draw[->, line width=0.5mm, to path={-| (\tikztotarget)}] (hmf_obs) edge (constraint);


\draw [line width=0.5mm,dotted, red, rounded corners=15pt]     ($(rho_eff.north west)+(-0.4,0.15)$) rectangle ($(mdyn_mtrue.south east)+(0.45,-0.1)$);
\node [above right = 0.10cm and -1.4cm of rho_eff] {\footnotesize{\color{red}\hypersetup{linkcolor=red}Chapter \ref{chapter:mdyn}}}; 
\draw[line width=0.5mm,dotted, blue, rounded corners=15pt]   ($(cat_true.north west)+(-0.4,0.15)$) rectangle ($(c_m.south east)+(0.4,-0.1)$);
\node [above left = 0.10cm and -1.6cm of cat_true] {\footnotesize{\color{blue}\hypersetup{linkcolor=blue}Chapters \ref{chapter:concentration}, \ref{chapter:DGP_clusters}}}; 
\draw[line width=0.5mm,dotted, brown, rounded corners=15pt]     ($(hmf_th.north west)+(-0.3,0.2)$) rectangle ($(constraint.south east -| hmf_obs.south east) +(0.25,-0.1)$);
\node [above right = 0.15cm and -1.2cm of hmf_obs] {\footnotesize{\color{brown}\hypersetup{linkcolor=brown}Chapter \ref{chapter:constraint_pipeline}}}; 
\draw[line width=0.5mm,dotted, black!60!green, rounded corners=15pt]     ($(scaling_relation.north west)+(-0.4,0.1)$) rectangle ($(scaling_relation.south east) +(0.4,-0.1)$);
\node [above right = 0.05cm and -2.3cm of scaling_relation] {\footnotesize{\color{black!60!green}\hypersetup{linkcolor=black!60!green}Chapters \ref{chapter:scaling_relations}, \ref{chapter:DGP_clusters}, \ref{chapter:baryonic_fine_tuning}}}; 
\end{tikzpicture}
\caption[Flowchart outlining our general framework for testing gravity using galaxy clusters.]{Flowchart outlining our general framework for testing gravity using cluster number counts. Using MG simulation data, we can calibrate models for the MG enhancements of the halo concentration (\textit{blue dotted box}) and the dynamical mass (\textit{red dotted box}). The concentration model can be used for converting between different halo mass definitions, which is required if the theoretical predictions and observations are initially defined using different spherical overdensities. The dynamical mass model can be used to convert the GR observable-mass scaling relation into a form that is consistent with the MG model of interest (\textit{green dotted box}). The latter is used to relate the observational mass function, ${\rm d}n/{\rm d}Y_{\rm obs}$ to the theoretical form ${\rm d}n/{\rm d}M$. Finally, MCMC sampling is used to constrain the MG model parameters using the theoretical predictions and observations (\textit{brown dotted box}). These components will be described for $f(R)$ gravity and nDGP in the annotated chapters. Chapter \ref{chapter:sz_power_spectrum}, which is not shown here, will focus on the thermal and kinetic SZ power spectra, which can potentially be used as alternative probes of gravity, and used in combination with the cluster number count constraint shown here.}
\label{fig:mg_flow_chart}
\end{figure}

The primary focus of our work is in developing a general framework for unbiased cluster tests of gravity. Fig.~\ref{fig:mg_flow_chart} provides a broad overview of this framework for the case cluster number count constraints of modified gravity. Most of the chapters of this thesis will focus on different parts of this flowchart.

In Chapter \ref{chapter:mdyn}, we will provide a more detailed overview of our framework and describe our modelling of the $f(R)$ enhancement of the dynamical mass of dark matter haloes using $f(R)$ gravity simulations. The dynamical mass is defined as the mass that is felt by a nearby massive test particle, and its enhancement is equivalent to the gravitational force enhancement. This model forms a core part of the framework for $f(R)$ gravity constraints and will be referred to in many of the subsequent chapters. 

Chapter \ref{chapter:concentration} will focus on our modelling of the enhancement of the halo concentration in $f(R)$ gravity, which can be used to model the density profiles of haloes. This will be required for converting between different halo mass definitions. Then, in Chapter \ref{chapter:scaling_relations}, we will demonstrate the effects of the fifth force on the scaling relations between the cluster mass and a number of cluster observables using $f(R)$ gravity simulations which include a full treatment of baryonic physics. Here, we will show how our model for the dynamical mass enhancement can be used to calibrate observable-mass scaling relations in $f(R)$ gravity using their GR counterparts. This is required in order to relate the cluster observable (provided in cluster catalogues) to the cluster mass.

In Chapter \ref{chapter:constraint_pipeline}, we will present our full $f(R)$ constraint pipeline which can be used to constrain $f_{R0}$ using cluster number counts. 
The pipeline uses MCMC sampling and it incorporates all of the fifth force effects described above.

The above chapters focus on HS $f(R)$ gravity, however our framework is designed to be easily extended to other gravity models and observables. In Chapter~\ref{chapter:DGP_clusters}, we will use nDGP simulations (again including a full baryonic treatment) to study and model the effect of the strengthened gravity in this model on observable-mass scaling relations, the halo concentration and the HMF. Then, in Chapter~\ref{chapter:sz_power_spectrum}, we will demonstrate the impact of $f(R)$ gravity and nDGP on the angular power spectrum of the SZ effect. This can potentially be used for precision constraints of gravity on large scales using data from future CMB experiments.

In Chapter \ref{chapter:baryonic_fine_tuning}, we will present a retuned baryonic model, which we have calibrated by running over $200$ test simulations, that can be used for full-physics simulations of screened modified gravity with much larger cosmological volumes. We have used this model to run a set of large-box simulations in $f(R)$ gravity and to revisit the $f(R)$ gravity observable-mass scaling relations using a mass range extending to much higher cluster masses.

Finally, in Chapter \ref{chapter:conclusions}, we will provide a summary of the main results of each chapter along with some considerations for future work.
\graphicspath{{./gfx/}}

\chapter{\boldmath Modelling the dynamical mass of haloes in \texorpdfstring{$f(R)$}{f(R)} gravity}
\label{chapter:mdyn}

\section{Introduction}

In observations, it is generally difficult to directly measure the masses of clusters. This is particularly the case for distant clusters, for which the required exposure time is prohibitively expensive.  Instead, one often has to infer them using mass proxies such as the X-ray temperature, luminosity and the SZ Compton $Y$-parameter. This, however, can lead to various sources of bias and uncertainty. For example, this can stem from the calibration procedures used to find the scaling relations linking these proxies to the masses, where observational uncertainty and various assumptions can lead to uncertain and possibly biased estimates of the mass. Unless these scaling relations are re-calibrated for any new cosmological models to be studied to remove any sources of bias, these will carry through to the predictions of properties that are dependent on the mass, such as the cluster abundance and the cluster gas fraction, which will therefore lead to biased constraints of the cosmological models and parameters. 

In practice, the calibration of the scaling relations can be achieved through different approaches. One way is to use full physics hydrodynamical simulations including radiative processes \citep[e.g.,][]{Fabjan:2011,Nagai:2007}. \citet{Fabjan:2011} employ this approach to calibrate the relations for three X-ray proxies. Another way is to use subsamples of a complete data-set as, e.g., in \cite{Vikhlinin:2009}, where Chandra observations are used to calibrate relations for X-ray proxies that can be cross-checked with weak lensing data. A third option is self-calibration, where the calibration is achieved with additional observables, for instance the clustering of clusters \citep{Schuecker:2003,Majumdar_2004}. In addition to these external calibrations, one can also calibrate data internally, e.g., by simultaneously constraining the scaling relations and cosmological models via a joint likelihood analysis \citep[e.g.,][]{Mantz:2010,Mantz:2015}.

The situation becomes even more complicated and largely unexplored when it comes to testing MG theories using galaxy cluster observations. A common effect of MG models which feature a fifth force is to enhance the dynamical mass of a galaxy cluster so that it becomes larger than the true (lensing) mass. This results from the additional gravitational forces. Tests which aim to measure both the dynamical and lensing masses to check for a disparity include recent works by \citet{Terukina:2013eqa,Wilcox:2015kna,Wilcox:2016guw,Pizzuti:2017diz}, which utilise actual measurements of the profiles of these two masses for massive clusters. Other probes include the cluster gas fraction \citep{Li:2015rva}, the clustering of clusters \citep{Arnalte-Mur:2016alq} and weak lensing \citep[e.g.,][]{Barreira:2015fpa} by clusters. The resulting weak lensing masses are only modified in some but not all MG models \citep{arnold:2014}. 

While earlier studies have pointed to a strong power of cluster observations in the tests of gravity, one potential issue that has so far not been given detailed attention is that the inferred cluster abundance, and other mass-dependent quantities, can change as a result of the enhancement of the dynamical mass with respect to the true mass, depending on which mass proxy is being used. If this enhancement is not accurately taken into account, the inferred abundance could be biased. In particular, scaling relations that are used to determine the cluster mass should first be calibrated in the contexts of specific MG models in order to incorporate this effect. Furthermore, these scaling relations are often derived using multiple probes, for example X-ray emission and weak lensing, which are affected by MG in different ways even in the same model. This adds more complexity and challenges for cosmological constraints. The main purpose of this chapter is to consider these complications and propose a suitable calibration method that is straightforward to implement in MG model tests.

As discussed in Chapter \ref{chapter:intro}, a primary focus of this thesis is to develop a framework to incorporate the various effects of MG on galaxy clusters in a self-consistent way. The aim is to have fully calibrated models which incorporates these effects into model predictions and allows for detailed MCMC searches of the parameter space to produce de-biased constraints of gravity. Of particular importance in this framework is the requirement to be able to make reliable model predictions for arbitrary model parameter values, as opposed to a very small number of model parameters that have been studied in detail in previous $N$-body simulations of MG (which are therefore not allowing for a continuous search of the large parameter space). To achieve this we will provide various simulation-calibrated fitting formulae that are essential for model predictions. In this chapter, we will focus on the relationship of the lensing and dynamical masses of galaxy clusters in $f(R)$ gravity. In this model, massive particles feel an extra force (the fifth force) mediated by an additional scalar field $f_{R}$. This field is redshift dependent, and, as described in Chapter \ref{chapter:intro}, its present day background value, $f_{R0}$, can be chosen as a model parameter. The enhancement of the dynamical mass therefore depends on the redshift and the background field strength at $z=0$.

Previous works analysing the dynamical mass and lensing mass in $f(R)$ gravity include \citet{Schmidt:2010jr,Zhao:2011cu,arnold:2014}. The studies were model specific, and they did not give a general formula that can be applied to arbitrary values of model parameters and redshifts. For example, the focus may only be on a particular present-day field strength at $z=0$: these results can be used for a qualitative understanding of particular models, but we really need a generic formula that is applicable to general models at all redshifts. In this chapter, we propose such a generic fitting formula which is based on a simple analytical model, the spherical thin-shell model \citep{Khoury:2003aq}. We check this fitting formula against simulations with different resolutions and find it to work very well across all tested field strengths. Although we use a specific choice of $f(R)$ gravity as our example, as discussed below, the results are expected to be applicable to or have useful implications for general chameleon gravity theories \citep[e.g.,][]{Gronke:2015}.

The chapter is organised as follows: Sec.~\ref{f(R)} discusses the key results of the thin-shell model, and defines the effective mass, which can be used interchangeably with the dynamical mass in simulations; Sec.~\ref{framework} discusses the background behind the use of galaxy clusters in constraining cosmological models, presents an outline of our proposed framework for $f(R)$ constraints, and proposes a method to account for the dynamical mass enhancement in scaling relations; Sec.~\ref{methods} summarises the properties of the simulations that are used and how we make use of them in our analyses, presents our fitting formula for the enhancement, and illustrates the method used to test this model; Sec.~\ref{results} presents the main results of our tests, including key formulae that have been fitted to the simulation data; and finally, Sec.~\ref{conclusions} summarises the key insights from this investigation and the implications for future work. Also, in Appendix \ref{appendix:mdyn}, we summarise the results obtained from using an alternative fitting procedure and show consistency tests to check for dispersions between the various data-sets used.

\section{Background} 
\label{f(R)}

In Sec.~\ref{thin_shell_modelling}, we describe the thin shell model, which provides a way of modelling the chameleon screening in $f(R)$ gravity. Then, in Sec.~\ref{dynamical_mass}, we define the dynamical and effective mass of dark matter haloes, and describe how to measure these from simulations.

\subsection{Thin-shell model} \label{thin_shell_modelling}

A useful way to model chameleon screening is via thin-shell modelling, which was first proposed in \cite{Khoury:2003rn} and has been used extensively in theoretical modelling, \citep[e.g.,][]{LE2012,Lombriser:2013wta,Lombriser:2013eza}. Consider a constant spherically symmetric top-hat matter density, $\rho_{\rm in}$, within a radius, $r_{\rm th}$, where $\phi_{\rm in}$ and $\phi_{\rm out}$ represent the scalar field inside and outside of $r_{\rm th}$ respectively.  Given this setup, one can make the following approximation:
\begin{equation}
\frac{\Delta r}{r_{\rm th}} \approx (3+2\omega)\frac{\phi_{\rm in}-\phi_{\rm out}}{6\Psi_{\rm N}} \approx -\frac{\phi_{\rm out}}{2\Psi_{\rm N}},
\label{r_ratio}
\end{equation}
where $\Delta r$ is the distance (from the boundary of the top-hat density distribution) necessary for the scalar field, $\phi$, to settle from $\phi_{\rm out}$ to $\phi_{\rm in}$, which to a good approximation is $\phi_{\rm in} \approx 0$. $\omega$ is the Brans-Dicke parameter, equal to zero for the $f(R)$ model under consideration. One can furthermore identify $\phi$ with $f_R$, and $\phi_{\rm out}$ with the background value $\bar{f}_R(z)$ for a given model and redshift. The depth of the Newtonian potential at the boundary, $\Psi_{\rm N}$, is given by,
\begin{equation}
\Psi_{\rm N} = \frac{GM}{r_{\rm th}},
\label{Newton}
\end{equation}
with $M$ the mass enclosed in the spherical top-hat. Using,
\begin{equation}
M \equiv \frac{4\pi}{3}\rho_{\rm in}r_{\rm th}^3,
\label{mass}
\end{equation}
we find that $\Psi_{\rm N} \propto M^{\frac{2}{3}}$ for a fixed density. 

In this chapter, we will focus on dark matter haloes found from $N$-body simulations. To make a connection between these haloes and the spherical top-hat densities described above which are used for thin-shell modelling, we make two approximations. First, dark matter structures in real simulations are not spherically symmetric, but we approximate them as spherical. Second, the radial density distribution of dark matter haloes are known to satisfy a Navarro-Frenk-White \citep[][NFW]{NFW} profile, 
\begin{equation}\label{eq:NFW}
\rho(r) = \frac{\rho_0}{(r/R_s)\left(1+r/R_s\right)^2},
\end{equation}
where $\rho_0$ is a parameter with the same unit as density, and $R_s$ is the scale radius. $\rho(r)$ scales like $r^{-1}$ ($r^{-3}$) in the inner (outer) part of a halo, and is not a constant within the halo radius, $R_{\Delta\rm c}$, which is determined as the distance from the halo centre within which the mean density is $\Delta$ times the critical density of the Universe, $\rho_{\rm crit}$, at the halo redshift. In our modelling, we treat the haloes as top-hats with density equal to $M_{\Delta\rm c}/\left(\frac{4}{3}\pi R^3_{\Delta\rm c}\right)$, where $M_{\Delta\rm c}$ is the halo mass, i.e., the mass enclosed in $R_{\Delta\rm c}$\footnote{For a more detailed and realistic modelling of chameleon screening, see, e.g., \citet{thin_shell,Lombriser:2013eza,Cataneo:2016iav}. However, as we show below, our simpler treatment works well and its predictions are in excellent agreement with simulations.}. It is furthermore shown in \cite{arnold:2016}, that the above scaling approach also works for ideal NFW haloes, validating our second assumption. The top-hat radius is given by $r_{\rm th}=R_{\Delta\rm c}$.

With the above approximations, we have 
\begin{equation}
\Psi_{\rm N} = \frac{\frac{4\pi G}{3}\rho_{\rm crit,0}
\Delta\left(1+z\right)^3\frac{r_{\rm th}^3}{\left(1+z\right)^3}}{\frac{r_{\rm th}}{1+z}}=\frac{GM}{r_{\rm th}}(1+z) \propto M^{\frac{2}{3}}(1+z),
\label{comoving}
\end{equation}
where $\rho_{\rm crit,0}$ is the critical density today, and so $\rho_{\rm crit,0}\Delta$ is the mean matter density in the halo today; the factor $(1+z)^3$ multiplying the density guarantees that we are using the physical density at redshift $z$, and the $(1+z)$ factors associated with $r_{\rm th}$ ensures that we use the physical radius (note that $R_{\Delta\rm c}=r_{\rm th}$ is the comoving radius of a halo).

With this setup, a qualitative argument can be made \citep[e.g.,][]{LE2012} that gravity is enhanced by the maximum factor 4/3 when $\Delta r\geq\frac{r_{\rm th}}{3}$. On the other hand, a small positive constant $\epsilon \ll 1$ can be defined such that one can assume no deviation from GR when $\Delta r \leq \epsilon r_{\rm th}$.

From the theoretical arguments discussed above, it is expected that the dynamical mass of a halo in $f(R)$ gravity varies in a range $M_{\rm true}\leq M_{\rm dyn}\leq \frac{4}{3}M_{\rm true}$ \citep{Schmidt:2010jr,Zhao:2011cu}. One can define the smallest true halo mass, $M_1$, for which there is no deviation from GR $(M_{\rm dyn}=M_{\rm true})$, and the highest true halo mass, $M_2$, for which there is no chameleon suppression of the scalar field $(M_{\rm dyn}=\frac{4}{3}M_{\rm true})$. From Eqs.~(\ref{r_ratio}) and (\ref{comoving}) and using the definitions for $M_1$ and $M_2$, these are respectively given by
\begin{equation}
M_1 = \kappa_1\left(\frac{1}{\epsilon}\frac{\bar{f}_R(z)}{1+z}\right)^{\frac{3}{2}} \propto \left(\frac{\bar{f}_R(z)}{1+z}\right)^{\frac{3}{2}},
\label{M_1}
\end{equation}
and
\begin{equation}
M_2 = \kappa_2\left(3\frac{\bar{f}_R(z)}{1+z}\right)^{\frac{3}{2}} \propto \left(\frac{\bar{f}_R(z)}{1+z}\right)^{\frac{3}{2}},
\label{M_2}
\end{equation}
where the constants $\kappa_1$ and $\kappa_2$ enclose Newton's gravitational constant along with some other constant factors from Eqs.~(\ref{r_ratio},\ref{Newton},\ref{mass}):
\begin{equation}\label{eq:kappa_12}
\kappa_1=\kappa_2=(2GH_0)^{-1}\Delta^{-1/2}.
\end{equation}
Both masses display power law fits as functions of $\frac{\bar{f}_R(z)}{1+z}$, and this is an important observation in this chapter: when comparing thin-shell model predictions against $N$-body simulations, both of them should be expressed as a function of $\bar{f}_R(z)/(1+z)$. An additional advantage is that this makes the dependence on the model parameter $f_{R0}$ implicit: two models, A and B, with different $f_{R0}$ values, should have the same $\bar{f}_R(z)/(1+z)$ value at some different redshifts $z_A$ and $z_B$. If the thin-shell model is generic enough, its predictions for model A at $z_A$ and model B at $z_B$ should be the same, irrespective of the fact that these are two  different models. We shall show below that this is indeed the case, and so promises a way to constrain general $f(R)$ models.

In reality, chameleon screening comes not only from a haloes own mass, but also from the matter that surrounds it. This can be considered as environmental screening. This is more important in F6 than in F4 and F5, because in the former the weak scalar field is more easily suppressed, occasionally resulting in total suppression of the field inside a low-mass halo if it is within a larger scale high-density environment. This means that the background field value at a halo, $\bar{f}_R(z)$, evaluated by Eq.~(\ref{eq:fR_background}), may often be incorrect if there is a surrounding high-density environment. Therefore a better approximation for the thin-shell modelling would be to replace $\Psi_{\rm N}$ in Eq.~(\ref{r_ratio}) with $\Psi_{\rm N}+\Psi_{\rm env}$ with $\Psi_{\rm env}$ the average Newtonian potential caused by the environment at the location of the halo \citep{He:2014eva,Shi:2017pyd}, which can be read from the simulation data. For the time being this will not be included in the modelling in this investigation, as it is not necessary to achieve such accuracy in the statistical treatment we aim for. Our approach will cover haloes which live in different environments so that the effects of $\Psi_{\rm env}$ largely cancel when looking at the median of all haloes (see below for further comments on this point).

\subsection{Dynamical mass and effective mass} \label{dynamical_mass}

The dynamical mass of a cluster or halo is the mass that massive test particles (e.g., stars or nearby galaxies) feel. It can be measured using the relationship between the gravitational potential energy and the kinetic energy of all of the constituent parts. In simulations it can be calculated for each halo, detected from the density field created by the dark matter particles. 

The formation of large-scale structures in $f(R)$ gravity is largely determined by the modified Poisson equation, which recall is given by Eq.~(\ref{eq:fr_poisson_eqn}). An effective density field, $\delta\rho_{\rm eff}$ \citep{effective_mass}, can be defined such that Eq.~(\ref{eq:fr_poisson_eqn}) can be cast into a form that is similar to Eq.~(\ref{eq:poisson_eqn}):
\begin{equation}
\nabla^2\Phi = 4\pi G\delta\rho_{\rm eff},
\label{effective_density_definition}
\end{equation}
where $\delta\rho_{\rm eff}$ and $\delta\rho$ are related via:
\begin{equation}
\delta\rho_{\rm eff} \equiv \left(\frac{4}{3}-\frac{\delta R}{24\pi G\delta\rho}\right)\delta\rho.
\label{eq:effective_density}
\end{equation}

The effective haloes are then identified from the effective density field, which is not necessarily the same as the true density field. In GR the two are seemingly the same but in MG they are different. It has been suggested in previous work by  \citet{effective_mass} that using the effective density field to describe haloes allows us to view the dynamical properties of haloes in an $f(R)$ model as in a $\Lambda$CDM cosmology. In this sense,  calculations of dynamical properties, such as the circular velocity of the halo, can be done assuming GR regardless of the model ($f(R)$ gravity or GR) that the simulation is actually run for, as long as the effective mass of a halo is known. Therefore the effective mass can be used as a proxy for $M_{\rm dyn}$. As is evident from Eq.~(\ref{eq:effective_density}), the maximum enhancement to the true density field is 4/3. Thus both the effective and the dynamical mass vary between $M_{\rm true}$ and $\frac{4}{3}M_{\rm true}$. In what follows we shall use the effective mass and dynamical mass interchangeably, regardless of the (minor) differences between them \citep{effective_mass}.

\section{A framework for gravity tests using clusters}
\label{framework}

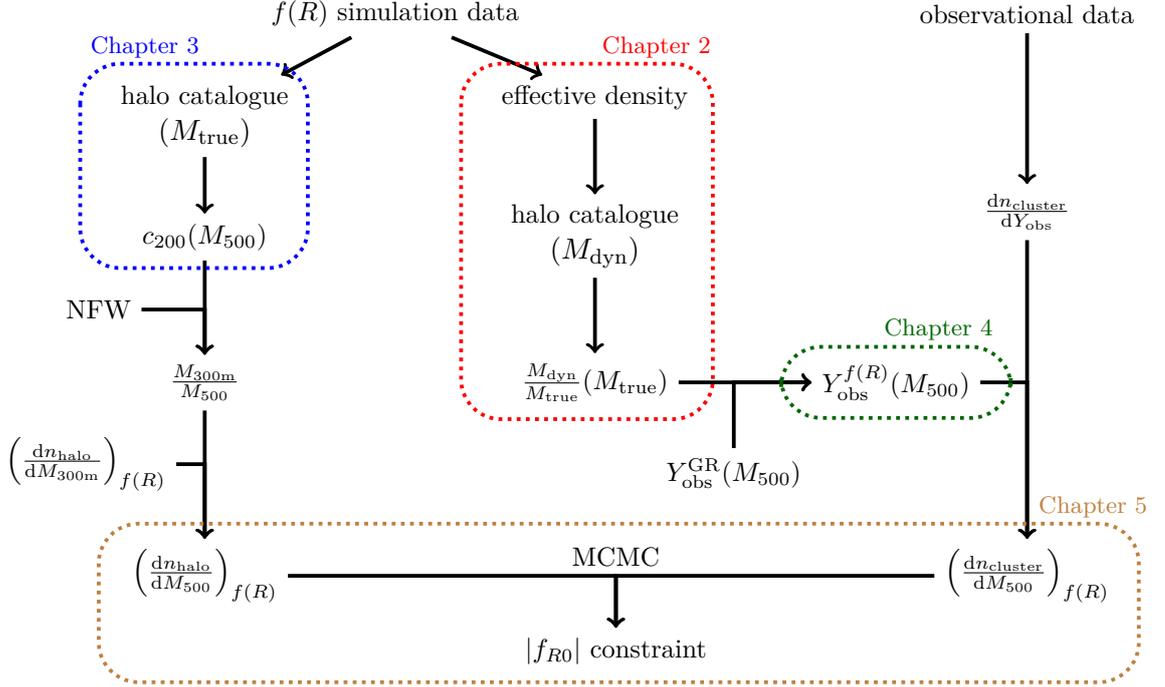
\begin{figure}
\centering
\begin{tikzpicture}
\tikzstyle{myarrow}=[line width=0.5mm,draw=black,-triangle 45,postaction={draw, line width=0.5mm, shorten >=4mm, -}]

\node    (simulations)    {\small $f(R)$ simulation data};
\node    (cat_true)    [below left = 0.5cm and -0.5cm of simulations]   [align=center]{\small halo catalogue \\ ($M_{\rm true}$)};
\node    (c_m)    [below = 0.75cm of cat_true] [align=center]   {\small $c_{\rm 200}(M_{\rm 500})$};
\node   (m300_m500)    [below = 1.25cm of c_m] [align=center]    {\small $\frac{M_{\rm 300m}}{M_{500}}$};
\node    (nfw)    [below left = 0.65cm and 1.0cm of c_m]  [anchor=west]   {\small NFW};
\node    (hmf_m300)    [below left = 2.7cm and 1.8cm of c_m]  [anchor=west]   {\small $\left( \frac{{\rm d} n_{\rm halo}}{{\rm d}M_{\rm 300m}} \right)_{f(R)}$};
\node    (hmf_th)    [below = 1.7cm of m300_m500]  [align=center]   {\small $\left( \frac{{\rm d} n_{\rm halo}}{{\rm d} M_{500}} \right)_{f(R)}$};
\node    (rho_eff)    [below right = 0.5cm and -0.5cm of simulations]   [align=center]  {\small effective density};
\node    (cat_mdyn)    [below = 1.0cm of rho_eff]    [align=center] [align=center]{\small halo catalogue \\ ($M_{\rm dyn}$)};
\node    (mdyn_mtrue)    [below = 1.0cm of cat_mdyn]  [align=center]   {\small $\frac{M_{\rm dyn}}{M_{\rm true}}(M_{\rm true})$};
\node    (observations)    [right = 5cm of simulations]    [align=center] {\small observational data};
\node    (n_Y)    [below = 2.0cm of observations]   [align=center]  {\small $\frac{{\rm d} n_{\rm cluster}}{{\rm d}Y_{\rm{obs}}}$};
\node    (hmf_obs)  at (hmf_th -| n_Y)  [ align=center]   {\small $\left(\frac{{\rm d} n_{\rm cluster}}{{\rm d} M_{\rm 500}}\right)_{f(R)}$};
\node    (scaling_relation)    [right = 1.75cm of mdyn_mtrue]    [align=center]{\small $Y_{\rm{obs}}^{f(R)}(M_{500})$};
\node    (scaling_relation_lcdm)    [below left= 0.5cm and -0cm of scaling_relation]    [align=center]{\small $Y_{\rm{obs}}^{\rm GR}(M_{500})$};

\node    (mcmc)    at ($(hmf_th)!0.5!(hmf_obs)+(0.0,0.25)$)   [align=center]  {\small MCMC};
\node    (constraint)  at ($(hmf_th)!0.5!(hmf_obs)+(0.0,-1.0)$) [align=center]  {\small $|f_{R0}|$ constraint};

\draw[->, line width=0.5mm] (simulations) -- (cat_true);
\draw[->, line width=0.5mm] (cat_true) -- (c_m);
\draw[->, line width=0.5mm] (c_m) -- (m300_m500);
\draw[->, line width=0.5mm, to path={-| (\tikztotarget)}] (nfw) edge (m300_m500);
\draw[->, line width=0.5mm] (m300_m500) -- (hmf_th);
\draw[->, line width=0.5mm, to path={-| (\tikztotarget)}] (hmf_m300) edge (hmf_th);
\draw[->, line width=0.5mm] (simulations) -- (rho_eff);
\draw[->, line width=0.5mm] (rho_eff) -- (cat_mdyn);
\draw[->, line width=0.5mm] (cat_mdyn) -- (mdyn_mtrue);

\draw[->, line width=0.5mm] (observations) -- (n_Y);
\draw[->, line width=0.5mm] (n_Y) -- (hmf_obs);
\draw[->, line width=0.5mm, to path={-| (\tikztotarget)}] (scaling_relation) edge (hmf_obs);
\draw[->, line width=0.5mm] (mdyn_mtrue) -- (scaling_relation);
\draw[->, line width=0.5mm, to path={|- (\tikztotarget)}] (scaling_relation_lcdm) edge (scaling_relation);

\draw[->, line width=0.5mm, to path={-| (\tikztotarget)}] (hmf_th) edge (constraint);
\draw[->, line width=0.5mm, to path={-| (\tikztotarget)}] (hmf_obs) edge (constraint);


\draw [line width=0.5mm,dotted, red, rounded corners=15pt]     ($(rho_eff.north west)+(-0.4,0.15)$) rectangle ($(mdyn_mtrue.south east)+(0.45,-0.1)$);
\node [above right = 0.10cm and -1.4cm of rho_eff] {\footnotesize{\color{red}\hypersetup{linkcolor=red}Chapter \ref{chapter:mdyn}}}; 
\draw[line width=0.5mm,dotted, blue, rounded corners=15pt]   ($(cat_true.north west)+(-0.4,0.15)$) rectangle ($(c_m.south east)+(0.4,-0.1)$);
\node [above left = 0.10cm and -1.3cm of cat_true] {\footnotesize{\color{blue}\hypersetup{linkcolor=blue}Chapter \ref{chapter:concentration}}}; 
\draw[line width=0.5mm,dotted, brown, rounded corners=15pt]     ($(hmf_th.north west)+(-0.3,0.2)$) rectangle ($(constraint.south east -| hmf_obs.south east) +(0.25,-0.1)$);
\node [above right = 0.15cm and -1.2cm of hmf_obs] {\footnotesize{\color{brown}\hypersetup{linkcolor=brown}Chapter \ref{chapter:constraint_pipeline}}}; 
\draw[line width=0.5mm,dotted, black!60!green, rounded corners=15pt]     ($(scaling_relation.north west)+(-0.4,0.1)$) rectangle ($(scaling_relation.south east) +(0.4,-0.1)$);
\node [above right = 0.05cm and -1.4cm of scaling_relation] {\footnotesize{\color{black!60!green}\hypersetup{linkcolor=black!60!green}Chapter \ref{chapter:scaling_relations}}}; 
\end{tikzpicture}
\caption[Flowchart outlining our pipeline for constraining $f(R)$ gravity using cluster number counts.]{Flowchart outlining our framework for constraining the present-day value of the background scalar field, $f_{R0}$, of HS $f(R)$ gravity. Using $f(R)$ simulation data, we have calibrated models for the $f(R)$ enhancements of the halo concentration (Chapter \ref{chapter:concentration}, \textit{blue dotted box}) and the dynamical mass (this chapter, \textit{red dotted box}). The concentration model is used to convert the mass definition of the theoretical HMF into a form that is consistent with the observational data. The dynamical mass model is used to convert the GR observable-mass scaling relation into a form that is consistent with $f(R)$ gravity (Chapter \ref{chapter:scaling_relations}, \textit{green dotted box}). The latter is used for relating the observational mass function, ${\rm d}n/{\rm d}Y_{\rm obs}$ to the theoretical form ${\rm d}n/{\rm d}M_{500}$. Finally, MCMC sampling is used to constrain $f_{R0}$ using the theoretical and observational predictions (Chapter \ref{chapter:constraint_pipeline}, \textit{brown dotted box}).}
\label{fig:fr_flow_chart}
\end{figure}

In this section, we focus on tests of HS $f(R)$ gravity using the galaxy cluster abundance \citep[see, e.g.,][for earlier works along this direction]{Schmidt:2009yyy,Mak_et_al.,PhysRevD.92.044009}. This is a specific case of the general framework shown in Fig.~\ref{fig:mg_flow_chart}, which is intended to be used for testing a range of MG models. Our proposed $f(R)$ framework is sketched in Fig.~\ref{fig:fr_flow_chart}. 

A fitting formula for the HMF is required to predict the halo abundance, and this can be obtained by using semi-analytical models calibrated by simulations. In this thesis, we adopt the model which has been proposed and calibrated by \cite{Cataneo:2016iav}, which itself is built upon earlier works \citep{LE2012,LiLam2012,LamLi2012,Lombriser:2013wta,Lombriser:2013eza} motivated by excursion set theory \citep{Bond:1990iw}; this will be discussed in Sec.~\ref{halo_abundance}. The \citet{Cataneo:2016iav} HMF has been calibrated using the halo mass definition $M_{300{\rm m}}$. To ensure generality, we will also require a mass conversion, $M_{\rm 300m}(M_{\Delta})$, to allow conversions to arbitrary mass definitions, which will require a concentration-mass relation (e.g., $c_{200}(M_{\Delta})$), of dark matter haloes in $f(R)$ gravity. This is discussed in Sec.~\ref{other_issues}, in addition to other work to be carried out. These ingredients will enable us to predict a theoretical cluster abundance for generic $f(R)$ models and mass definitions.

On the observational side, a key observable to be used in our test framework is the cluster abundance derived from SZ and X-ray surveys, such as Planck's SZ cluster abundance \citep{Planck_SZ_cluster}. As discussed in Sec.~\ref{mg_scaling_relations}, converting from cluster observables to the cluster mass typically involves the use of a scaling relation, however the most accurate scaling relations that are currently available are observational and/or derived for $\Lambda$CDM. We propose a method for converting these relations from $\Lambda$CDM to $f(R)$ gravity, based on the findings of \cite{He:2015mva}. We discuss this point in more detail in Sec.~\ref{mg_scaling_relations}. The conversion requires a formula for the ratio $M_{\rm dyn} / M_{\rm true}$, which is the focus of this chapter. Our procedure to measure $M_{\rm dyn} / M_{\rm true}$ as a function of $M_{\rm true}$, $z$ and $\bar{f}_{R}$ is discussed in Sec.~\ref{methods} and our results are presented in Sec.~\ref{results}. We show that a simple fitting formula for $M_{\rm dyn}/M_{\rm true}$ motivated by the theoretical modelling of Sec.~\ref{thin_shell_modelling} works very well in describing the results of a large suite of simulations. The simulations are introduced in Sec.~\ref{simulations}.

Following the corrections described above, the predicted and observed abundances can be combined to constrain $|f_{R0}|$ by confronting theoretical predictions for models with an arbitrary value of  $f_{R0}$ with observations. A continuous parameter space search can be carried out using techniques such as MCMC, which accounts for relevant covariances between data. The fitting formulae for various quantities, with corresponding errors, can be used to construct mock cluster catalogues to validate the model constraint pipeline. In Sec.~\ref{other_observables}, we will also mention some other possible observables which can be included in this framework and which will also require a knowledge of $M_{\rm dyn}/M_{\rm true}$ which we focus on in this chapter.

\subsection{Cluster abundance tests}
\label{cluster_abundance}

One of the frequently used probes of cosmological models and the underlying theory of gravity is the cluster abundance, defined as the number density of galaxy clusters per unit mass interval, $\frac{{\rm d}n_{\rm cluster}}{{\rm d}\log_{10}M}$. This depends sensitively on the cluster mass, $M$, which means that model tests using the cluster abundance require an accurate measurement of the cluster mass. We have seen that the term `mass' can be ambiguous in MG theories because different observables depend on different masses, e.g., dynamical versus lensing mass. Therefore, any effects of $f(R)$ gravity on the mass should be accounted for to prevent a biased prediction of the abundance. 

The theoretical counterparts of galaxy clusters in $N$-body simulations are massive dark matter haloes 
($\gtrsim10^{14}h^{-1}M_{\odot}$). A prediction of the cluster abundance can be obtained by measuring the abundance of haloes. Some efforts must also be made to account for the limitations of an observational survey, for example the blocking of many clusters by foreground stars and the galactic plane, and the rejection of low signal-to-noise sources. These effects are specific to the survey under consideration. {In summary, the following quantities are required:}
\begin{itemize}
\item An HMF which evaluates the number density of dark matter haloes per unit mass interval;
\item A scaling relation to predict the cluster observable, given the mass of the dark matter halo;
\item The selection function of the survey, which evaluates the probability of a cluster being detected and included in the resulting data-set, as a function of the observable flux, redshift, etc.;
\item The likelihood of the measurements, which would be produced along with the observed data itself.
\end{itemize}

These corrections will ensure that the prediction of the cluster abundance is consistent with measurements taken in the real Universe using detectors with finite precision. However, the HMF and scaling relations are generally more challenging to implement in $f(R)$ gravity tests without inducing sources of bias. This can stem from effects like the chameleon screening mechanism and the enhancement of the dynamical mass, which are complicated to model exactly. Secs.~\ref{halo_abundance}-\ref{mg_scaling_relations} illustrate our proposed methods to tackle these difficulties, and Sec.~\ref{other_issues} discusses other current issues in using the cluster abundance to test $f(R)$ gravity which we seek to address in this thesis.

\subsubsection{Halo abundance}
\label{halo_abundance}

The abundance of dark matter haloes can be predicted using semi-analytical models, such as excursion set theory \citep{Bond:1990iw}, which generally show reasonable qualitative agreement with simulations. These models connect high peaks in the initial density field to the late-time massive dark matter haloes by assuming spherical collapse. However quantitative agreements with simulations are not great, which has motivated models with more physical assumptions, such as the ellipsoidal collapse model \citep{Sheth:1999mn,Sheth:2001dp,Sheth:1999su} which gives up the sphericity assumption above. These efforts have led to various fitting formulae of the HMF in standard $\Lambda$CDM, whose parameters can be calibrated using simulations \citep[e.g.,][]{Jenkins:2000bv,Warren:2005ey,Reed:2006rw,Tinker:2008ff}.

In MG theories, excursion set theory still applies but the connection between initial density peaks and late-time dark matter haloes becomes more complicated. In some scenarios, such as the Galileon model \citep[e.g.,][]{Nicolis:2008in,Deffayet:2009wt}, as in $\Lambda$CDM, the spherical collapse of an initial top-hat overdensity does not depend on the environment, and analytical solutions can be obtained for their HMFs \citep{2010PhRvD..81f3005S,Barreira:2013xea,Barreira:2014zza}. In $f(R)$ gravity and general chameleon models, however, the behaviour of the fifth fore is more complicated and the spherical collapse becomes environment-dependent. Theoretical models of HMFs in these theories have been studied in \citet{LE2012,LiLam2012,LamLi2012,Lombriser:2013wta,Lombriser:2013eza,Kopp:2013lea}, and qualitative agreement with simulations is reasonable.

In this thesis, we adopt the HMF as proposed in \cite{Cataneo:2016iav}, which is based on an extension of the theoretical modelling described in \cite{Lombriser:2013wta,Lombriser:2013eza} by adding free parameters to the theoretical HMF to account for the chameleon screening mechanism and allow a better match with simulations. These parameters have been fitted using a subset (Crystal, see Sec.~\ref{simulations}) of our DMO $f(R)$ gravity simulations which have been run for F4, F5 and F6, but they work for general values of $|f_{R0}|$ within $[10^{-6},10^{-4}]$. \cite{Cataneo:2016iav} show that their HMF fitting formula agrees with simulation results to within $5\%$.

We note that the HMF fitting formula is an independent ingredient in our framework as depicted in Fig.~\ref{fig:fr_flow_chart}, by virtue of which we can always use the latest and most accurate in our analysis.

\subsubsection{Scaling relations in \texorpdfstring{$f(R)$}{f(R)} gravity}
\label{mg_scaling_relations}

The cluster mass is difficult to measure via direct observations, and a scaling relation is usually used to connect the cluster mass to some more readily observable quantities, such as the {average} temperature, $T_{\rm gas}$, of the intra-cluster gas. This relates to the {total mass, $M$,} via the virial theorem which leads to:
\begin{equation}
\frac{GM}{R} = \frac{3}{2}\frac{k_{\rm B}T_{\rm gas}}{\mu m_{\rm p}},
\label{M(T)}
\end{equation}
where 
$R$ is the cluster radius, $m_{\rm p}$ is the proton mass, $k_{\rm B}$ is the Boltzmann constant and $\mu$ is the molecular weight.

We are interested in cluster abundances measured from X-ray emission, the SZ effect and weak lensing. The X-ray radiation by a cluster is generated by the bremsstrahlung process, and the SZ effect is due to the inverse-Compton scattering of cosmic microwave background photons off electrons in the intra-cluster medium. Both of these effects depend on $T_{\rm gas}$. Therefore, several related and easily observable quantities can be used as mass proxies, such as the integrated SZ Compton $Y$-parameter, $Y_{\rm SZ}$, the X-ray equivalent of the integrated SZ flux, $Y_{\rm X}$, and the X-ray luminosity, $L_{\rm X}$. For each of these observables the cluster mass can be inferred through a scaling relation. 

In $\Lambda$CDM, such scaling relations can be obtained in different ways, such as by using hydrodynamical simulations \citep[e.g.,][]{Fabjan:2011,Nagai:2007} or from subsets of observed clusters whose masses can be measured in other means, e.g., weak lensing \citep{Vikhlinin:2009}. An example is the $Y_{\rm SZ}-M$ scaling relation calibrated by the Planck Collaboration \citep{Planck_SZ_cluster}, which incorporates the results from various observational surveys and simulations, and where rigorous methods have been used to prevent various sources of bias, including Malmquist bias and hydrostatic equilibrium bias. 

 

In $f(R)$ gravity, and in general for any new gravity theory, the scaling relations calibrated for $\Lambda$CDM are unlikely to still apply. It is impractical to calibrate these relations by using hydrodynamical simulations, since they are expensive even for a single specific $f(R)$ model, let alone the whole $f_{R0}$ parameter range. Calibrations using a subset of data or using other observables should be treated with caution as well. For example, the scaling relations may be different between the subset of data and the whole sample, due to the environmental dependence of the MG effect, and different observables are proxies of different masses in $f(R)$ gravity, and so the combined use of different observations is tricky. It is therefore highly desirable to have a physically motivated model for obtaining (certain) scaling relations for arbitrary values of the $f(R)$ parameter $f_{R0}$ with good precision and minimal effort.

Along this line and based on the use of the so-called effective mass (Sec.~\ref{dynamical_mass}), a procedure for correcting for the effect of MG on the physical properties of clusters, such as their various observable-mass scaling relations, has been proposed by \citet{He:2015mva}. This method avoids direct calibration of the cluster mass using full hydrodynamical simulations in the $f(R)$ model, and instead calculates the scaling relations in $f(R)$ gravity by using the corresponding ones in standard $\Lambda$CDM (which are better known) with a rescaled baryon-to-total mass ratio. Its results are found to agree very well with $f(R)$ simulations. 

\cite{He:2015mva} discussed the cluster mass proxies $L_{\rm X}$, $Y_{\rm SZ}$ and $Y_{\rm X}$, and here we describe the result for $Y_{\rm SZ}$ as an example. Using a non-radiative approximation, in which the baryonic content of the hydrodynamical simulations behaves as an ideal gas satisfying Eq.~(\ref{M(T)}), $Y_{\rm SZ}$ is given by:
\begin{equation}
Y_{\rm SZ} = \frac{\sigma_{\rm T}}{m_{\rm e}c^2}\int_0^r {\rm d} r 4\pi r^2P_{\rm e},
\label{Y_SZ}
\end{equation}
where $\sigma_{\rm T}$ is the Thomson cross section and $m_{\rm e}$ is the electron mass. The electron pressure, $P_{\rm e}$, is given by $P_{\rm e}=\frac{2+\mu}{5}n_{\rm gas}k_{\rm B}T_{\rm gas}$, where $n_{\rm gas}$ is the number density of gas particles. From the simulations it was found that the $T_{\rm gas}$-$M$ relations for the effective haloes in $f(R)$ gravity and the haloes in $\Lambda$CDM agree very well:
\begin{equation}
T_{\rm gas}^{f(R)}\left(M_{\rm dyn}^{f(R)}\right) = T_{\rm gas}^{\Lambda {\rm CDM}}\left(M^{\Lambda {\rm CDM}}\right).
\label{T_agreement}
\end{equation}
This is as expected given that the temperature and the gravitational potential of a halo are intrinsically linked through the virial theorem.

Using a suite of non-radiative hydrodynamical simulations, it was found that outside the core regions, the profiles of effective haloes in $f(R)$ gravity closely resemble those in $\Lambda$CDM, with a rescaled gas mass fraction: 
\begin{equation}
\rho_{\rm gas}^{f(R)}(r) \approx \frac{M^{f(R)}}{M_{\rm dyn}^{f(R)}}\rho_{\rm gas}^{\Lambda {\rm CDM}}(r) \propto \frac{M^{f(R)}}{M_{\rm dyn}^{f(R)}}\frac{\Omega_{\rm b}}{\Omega_{\rm m}}\left(r^2+r_{\rm core}^2\right)^{-\frac{3\beta}{2}},
\label{gas_density_eff}
\end{equation}
where $r_{\rm core}$ is the core radius and $\beta$ is the ratio between the specific kinetic energy (kinetic energy per unit mass) of cold dark matter and the specific internal energy (internal energy per unit mass) of gas. For an effective halo in $f(R)$ gravity with an effective mass that is equal to the true mass of a $\Lambda$CDM halo, $M_{\rm dyn}^{f(R)} = M^{\Lambda {\rm CDM}}$, it follows from Eqs.~(\ref{T_agreement}) and (\ref{gas_density_eff}) that:
\begin{equation}
\begin{aligned}
& \int_0^r {\rm d} r4\pi r^2\left(\rho_{\rm gas}^{f(R)}\right)^a\left(T_{\rm gas}^{f(R)}\right)^b \\
& \approx \left(\frac{M^{f(R)}}{M_{\rm dyn}^{f(R)}}\right)^a\int_0^r {\rm d} r4\pi r^2 \left(\rho_{\rm gas}^{\Lambda {\rm CDM}}\right)^a\left(T_{\rm gas}^{\Lambda {\rm CDM}}\right)^b,
\end{aligned}
\label{dens_and_temp}
\end{equation}
where $a$ and $b$ are indices of power. By combining this result with Eq.~(\ref{Y_SZ}) it follows that the $Y_{\rm SZ}$-$M$ scaling relations in these two models can be related by:
\begin{equation}
\frac{M_{\rm dyn}^{f(R)}}{M_{\rm true}^{f(R)}}Y_{\rm SZ}^{f(R)}\left(M_{\rm dyn}^{f(R)}\right) \approx Y_{\rm SZ}^{\Lambda {\rm CDM}}\left(M^{\Lambda {\rm CDM}}=M_{\rm dyn}^{f(R)}\right).
\label{Li and He}
\end{equation}
As mentioned previously, this relation has been verified by a suite of non-radiative hydrodynamical simulations. Similar results have been obtained and verified for the other two proxies ($Y_{\rm X}$ and $L_{\rm X}$) as well, and are particularly accurate for $Y_{\rm SZ}$ and $Y_{\rm X}$ with the error just slightly over 3\%.

As the scaling relations in $\Lambda$CDM are much better understood than in $f(R)$ gravity, Eq.~(\ref{Li and He}) can potentially be used to re-calibrate a scaling relation obtained for $\Lambda$CDM, into a form linking $Y_{\rm SZ}$ to the cluster dynamical mass in $f(R)$ gravity.

\subsubsection{Other issues}
\label{other_issues}

The mass of a galaxy cluster or dark matter halo is usually defined as the mass enclosed in some radius centred around the cluster or halo centre. This is the radius in which the average matter density is $\Delta$ times the mean matter density or the critical density at the halo redshift. In the literature, different values of $\Delta$ are commonly used, and so it is essential to be able to convert between them. As an example, \cite{Cataneo:2016iav}, whose $f(R)$ gravity HMF fitting formula we use by default in our framework, {work with $M_{\rm 300m}$. As another example, in the literature $M_{\rm 200c}$ is very commonly used.}

It is straightforward to convert between the different masses by noting that the different definitions only differ in where the halo boundary lies. Therefore, all we need is the density profile $\rho(r)$ of a halo. In $\Lambda$CDM, dark matter haloes are well described by the NFW density profile given by Eq.~(\ref{eq:NFW}), which has two free parameters, $\rho_0$ and $R_s$. The NFW profile has also been shown to work well for haloes in $f(R)$ gravity \citep[][]{thin_shell,Shi:2015aya}. Of the two NFW parameters, the scale radius, $R_s$, can be expressed by using the halo concentration, $c_{\Delta}\equiv R_{\Delta}/R_s$, and $\rho_0$ can be further fixed using the halo mass, $M_\Delta\equiv M(\leq R_{\Delta})$. Therefore, to convert between the different mass definitions requires an understanding of the concentration-mass relation, $c_{\Delta}(M_{\Delta})$. We note that, as long as the concentration is known for one overdensity $\Delta$, it can be inferred for any other overdensity (for example, see Appendix \ref{sec:appendix:pipeline:mass_conversions}). We have studied the $c_{\rm 200}(M_{500})$ concentration-mass relation in both screened and unscreened regimes, using data from various $f(R)$ simulations, and the results will be presented in Chapter \ref{chapter:concentration}.

Another issue that merits further investigation is a check of the method by \cite{He:2015mva} against full-physics hydrodynamical simulations including baryonic feedback processes, which go beyond the non-radiative approximations originally used. Studies in $\Lambda$CDM \citep[e.g.,][]{Fabjan:2011} have found that, for certain quantities such as $Y_{\rm X}$, the resulting scaling relation is insensitive to baryonic processes, such as cooling, star formation and black hole feedback, in galaxy formation if the data from the very inner part of a cluster is excluded. We expect the same to apply in $f(R)$ gravity, but in order to be certain we have conducted an analysis using full-physics hydrodynamical simulations for HS $f(R)$ gravity, and will present the results in Chapter \ref{chapter:scaling_relations}.

Such simulations will also be useful to better understand the impact of galaxy formation on the HMF in $f(R)$ gravity, though we expect it to be small. We also note that the fitting formula by \cite{Cataneo:2016iav}, which has a $3$-$5\%$ accuracy with the simulation data for F4-F6 and halo masses above $10^{13}h^{-1}M_{\odot}$, was calibrated using DMO simulations (Crystal, see Sec.~\ref{simulations}).

\subsection{Other observables}
\label{other_observables}

As mentioned above, the focus of the remainder of this chapter is a fitting function for the relationship between the dynamical and true masses of dark matter haloes, which would be useful for deriving cluster scaling relations in $f(R)$ gravity. But the use of this relation is certainly not restricted to this.

A direct use of the $M_{\rm dyn}/M_{\rm true}$ relation is to constrain the fifth force by comparing measurements of $M_{\rm dyn}$ and $M_{\rm true}$. In observations, the profiles of these masses can be obtained using the X-ray surface brightness profile and lensing tangential shear profile of a cluster respectively. The measurements can be done for massive clusters for which high-quality X-ray and lensing data are available. \cite{Terukina:2013eqa,Wilcox:2015kna,Wilcox:2016guw} performed the first analyses using this method and found constraints on general chameleon gravity theories. A more recent analysis can be found in \citet{Pizzuti:2017diz}. The dynamical mass or potential can also be inferred from the escape velocity edges in the radius/velocity phase space, which can be compared with the lensing-inferred mass profile, or the gravitational potential profiles for samples of low- and high-mass haloes, which would feel different effects of gravity due to the chameleon screening, can be compared \citep{Stark:2016mrr}.

Another potentially powerful probe in cluster cosmology is the cluster gas fraction \citep[e.g.,][]{fgas}, $f_{\rm gas} = {M_{\rm gas}}/{M_{\rm halo}}$, where $M_{\rm gas}$ is the mass of baryons (or hot gas) in the intra-cluster medium and $M_{\rm halo}$ is the total halo mass. In massive clusters, the mass of the hot intra-cluster gas dominates over that in cold gas and stars, and thus $f_{\rm gas}$ is expected to approximately match the cosmic baryon fraction, ${\Omega_{\rm b}}/{\Omega_{\rm M}}$. However, measurements of $f_{\rm gas}$ involve measuring $M_{\rm halo}$, which is the dynamical rather than the true mass of the halo. Constraints from $f_{\rm gas}$ on $f(R)$ gravity are therefore likely to be biased \citep{Li:2015rva}. To make amends for this we will require a general formula for the ratio $M_{\rm dyn}/M_{\rm true}$, which is presented in Sec.~\ref{results}. 

Our framework is sufficiently flexible to include these, among other, observables in the ultimate cluster constraints, though certain generalisations may be needed, such as the concentration-mass relations for not only the true but also the effective haloes.

\section{Simulations and methods} 
\label{methods}

The specifications of the $f(R)$ gravity simulations used in this chapter are presented in Sec.~\ref{simulations}. The procedure to measure the dynamical mass enhancement from this data is discussed in Sec.~\ref{measure_m_dyn}, along with the details for the modelling of this enhancement and its parameters.

\subsection{Simulations} 
\label{simulations}

Our collisionless simulations are run using the \textsc{ecosmog} code \citep{Li:2011vk}, a code based on the publicly-available $N$-body and hydrodynamical code \textsc{ramses} \citep{Teyssier:2001cp}, and which can be used to run $N$-body simulations for a wide range of MG and dynamical dark energy scenarios. The code is 
efficiently parallelised, and uses adaptive mesh refinement to ensure accuracy of the fifth force solution in high-density regions. {In order to reliably fit the dynamical mass enhancement as a function of the halo mass, an appropriate range of  halo true mass which covers the transition between $M_{\rm dyn}=M_{\rm true}$ and $M_{\rm dyn}=\frac{4}{3}M_{\rm true}$ would be required.  For this reason, three different simulations of varying resolutions were utilised. For the purposes of clarification, }these are listed as the {\it Crystal}, {\it Jade} and {\it Diamond} simulations with increasing resolutions. 

\begin{table}
\centering

\caption[Specifications of the \textsc{ecosmog} simulations used to study the ratio of the dynamical mass to the true mass.]{Specifications of the three \textsc{ecosmog} simulations used to study the $f(R)$ enhancement of the dynamical mass, labelled Diamond, Jade and Crystal for convenience. The gold data is defined as having been generated by $f(R)$ gravity simulations, whereas the silver data comes from effective density data generated from $\Lambda {\rm CDM}$ simulations. The Hubble constant, $H_0$, is set to 69.7 kms$^{-1}$Mpc$^{-1}$ in all simulations.}
\label{table:simulations}

\small
\begin{tabular}{ cccc } 
 \toprule
 
 Parameters and & & Simulations & \\
 data types & Diamond & Jade & Crystal \\

 \midrule

 box size / $h^{-1}$Mpc & 64 & 450 & 1024 \\ 
 particle number & $512^3$ & $1024^3$ & $1024^3$ \\ 
 particle mass / $h^{-1}M_{\odot}$ & $1.52\times10^8$ & $6.60\times10^9$ & $7.80\times10^{10}$ \\
  & & &  \\
 $\Omega_{\rm M}$ & 0.281 & 0.282 & 0.281 \\ 
 $\Omega_{\Lambda}=1-\Omega_{\rm M}$ & 0.719 & 0.718 & 0.719 \\
  & & & \\
 gold & F6 & F5 & F4, F5, F6 \\
 silver & F5.5, F6.5 & F4.5, F5.5, F6.5 & F4.5, F5.5 \\
 
 \bottomrule
 
\end{tabular}

\end{table}

The parameters and technical specifications of the simulations are listed in Table \ref{table:simulations}.  {The Hubble expansion rate, $H_0$, is set to 69.7 kms$^{-1}$Mpc$^{-1}$.} Diamond is the highest resolution simulation, and its small particle mass allows lower-mass haloes to be investigated. While Crystal is the lowest resolution, its large volume and particle number mean that higher-mass haloes can be included. Jade is needed in order to provide bridging halo mass regimes with both Crystal and Diamond to ensure that a complete range of masses is tested and to verify that the different simulations agree well in the overlapping regions (see Appendix \ref{appendix:mdyn:consistency}). Because the results of this investigation are intended to be used with the Planck 2015 data, which only covers up to redshift $z=1$, only simulation snapshots with $z<1$ are used. This includes 19 snapshots from both Crystal and Diamond, and 33 from Jade. The use of data from only $z<1$ also means that we can avoid using high-$z$ data from the Crystal simulations, which suffer from poor resolutions.

Halo catalogues for these simulations are constructed in two steps. First a modified \textsc{ecosmog} code is run to generate effective density data from the particle data for all of the snapshots. After that, \textsc{ahf} \citep{AHF1,AHF2}, a halo finder which is properly modified to read the effective density data, is run to identify effective haloes. {\sc ahf} is run with the $M_{500c}$ mass definition, and the outputted halo catalogues include the ratio $M_{\rm dyn}/M_{\rm true}$ for each halo, as well as the lensing mass which can be treated as $M_{\rm true}$.

Given the expensive cost of  full MG simulations, our $f(R)$ simulation suite only includes a limited number of models. The Crystal simulations have only been run for F4, F5 and F6, Jade has been run for F5 only and Diamond for F6 only. From Fig.~\ref{fig:fR_z_dependence}, we can see that up to $z=1$ (the redshift limit in the simulation data for our analysis) the three simulated models --- F4, F5, F6 --- do not cover all possible values of $\bar{f}_R(z)$ continuously but leave gaps in between. In order to test the proposed model for the dynamical mass enhancement over the greatest possible range of field values, without making too much effort in running full $f(R)$ simulations for other $f_{R0}$ values, {we propose a simpler approach}. At any desired redshift $z$, {the MG solver in the} \textsc{ecosmog} code was run on the particle data of $\Lambda$CDM simulations to generate further effective density data by assuming these were actually $f(R)$ gravity calculations with strengths F4.5 ($|f_{R0}|=10^{-4.5}$), F5.5 ($|f_{R0}|=10^{-5.5}$) and F6.5 ($|f_{R0}|=10^{-6.5}$). Because these calculations involve running the {\sc ecosmog} only for one step (for each $f_{R0}$ and $z$), they are much less expensive than a full simulation which means that we can afford to run many of them. Indeed, we could repeat this for any other values of $f_{R0}$, but found that the above three additional values already give decent overlapping in the halo mass ranges (see below). 

\textsc{ahf} effective halo catalogues were then generated for the additional $f_{R0}$ values using the effective density field from these `approximate simulations', 
the latter neglecting effects from the different structure formations under these models which could lead to different internal structure and large-scale environments of haloes.
For this reason, this additional data is labelled `silver' data, and it was used in addition to the `gold' data which was generated from the actual full $f(R)$ gravity simulations. We justify the use of silver data by noticing that our thin-shell modelling (see above) treats haloes as spherical top-hats by averaging the mass distribution within $R_{500\rm c}$ (the same can be done for other halo mass definitions, although in this study we use $M_{500\rm c}$ when studying $M_{\rm dyn}/M_{\rm true}$) and therefore {is not sensitive to} the actual subtle differences in the halo density profiles from the full and approximate simulations. In addition, we have checked the validity of using silver data by doing the same analysis for $|f_{R0}|=10^{-5}$, for which we have gold data to compare to: as is shown in Appendix \ref{appendix:mdyn:consistency}, in this case the gold and silver data of F5 are in excellent agreement.

\subsection{Measuring the dynamical mass enhancement}
\label{measure_m_dyn}

The ratio of the dynamical mass to the true mass of a halo depends on the mass of the halo, the background scalar field of the Universe and the redshift. Because the field is a redshift-dependent quantity, the different snapshots for a given model all have different field values with which to investigate the dynamical mass enhancement. The ratio $M_{\rm dyn}/M_{\rm true}$ is described by two parameters $p_1, p_2$ (as will be discussed below), which vary with the background field value $\bar{f}_{R}(z)$ and redshift $z$. In Sec.~\ref{thin_shell_modelling} it was shown that, according to our thin-shell modelling, the screening effect can be described by a specific combination of $\bar{f}_R(z)$ and $z$, $\bar{f}_R(z)/(1+z)$, and so we expect that both $p_1$ and $p_2$ can be fitted as functions of $\bar{f}_R(z)/(1+z)$ using their values at the snapshots. In this subsection we describe how this fitting process was carried out in our analysis.

\subsubsection{tanh function fit to \texorpdfstring{$M_{\rm dyn}/M_{\rm true}$}{Mdyn/Mtrue}}

In this step, the \textsc{ahf} halo catalogues were first sifted to keep only haloes made up of a sufficient number of dark matter particles and to exclude sub-haloes. The mass criteria for the sifting of Crystal, Jade and Diamond was, respectively, $M_{500}>(4\times10^{13},3\times10^{12},6.5\times10^{10})h^{-1}M_{\odot}$, which correspond to a minimum number of particles per halo of 513, 454 and 428. These numbers were chosen conservatively to ensure that the $\Lambda$CDM halo catalogues are complete down to those masses, which in practice was done by requiring that the HMF is in good agreement with the \citet{Tinker:2010ff} analytical fitting formula.

\begin{figure*}
\centering
\includegraphics[width=1.0\textwidth]{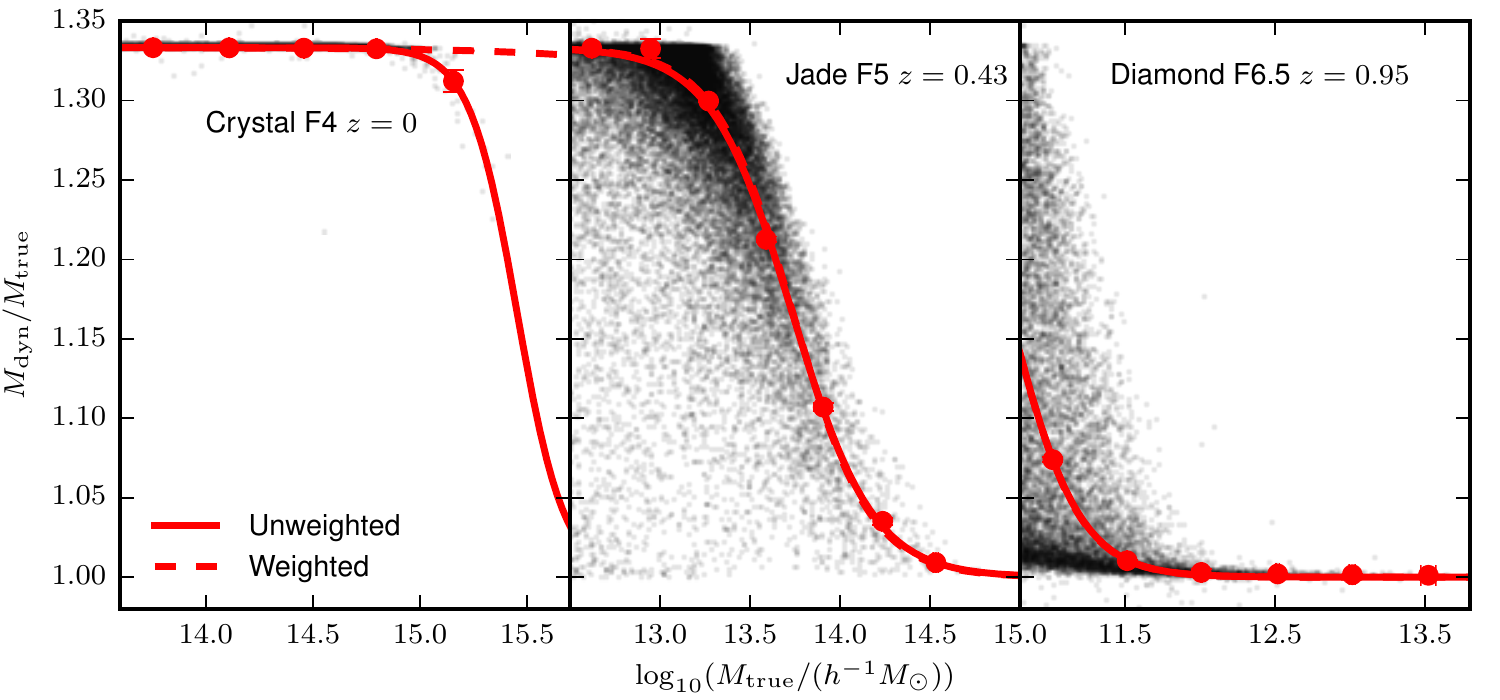} 
\caption[The ratio of the dynamical mass to the true mass as a function of the true mass for dark matter haloes in HS $f(R)$ gravity.]{The ratio of the dynamical mass to the lensing mass versus the lensing mass for the dark matter haloes generated from $N$-body simulations run with modified \textsc{ecosmog} simulations for HS $f(R)$ gravity. \textit{From left to right}: Crystal simulation with $z=0$ for F4; Jade simulation with $z=0.43$ for F5; Diamond simulation with $z=0.95$ for F6.5. The simulation specifications are provided in Table \ref{table:simulations}. Unweighted (\textit{solid}) and weighted (\textit{dashed}) least squares fits of Eq.~(\ref{eq:mdyn_enhancement}) are plotted over the data. These are generated using mass bins represented by the mean mass and the median ratio, shown by the red points. These points and their one standard deviation error bars are produced using jackknife resampling. For jackknife errors less than $10^{-4}$, we replace these with half of the 68\% width of the data, between the 16th and 84th percentiles (see main text, below).}
\label{fig:raw_fits}
\end{figure*}

Three plots of the mass ratio $M_{\rm dyn}/M_{\rm true}$ as a function of the halo mass $M_{\rm true}$ are shown in Fig.~\ref{fig:raw_fits}, for the sifted Crystal F4, Jade F5 and Diamond F6.5 data for redshifts $0$, $0.43$ and $0.95$ respectively. These include the extremes in both field strength and redshift. Each black data point corresponds to an individual halo. In each plot a majority of the haloes lie along a dark band of points that is asymptotic at ratios 4/3 and 1. The asymptote at ratio 1 corresponds to $M_{\rm dyn}=M_{\rm true}$, which holds for higher-mass haloes whose self-screening is sufficient to completely remove the enhancement due to the fifth force. The asymptote at 4/3 represents the maximum possible enhancement to $M_{\rm dyn}$, and therefore results for haloes in a relatively empty environment and with mass low enough that there is effectively no self-screening of the fifth force. 

For F5 many points are found below the dark band. These correspond to haloes that have most likely experienced environmental screening due to nearby more massive haloes, such that chameleon suppression of the fifth force is active even though the halo mass itself might not be great enough for self-screening. The effect of environmental screening in F5 is weak enough that the dark band of data only traces haloes for which self-screening dominates over environmental screening. In F4, few data points are observed below the band because environmental screening is less effective in stronger background fields. For F4 and F5, apart from numerical noise, no data points are found to lie above $4/3$ which is the maximally-allowed dynamical mass enhancement in $f(R)$ gravity. 
In F6.5 the dark band of data is observed to have lower enhancement, with many data points found above it, particularly at $M_{\rm true}\leq10^{11.5}h^{-1}M_\odot$. With such low field values and halo masses in this mass range in F6.5, environmental screening is now able to begin to dominate over self-screening, which means the dark band of data no longer traces the haloes with self-screening only, as it did for F4 and F5. This is why it is now possible to find haloes above the main trend, as these simply correspond to haloes in emptier environments. Note that the upper bound of $4/3$ applies also in this case.

In order to extract a trend for this data, the haloes are grouped into a set of equally-spaced logarithmic mass bins, 
which effectively cover the full range of halo masses under consideration for a given model and snapshot. For each bin, the mean halo mass is measured along with the median ratio $M_{\rm dyn}/M_{\rm true}$ among all haloes. The data in each bin approximately follows a lognormal distribution, and the median is expected to yield an appropriate ratio from within the main band of data. While we do not provide a detailed modelling of the distribution of $M_{\rm dyn}/M_{\rm true}$ here, we have provided a simple model for the root-mean-square scatter in Chapter \ref{chapter:constraint_pipeline}. 

In the absence of multiple realisations of the data, the errors on the mean halo mass and median $M_{\rm dyn}/M_{\rm true}$ in the bins are evaluated using jackknife resampling, in which the data is randomly split into 150 sub-volumes at each snapshot. By systematically excluding one sub-volume at a time, 150 resamples are created. For each resample, the haloes are split into the same set of mass bins, and 150 median ratios $M_{\rm dyn}/M_{\rm true}$ and 150 mean masses are measured for each bin. Following the procedure outlined by \cite{Norberg:2008tg}, the errors in the median ratio and mean mass are generated by taking the square root of the variance of the 150 values, which has to be rescaled by a factor 149 to account for the lack of independence of the resamples.

The mass ratio data is quoted to 4 decimal places in the \textsc{ahf} output. Such precision can result in zero, or an unphysically small, variance being measured by the jackknife method. This can happen in unscreened or completely screened regimes where most of the data in the bin spans only a small range of ratios. Using the argument that the ratio errors must at least equal $10^{-4}$, any errors generated by jackknife which are less than this value are replaced with half of the width of the $68\%$ range (in the bin under consideration), which spans from the 16th percentile to the 84th percentile. The percentile spread is most often used for lower-mass bins in strongly unscreened regimes, where the ratio data spans only a very small range. This ensures that the errors for these bins become a reasonable size relative to the errors of the other bins, which are estimated by jackknife, though rigorously speaking the $68\%$ range is more of a description of the spread of the mass ratio rather than sample variation of the median ratio as jackknife gives. As discussed below, in the main results of this chapter we do not use the error bars estimated using this combination of jackknife and the $68\%$ range.

The results for these bins are shown in Fig.~\ref{fig:raw_fits}, plotted over the raw data. To account for the asymptotic nature of the data, we fit the following tanh curve:
\begin{equation}
\frac{M_{\rm dyn}}{M_{\rm true}} = \frac{7}{6}-\frac{1}{6}\tanh\left(p_1\left[\log_{10}\left(M_{\rm true}M_{\odot}^{-1}h\right)-p_2\right]\right).
\label{eq:mdyn_enhancement}
\end{equation}
The two constants 7/6 and 1/6 are used to ensure the function remains between fixed asymptotes at ratios $4/3$ and $1$. The parameters $p_1$ and $p_2$ represent, respectively, the inverse width of the mass transition and the mass logarithm at the centre of the transition. 


\begin{figure*}
\centering
\includegraphics[width=0.86\textwidth]{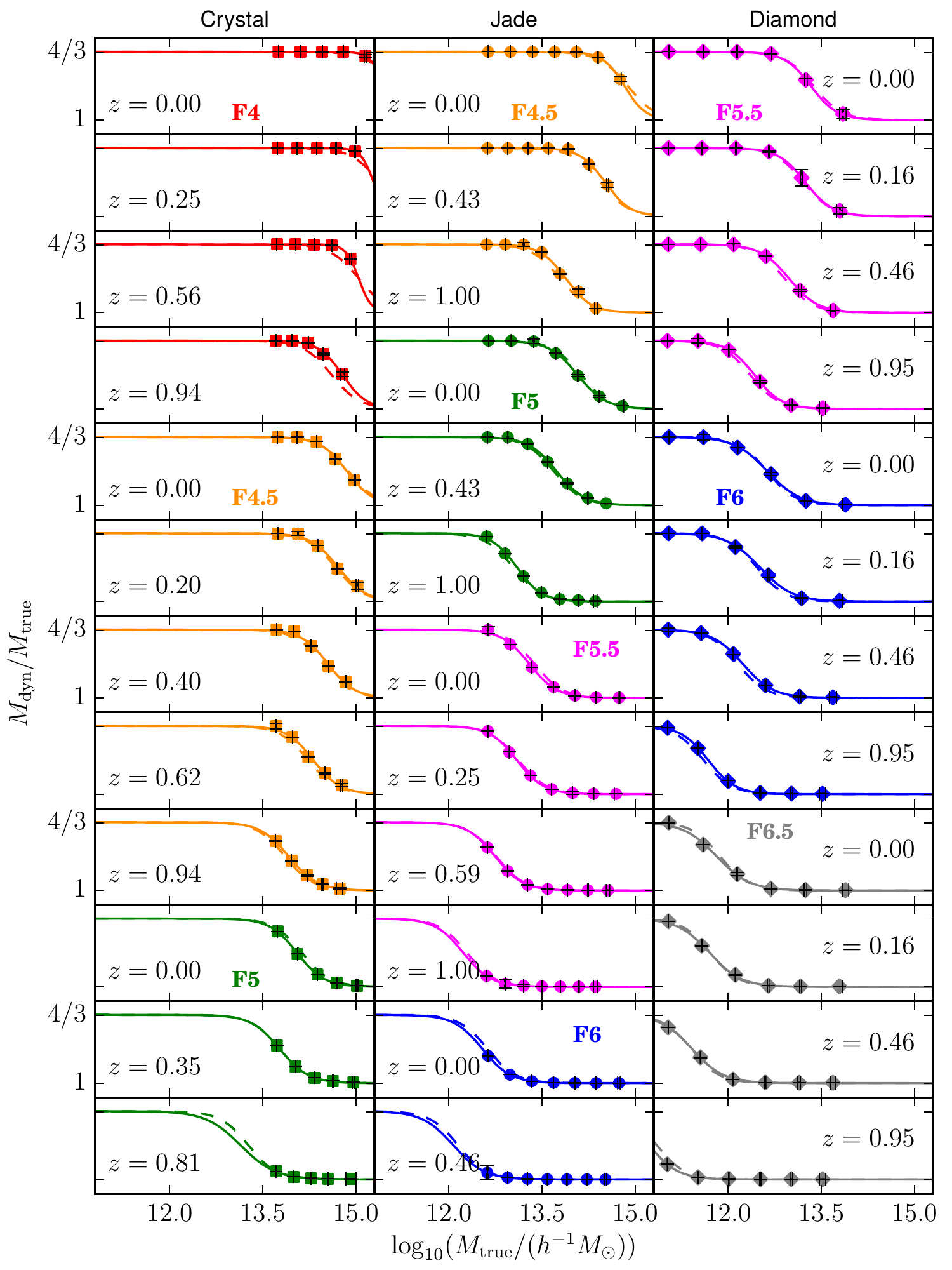}
\caption[Dynamical mass to true mass ratio as a function of the true mass for several present-day field strengths $f_{R0}$ of HS $f(R)$ gravity at various redshifts.]{Dynamical mass to lensing mass ratio as a function of the lensing mass for F4 (\textit{red}), F4.5 (\textit{orange}), F5 (\textit{green}), F5.5 (\textit{magenta}), F6 (\textit{blue}) and F6.5 (\textit{grey}) at various redshifts as annotated. The halo data comes from the Crystal (\textit{left column}), Jade (\textit{middle column}) and Diamond (\textit{right column}) simulations (see Table \ref{table:simulations}). The data points, corresponding to mass bins represented by their median ratio and mean mass, and their one standard deviation error bars are produced using jackknife resampling. Jackknife errors less than $10^{-4}$ are replaced with half of the range between the 16th and 84th percentiles. \textit{Solid line}: Eq.~(\ref{eq:mdyn_enhancement}) with $p_1$ and $p_2$ determined by unweighted least squares fitting for the given snapshot; \textit{Dashed line}: Eq.~(\ref{eq:mdyn_enhancement}) with best-fit constant $p_1$ result ($p_1=2.21$) and linear $p_2$ result (Eq.~(\ref{eq:p2})) from Fig.~\ref{fig:unweighted_p_1} and Fig.~\ref{fig:unweighted_p_2}, respectively.}
\label{fig:unweighted_matrix}
\end{figure*}


For the dashed line, the parameters have been optimised through {weighted least squares: the} minimisation of the sum of the squared normalised residuals, where the normalisation is equal to the size of the error bars. For F5 and F6.5, this {fit of} Eq.~(\ref{eq:mdyn_enhancement}) shows excellent agreement with the bin data, however for F4 the fit shows poor agreement with the result for the highest mass bin. This is because the error bar of this bin is substantially greater than those of the lower-mass bins, and it contributes very little weight in the optimisation. Weighted least squares therefore over-estimates the value of $p_2$ for this snapshot, as the $\tanh$ curve starts to drop at a higher mass than the raw data. In contrast, the data point in the second highest mass bin has a much smaller error and it slightly overshoots a $\tanh$ curve which would perfectly go through the highest mass data point (the solid line, see below). Note that the same happens to the second and third lowest-mass data points for F5 (the middle panel of Fig.~\ref{fig:raw_fits}), but in this case there are four other data points at higher masses which dominate the optimisation, resulting in a good visual agreement between the dashed curve and the data points. This indeed highlights the importance of having data points which cover the full transition of the $\tanh$ curve in order to fit $p_1$ and $p_2$ accurately. Furthermore, the observation that the second lowest mass point for F5 lies above the $\tanh$ curve is quite generic and happens in most other plots where the curve starts to deviate from $4/3$, implying a slight insufficiency in the $\tanh$ fitting (we will comment on how this affects the fitted values of $p_1$ and $p_2$ below). 

On the other hand, for the solid line in Fig.~\ref{fig:raw_fits} the parameters have been optimised via unweighted least squares: the minimisation of the sum of the squared residuals, which have equal weights for all bins now. Since it does not suffer from the same issues as described above for the weighted fitting, this fit shows better agreement with the data point of the highest mass bin of F4, while elsewhere shows equally good agreement as weighted least squares. 

As discussed above, in the completely screened or unscreened regimes there is very little variation of the mass ratio and therefore the resulting uncertainties --- by using either Jackknife resampling or the 68\% range --- for mass bins in those regimes are extremely small. Together with the facts that in many snapshots (e.g., the left panel of Fig.~\ref{fig:raw_fits}) the data points only cover part of the transition of the tanh curve and that the lower-mass bins can contain around three orders of magnitude more haloes than the higher-mass bins, this makes it challenging to find a consistent way to estimate uncertainties in all mass bins across all models/snapshots. Since the inhomogeneous sizes of error bars in the data points can lead to clearly unphysical fitting results, as shown in the dashed lines of the left panel of Fig.~\ref{fig:raw_fits}, the main results of this chapter shall be given using the unweighted least squares approach. We have tried a number of different ways to assign data error bars, including setting a lower limit such as $10^{-4}$ to the individual errors, which all involve certain degrees of arbitrariness (for example, the 68\% range to get error bars in Fig.~\ref{fig:raw_fits} is really a characterisation of the spread of the data rather than an uncertainty of the median, and it is used solely to avoid very small uncertainties for some mass bins). Perhaps more importantly, the different ways of estimating uncertainties for the weighted least squares approach that we have tried all lead to similar fitting results of $p1, p2$ as functions of $\bar{f}_R(z)/(1+z)$ (the topic of the next sub-subsection), and the situation depicted in the left panel of Fig.~\ref{fig:raw_fits} happens only for a few snapshots. As an example for reassurance, in Appendix \ref{appendix:mdyn:weighted_fitting} we present fitting results of $p_1$ and $p_2$ using the weighted least squares approach with the error bars estimated as in Fig.~\ref{fig:raw_fits}, which confirms that this different approach does not significantly affect the final result.

For each snapshot in the investigation, five mass bins were used for Crystal, seven for Jade and six for Diamond, as these are the maximum possible numbers of bins such that there are a minimum of five haloes in almost all bins. We have checked different bin numbers, and this combination of bin numbers was also found to yield the smoothest results.

\subsubsection{Fitting of \texorpdfstring{$p_1, p_2$}{p1,p2} as functions of \texorpdfstring{$\bar{f}_R(z)/(1+z)$}{fR/(1+z)}}

By carrying out a fitting of Eq.~(\ref{eq:mdyn_enhancement}) for all snapshots of all models, the field and redshift dependence of $p_1$ and $p_2$ can be tested. 
To understand what should be plotted, Sec.~\ref{f(R)} and in particular the approximations for $M_1$ and $M_2$, given by Eqs.~(\ref{M_1}) and (\ref{M_2}), are used. From the way that $p_1$ and $p_2$ have been defined, the following can be shown:
\begin{equation}
p_1(z,\bar{f}_R) \propto \frac{1}{\log_{10}\left(M_1\right)-\log_{10}\left(M_2\right)} = {\rm const};
\label{p_1}
\end{equation}
\begin{equation}
p_2(z,\bar{f}_R) = \frac{\log_{10}\left(M_1\right)+\log_{10}\left(M_{2}\right)}{2} = \frac{3}{2}\log_{10}\left(\frac{|\bar{f}_R|}{1+z}\right)+{\rm const}.
\label{p_2}
\end{equation}
Eqs.~(\ref{M_1}) and (\ref{M_2}) have been used to bring in the $z$ and $\bar{f}_R(z)$ dependences. Eq.~(\ref{p_2}) implies $p_2$ should have a linear trend as a function of $\log_{10}\left(\frac{\bar{f}_R}{1+z}\right)$ with a slope of $1.5$. This comes from the power $3/2$ in Eqs.~(\ref{M_1}) and (\ref{M_2}), where it in turn stems from the $2/3$ power in $\Psi_{\rm N} \propto M^{\frac{2}{3}}$ for the Newtonian potential given by Eq.~(\ref{Newton}). On the other hand Eq.~(\ref{p_1}) implies $p_1$ has no dependence on $z$ and $\bar{f}_R$ {apart from through higher order effects,} such as the non-sphericity of haloes, non-uniformity of the mass distributions within haloes, environmental screening, etc. Due to the simplicity of our thin-shell modelling, here we shall not attempt to include these higher-order effects. {Indeed, under the thin-shell approximation, using Eqs.~(\ref{M_1}, \ref{M_2}, \ref{eq:kappa_12}), it is found that the intercept of $p_2$ in Eq.~(\ref{p_2}) only depends on $\epsilon$, $G$ (there is no dependence on $H_0=100h$~kms$^{-1}$Mpc$^{-1}$ since the $h$ is absorbed into the unit of $10^{p_2}$, $h^{-1}M_\odot$) and $\Delta$, and $p_1$ depends only on $\epsilon$; neither depends on the cosmological parameters, whose effects are completely in determining $\bar{f}_R(z)$.} We will find later that $p_1$ is indeed very weakly dependent on $\bar{f}_R(z)/(1+z)$. We also show that this dependency can be safely ignored without significantly affecting the value of the ratio $M_{\rm dyn}/M_{\rm true}$.


A potential issue arises from the limitations of the mass range covered by a particular set of data. As can be seen from Fig.~\ref{fig:unweighted_matrix}, the mass bins are located almost entirely in the unscreened regime for F4 at low redshifts, while for high redshift Crystal F5, Jade F6 and Diamond F6.5 the mass bins are mostly found in the completely screened regime. As will be discussed in Figs.~\ref{fig:unweighted_p_2} and \ref{fig:unweighted_p_1} of Sec.~\ref{results}, the latter can result in under-estimation of the $p_1$ and $p_2$ values, and we have already seen in Fig.~\ref{fig:raw_fits} how, depending on the choice of fitting procedure, $p_2$ can be over-estimated for F4 at low redshift.

To understand why the parameters are affected in such a manner, consider the scenario where all mass bins are located at ratio $4/3$. As can be seen in the F4 and F5 panels of Fig.~\ref{fig:raw_fits}, the median ratio data from the simulations in this regime is almost completely flat, so a $\tanh$ fit will predict a turning point at a mass higher than is actually the case, and so $p_2$ will be over-estimated. This flatness of the raw data in the unscreened regime is particularly evident in the F5 panel, where the second data point from the left ends up above the trend line, despite having a negligible error, at the same height as the first data point (this suggests that this region of the data cannot be fitted perfectly by a tanh curve). On the other hand, for mass bins at high-redshift snapshots and for low field strengths (F6.5 - F5), where almost all of the data points lie at a ratio of $1$, because the data here is flatter than predicted by Eq.~(\ref{eq:mdyn_enhancement}) the turning point at ratio 1 will thus be predicted at lower mass, leading to an under-estimation of $p_2$. The effect on $p_1$ turns out to be similar to $p_2$, but is even more sensitive to these limitations.

The issues presented here were the main motivation for using data from simulations with differing resolutions. To prevent such dubious estimations of $p_1$ and $p_2$ from adversely affecting the main results, a strict criterion is enforced: we only trust $p_1$ and $p_2$ values that have been calculated using snapshots for which the mass bins enclose at least half of the height of the mass ratio transition (a median ratio range of 1/6 or greater).

\section{Results} 
\label{results}

As mentioned above, a fitting function for the ratio $M_{\rm dyn}/M_{\rm true}$ that works for general scalar field strength $f_{R0}$ and redshift $z$ should  be calibrated and validated against full numerical simulations with a large dynamical range of halo masses in order to maximally cover the transition between screened and unscreened regimes, which itself varies strongly with $z$ and $f_{R0}$. However, $N$-body simulations are known to have a limited dynamical range and it is also too expensive to run full simulations for too many $f_{R0}$ values. Our recipe to tackle the former challenge is to combine a suite of simulations with varying resolutions (Crystal, Jade and Diamond) to increase the halo mass range, while for the latter issue we have introduced the low-cost `silver' simulations (see Sec.~\ref{simulations}). Both approaches need to be explicitly checked to guarantee validity and consistency. Furthermore, in Sec.~\ref{measure_m_dyn} we have discussed subtleties in the $\tanh$ curve fitting such as the weighted and unweighted least squares approaches. In this section we give the main results on $p_1$ and $p_2$ from using this methodology, for unweighted least squares, and leave various consistency checks to the Appendices. In Appendix \ref{appendix:mdyn:weighted_fitting} we compare with results from using the weighted least squares approach as a double check, and in Appendix \ref{appendix:mdyn:consistency} we check the use of `silver' data and the combination of the Crystal, Jade and Diamond simulations. 

\begin{figure*}
\centering
\includegraphics[width=0.85\textwidth]{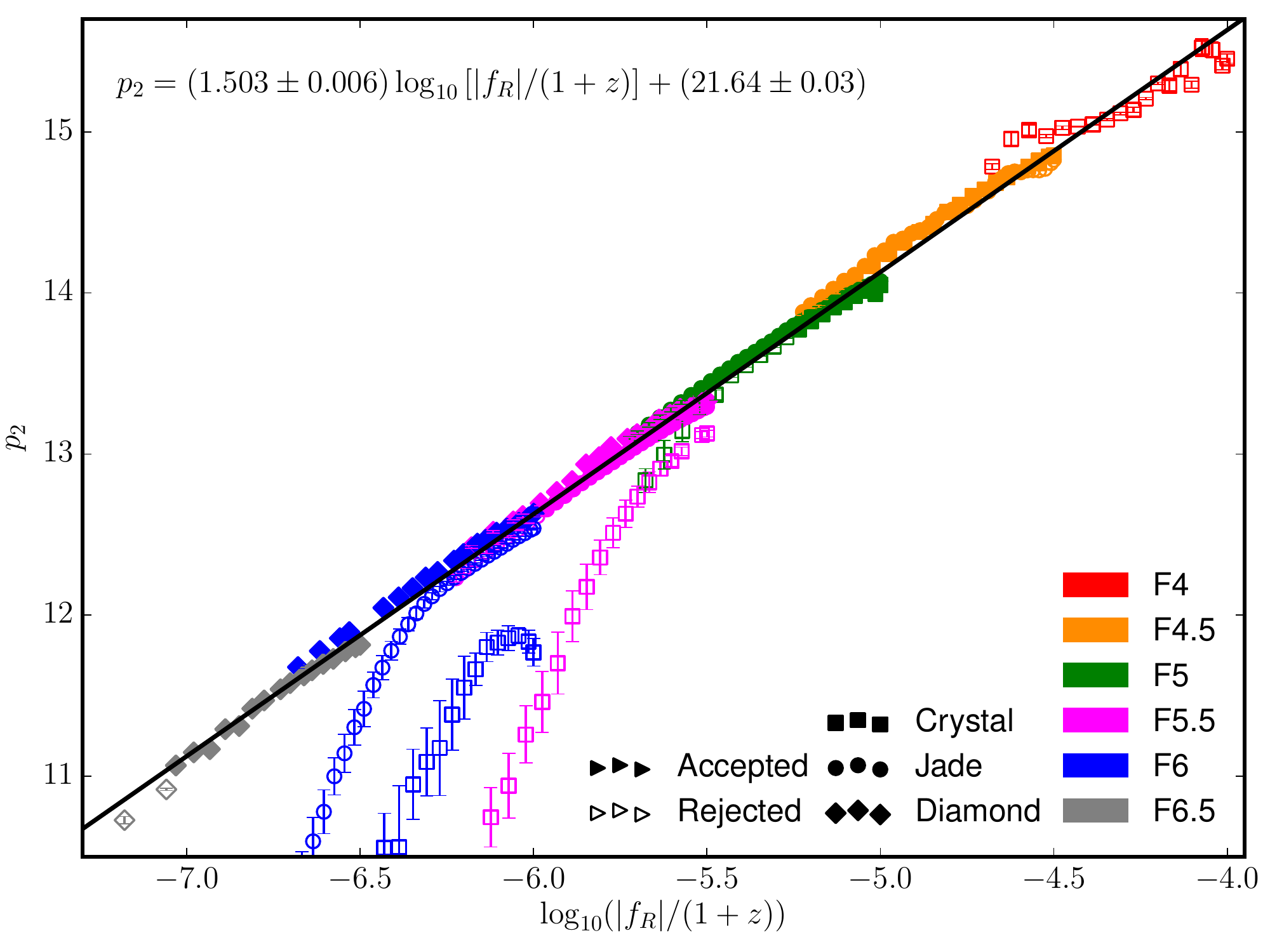}
\caption[Parameter $p_2$ in Eq.~(\ref{eq:mdyn_enhancement}) as a function of the background scalar field at redshift $z$, $\bar{f}_R(z)$, divided by $(1+z)$.]{Parameter $p_2$ in Eq.~(\ref{eq:mdyn_enhancement}) plotted as a function of the background scalar field at redshift $z$, $\bar{f}_R(z)$, divided by $(1+z)$, for several present day field strengths $f_{R0}$ (see legends) of HS $f(R)$ gravity. $p_2$ is measured via an unweighted least squares optimization of Eq.~(\ref{eq:mdyn_enhancement}) to data from modified \textsc{ecosmog} simulations, described by Table \ref{table:simulations}, at simulation snapshots with redshift $z<1$. $\bar{f}_R(z)$ is calculated for each snapshot using Eq.~(\ref{eq:fR_background}). The trend line has been produced via a weighted least squares linear fit, using the one standard deviation error bars, of the solid data points, which correspond to snapshots for which the mass bins contain at least half of the median mass ratio range 1 to 4/3. The hollow data does not meet this criterion, so is deemed unreliable and neglected from the fit, which is given by Eq.~(\ref{eq:p2}).}
\label{fig:unweighted_p_2}
\end{figure*}

A plot of $p_2$ as a function of $\log_{10}\left(\frac{|\bar{f}_R|}{1+z}\right)$ is shown in Fig.~\ref{fig:unweighted_p_2}. A linear trend is fitted using the filled data points, which correspond to snapshots for which the mass bins enclose a median ratio range of $1/6$ or greater. The motivation for this criterion is discussed in Sec.~\ref{measure_m_dyn}. The filled data points are expected to give a reasonable estimate for the logarithm of the mass at the centre of the transition, and they all turn out to lie along a clear linear trend in Fig.~\ref{fig:unweighted_p_2}. The result of the linear fit, found using the one standard deviation error bars, is: 
\begin{equation}
    p_2=(1.503\pm0.006)\log_{10}\left(\frac{|\bar{f}_R|}{1+z}\right)+(21.64\pm0.03). 
    \label{eq:p2}
\end{equation}
The gradient of $1.503\pm0.006$ shows excellent agreement with the theoretical prediction of 1.5 from Eq.~(\ref{p_2}).

Many of the hollow data points are observed to be peeling off the trend, particularly in the F6 and F5.5 models. These snapshots correspond to cases in which all mass bins are found in the totally screened regime, resulting in an under-estimation of the centre of the transition as discussed in the previous section. This behaviour provides no useful information about the dynamical mass enhancement, but rather it tells us that a higher resolution simulation, with lower-mass particles to probe haloes of lower mass, is required. For F5.5 the peeling-off corresponds to Crystal data, whereas the higher resolution Jade and Diamond simulations produce linear data. For F6 both the Crystal and Jade data peel off from the linear trend, as only Diamond has a high enough resolution to probe unscreened haloes in F6. Diamond turns out to have a sufficient resolution to effectively examine F6.5 as well, although a couple of high redshift snapshots do not get used in the linear fit, suggesting these are on the boundary between reliable and  untrustworthy data. F6.5 nevertheless agrees with the linear behaviour of the rest of the filled data. 

A relatively noisy trend is observed in the F4 data (though the data points all reasonably follow the linear trend), probably because each snapshot only has one or two mass bins lying within the mass range where the ratio $M_{\rm dyn}/M_{\rm true}$ undergoes a transition between $1$ and $4/3$. Most bins lie in the unscreened regime, such that none of the snapshots in F4 satisfy the selection criterion to be included in the linear fit -- all data points for F4 are hollow in Fig.~\ref{fig:unweighted_p_2}. An improvement of this result would require a simulation with a sufficiently large box size to include more haloes at the higher masses necessary to properly examine screening in F4.

\begin{figure*}
\centering
\includegraphics[width = 0.85\textwidth]{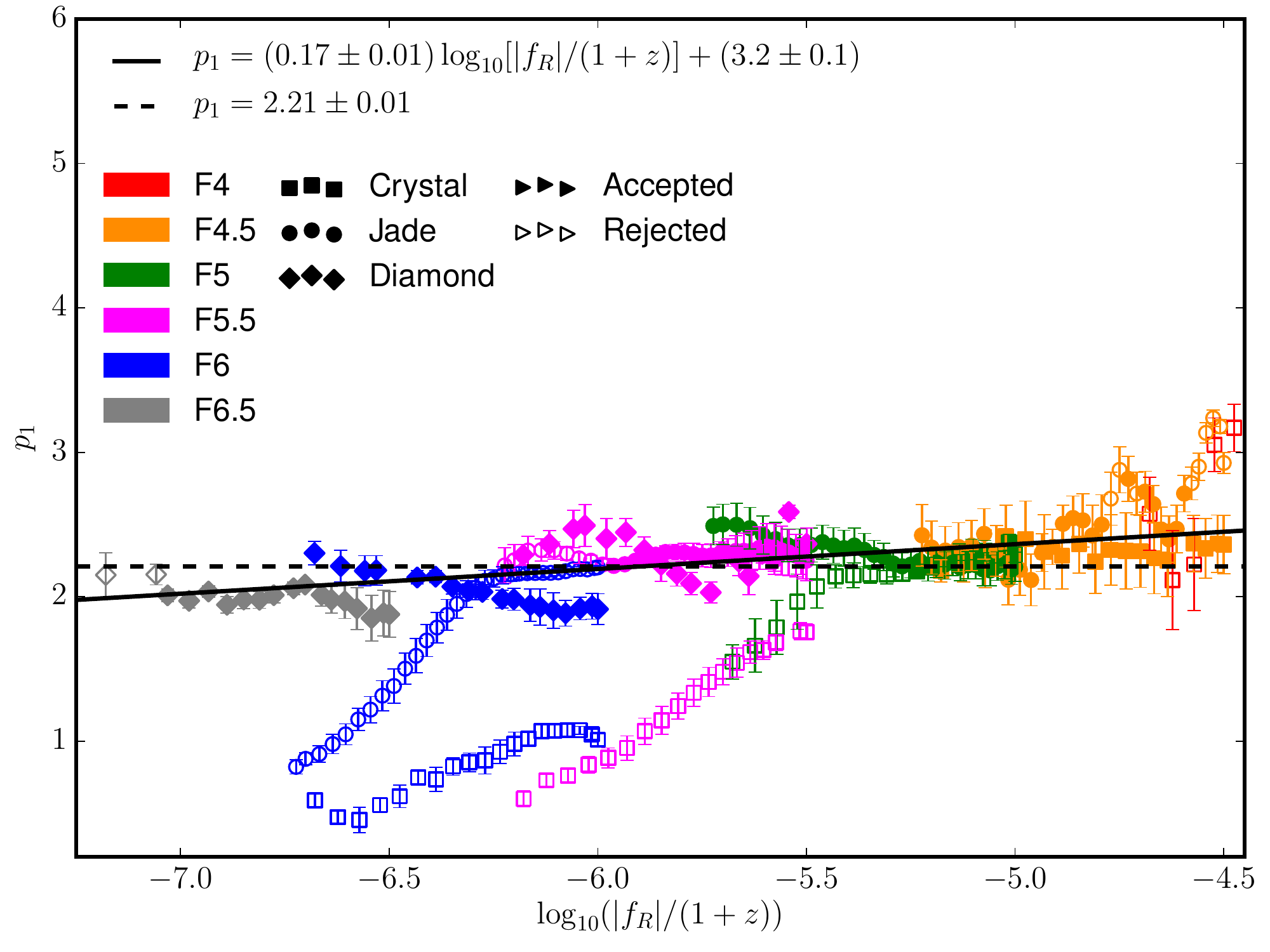}
\caption[Parameter $p_1$ in Eq.~(\ref{eq:mdyn_enhancement}) as a function of the background scalar field at redshift $z$, $\bar{f}_R(z)$, divided by $(1+z)$.]{Parameter $p_1$ in Eq.~(\ref{eq:mdyn_enhancement}) plotted as a function of the background scalar field at redshift $z$, $\bar{f}_R(z)$, divided by $(1+z)$, for several present day field strengths $f_{R0}$ (see legends) of HS $f(R)$ gravity. $p_1$ is measured via an unweighted least squares optimisation of Eq.~(\ref{eq:mdyn_enhancement}) to data from modified \textsc{ecosmog} simulations, described by Table \ref{table:simulations}, at simulation snapshots with redshift $z<1$. $\bar{f}_R(z)$ is calculated for each snapshot using Eq.~(\ref{eq:fR_background}). Weighted least squares linear (\textit{solid line}) and constant (\textit{dashed line}) fits, {using the one standard deviation error bars}, of the solid data points, which correspond to snapshots for which the mass bins contain at least half of the median mass ratio range $1$ to $4/3$, are shown. The hollow data points do not meet this selection criterion, and therefore are deemed unreliable and neglected from the fits, which are given by Eq.~(\ref{eq:p1}) and $p_1=(2.21\pm0.01)$, respectively.}
\label{fig:unweighted_p_1}
\end{figure*}

The corresponding plot for $p_1$ is shown in Fig.~\ref{fig:unweighted_p_1}. The {trend} is more complicated than that of $p_2$, partly because the thin-shell model {result described in} Eq.~(\ref{p_1}) predicts no dependence of $p_1$ on $\bar{f}_R$ and $z$, while dependence can still be introduced through effects such as environmental screening which are harder to model. However, we expect that these effects have a relatively small impact, and indeed, an approximately flat trend of $p_1$ is observed. The results are noisier here than in Fig.~\ref{fig:unweighted_p_2} for $p_2$, because the width of the mass transition requires a greater range of halo masses for a $\tanh$ fit to be reliable. 
The criterion for selecting snapshots in the fit of $p_1$ is the same as for $p_2$, and again only the solid data points which satisfy this criterion are fitted. This rules out all of the data from the F4 model {(which produces a wild trend here that is left out of the plot area)}, and several snapshots from other models. 

The result for the constant $p_1$ fit, as predicted by Eq.~(\ref{p_1}), is $p_1=(2.21\pm0.01)$ and is shown by the dashed line in Fig.~\ref{fig:unweighted_p_1}. A linear model was also fitted, shown by the solid line, yielding the following result:
\begin{equation}
    p_1=(0.17\pm0.01)\log_{10}\left(\frac{|\bar{f}_R|}{1+z}\right)+(3.2\pm0.1). 
    \label{eq:p1}
\end{equation}
These trends have been fitted using the one standard deviation error bars. The gradient of $0.17\pm0.01$ is small, though not in agreement with the prediction of a flat trend. With a theoretical modelling which neglects effects such as environmental screening, a small gradient nevertheless seems like a reasonable result. Being able to accurately predict the width of the mass transition is not as important as being able to predict the central mass of the transition, because the $\tanh$ curve is less sensitive to $p_1$ than to $p_2$ (which can be easily checked). Almost all the data points observed to be significantly peeling off from the horizontal band of data in Fig.~\ref{fig:unweighted_p_1} (including Jade and Crystal F6, Crystal F5.5 and some of Jade F4.5) fail to satisfy the selection criterion. This is further evidence that these particular trends are indeed caused by the limitations of the simulation resolution. Also, a comparison of Figs.~\ref{fig:unweighted_p_1} and \ref{fig:p_1} shows that the use of an unweighted approach to measure $p_1$ produces the smoother trend in the $p_1$ data.


The quality of the above fits for $p_1$ and $p_2$ as well as the validity of the theoretical predictions, given by Eqs.~(\ref{p_1}) and (\ref{p_2}), can be assessed by examining Fig.~\ref{fig:unweighted_matrix}. The solid lines represent the exact fits produced in the {unweighted} least squares optimisation of Eq.~(\ref{eq:mdyn_enhancement}) to each snapshot of data. The dashed lines are plotted using Eq.~(\ref{eq:mdyn_enhancement}) and the $p_1$ and $p_2$ values that are predicted using the constant fit of Fig.~\ref{fig:unweighted_p_1} {(dashed line)} and the linear fit of Fig.~\ref{fig:unweighted_p_2} {(solid line)} respectively. Noticeable disparities between the dashed line and solid line fits are observed in the F4 data, resulting from the relatively flat trend produced by the raw data in unscreened regimes {and the limited number of haloes in Crystal covering the high masses necessary for properly examining the transition to complete screening in F4.} 
The agreement between the dashed and solid lines in Fig.~\ref{fig:unweighted_matrix} generally improves if one uses the linear fit predictions for $p_1$, {although we only use the constant fit here, which is motivated by our theoretical modelling. Nevertheless, in general the dashed line fits show excellent agreement with the simulation data over the full range of redshifts and models that are plotted in Fig.~\ref{fig:unweighted_matrix}, implying that Eq.~(\ref{eq:mdyn_enhancement}) can be treated as a general formula when using our constant and linear fits of $p_1$ and $p_2$ respectively.}

\subsection{Potential implications}

Although they are not directly related to the preparation for cluster constraints, we make the following interesting observations in the results of this section, mainly Fig.~\ref{fig:unweighted_p_2}.

First, the solid straight line in Fig.~\ref{fig:unweighted_p_2} represents the logarithm of the halo mass, $\log_{10}M_{\rm true}$, at the centre of the transition of the median of $M_{\rm dyn}/M_{\rm true}$, and it roughly separates the haloes into two parts -- a screened sample ($\log_{10}M_{\rm true}$ well above the line) and an unscreened sample ($\log_{10}M_{\rm true}$ well below the line). From Figs.~\ref{fig:raw_fits} and \ref{fig:unweighted_p_2} we notice that even at $|\bar{f}_R(z)|/(1+z)=10^{-7}$, corresponding to a strongly screened model, about half of the haloes (with high ratio $M_{\rm dyn}/M_{\rm true}$) with mass $M_{\rm true}\sim10^{11}h^{-1}M_\odot$ are unscreened, and these are haloes which are likely to reside in under-dense regions. The other half of these haloes (with low ratio $M_{\rm dyn}/M_{\rm true}$) are screened, aided by their environments, implying the importance of environmental screening. It would certainly be interesting to see if this linear trend goes to even smaller values of $|f_{R}(z)|/(1+z)$, which will tell us whether dwarf galaxy haloes can be environmentally screened for those field values. This will be relevant for astrophysical tests of $f(R)$ gravity \citep[e.g.,][]{Jain:2012tn,Vikram:2013uba,Sakstein:2014nfa}.

Second, it is interesting that the screening of haloes in models with different $f_{R0}$ can be well described by a single parameter: $\bar{f}_R(z)/(1+z)$. This implies that the theoretical modelling of various other properties in $f(R)$ gravity can perhaps be simplified into a one-parameter family of description and therefore may have profound theoretical and practical implications.

\section{Summary, discussion and conclusions}
\label{conclusions}

The global properties of galaxy clusters, such as their abundance and clustering on large scales, are sensitive to the strength of gravity and can be predicted accurately using cosmological simulations. They therefore offer a powerful means of testing alternative models of gravity, including $f(R)$ gravity, on large scales. In order to utilise the wealth of information being made available through current and upcoming galaxy cluster surveys, it is important to ensure that numerical predictions are prepared that can be directly confronted to the observational data. This includes accounting for various sources of theoretical bias, such as the enhancement of the dynamical mass of galaxy clusters  resulting from the presence of the fifth force in unscreened $f(R)$ gravity. This effect is currently not included in the derivations of scaling relations used to determine the cluster mass. The best means of correcting this would be through a re-calibration of the scaling relations which are better understood in $\Lambda$CDM, and make them work in the context of MG, which requires an understanding of the relationship between the dynamical mass and lensing mass. However, previous studies of this relationship in the literature are specific and do not include a general formula that can be applied to arbitrary model parameters and redshifts.

We have found a simple model to describe the relationship between the dynamical mass and lensing mass of dark matter haloes in the HS $f(R)$ model. As shown by the solid line fits of Fig.~\ref{fig:unweighted_matrix}, the $\tanh$ fitting formula of Eq.~(\ref{eq:mdyn_enhancement}) has generally shown excellent agreement with \textsc{ahf} halo data, for $z<1$, from three \textsc{ecosmog} DMO simulations, which are summarised in Table \ref{table:simulations}. By taking advantage of the variety of resolutions offered by these simulations, and using $\Lambda$CDM simulations to produce approximate data for field strengths not covered by the $f(R)$ gravity simulations, the validity of Eq.~(\ref{eq:mdyn_enhancement}) has been probed vigorously across a wide and continuous range of field values that cover $10^{-6.5}<|f_{R0}|<10^{-4}$ within $z<1$.

In addition, we have used a simple thin-shell model (Sec.~\ref{thin_shell_modelling}) to predict the behaviours of free parameters $p_1$ and $p_2$ in Eq.~(\ref{eq:mdyn_enhancement}), which characterise the inverse width and the central logarithmic mass of the tanh-like transition respectively. The predictions, which neglect the effects of environmental screening due to nearby dark matter haloes, are given by Eqs.~(\ref{p_1}, \ref{p_2}). Using a stringent criterion to exclude unreliable snapshots in the fitting, the result for $p_2$, shown in Fig.~\ref{fig:unweighted_p_2}, is given by Eq.~(\ref{eq:p2}). The slope value of $1.503\pm0.006$ shows excellent agreement with the prediction of $1.5$ by Eq.~(\ref{p_2}), and the data of Fig.~\ref{fig:unweighted_p_2} shows a clear linear trend as predicted. As shown by Fig.~\ref{fig:unweighted_p_1}, the $p_1$ data is more scattered, but given the size of the one standard deviation error bars, the constant trend predicted by Eq.~(\ref{p_1}) is not unreasonable, resulting in $p_1=(2.21\pm0.01)$. As shown by the dashed line fits of Fig.~\ref{fig:unweighted_matrix}, these results for $p_1$ and $p_2$ show good agreement with the simulation data across the full range of field values and redshifts. We have also repeated the analysis using a different approach to utilise the errors in the simulation data, and the results, shown in Appendix \ref{appendix:mdyn:weighted_fitting}, also agree with the thin-shell model prediction very well. In Appendix \ref{appendix:mdyn:consistency} we further argue that the results in this chapter apply to models with different cosmological parameters such as $\sigma_8$ and $\Omega_{\rm M}$.

On the other hand, although we make a very specific choice of $f(R)$ gravity in this chapter, the theoretical model and the procedure we followed to calibrate it are expected to be applicable to general chameleon gravity theories \citep{Gronke:2015,Gronke:2016}. As discussed briefly in Appendix \ref{appendix:mdyn:consistency}, in other $f(R)$ models the transition between screened and unscreened regimes can be different from the \citet{Hu:2007nk} model with $n=1$, which may cause the exact fitted values of $p_i$ to differ from what we presented in the above. Therefore, other $f(R)$ models may require a re-calibration based on simulations. However, given that all $f(R)$ models are phenomenological, it is perhaps more sensible to focus on a representative example, such as that by \citet{Hu:2007nk}, to make precise observational constraints. The pipeline and methodology can then be applied to any other models following general parameterisation schemes \citep[e.g.,][]{Brax:2012gr,Brax:2011aw,Lombriser:2016zfz}, which are useful for capturing the essential features of large classes of models using a few parameters. Should a preferred one emerge, the conclusion for the HS model can serve as a rough guideline as to what level future cluster observations can constrain scalar-tensor-type screened theories. For this reason we decide not to explore other forms of $f(R)$ in this thesis.

A generic fitting function for the relationship between the dynamical and lensing masses of dark matter haloes is an essential ingredient of the new framework proposed in this thesis, to carry out cosmological tests of gravity in an unbiased way. Taking Eq.~(\ref{Li and He}) as an example, our general formula for the dynamical mass enhancement allows us to incorporate this particular effect of $f(R)$ gravity into galaxy cluster scaling relations in a self-consistent way. A key benefit of a fitting function is that it allows a continuous search through the model parameter space without having to run full simulations for every parameter point sampled in MCMC. The results will also be useful for other cluster tests of gravity that employ the difference between dynamical and lensing masses, such as by comparing cluster dynamical and lensing mass profiles, or by looking at measured cluster gas fractions.

The results presented in this chapter indicate that a simple model sometimes works surprisingly well despite the greatly simplified treatment of the complicated nonlinear physics of (modified) gravity. It naturally raises the following question: can other theoretical or observational properties of dark matter haloes also be modelled accurately, based on a simplified physical picture and calibrated by numerical simulations? An example is the relationship between the masses and density profiles of haloes, as mentioned in Sec.~\ref{framework}. This concentration-mass relation is critical for converting between the different halo mass definitions commonly used in different communities, and a great deal of effort has been made to explain it in the standard $\Lambda$CDM model, while in MG models, such as $f(R)$ gravity, the understanding is still purely numerical and confined to a limited few cases. We will explore this issue in Chapter \ref{chapter:concentration}.

Throughout the analysis in this chapter, we used DMO simulations. The method to rescale the $\Lambda$CDM cluster scaling relations to get scaling relations that apply to MG \citep{He:2015mva}, has been tested and validated using non-radiative hydrodynamical simulations. In Sec.~\ref{framework} we argued that adding the full baryonic physics in the simulations will not substantially change the conclusion, based on previous work on $\Lambda$CDM full physics simulations. We will check this using full hydrodynamical simulations for HS $f(R)$ gravity in Chapter \ref{chapter:scaling_relations}.

Finally, we note again that a key ingredient of any test of gravity using the cluster abundance is the ability to predict the HMF for arbitrary model parameters. In this thesis, we have used the recently-developed HMF fitting formula by \citet{Cataneo:2016iav}, which was calibrated using a subset of simulations (Crystal) used in this chapter. This formula has 3-5\% accuracy for a range of $f_{R0}$ values between F4 and F6 and for halo masses above $10^{13}h^{-1}M_\odot$, making it ideal for comparing with observed cluster abundances. A full hydrodynamical simulation can also be useful in understanding how the predicted abundance of dark matter haloes can change with the inclusion of baryonic physics.

We will test the above-mentioned framework, which incorporates these effects into model predictions and allows for detailed MCMC searches of the parameter space, in Chapter \ref{chapter:constraint_pipeline}. Here, the fitting function for $M_{\rm dyn}/M_{\rm true}$ is also useful for constructing mock observational data that are used to validate the MCMC model constraint pipelines.

\graphicspath{{./gfx/}}

\chapter{\boldmath A universal model for the halo concentration in \texorpdfstring{$f(R)$}{f(R)} gravity}
\label{chapter:concentration}

\section{Introduction}
\label{sec:introduction}

In Chapter \ref{chapter:mdyn}, we presented a general analytical model for the enhancement of the dynamical mass. This is an important component of our framework for $f(R)$ gravity constraints using cluster number counts (Fig.~\ref{fig:fr_flow_chart}): it is used to predict the dynamical masses of dark matter haloes in mock catalogues and to convert cluster observable-mass scaling relations that have been calibrated in $\Lambda$CDM into a form that works in $f(R)$ gravity. Another important aspect of our framework is the ability to make conversions between halo masses corresponding to different overdensities $\Delta$. For example, in order to constrain $f(R)$ gravity using the cluster abundance, we require a model-dependent calibration of the HMF, which quantifies the number density of dark matter haloes per unit mass interval, ${\rm d}n_{\rm halo}/{\rm d}M_{\Delta}$. Our choice of HMF \citep{Cataneo:2016iav} has been calibrated for overdensity $\Delta=300\Omega_{\rm M}(z)$, while overdensity $\Delta=500$ is more generally used in cluster surveys. Therefore in order to make constraints using observational data that has been calibrated for overdensity $\Delta=500$, it will be necessary for us to apply the conversion $M_{\rm 300m} \rightarrow M_{500}$ to the HMF. It is also likely that conversions to other overdensities will be required. For example $\Delta=2500$ is also sometimes used in observational surveys, and $\Delta=200$ is often used in theoretical studies. Therefore, a prediction for the conversion between halo masses corresponding to arbitrary values of $\Delta$ is essential. 

The halo mass measured for different overdensities corresponds to the total mass enclosed by different halo radii $R_{\Delta}$, where $R_{\Delta}$ is larger for lower values of $\Delta$. Therefore conversions between the halo mass at different overdensities can be estimated if the density profile of dark matter haloes can be predicted. Typically the universal NFW density profile \citep{NFW} is assumed. This is a 2-parameter profile, but can be written with one parameter if the mass (or radius) for a particular overdensity is known. This parameter can be the halo concentration, and predicting it in $\Lambda$CDM as a function of the cluster mass and redshift has been the subject of much work over the two decades since it was first introduced \citep[e.g.,][]{Bullock:1999he,Neto:2007vq,Duffy:2008pz,Dutton:2014xda,Ludlow:2013vxa}.

In addition to facilitating mass conversions the concentration is also important in studies of the non-linear matter power spectrum \citep[e.g.,][]{Brax:2013fna,Lombriser:2013eza,2016PhRvD..93j3522A,Hu:2017aei,2018arXiv181205594C}, which, like the cluster abundance, can also be used to probe dark energy and MG theories. The large-scale part of the matter power spectrum can, for example, be predicted using linear perturbation theory by incorporating the linear halo bias. On the other hand, the small-scale part of the matter power spectrum can be assembled using the HMF and the halo concentration. The concentration is necessary in order to predict the density profile, which is required, for example, in order to model the size of haloes.

In $f(R)$ gravity, the concentration can become enhanced due to the effects of the fifth force on the density profile. For example, for an unscreened halo the in-falling particles experience a greater acceleration due to the stronger gravitational force, and this can alter the profile such that the density is raised at the inner regions and lowered at the outer regions. Therefore, the $\Lambda$CDM predictions of the concentration are unlikely to apply for lower-mass unscreened haloes in $f(R)$ gravity. Yet there is no general quantitative model for the concentration in $f(R)$ gravity that is discussed in the literature, which instead tends to focus on a more qualitative understanding of the effects of the fifth force on the concentration and on the density profile \citep[e.g.,][]{Zhao:2010qy,thin_shell,Shi:2015aya,arnold:2016,Arnold:2018nmv}. Therefore, a $\Lambda$CDM relation for the concentration is often used in the literature. For example, the modellings of the non-linear matter power spectrum in HS $f(R)$ gravity by \citet{Brax:2013fna,Hu:2017aei,2018arXiv181205594C} use prescriptions for the concentration-mass relation that have been calibrated in $\Lambda$CDM. Also, due to the large scatter of the concentration-mass relation, some works \citep[e.g.,][]{PhysRevD.92.044009} argue that it is fine to assume a fixed concentration for a sample of clusters that covers a sufficiently narrow mass range. 

In order to prevent potential biases resulting from a simplified treatment of the concentration, the focus of this chapter is to produce a general model for the concentration in HS $f(R)$ gravity (blue dotted box of Fig.~\ref{fig:fr_flow_chart}) that may be applied in future studies. Rather than calibrating a relation for the absolute concentration, we decided to focus on finding a universal model for the enhancement of the concentration as a function of the halo mass and redshift. This has been achieved using data from a suite of DMO $N$-body simulations run for three models of HS $f(R)$ gravity. Note that we define the enhancement as the ratio of the $f(R)$ concentration to the concentration in GR. This means that one can select a $\Lambda$CDM concentration-mass-redshift relation from the literature \citep[e.g.,][]{Bullock:1999he,Neto:2007vq,Duffy:2008pz,Dutton:2014xda,Ludlow:2013vxa} that they wish to use, then this can be converted into a form in HS $f(R)$ gravity. Our model includes a dependence on the cosmological density parameters $\Omega_{\rm M}$ and $\Omega_{\Lambda}$, so any $\Lambda$CDM relation can be used regardless of the values of these parameters. Our model depends on the particular combination $\bar{f_R}(z)/(1+z)$ where $\bar{f_R}(z)$ is the background scalar field at redshift $z$, and does not explicitly depend on the model parameter $f_{R0}$, i.e., the present-day background scalar field value, as one would naively expect. This has the implication that predictions may be made for arbitrary values of $f_{R0}$ and $z$ (as long as the above combination is within the range of validity of our fitting). This generality of our model was achieved by combining data from the different $f(R)$ gravity models by applying a simple transformation to the halo mass using the $p_2$ parameter defined in Chapter \ref{chapter:mdyn}, where $M_{500}=10^{p_2}h^{-1}M_{\odot}$ can be considered as the mass above which haloes are screened and below which haloes are unscreened.

This chapter is arranged as follows: Sec.~\ref{sec:simulations_and_methods} provides an overview of the DMO $N$-body simulations that are used in this chapter, along with an outline of the methods used to measure the concentration and its enhancement; Sec.~\ref{sec:results} discusses the results of this chapter, including the general model for the concentration enhancement; and, finally, Sec.~\ref{sec:conclusions} summarises the main conclusions from this chapter, and outlines the next steps of our framework.

\section{Simulations and methods}
\label{sec:simulations_and_methods}

The simulations that we have used to study the concentration are presented in Sec.~\ref{sec:simulations}, along with the methods that we use to extract their halo data. Our methods to measure the concentration and its enhancement and a useful technique of rescaling the halo mass are discussed in Sec.~\ref{sec:methods}.

\subsection{Simulations}
\label{sec:simulations}

\begin{table*}
\centering

\small
\begin{tabular}{ ccccc } 
 \toprule
 
 Parameters and & \multicolumn{4}{c}{Simulations} \\
 models & Diamond & Jade & Crystal & \textsc{arepo} \\

 \midrule

 box size / $h^{-1}$Mpc & 64 & 450 & 1024 & 62 \\ 
 particle number & $512^3$ & $1024^3$ & $1024^3$ & $512^3$ \\ 
 particle mass / $h^{-1}M_{\odot}$ & $1.52\times10^8$ & $6.64\times10^9$ & $7.78\times10^{10}$ & $1.52\times10^8$ \\
 & & & & \\
 $\Omega_{\rm M}$ & 0.281 & 0.2819 & 0.281 & 0.3089 \\ 
 $\Omega_{\Lambda}=1-\Omega_{\rm M}$ & 0.719 & 0.7181 & 0.719 & 0.6911 \\
 $h$ & 0.697 & 0.697 & 0.697 & 0.6774 \\
 $f(R)$ models & F6 & F5 & F4, F5, F6 & F4, F5, F6\\
 
 \bottomrule
 
\end{tabular}

\caption[Specifications of the \textsc{ecosmog} and \textsc{arepo} simulation used to study the $f(R)$ halo concentration.]{Specifications of the three \textsc{ecosmog} simulations and the \textsc{arepo} simulation used in this investigation. The \textsc{ecosmog} simulations are labelled Diamond, Jade and Crystal for convenience. All simulations have been run for $\Lambda$CDM in addition to the HS $f(R)$ gravity models listed. The Hubble constant, $H_0$, is equal to $100h$ kms$^{-1}$Mpc$^{-1}$ for each simulation.}
\label{table:simulations:concentration}

\end{table*}

Our DMO simulations are shown in Table \ref{table:simulations:concentration}. Three of these --- the Crystal, Jade and Diamond simulations --- were also used in Chapter \ref{chapter:mdyn}. We also include a simulation that has been run using the \textsc{arepo} code \citep{2010MNRAS.401..791S} with its MG solver. 

The cosmological parameters and technical specifications of the simulations are listed in Table~\ref{table:simulations:concentration}. The \textsc{arepo} simulation, which is the DM-only subset of the \textsc{shybone} simulation suite \citep{Arnold:2019vpg}, has the same mass resolution as the Diamond simulation, with particle mass $1.52\times10^{8}h^{-1}M_{\odot}$, as well as a similar box size of $62h^{-1}{\rm Mpc}$. Together, the four simulations cover a wide range of halo masses, which is essential in order to comprehensively study the halo concentration across the full transition between the screened and the unscreened regimes. The \textsc{arepo} data is particularly useful because it has been run for all three $f(R)$ gravity models examined in this chapter. Its low particle mass allows low-mass, unscreened haloes to be studied in all three models, ensuring a more detailed exploration of the transition between the screened and unscreened regimes. In addition to this, the similar resolutions of the \textsc{arepo} simulation and Diamond allow a consistency test of the \textsc{ecosmog} and \textsc{arepo} simulations, which is necessary due to the potential disparities between the results from these two codes which employ different algorithms and assume different cosmological parameters.

The simulation data covers redshifts up to at least $z=1$ for all simulations. However redshifts $z<2$ and $z<3$ have been included for \textsc{arepo} F5 and F4, respectively, as otherwise the data from these models would only cover the unscreened regime. The dataset consists of 19 snapshots from Crystal, 33 from Jade and 44 from Diamond. For \textsc{arepo} there are 46, 37 and 24 snapshots from F4, F5 and F6, respectively.

The halo catalogues that we construct consist of dark matter haloes identified using the \textsc{subfind} code \citep{springel2001} implemented in \textsc{arepo}. This employs a standard friends-of-friends (FOF) algorithm to identify FOF groups (haloes) and a gravitational un-binding method to locate the bound substructures (subhaloes) within each group. The mass and radii of the haloes have been measured for overdensities $\Delta=500$ and $200$ with respect to the critical density of the Universe. Two methods that can be used to calculate the halo concentration (see Sec.~\ref{sec:c_measurement}) also require measurements of the maximum circular velocity, $V_{\rm max}$, and the corresponding orbital radius, $R_{\rm max}$. The \textsc{subfind} code calculates these quantities using just the bound particles of the central, dominant subhalo. To a good approximation, these measurements can be used to represent the maximum circular velocity and the corresponding orbital radius for the entire FOF group. We have checked that within $R_{\rm max}$ almost all particles are bound regardless of whether the fifth force is felt or not.

In order to accurately measure the halo concentration, it is important that the halo consists of enough particles so that it is well-resolved at both the inner and outer regions. Therefore in all of our analyses in this chapter we only use haloes with more than 1000 particles contained within $R_{500}$. This corresponds to minimum halo masses of $M_{500}=(1.52\times10^{11}$,$6.64\times10^{12}$,$7.78\times10^{13}$,$1.52\times10^{11})h^{-1}M_{\odot}$ for Diamond, Jade, Crystal and \textsc{arepo} respectively.

We bin our haloes by $M_{500}$ (see Sec.~\ref{sec:c_enhancement_measurement}), and our model for the enhancement of the concentration is designed to predict the concentration in $f(R)$ gravity as a function of $M_{500}$ (see Sec.~\ref{sec:general_model}). The reason for choosing overdensity $\Delta=500$ is to be consistent with Chapter \ref{chapter:mdyn}, in which $M_{500}$ was used to study the enhancement of the dynamical mass and, crucially, to define the parameter $p_2$. As will be discussed later (Sec.~\ref{sec:rescaled_mass}), the halo mass can be rescaled by this parameter in order to combine data from snapshots with different values of $|\bar{f}_R|/(1+z)$.

Various works in the literature which study the halo evolution and aim to model the concentration as a function of redshift and mass often use relaxed samples \citep[e.g.,][]{Neto:2007vq}. For example, haloes which have undergone recent mergers are unlikely to give reliable estimates of the concentration. However, we have decided to include all haloes (that satisfy the above mass criteria) since one of the applications of our results will include matter power spectrum predictions, which requires the use of all haloes.

\subsection{Methods}
\label{sec:methods}

In this section we present the three approaches for measuring the halo concentration that are used in this chapter (Sec.~\ref{sec:c_measurement}), a useful rescaling of the halo mass (Sec.~\ref{sec:rescaled_mass}) and our method of binning the concentration and evaluating its enhancement (Sec.~\ref{sec:c_enhancement_measurement}).

\subsubsection{Concentration measurement}
\label{sec:c_measurement}

Three methods were considered for the measurement of the halo concentration. The most accurate is to directly fit the profiles of the haloes using the NFW profile \citep{NFW}:
\begin{equation}
\rho(r) = \frac{\rho_{\rm s}}{(r/R_{\rm s})(1+r/R_{\rm s})^2},
\label{eq:nfw}
\end{equation}
where $\rho_{\rm s}$ is the characteristic density and $R_{\rm s}$ is the scale radius. The concentration is defined as $c_{200}=R_{200}/R_{\rm s}$. Note that the convention to define the concentration with respect to overdensity 200 is frequently used by the literature. Therefore we elected to use this definition, even though our haloes are required to be binned by $M_{500}$ (see above). This approach is still consistent because as long as the concentration can be predicted for one overdensity, it can also be predicted for overdensity 500.

Radial bins that are equally spaced in logarithmic distance from the halo centre were used to ensure that both the inner and outer regions were equally well-fitted. As shown by \cite{Neto:2007vq}, the choice of radial range over which the profile is fitted can be important. Resolution effects can occur at the outer less-dense regions of the halo or the innermost regions where, due to the limited number of particles, the density can be underestimated. In order to avoid these effects, we chose to calculate the densities of 20 logarithmic radial bins, spanning distances $0.05R_{200}$ to $R_{200}$ from the halo centre, which is consistent with the range used by \cite{Neto:2007vq}. These densities were fitted using the formula,
\begin{equation}
\log_{10}(\rho) = \log_{10}(\rho_{\rm s})-\log_{10}(xc_{200})-2\log_{10}(1+xc_{200}),
\label{eq:nfw_fit}
\end{equation}
where $x=r/R_{200}$. This was achieved via unweighted least squares, where $\rho_{\rm s}$ and $c_{200}$ are allowed to vary independently. The halo concentration was set equal to the optimal value of $c_{200}$.

The concentration was originally defined by \cite{NFW} as a parameter of the NFW profile. Therefore, fitting this profile to individual haloes is the only means of accurately measuring the concentration in a way that is true to its definition\footnote{We note that for the full NFW fitting one can choose to fit the mass profiles instead of the density profiles of haloes \citep[e.g.,][]{2013ApJ...768..123K}. We consider both cases as full NFW fitting, because they make use of the whole range of halo radius (neglecting certain regions excluded from the fitting).}. Even for an unscreened halo in $f(R)$ gravity, where the density profile can in principle deviate from an NFW profile, the concentration should still be measured in the same way. However there are a number of simplified methods that have been adopted in the literature, including those presented by \cite{Prada:2011jf} (P12) and \cite{Springel:2008cc} (S08), which we also consider in this chapter. Both methods assume that the halo is well-characterised by the NFW profile without performing a direct fitting. This allows the concentration to be predicted with more limited information, which can save time. 

The P12 method uses the relation between $V_{\rm max}$ and the circular velocity at the halo radius $R_{200}$:
\begin{equation}
V_{200} = \left(\frac{GM_{200}}{R_{200}}\right)^{1/2}.
\label{eq:circular_velocity}
\end{equation}
For the NFW profile the ratio $V_{\rm max}/V_{200}$ is directly related to the halo concentration, $c_{200}$, by,
\begin{equation}
\frac{V_{\rm max}}{V_{200}} = \left(\frac{0.216c_{200}}{f(c_{200})}\right)^{1/2},
\label{eq:prada}
\end{equation}
where the function $f(c)$ is given by the following:
\begin{equation}
f(c) = \ln(1+c)-\frac{c}{(1+c)}.
\label{eq:c_function}
\end{equation}
Using only the measurements of $V_{\rm max}$ and $R_{200}$ (or $M_{200}$), Eqs.~(\ref{eq:circular_velocity})-(\ref{eq:c_function}) can be combined and solved numerically to estimate the concentration.

One challenge of this method that we encountered was that for some haloes in our sample, $V_{200}>V_{\rm max}$. Our estimate of $V_{\rm max}$ includes only particles bound to the central, dominant subhalo of the FOF group, while $V_{200}$ has been calculated using all particles, bound or unbound, contained within $R_{200}$. Therefore it is possible that some large substructure found in the outer regions of the halo or some additional unbound particles flying past the halo could result in this inequality. We did not include P12 measurements of the concentration for haloes with $V_{200}>V_{\rm max}$.

Meanwhile the S08 method involves measuring the mean overdensity within $R_{\rm max}$ in units of the present-day critical density, $\rho_{\rm crit}$:
\begin{equation}
\delta_{\rm V} = \frac{\bar{\rho}(R_{\rm max})}{\rho_{\rm crit}} = 2\left(\frac{V_{\rm max}}{H_0R_{\rm max}}\right)^2.
\label{eq:rmax_density}
\end{equation}
The characteristic NFW overdensity, $\delta_{\rm c}$, is then calculated using,
\begin{equation}
\delta_{\rm c} = \frac{\rho_{\rm s}}{\rho_{\rm crit}} = 7.213\delta_{\rm V},
\label{eq:characteristic_overdensity}
\end{equation}
which is related to $c_{200}$ as follows:
\begin{equation}
\delta_{\rm c} = \frac{200}{3}\frac{c_{200}^3}{f(c_{200})}.
\label{eq:c_calculation}
\end{equation}
By starting with a measurement of $V_{\rm max}$ (or $R_{\rm max}$), the concentration can be estimated by combining Eqs.~(\ref{eq:rmax_density})-(\ref{eq:c_calculation}). Note that this method can be used for all haloes in our sample, including haloes with $V_{200}>V_{\rm max}$.

The main distinction between these two simplified methods is that the S08 method can give a more stable result which is independent of deviations from an NFW profile. This can be useful for simulations of a low resolution, since the S08 method is only sensitive to density changes around $R_{\rm max}$ and can therefore reproduce the density profile without too much noise. On the other hand, the P12 method uses information at $R_{200}$ as well as $R_{\rm max}$, so if the density profile deviates from an NFW profile outside $R_{\rm max}$ then this method can pick up this effect and yield a different result. So the P12 concentration estimate is expected to be more likely to agree with a measurement from a full fit of the NFW profile.

Because the density profile is slightly changed by $f(R)$ gravity for unscreened haloes, the differences between the methods will prove to be important for the results (see Sec.~\ref{sec:measurement_comparison}).

\subsubsection{Rescaled mass}
\label{sec:rescaled_mass}

A key aim of this chapter is to be able to predict how the halo concentration is affected in the screened and unscreened regimes of $f(R)$ gravity. The mass of the transition between these two regimes can be predicted using the $p_2$ parameter defined in Chapter \ref{chapter:mdyn}. Therefore it is useful to rescale the mass by the transformation $\log_{10}(M_{500}M_{\odot}^{-1}h) \rightarrow \log_{10}(M_{500}M_{\odot}^{-1}h)-p_2\equiv\log_{10}(M_{500}/10^{p_2})$, such that negative values correspond to the unscreened regime and positive values correspond to the screened regime. It is essential to use a halo mass overdensity of 500 here, as discussed in Sec.~\ref{sec:simulations}. Note that $p_2$ can be measured using Eq.~(\ref{eq:p2}), together with Eq.~(\ref{eq:fR_background}) which is used to evaluate the background scalar field.

\begin{figure*}
\centering
\includegraphics[width=1.0\textwidth]{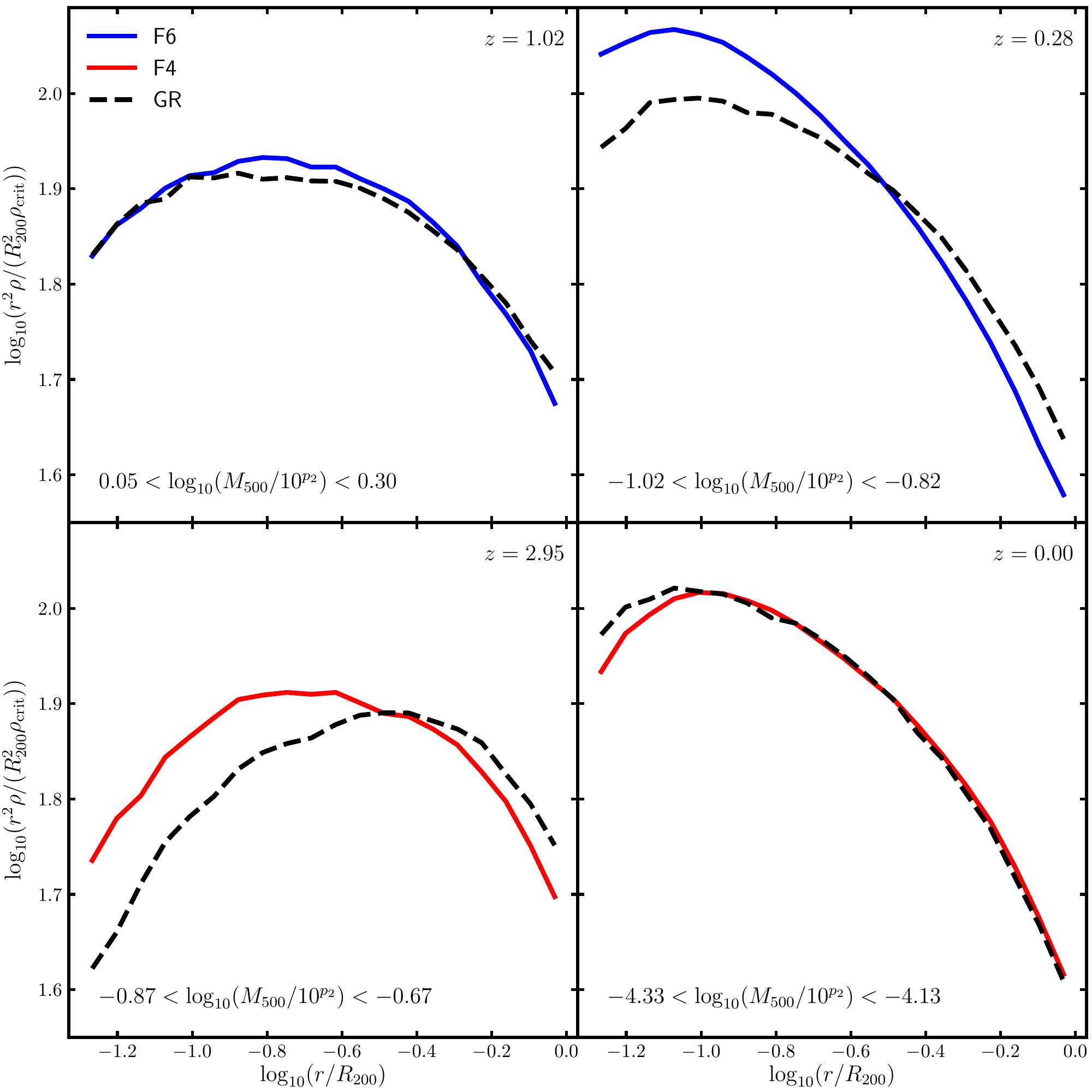}
\caption[Median density profiles of simulated dark matter haloes within four bins of rescaled mass $log_{10}(M_{500}/10^{p_2})$.]{Stacked density profile, scaled by $r^2$, measured using the median density profile of the haloes found within a particular mass bin. The data is from the \textsc{arepo} simulation (see Table \ref{table:simulations:concentration}), run for Hu-Sawicki $f(R)$ gravity with $n=1$ (\textit{solid line}) and GR (\textit{dashed line}). Each panel corresponds to a particular value of the present-day scalar field $f_{R0}$, where F6 and F4 correspond to $|f_{R0}|=10^{-6}$ and $|f_{R0}|=10^{-4}$, respectively, and a particular redshift $z$. The mass bins used cover the same range of halo masses, $M_{500}$, for the $f(R)$ and GR data, and each panel corresponds to a unique range of $\log_{10}(M_{500}/10^{p_2})$ values (annotated), where $p_2$ is evaluated with Eq.~(\ref{eq:p2}) using the values $|f_{R0}|$ and $z$ of that panel.}
\label{fig:stacked_profiles}
\end{figure*}

We motivate the rescaling approach in Fig.~\ref{fig:stacked_profiles}, which shows the stacked density profiles of the $f(R)$ haloes for four bins of $\log_{10}(M_{500}/10^{p_2})$, shown across four panels. The GR data for the same redshifts is also shown in each panel for a comparison. The density profile has been scaled by $r^{2}$ such that it peaks at distance $r=R_{\rm s}$ from the halo centre. This allows the concentration to be easily read off from the peak of the data, given that this is equal to $R_{200}/R_{\rm s}$.

The $f(R)$ data in the panel at the top-left corresponds to haloes that are partially screened. This is because the values of $\log_{10}(M_{500}/10^{p_2})$ used are just slightly above zero, and because the transition from screened to unscreened is smooth and gradual these haloes are therefore unscreened in the outer, less-dense regions but still screened in the inner, denser regions. This means that the particles falling towards the intermediate regions from the outer regions of the halo feel a stronger gravitational pull. However, the particle motions in the innermost regions of the halo are not affected by the fifth force, since this region is still screened. Therefore in $f(R)$ gravity the density at the outer regions is lower and the density at the intermediate regions is greater than in GR, but the density at the innermost regions is unaffected. This can actually lead to a larger scale radius $R_{\rm s}$ in $f(R)$ gravity than in GR, resulting in a lower value of the concentration. However the deviation between the $f(R)$ and GR density profiles in this regime of $\log_{10}(M_{500}/10^{p_2})$ is still quite small.

The top-right and bottom-left panels show regimes in which the entire halo has become unscreened. It is likely that these haloes have only recently entered the unscreened regime, since the values of $\log_{10}(M_{500}/10^{p_2})$ used are negative but still quite close to zero. Because the entire halo is now unscreened, the particles within both the inner and outer regions of the halo feel a stronger pull of gravity and thus fall towards the halo centre with a greater acceleration. Therefore the density is greater at the inner regions and lower at the outer regions than in GR. This results in a scale radius $R_{\rm s}$ that is smaller in $f(R)$ gravity, and so the concentration is greater than in GR. This regime of $\log_{10}(M_{500}/10^{p_2})$ has the greatest deviation between the $f(R)$ and GR density profiles.

Finally, the bottom-right panel shows data that is deep within the unscreened regime. This is because the values of $\log_{10}(M_{500}/10^{p_2})$ used are negative and much lower than the values spanned by the top-right and bottom-left panels. Therefore the haloes in this bin are likely to have been unscreened for a significant period of time. Interestingly, the density profiles in GR and $f(R)$ gravity are in reasonable agreement in all regions apart from the innermost region, in which the GR haloes are more dense. One possibility is that in $f(R)$ gravity the particles that initially fall into the halo centre gain a higher velocity during this substantial period of enhanced gravitational acceleration \citep{Shi:2015aya}, such that they are unable to settle into orbits at the innermost regions. As a result the scale radius is larger in $f(R)$ gravity than in GR, such that the concentration is greater in GR.

From these results, it is clear that the variable $\log_{10}(M_{500}/10^{p_2})$ can be a useful measure of the amount of screening of a halo, and so the model used to predict the enhancement of the concentration (Sec.~\ref{sec:general_model}) has been measured with respect to this variable. Plotting all results as a function of this variable also allows the combination of data with different values of $f_{R0}$ and $z$, since these are encapsulated by $p_2$ (see Sec.~\ref{sec:results}). 

\subsubsection{Concentration enhancement}
\label{sec:c_enhancement_measurement}

The GR data of each simulation has been outputted at snapshots with the same redshifts as the $f(R)$ data. Therefore the concentration enhancement of the haloes in a particular mass bin can be evaluated by first computing the median concentrations using the $f(R)$ and GR data from that bin, and then taking a ratio of these quantities. The choice of binning, the measurement of the median concentration and its error, and the evaluation of the concentration enhancement and its error is discussed in this section.

In order to evaluate the concentration enhancement for the haloes of a particular snapshot, the absolute concentrations of all haloes in $f(R)$ gravity and GR were measured using the methods discussed in Sec.~\ref{sec:c_measurement}. The haloes in each model were then binned by the halo mass $M_{500}$, with the binning chosen such that all bins are of equal width when viewed on a logarithmic axis apart from the highest-mass bin, which was allowed to be wide enough to contain at least 75 haloes in both $f(R)$ gravity and GR. For a given snapshot, the same bins are used for both the GR and $f(R)$ gravity datasets, and the number of bins used is the greatest possible number such that each bin contains at least 75 haloes for both models.

The decision to use a wider highest-mass bin was taken due to the much smaller halo count at higher masses. The choice of having at least 75 haloes in each bin ensured that a balance was found between having enough bins with which to fit a reliable trend and keeping the errors low for each bin. This decision followed some tests which found that a minimum of 50 haloes per bin generates bins with a scatter that is too large, while a minimum of 100 haloes per bin results in very few bins, particularly in the \textsc{arepo} and Diamond data which do not contain as many haloes as the other simulations.

In each bin, the mean logarithm of the halo mass and the median concentration was measured for both $f(R)$ gravity and GR. The average of the two mean mass logarithms has been used to represent the mass of each bin. The error of the median concentration was evaluated using jackknife resampling (see Chapter~\ref{chapter:mdyn}), with 20 resamples. 


\begin{figure*}
\centering
\includegraphics[width=0.90\textwidth]{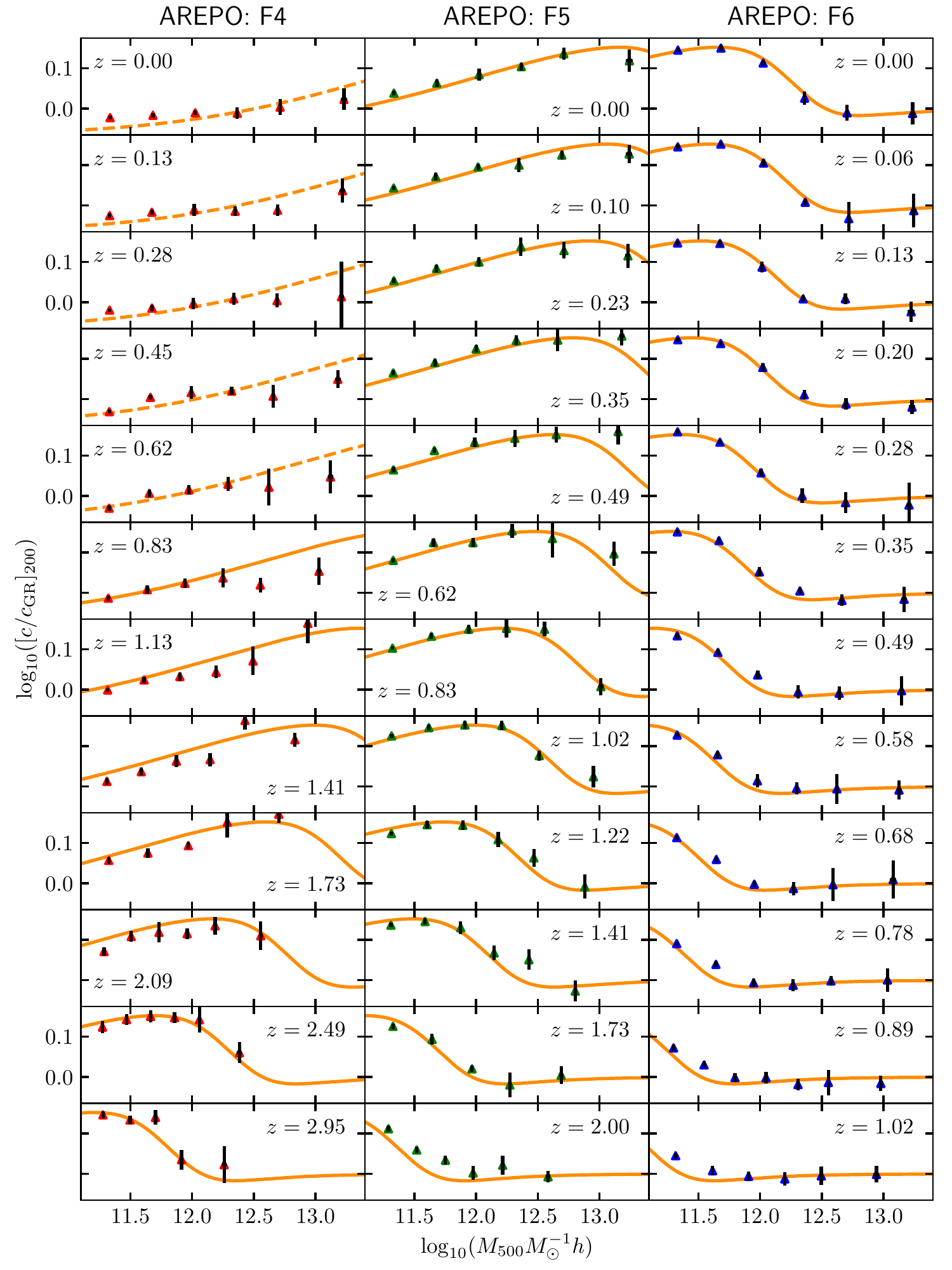}
\caption[Enhancement of the halo concentration in three $f(R)$ models as a function of the halo mass, generated using \textsc{arepo} data at various redshifts.]{Ratio of the median concentrations of $f(R)$ gravity and GR as a function of the halo mass, for F4 (\textit{left column}), F5 (\textit{middle column}) and F6 (\textit{right column}) at various redshifts, $z$, as annotated. The parameter $p_2$ is evaluated using the values $|f_{R0}|$ and $z$ of each panel. The data has been generated from the \textsc{arepo} simulation (see Table \ref{table:simulations:concentration}). The one standard deviation error bars are shown. Predictions have been plotted (\textit{solid line}) for most snapshots, and are measured using the fit of Eq.~(\ref{eq:skewtanh}) to the data of Fig.~\ref{fig:skewtanh_fit}. In panels corresponding to data excluded from the fit, the predictions are shown with dashed lines.}
\label{fig:arepo_matrix}
\end{figure*}



\begin{figure*}
\centering
\includegraphics[width=0.90\textwidth]{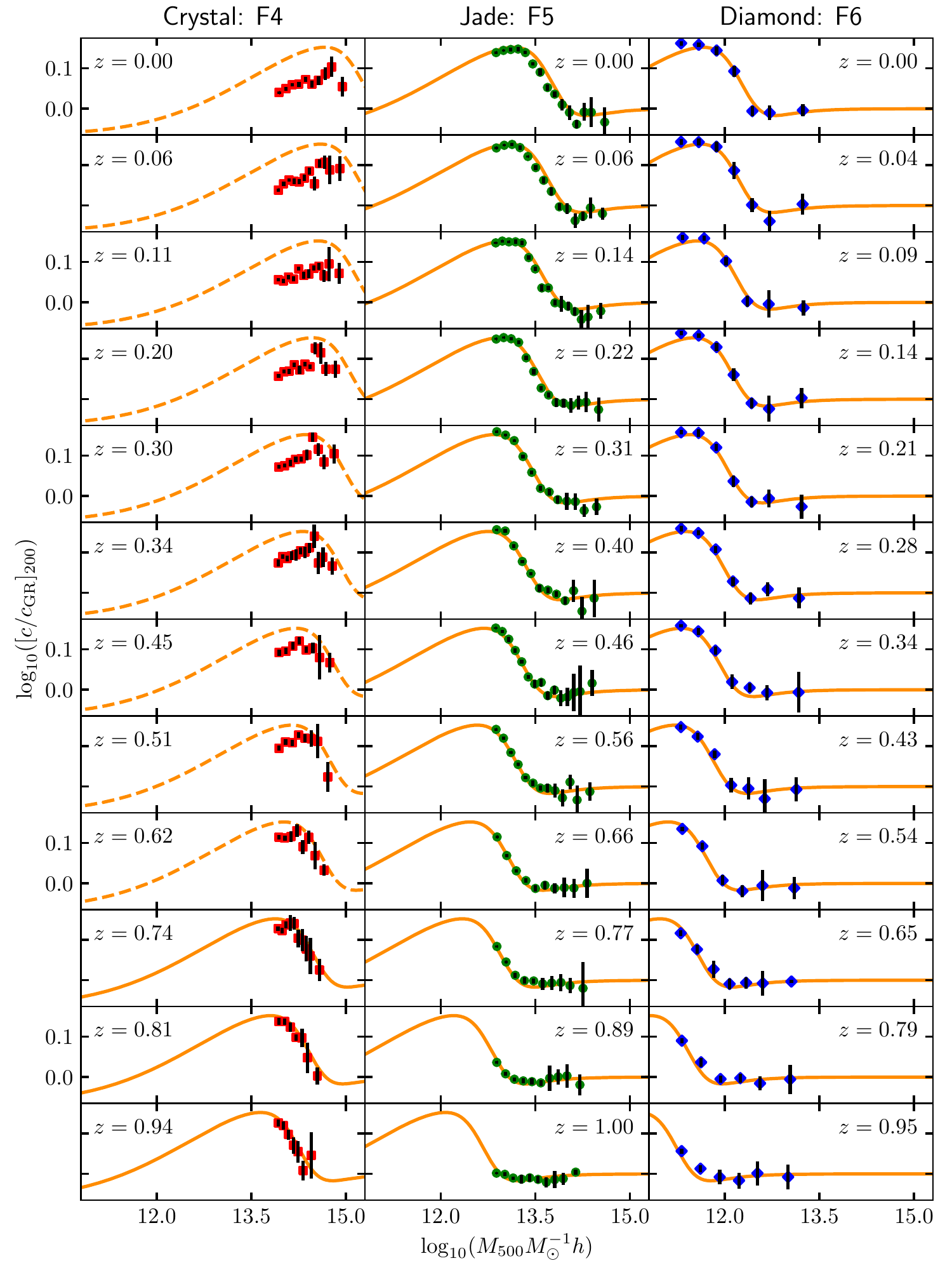}
\caption[Enhancement of the halo concentration in three $f(R)$ models as a function of the halo mass, generated using \textsc{ecosmog} data at various redshifts.]{Ratio of the median concentrations of $f(R)$ gravity and GR as a function of the halo mass, for F4 (\textit{left column}), F5 (\textit{middle column}) and F6 (\textit{right column}) at various redshifts, $z$, as annotated. The parameter $p_2$ is evaluated using the values $|f_{R0}|$ and $z$ of each panel. The data has been generated using the Crystal, Jade and Diamond \textsc{ecosmog} simulations (see Table \ref{table:simulations:concentration}), for F4, F5 and F6, respectively. The one standard deviation error bars are shown. Predictions have been plotted (\textit{solid lines}) for most snapshots, and are measured using the fit of Eq.~(\ref{eq:skewtanh}) to the data of Fig.~\ref{fig:skewtanh_fit}. In panels corresponding to data excluded from the fit, the predictions are shown with dashed lines.}
\label{fig:ecosmog_matrix}
\end{figure*}


Finally, the logarithm of the ratio of the median concentration values in $f(R)$ gravity and GR, $\log_{10}\left(\left[c/c_{\rm GR}\right]_{200}\right)$, was evaluated to obtain the concentration enhancement. The error of the enhancement was measured by combining the median concentration errors in quadrature to find the error of the ratio, then the error of the logarithm of this ratio. Treating the errors as independent is justified here because the particle positions in collapsed structures are uncorrelated in these two simulations even though they start from the same initial conditions. This data is shown by the symbols in Figs.~\ref{fig:arepo_matrix} and \ref{fig:ecosmog_matrix} for an arbitrary selection of snapshots and models for each simulation. Note that the lines plotted in each panel correspond to predictions from our general model, which is discussed in Sec.~\ref{sec:general_model}. For each column the snapshots that are shown span the full range of available redshifts (see Table \ref{table:simulations:concentration}). For each bin the concentration enhancement and its error bar has been measured using the methods discussed above. The use of a wider highest-mass bin in each snapshot, as discussed above, can also be seen in each panel. Note that the same GR data is used when measuring the concentration enhancement for different $f(R)$ gravity models of the same simulation, although the binning scheme that is used may vary slightly.

\section{Results}
\label{sec:results}

The results shown in Figs.~\ref{fig:arepo_matrix} and \ref{fig:ecosmog_matrix} are plotted against the halo mass $M_{500}$. However, as discussed in Sec.~\ref{sec:rescaled_mass}, performing the transformation $\log_{10}(M_{500}M_{\odot}^{-1}h) \rightarrow \log_{10}(M_{500}/10^{p_2})$ converts the mass into a rescaled form in which negative values roughly correspond to the unscreened regime of halo mass and positive values roughly correspond to the screened regime. One advantage of this mass rescaling is that the $f(R)$ gravity model, redshift and cosmological parameters $\Omega_{\rm M}$ and $\Omega_{\Lambda}$ are all encapsulated by the $p_2$ parameter (Eq.~\ref{eq:p2}), so plotting against this rescaled mass can allow data from different $f(R)$ gravity models and even from simulations run for different cosmologies to be combined and plotted together in order to extract general trends. Note that there is a unique $p_2$ value for every combination of $z$, $f_{R0}$, $\Omega_{\rm M}$ and $\Omega_{\rm \Lambda}$, so, for example, all of the data in a particular panel of Fig.~\ref{fig:arepo_matrix} or \ref{fig:ecosmog_matrix} would be shifted by the same amount along the $\log_{10}(M_{500}M_{\odot}^{-1}h)$ axis when applying the transformation to the mass. 

We first used this plotting scheme to look into the three methods for measuring the concentration (see Sec.~\ref{sec:c_measurement}) to see how the choice of measurement can affect the results. This is discussed in Sec.~\ref{sec:measurement_comparison}. Then, focusing on the data produced from performing direct fitting of the NFW profile to individual haloes, a general model to describe the enhancement of the halo concentration in $f(R)$ gravity was sought and is discussed in Sec.~\ref{sec:general_model}.

\subsection{Concentration measurement comparison}
\label{sec:measurement_comparison}

\begin{figure*}
\centering
\includegraphics[width=1.0\textwidth]{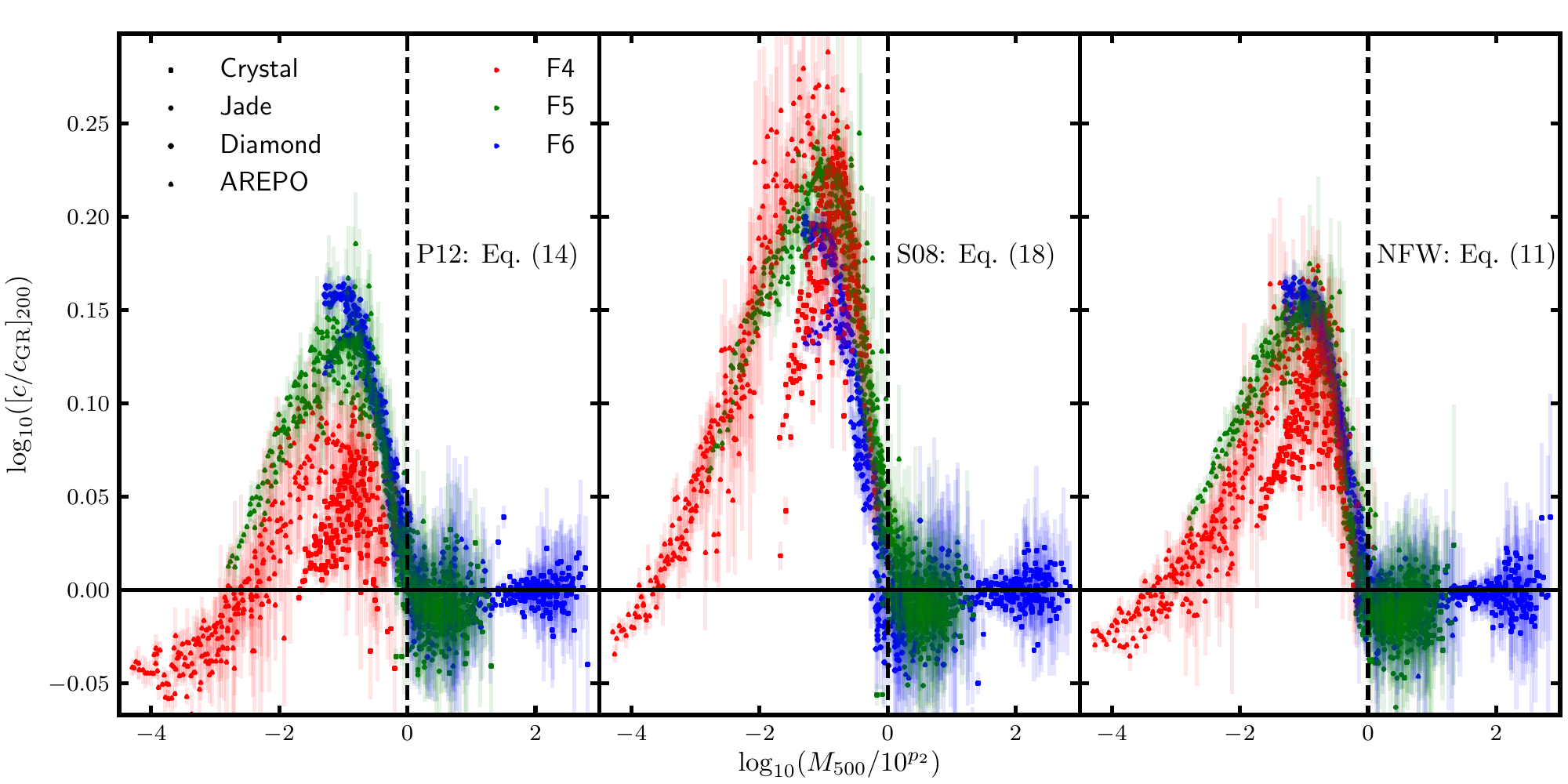}
\caption[Ratio of the median concentrations in $f(R)$ gravity and GR, computed using three different methods, as a function of the rescaled mass $\log_{10}(M_{500}/10^{p_2})$.]{Ratio of the median concentrations of $f(R)$ gravity and GR as a function of the rescaled mass, $\log_{10}(M_{500}/10^{p_2})$, for F4 (\textit{red}), F5 (\textit{green}) and F6 (\textit{blue}). The plotted data is from the simulations summarised by Table \ref{table:simulations:concentration}, sifted so that only haloes with more than 1000 particles enclosed within $R_{500}$ are included. The concentration of each halo has been calculated using the methods discussed in Sec.~\ref{sec:c_measurement}, namely the methods discussed by P12 (\textit{left}) and S08 (\textit{middle}) and by performing a direct fitting of the NFW profile to the halo density profiles (\textit{right}). The one standard deviation error bars are shown.}
\label{fig:3_panel}
\end{figure*}

The concentration of each halo was measured using each of the three methods presented in Sec.~\ref{sec:c_measurement}. For each of these measures, the data was binned for every snapshot of each $f(R)$ model and the concentration enhancement and its error was measured for each bin using the method discussed in Sec.~\ref{sec:c_enhancement_measurement}. All of this data was plotted together against $\log_{10}(M_{500}/10^{p_2})$ to yield Fig.~\ref{fig:3_panel}, in which the three panels correspond to the three methods of measuring $c_{200}$.

The data from each panel follows a similar general trend. There is approximately zero enhancement of the concentration in the screened regime, then at lower mass (entering the unscreened regime) the enhancement rises to a peak at $\log_{10}(M_{500}/10^{p_2}) \approx -1$, before dropping to a negative enhancement at $\log_{10}(M_{500}/10^{p_2}) \lesssim -3$, where the GR concentration exceeds the $f(R)$ concentration. There is also a small dip in the concentration enhancement for $0 < \log_{10}(M_{500}/10^{p_2}) < 1$. The stacked profiles of Fig.~\ref{fig:stacked_profiles} and the discussion in Sec.~\ref{sec:rescaled_mass} can provide a physical interpretation of this behaviour.

The most accurate measurement of $c_{200}$ is by performing a direct fitting of the NFW profile to the halo profiles, so the panel on the right in Fig.~\ref{fig:3_panel} is expected to give the most reliable result. Here, the three $f(R)$ gravity models all show a similar behaviour and even peak at the same enhancement, which is approximately 0.15. Only the F4 data from the Crystal simulation shows any deviation from this behaviour, as it appears to have a lower concentration enhancement than the rest of the data at $\log_{10}(M_{500}/10^{p_2}) \approx -1$. However the good agreement of most of the data means that a general model can be fitted using a portion of the data, as discussed in Sec.~\ref{sec:general_model}.

The P12 data reaches the same maximum enhancement of approximately 0.15. However, the disparity between the three $f(R)$ models is now greater, with the models each peaking at a different enhancement. F6 has the highest peak enhancement and F4 has the lowest peak. Note that the P12 data has a reduced sample which excludes all haloes with $V_{200}<V_{\rm max}$ (see Sec.~\ref{sec:c_measurement}). The S08 data reaches a greater maximum enhancement of approximately 0.25, and shows an opposite trend in terms of the order of the models: F4 now has the highest peak enhancement and F6 has the lowest peak.

The difference in the results from the P12 and S08 methods is not surprising, yet it is significant. As shown in Fig.~\ref{fig:stacked_profiles}, data at $\log_{10}(M_{500}/10^{p_2}) \approx -1$ has a greater density at the inner regions of the halo and a lower density at the outer regions compared with GR, due to the enhanced acceleration of the in-falling particles. The S08 method only uses data from $R_{\rm max}$ to measure $c_{200}$, whereas the P12 method uses data from $R_{\rm max}$ and $R_{200}$. Being at a smaller distance from the halo centre, the mass enclosed by $R_{\rm max}$ is more affected by the in-fall of particles than the mass enclosed by $R_{200}$. So $V_{\rm max}$ is more affected than $V_{200}$, with the result that the S08 method measures a higher $f(R)$ concentration than the other methods. The P12 method is closer to actually fitting a density profile to the full extent of the halo, and so it yields a closer result to the full NFW fitting. 

These results indicate that the choice of method for measuring the concentration can be very important in MG models. The three methods discussed in this chapter only agree perfectly for ideal NFW profiles, and therefore only the direct NFW fitting should be used for realistic cases. Note that this statement assumes that the halo density profiles in $f(R)$ gravity can still be well described by the NFW profile, which needs to be checked explicitly (see below). However, even if NFW is no longer valid, approximate methods such as P12 and S08 should not be used instead of full NFW fitting either as they are derivatives of the latter. Indeed, using the S08 method in $f(R)$ gravity could lead to a measurement of the halo concentration that is up to 26\% greater than from performing a fit. We therefore elected to use the direct NFW fitting method for all other results in this chapter. Note that in Sec.~\ref{sec:general_model} we include a check of the validity of the NFW profile in $f(R)$ gravity.

\subsection{General model for the concentration enhancement}
\label{sec:general_model}

\begin{figure*}
\centering
\includegraphics[width=1.0\textwidth]{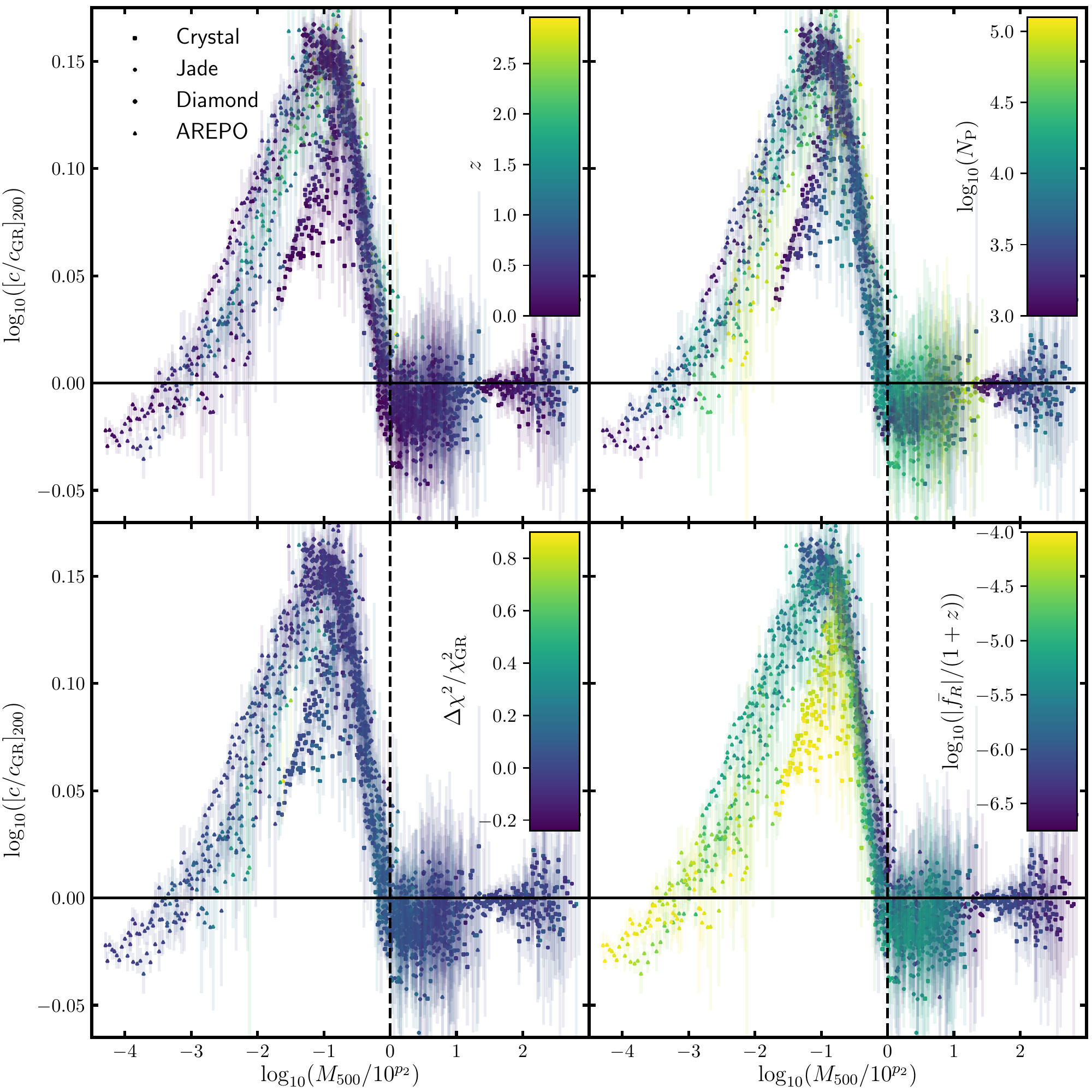}
\caption[Ratio of the median concentrations in $f(R)$ gravity and GR as a function of the rescaled mass, with colouring to indicate redshift, particle resolution, $|\bar{f}_R|/(1+z)$, and the goodness of the NFW fitting compared to GR.]{Ratio of the median concentrations of $f(R)$ gravity and GR as a function of the rescaled mass, $\log_{10}(M_{500}/10^{p_2})$, for F4, F5 and F6. The plotted data is from the simulations summarised by Table \ref{table:simulations:concentration}, sifted so that only haloes with more than 1000 particles enclosed within $R_{500}$ are included. The one standard deviation error bars are shown. The data is coloured by (\textit{clockwise from top left}) the redshift, the mean number of particles within $R_{500}$, the logarithm of a combination of the scalar field and redshift, $|\bar{f}_R|/(1+z)$, and the fractional difference of chi squared measures generated by NFW fits of the halo profiles.}
\label{fig:4_panel}
\end{figure*}

Before applying a fitting formula to the data it was useful to check whether certain factors could be affecting the shape of the trend, therefore the data was coloured via various schemes which are shown in Fig.~\ref{fig:4_panel}.

The top-left and top-right panels show the colourings by redshift and the particle number within each halo, respectively. These can both be viewed as tests of the effect of the halo resolution. For example, haloes with fewer particles can be prone to resolution effects at the innermost and outer regions, where the density can be underestimated. Haloes are less dense and more diffuse at higher redshift, which therefore leads to a greater exposure to these effects. As discussed in Sec.~\ref{sec:c_measurement}, in an effort to prevent these effects we restricted the fitting of the NFW profile to the radial range $0.05R_{200}$ to $R_{200}$. Furthermore we only include haloes that contain at least 1000 particles within $R_{500}$ in our sample. The coloured data of Fig.~\ref{fig:4_panel} suggests that these measures were sufficient, as it can be seen that even data at $z \approx 3$ agrees with the main trend and every part of the trend consists of haloes with both low and high particle numbers. Therefore even for the F4 Crystal data, low resolution is unlikely to be the reason for any disparity with the main trend.

It is significant that we are able to use redshifts up to $z=3$ for the \textsc{arepo} data. For a given $f_{R0}$ value, haloes at high $z$ are more screened than haloes at low $z$. This is because the magnitude of the background scalar field $\bar{f}_R$ grows as a function of time, such that haloes of a given mass will eventually go from being screened to unscreened. For $f(R)$ gravity models with a stronger scalar field (greater $|f_{R0}|$), haloes at a given redshift are more unscreened and therefore have a lower $\log_{10}(M_{500}/10^{p_2})$ value. However, models that are stronger than F4 are infeasible given current constraints on $f(R)$ gravity, and the minimum redshift that is available is $z=0$. Therefore the minimum value of $\log_{10}(M_{500}/10^{p_2})$ that can be plotted is only limited by the simulation resolution, as only haloes with lower mass can exist at lower values of this rescaled mass; similarly, the maximum value of $\log_{10}(M_{500}/10^{p_2})$ is limited by the box size. For each of the $f(R)$ gravity models tested in this analysis, the range of redshifts used provides a range of $\log_{10}(M_{500}/10^{p_2})$ that extends from the lowest value that is possible at halo mass $M_{500}=1.52\times10^{11}h^{-1}M_{\odot}$ to values in the screened regime, at which there is approximately no enhancement of the concentration compared with GR and so the concentration is much easier to predict. A weaker model of $f(R)$ gravity would likely exist close to or within the screened regime for $M_{500} \geq 1.52\times10^{11}h^{-1}M_{\odot}$ at $z=0$. Therefore, given that all three models tested in this chapter show excellent agreement for $-0.5 \leq \log_{10}(M_{500}/10^{p_2}) \leq 0.0$, it seems that a fit of this trend should be general for $M_{500} \geq 1.52\times10^{11}h^{-1}M_{\odot}$ for arbitrary values of $f_{R0}$ that are allowed by current constraints.

For every NFW fit, we stored the $\chi^2$ value, which is measured by summing the squared residuals of the 20 radial bins. Storing the median $\chi^2$ value for every mass bin, the GR and $f(R)$ values of the latter were then combined to generate the fractional $\chi^2$ difference. The bottom-left panel of Fig.~\ref{fig:4_panel} shows the data coloured by this measure, and can therefore be seen as a test of the validity of the NFW profile in $f(R)$ gravity. The colouring shows that the goodness-of-fit of the NFW profile for most haloes in $f(R)$ gravity is within 20\% of the goodness-of-fit in GR. The colour-bar here shows the full range of fractional differences that were observed in the data, and we note that only a very small minority of haloes have a $\chi^2$ that is almost 90\% higher than in GR. These results are promising, and imply that systematics induced through the fitting of the NFW profile are unlikely to impact on the scatter of the halo concentration in $f(R)$ gravity.

Finally, the data was coloured by the logarithm of $|\bar{f_R}|/(1+z)$, and this is shown in the bottom-right panel of Fig.~\ref{fig:4_panel}. In Chapter \ref{chapter:mdyn} we found that complicated $f(R)$ gravity effects, including screening, can effectively be described by this useful parameter. It is therefore useful to see how the enhancement of the halo concentration at screened and unscreened regimes can depend on this. An interesting observation is that bins with $|\bar{f}_R|/(1+z) \lesssim 10^{-4.5}$ are in excellent agreement with a smooth trend for $-4 \lesssim \log_{10}(M_{500}/10^{p_2}) \lesssim 3$. The F6 data, which reaches a peak enhancement at $z \approx 0$, shows very good agreement with the F5 data, and both models agree well with the higher-$z$ F4 data. Therefore if a cut is made so that only data with $|\bar{f}_R|/(1+z) \leq 10^{-4.5}$ is used in the fitting, then, at least for halo masses $M_{500}\geq1.52\times10^{11}h^{-1}M_{\odot}$, a general model can be created that applies to arbitrary $f_{R0}$ values provided $|\bar{f}_R|/(1+z) \leq 10^{-4.5}$. For models with $|f_{R0}|>10^{-4.5}$ the concentration enhancement does not follow the same trend, and therefore cannot be described by the universal fitting formula below. However, we note that models with $|f_{R0}|>10^{-4.5}$ have already been strongly disfavoured by observations. 

\begin{figure*}
\centering
\includegraphics[width=0.7\textwidth]{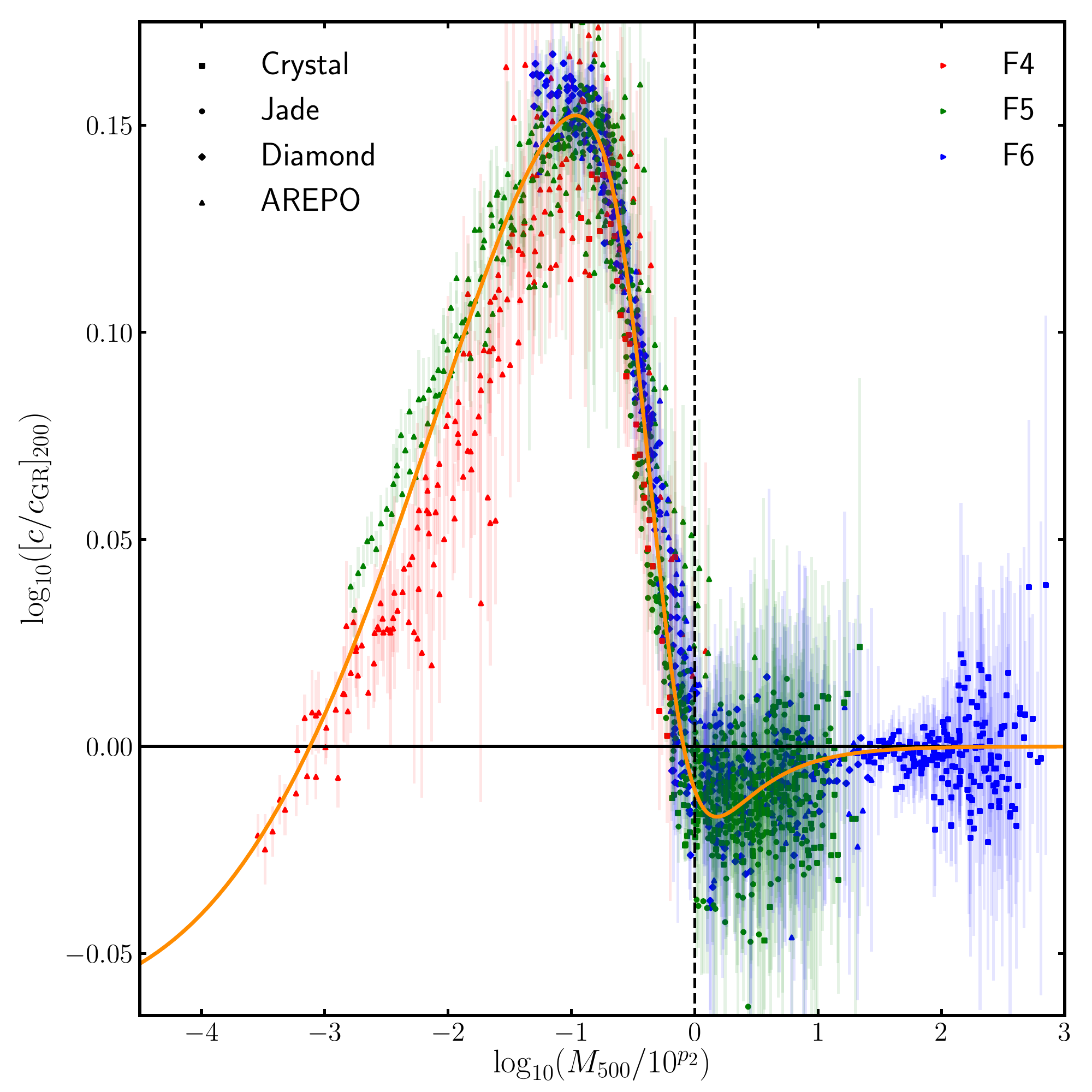}
\caption[Ratio of the median concentrations in $f(R)$ gravity and GR as a function of the rescaled mass $\log_{10}(M_{500}/10^{p_2})$, plotted along with our best-fit universal model.]{Ratio of the median concentrations of $f(R)$ gravity and GR as a function of the rescaled mass, $\log_{10}(M_{500}/10^{p_2})$, for F4 (\textit{red}), F5 (\textit{green}) and F6 (\textit{blue}). Only data with $|\bar{f}_R|/(1+z) \leq 10^{-4.5}$ has been included and fitted with Eq.~(\ref{eq:skewtanh}). This fit is shown by the trend-line, and the optimised parameter values are listed in Table \ref{table:fitting}. The plotted data is from the simulations summarised by Table \ref{table:simulations:concentration}, sifted so that all haloes have at least 1000 particles enclosed within $R_{500}$. The one standard deviation error bars are shown.}
\label{fig:skewtanh_fit}
\end{figure*}

Cleaning the data to remove bins with $|\bar{f}_R|/(1+z) > 10^{-4.5}$ and re-plotting yields Fig.~\ref{fig:skewtanh_fit}. This shows a clear trend. As $\log_{10}(M_{500}/10^{p_2})$ is reduced from value 3, the concentration enhancement, which is initially zero, drops slightly to a negative enhancement. Continuing into the unscreened regime the enhancement then rises to a distinct peak with value $\approx0.15$ at $\log_{10}(M_{500}/10^{p_2})\approx-1$, before dropping down to negative values again for $\log_{10}(M_{500}/10^{p_2})\lesssim-3$. From the discussion in Sec.~\ref{sec:rescaled_mass} and the results in Fig.~\ref{fig:stacked_profiles}, the above behaviour is physical and should therefore be fully included in the fitted model. 

In selecting a suitable fitting formula, both the screened and unscreened regimes of the data were considered. The data in the unscreened part of Fig.~\ref{fig:skewtanh_fit} shows good agreement with a skewed Gaussian curve, whose steepness is different on the two sides of the peak. This requires five parameters: a scaling $\lambda$ of the height of the curve, a shift $\gamma$ along the $\log_{10}\left(\left[c/c_{\rm GR}\right]_{200}\right)$ axis, a width $\omega_{\rm s}$, a skewness parameter $\alpha$ and a parameter $\xi_{\rm s}$ to describe the location with respect to the $\log_{10}(M_{500}/10^{p_2})$ axis. As discussed in Sec.~\ref{sec:rescaled_mass} and from examining the top-left panel of Fig.~\ref{fig:stacked_profiles}, there is a physical motivation that the concentration should dip slightly at halo masses just greater than $10^{p_2}h^{-1}M_{\odot}$. Therefore the model would have to include a minimum in this regime. This was achieved by multiplying the skewed Gaussian with a $\tanh$ curve, which takes value 1 at low values of $\log_{10}(M_{500}/10^{p_2})$ and drops to value 0 at high values. This induces a further two parameters: a location $\xi_{\rm t}$ and a width $\omega_{\rm t}$ with respect to the $\log_{10}(M_{500}/10^{p_2})$ axis. The $\tanh$ curve also ensures that the model tends to zero at higher $\log_{10}(M_{500}/10^{p_2})$. This model would gradually level out at $\log_{10}(M_{500}/10^{p_2}) < -4$, however this is not necessarily how the concentration enhancement would behave in this regime. The only way to understand this would be to run higher-resolution simulations so that lower-mass haloes can be investigated.

By taking the above considerations into account, we arrive at a 7-parameter fitting formula, which is given by,
\begin{equation}
y(x) = \frac{1}{2}\left(\frac{\lambda}{\omega_{\rm s}}\phi(x')\left[1+\rm{erf}\left(\frac{\alpha x'}{\sqrt[]{2}}\right)\right]+\gamma\right)\left(1-\tanh\left(\omega_{\rm t}\left[x+\xi_{\rm t}\right]\right)\right),
\label{eq:skewtanh}
\end{equation}
where $y=\log_{10}\left(\left[c/c_{\rm GR}\right]_{200}\right)$, $x'=(x-\xi_{\rm s})/\omega_{\rm s}$ and $x=\log_{10}(M_{500}/10^{p_2})$. The left-hand bracket of Eq.~(\ref{eq:skewtanh}) represents the skewed Gaussian curve, where $\phi(x)$ is the normal distribution:
\begin{equation}
\phi(x) = \frac{1}{\sqrt[]{2\pi}}\exp\left(-\frac{x^2}{2}\right).
\label{eq:normal_dist}
\end{equation}
This also includes a multiplication with the error function ${\rm erf}(x')$ in order to generate a skewed curve. The error function is given by,
\begin{equation}
{\rm erf}(x) = \frac{2}{\sqrt[]{\pi}}\int_0^xe^{-t^2}{\rm d}t.
\end{equation}
The fit of Eq.~(\ref{eq:skewtanh}) to the data of Fig.~\ref{fig:skewtanh_fit} was carried out by minimising the sum of the squared normalised residuals of the data points, where the residuals have been normalised by the one standard deviation error bars shown. The sum is evaluated in a way that treats different parts of the $\log_{10}(M_{500}/10^{p_2})$ range equally. This has been achieved by splitting this range into 13 equal-width bins. The squared normalised residuals are then weighted by the reciprocal of the number of data points in the current bin, prior to minimising the sum through varying the parameters. The optimal parameters are listed in Table \ref{table:fitting} and the corresponding fit is shown in Fig.~\ref{fig:skewtanh_fit}.

\begin{table*}
\centering

\small
\begin{tabular}{ ccccccc } 
 \toprule
 
 $\lambda$ & $\xi_{\rm s}$ & $\omega_{\rm s}$ & $\alpha$ & $\gamma$ & $\omega_{\rm t}$ & $\xi_{\rm t}$ \\

 \midrule

 $0.55\pm0.18$ & $-0.27\pm0.09$ & $1.7\pm0.4$ & $-6.5\pm2.4$ & $-0.07\pm0.04$ & $1.3\pm1.0$ & $0.1\pm0.3$ \\ 
 
 \bottomrule
 
\end{tabular}

\caption[Optimal parameter values and errors from the fit of Eq.~(\ref{eq:skewtanh}) to the concentration enhancement data in Fig.~\ref{fig:skewtanh_fit}.]{Optimal parameter values and errors from the fit of Eq.~(\ref{eq:skewtanh}) to the data of Fig.~\ref{fig:skewtanh_fit}. The fit is carried out by first splitting the range of $\log_{10}(M_{500}/10^{p_2})$ into 13 equal-width bins. The squared normalised residuals of the data points are then weighted by the reciprocal of the number of data points in the current bin. The sum of these is minimised by varying the parameters.}
\label{table:fitting}

\end{table*}

We have also considered a weighted least squares fit which neglects the weighting of the squared normalised residuals described above. This results in a model that produces nearly identical predictions to the model shown in Fig.~\ref{fig:skewtanh_fit}. However, neglecting the weighting of the squared normalised residuals disfavours parts of the $\log_{10}(M_{500}/10^{p_2})$ range that contain fewer data points, including the \textsc{arepo} F4 data at $\log_{10}(M_{500}/10^{p_2})\lesssim-3$. Therefore, we only include results from the fitting described above.

In order to test our model, its predictions were generated for the data shown in Figs.~\ref{fig:arepo_matrix} and \ref{fig:ecosmog_matrix}. The predictions are shown by the plotted lines. Solid lines are used in snapshots which were used to generate the fit in Fig.~\ref{fig:skewtanh_fit} and dashed lines are used in snapshots excluded from the fit (snapshots with $|\bar{f}_R|/(1+z) > 10^{-4.5}$). Agreement is generally excellent between the data and the predictions in both figures. Agreement is reasonable even for the low-$z$ F4 snapshots of \textsc{arepo} that were not used to generate the model, as can be seen in Fig.~\ref{fig:arepo_matrix}. Some small disparity exists in the higher-$z$ F5 and F6 snapshots, where the data does not appear to agree with the predicted minimum in the screened regime. Again, there are probably some physical effects that result in subtly different trends at different redshifts. However, given the complexity of the behaviour of the halo concentration in chameleon $f(R)$ gravity and the simplicity of our modelling, the amount of agreement shown in these figures is indeed very good.

\section{Summary, Discussion and Conclusions}
\label{sec:conclusions}

In this chapter, a model has been developed for the enhancement of the halo concentration in HS $f(R)$ gravity with $n=1$ using a suite of simulations that are summarised by Table~\ref{table:simulations:concentration}. The model is shown in Fig.~\ref{fig:skewtanh_fit}, and is given by Eq.~(\ref{eq:skewtanh}) with the parameter values listed in Table \ref{table:fitting}. It has been defined in terms of a useful rescaling of the halo mass, $M_{500}/10^{p_2}$, with $p_2$ calculated using Eq.~(\ref{eq:p2}), such that data from three different $f(R)$ gravity models can satisfy a universal description. These models have $\log_{10}(|f_{R0}|)=(-4,-5,-6)$, and the fitting was carried out using data from all simulation snapshots with $\log_{10}\left(|\bar{f}_R|/(1+z)\right)\leq-4.5$. This universal description is shown to have very good agreement with simulations for $M_{500}/10^{p_2}$ covering nearly 7 orders of magnitude, and covering five decades of the halo mass.

Our model has been tested by comparing its predictions of the enhancement of the concentration with an arbitrarily chosen set of snapshots from our simulations, as shown by the lines plotted in Figs.~\ref{fig:arepo_matrix} and \ref{fig:ecosmog_matrix}. These predictions show excellent agreement with the data for all snapshots, apart from the Crystal snapshots with $\log_{10}\left(|\bar{f}_R|/(1+z)\right)>-4.5$. This is not surprising given that this data was not used in the fit of the model. Having a general model that works for $\log_{10}\left(|\bar{f}_R|/(1+z)\right)\leq-4.5$ will prove very useful, particularly given that an analytical theoretical modelling was not available.

The data of Fig.~\ref{fig:skewtanh_fit} shows that  in the unscreened regime the enhancement of the concentration reaches a distinct peak as a function of the halo mass, but drops to negative values at lower mass, where the $f(R)$ concentration is less than the GR concentration. As shown by Fig.~\ref{fig:stacked_profiles}, such negative enhancement occurs because the innermost regions of the haloes are less dense in $f(R)$ gravity than in GR. This could be caused by the velocity gained by particles in haloes, which makes it difficult for them to settle into orbits at the central regions of the halo. Meanwhile in the screened regime of Fig.~\ref{fig:skewtanh_fit} there is a small dip in the concentration. Fig.~\ref{fig:stacked_profiles} suggests that this is caused by the halo being only partially screened, so that outer particles are moved further towards the centre of the halo while the inner regions remain screened. The density profile is therefore unaffected at the innermost regions but is greater at intermediate radii. Therefore the scale radius becomes greater, and fitting an NFW profile would then result in an estimate for the concentration that is lower in $f(R)$ gravity than in GR. All of these effects are incorporated by the fitted model of Eq.~(\ref{eq:skewtanh}).

Some further investigations were carried out which can be useful for further studies of the concentration in $f(R)$ gravity, and in other similar MG theories. Firstly, in addition to applying a direct NFW profile fitting to each of the haloes to measure the concentration, two simplified approaches were also used, namely the methods that are used by \cite{Prada:2011jf} and \cite{Springel:2008cc}. The resulting enhancement of the concentration from using these two methods (shown in Fig.~\ref{fig:3_panel}) shows a difference from direct NFW fitting. This is due to the effects of $f(R)$ gravity on the internal density profile, which means that the choice of regions of the halo to use in measuring the concentration becomes important. The method used by \cite{Springel:2008cc} only requires the mass enclosed by the orbital radius with the maximum circular velocity. Being found at the inner regions of a halo, which become more dense as the halo becomes unscreened, this results in the concentration being overestimated by up to 26\%. From this, we conclude that only the direct NFW fitting should be used in $f(R)$ studies. Secondly, we looked at the validity of the NFW profile fitting in $f(R)$ gravity and found that, as shown by the bottom-left panel of Fig.~\ref{fig:4_panel}, for most haloes the $\chi^2$ measure for the fit is within 20\% of the GR measure, and for some haloes the fit is even better. Therefore the systematic effects caused by fitting the NFW profile in $f(R)$ gravity are unlikely to have a significant effect on the scatter of the concentration measure.

The results of this chapter show that the $p_2$ parameter defined in Chapter \ref{chapter:mdyn} can indeed be very useful in the description and modelling of complicated effects in $f(R)$ gravity. In addition to its relatively simple one-parameter definition, it may also allow the combining of data generated by simulations run for different cosmological parameters, as $p_2$ encapsulates the values of $\Omega_{\rm M}$ and $\Omega_{\Lambda}$. Indeed, the data for the concentration enhancement from \textsc{arepo} and Diamond F6 shows excellent agreement (see Fig.~\ref{fig:skewtanh_fit}), even though these two simulations were run for different cosmological parameters and using very different codes. It will be interesting to see where else $p_2$ can be used in $f(R)$ studies. Of particular interest would be to see how it can simplify the modelling of the HMF. The enhancement of the HMF in $f(R)$ gravity peaks at a particular halo mass which depends on the strength of the scalar field. A stronger scalar field allows higher-mass haloes to be unscreened, and therefore results in an enhancement of the HMF at a higher mass. At the very least, the mass of the peak enhancement of the HMF can be expected to be strongly correlated to $p_2$. The enhancement of the matter power spectrum could also be investigated via a similar treatment.

For the results of this chapter, we used data from four different simulations, allowing a wide range of resolutions to be used. However, one potential drawback is that these are run for dark matter only. It will therefore be important to test these results using cluster data from full-physics hydrodynamical simulations run for $f(R)$ gravity. 
\graphicspath{{./gfx/}}

\chapter{\boldmath Observable-mass scaling relations in \texorpdfstring{$f(R)$}{f(R)} gravity}
\label{chapter:scaling_relations}

\section{Introduction}
\label{sec:introduction:scaling_relations}

In this chapter, we analyse the effects of $f(R)$ gravity on cluster observable-mass scaling relations (green dotted box of Fig.~\ref{fig:fr_flow_chart}). These are modified by the effects of the fifth force on the gravitational potentials of haloes, which are intrinsically linked to the dynamical mass and gas temperature. According to \citet{He:2015mva}, who employed a suite of non-radiative simulations to investigate the effect of $f(R)$ gravity on a number of mass proxies, it is possible to map between the scaling relations in $f(R)$ gravity and GR using only the relation between the dynamical and true (or lensing) masses of a halo, which is accurately captured by our tanh fitting formula (Eq.~(\ref{eq:mdyn_enhancement})). Here, we test these predictions using non-radiative hydrodynamic simulations with much higher resolutions, and build upon this by checking how the addition of full-physics effects such as cooling, star formation and feedback impact the accuracy. To this end, we make use of the first simulations that simultaneously incorporate both full-physics and $f(R)$ gravity \citep{Arnold:2019vpg}. In addition, we propose and test a set of alternative mappings from the scaling relations in GR to their $f(R)$ counterparts, which again require only our tanh fitting formula. We note that, for this chapter, our simulations only cover galaxy groups and low-mass clusters ($M\lesssim10^{14.5}M_\odot$). However, in Chapter \ref{chapter:baryonic_fine_tuning} we will use larger simulations (run using a retuned model for the baryonic physics) to extend these tests to $\sim10^{15}M_{\odot}$.

This chapter is organised as follows: in Sec.~\ref{sec:background:scaling_relations}, we summarise the scaling relation mappings proposed by \citet{He:2015mva} and our novel alternative mappings; in Sec.~\ref{sec:methods:scaling_relations}, we describe our simulations and calculations of the halo masses and observable proxies; then, in Sec.~\ref{sec:results:scaling_relations}, we present our results for the scaling relations of four mass proxies; finally, in Sec.~\ref{sec:conclusions:scaling_relations}, we summarise the results of this chapter and their significance for our framework.

\section{Background}
\label{sec:background:scaling_relations}


We will study the effects of HS $f(R)$ gravity on the cluster scaling relations for four mass proxies: the gas temperature $T_{\rm gas}$, the X-ray luminosity $L_{\rm X}$, the integrated SZ flux, given by the Compton $Y$-parameter $Y_{\rm SZ}$, and its X-ray analogue $Y_{\rm X}$. These observables have a one-to-one mapping with the mass because of the link between the gravitational potential of a cluster and its temperature. During cluster formation, baryonic matter is accreted onto the dark matter halo from its surroundings. The gravitational potential energy of the gas is converted into kinetic energy as it falls in. During accretion, the in-falling gas undergoes shock heating, resulting in the conversion of its kinetic energy into thermal energy. The resulting self-similar model for cluster mass scaling relations predicts that the gravitational potential alone can determine the thermodynamical properties of a cluster \citep{1986MNRAS.222..323K}. 

The X-ray luminosity within radius $r$ from the cluster centre is given by,
\begin{equation}
    L_{\rm X}(<r) = \int^r_0{\rm d}r'4\pi r'^2\rho_{\rm gas}^2(r')T_{\rm gas}^{1/2}(r'),
    \label{eq:lx_obs}
\end{equation}
where $\rho_{\rm gas}(r)$ is the gas density profile. The $Y_{\rm SZ}$ parameter is related to the integrated electron pressure of the cluster gas, and is given by,
\begin{equation}
     Y_{\rm SZ}(<r) = \frac{\sigma_{\rm T}}{m_{\rm e}c^2}\int^r_0{\rm d}r'4\pi r'^2n_{\rm e}(r')T_{\rm gas}(r'),
     \label{eq:ysz_obs}
\end{equation}
where $n_{\rm e}$ is the electron number density. Meanwhile, the $Y_{\rm X}$ parameter \citep{Kravtsov:2006db} is equivalent to the product of the gas mass and the mass-weighted gas temperature, $\bar{T}_{\rm gas}(<r)$:
\begin{equation}
    Y_{\rm X}(<r) = \bar{T}_{\rm gas}(<r)\int^r_0{\rm d}r'4\pi r'^2\rho_{\rm gas}(r').
    \label{eq:yx_obs}
\end{equation}

It has been shown in previous studies \citep[e.g.,][]{Fabjan:2011} that $Y_{\rm X}$ and $Y_{\rm SZ}$ have comparatively low scatter as mass proxies and are relatively insensitive to dynamical processes including cluster mergers. It has also been found that their scaling relations with the mass show good agreement with the self-similar model predictions even after the inclusion of full-physics effects such as feedbacks, which can heat up and blow gas out from the central regions \citep[e.g.,][]{Fabjan:2011,Truong:2016egq,2018MNRAS.480.2898C}. 

Because these cluster mass proxies are closely related to the gravitational potential (and hence the dynamical mass) of clusters, and because in $f(R)$ gravity the dynamical mass can be cleanly modelled (see Chapter \ref{chapter:mdyn}), it is natural to expect that cluster observable-mass scaling relations in $f(R)$ gravity can be modelled given their counterparts in GR. To study the effects of $f(R)$ gravity on the scaling relations for these proxies, we adopt two methods which are described in the sections that follow.

\subsection{Effective density approach}
\label{sec:eff_approach}

In Chapter \ref{chapter:mdyn}, we introduced the $f(R)$ gravity effective density field $\rho_{\rm eff}$ \citep{He:2014eva}, which can be expressed in terms of the true density field $\rho$ using Eq.~(\ref{eq:effective_density}). The true density field corresponds to the intrinsic mass of simulation particles. As we discussed, the mass of haloes computed using the effective density field is equivalent to the dynamical mass. 

Using non-radiative simulations run for the F5 model and GR, \citet{He:2015mva} generated halo catalogues using the effective density field. The radius, $R_{\rm 500}^{\rm eff}$, of these haloes enclosed an average \textit{effective} density of 500 times the ({\it true}) critical density of the Universe. Both the total true and dynamical mass were computed within this radius, and the cluster observables were computed using all enclosed gas cells.

Analysing these data, \citet{He:2015mva} found that haloes in GR and $f(R)$ gravity with the same dynamical mass, $M^{\rm GR}=M_{\rm dyn}^{f(R)}=M_{\rm dyn}$, also have the same gas temperature:
\begin{equation}
    T^{f(R)}_{\rm gas}\left(M^{f(R)}_{\rm dyn}\right) = T^{\rm GR}_{\rm gas}\left(M^{\rm GR}=M^{f(R)}_{\rm dyn}\right).
    \label{eq:temp_equiv_eff}
\end{equation}
The physical origin of this result is the intrinsic relationship between the gravitational potential and the gas temperature (see above). Two haloes with the same dynamical mass $M_{\rm dyn}$ (which we recall has also been computed within the same radius $R_{500}^{\rm eff}$), would also have the same gravitational potential, $\phi=(GM_{\rm dyn})/R_{500}^{\rm eff}$, and are therefore expected to have the same temperature. The authors also found that, outside the core region, the gas density profiles of haloes in GR are enhanced by a factor $M^{f(R)}_{\rm dyn}/M^{f(R)}_{\rm true}$ with respect to haloes in $f(R)$ gravity which have the same dynamical mass. This is because the gas density profile follows the true density profile more closely than the effective density profile, which itself is a result of the fact that clusters form from very large regions in the Lagrangian space, so that the ratio between the baryonic and total masses within clusters resembles the cosmic mean, $\Omega_{\rm b}/\Omega_{\rm M}$ \citep{1993Natur.366..429W}. 
The extra forces in MG theories and feedbacks from galaxy formation can add further complications to this through their effects on the gas density profiles, especially in the inner regions; however, as we will show in the following paragraph, the good agreement between the GR and rescaled $f(R)$ gas density profiles still holds in the outer halo regions.

\begin{figure*}
\centering
\includegraphics[width=1.0\textwidth]{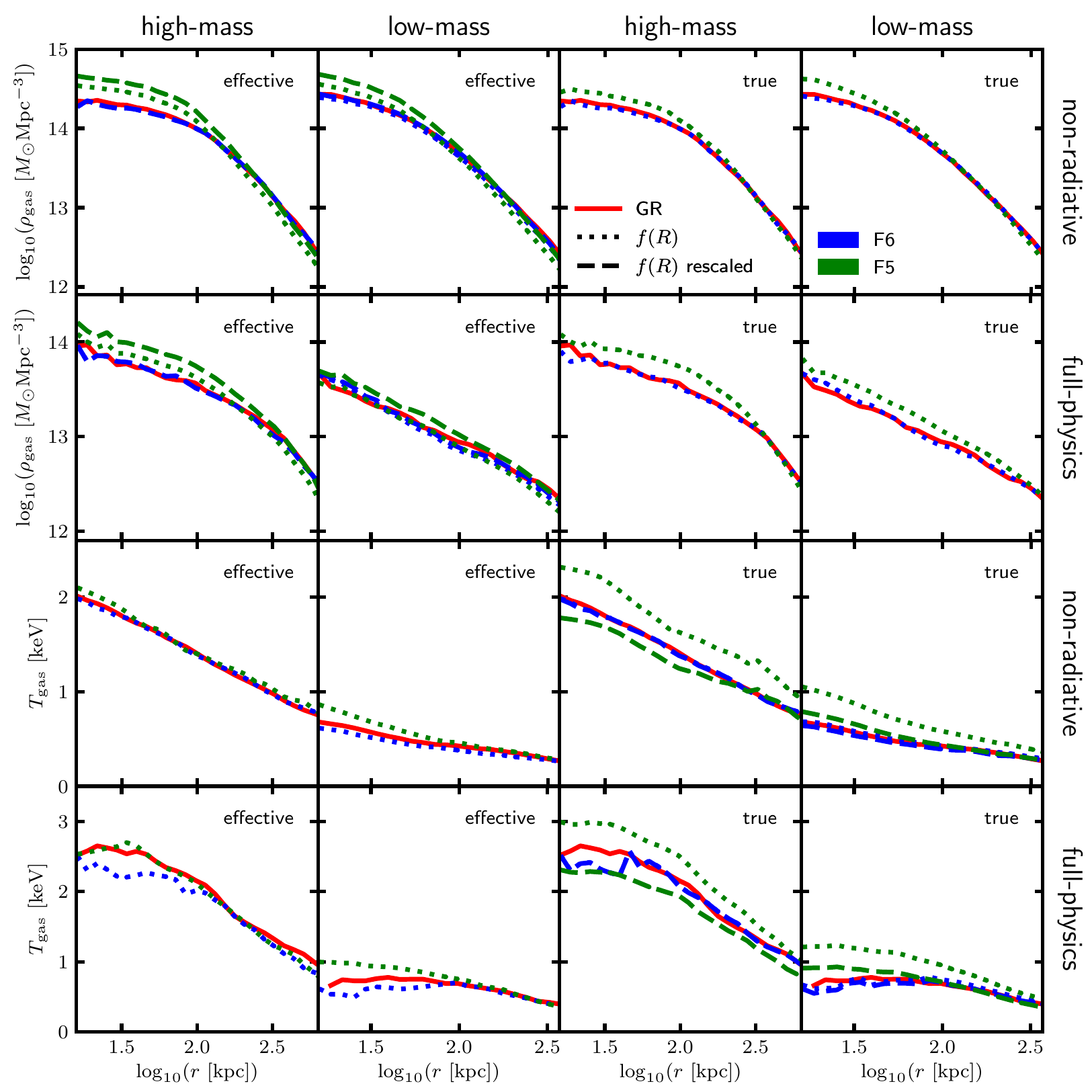}
\caption[Median gas density and temperature profiles of haloes in $f(R)$ gravity and GR, using the full-physics and non-radiative \textsc{shybone} simulations.]{Median gas density profiles (\textit{top two rows}) and median temperature profiles (\textit{bottom two rows}) of FOF groups from two mass bins: $13.7<\log_{10}\left(M/M_{\odot}\right)<14.0$ (\textit{high-mass}) and $13.0<\log_{10}\left(M/M_{\odot}\right)<13.3$ (\textit{low-mass}). The group data from the non-radiative and full-physics \textsc{shybone} simulations (see Sec.~\ref{sec:simulations:scaling_relations}) has been used. In addition to GR (\textit{red solid lines}), the profiles for F6 (\textit{blue lines}) and F5 (\textit{green lines}) are shown. Rescaled $f(R)$ gravity profiles (\textit{dashed lines}) are shown, along with the unaltered profiles (\textit{dotted curves}). The rescalings correspond to the effective density (\textit{left two columns}) and true density (\textit{right two columns}) approaches discussed in Sec.~\ref{sec:background:scaling_relations}. For the effective approach, the maximum radius shown is $R_{500}^{\rm eff}$ (see Sec.~\ref{sec:eff_approach}) and the halo mass $M$ is the total dynamical mass within this radius. For the true approach, the maximum radius shown is $R_{500}^{\rm true}$ (see Sec.~\ref{sec:true_approach}) and the halo mass $M$ is the total true mass within this radius.}
\label{fig:profiles}
\end{figure*}

We have replicated the procedure adopted by \citet{He:2015mva} using our full-physics and non-radiative simulations (for full details, see Sec.~\ref{sec:methods:scaling_relations}). In Fig.~\ref{fig:profiles}, the stacked temperature and gas density profiles of haloes from mass bins $10^{13.7}M_{\odot}<M_{\rm dyn}(<R_{500}^{\rm eff})<10^{14.0}M_{\odot}$ and $10^{13.0}M_{\odot}<M_{\rm dyn}(<R_{500}^{\rm eff})<10^{13.3}M_{\odot}$ are shown in the first and second columns from the left, respectively. The radial range is shown up to the mean logarithm of $R_{500}^{\rm eff}$ (which is almost exactly the same for GR, F6 and F5). For the non-radiative temperature profiles, shown in the third row, it is clear that the F6 and F5 predictions agree very well with GR in the outer regions. There is also encouraging agreement for the full-physics data, although there is a small disparity between F5 and GR in the outer regions for the higher-mass bin. For the $f(R)$ gravity gas density profiles, the results both with and without the $M^{f(R)}_{\rm dyn}/M^{f(R)}_{\rm true}$ rescaling are shown. As was found by \citet{He:2015mva}, the rescaled $f(R)$ gravity profiles (shown by the dashed curves) agree very well with GR in the outer regions. Again, there is also promising agreement for the full-physics profiles.

These results suggest that for haloes in $f(R)$ gravity and GR which have the same dynamical mass, $M^{\rm GR}=M^{f(R)}_{\rm dyn}$, the following relation is expected to apply:
\begin{equation}
\begin{aligned}
& \int_0^r {\rm d} r'4\pi r'^2\left(\rho_{\rm gas}^{f(R)}\right)^a\left(T_{\rm gas}^{f(R)}\right)^b \\
& \approx \left(\frac{M_{\rm true}^{f(R)}}{M_{\rm dyn}^{f(R)}}\right)^a\int_0^r {\rm d} r'4\pi r'^2 \left(\rho_{\rm gas}^{\rm GR}\right)^a\left(T_{\rm gas}^{\rm GR}\right)^b,
\end{aligned}
\label{eq:eff_rescaling}
\end{equation}
where $a$ and $b$ represent indices of power, and we note that $\rho_{\rm gas}$ represents the intrinsic (not effective) gas density. Using Eqs.~(\ref{eq:lx_obs})-(\ref{eq:yx_obs}) for the mass proxies, this relation can be applied to derive the following mappings between the respective mass scaling relations in GR and $f(R)$ gravity:
\begin{equation}
    \frac{M_{\rm dyn}^{f(R)}}{M_{\rm true}^{f(R)}}Y_{\rm SZ}^{f(R)}\left(M_{\rm dyn}^{f(R)}\right) \approx Y_{\rm SZ}^{\rm GR}\left(M^{\rm GR}=M_{\rm dyn}^{f(R)}\right),
    \label{eq:ysz_mapping}
\end{equation}
\begin{equation}
    \frac{M_{\rm dyn}^{f(R)}}{M_{\rm true}^{f(R)}}Y_{\rm X}^{f(R)}\left(M_{\rm dyn}^{f(R)}\right) \approx Y_{\rm X}^{\rm GR}\left(M^{\rm GR}=M_{\rm dyn}^{f(R)}\right),
    \label{eq:yx_mapping}
\end{equation}
\begin{equation}
    \left(\frac{M_{\rm dyn}^{f(R)}}{M_{\rm true}^{f(R)}}\right)^2L_{\rm X}^{f(R)}\left(M_{\rm dyn}^{f(R)}\right) \approx L_{\rm X}^{\rm GR}\left(M^{\rm GR}=M_{\rm dyn}^{f(R)}\right).
    \label{eq:lx_mapping}
\end{equation}
Note that to obtain these relations, the two integrations in Eq.~\eqref{eq:eff_rescaling} have used the same upper limit, $r=R^{\rm eff}_{500}$, for GR and $f(R)$ gravity, as mentioned above.

\citet{He:2015mva} demonstrated an accuracy $\approx3\%$ for the $Y_{\rm SZ}$ and $Y_{\rm X}$ relations and $\approx13\%$ for $L_{\rm X}$. These quantities are all cumulative: they are computed by summing over the entire volume of a halo up to some maximum radius (in this case $R_{500}^{\rm eff}$). Cumulative quantities typically depend more on the outer regions, which contain most of the volume and mass, than on the inner regions. Therefore, even though the $f(R)$ gravity profiles (with appropriate rescaling) do not agree with GR for the inner regions (see Fig.~\ref{fig:profiles}), this is expected to have a negligible contribution overall to these mass proxies and explains why \citet{He:2015mva} found such a high accuracy for these mappings. 

In this chapter, we will expand on these tests by using full-physics hydrodynamical simulations to check how the addition of effects such as cooling and feedback can alter the accuracy of the mappings defined by Eqs.~(\ref{eq:ysz_mapping})-(\ref{eq:lx_mapping}) and the temperature equivalence given by Eq.~(\ref{eq:temp_equiv_eff}). Our tests with the non-radiative runs can also be used as a check for consistency with \citet{He:2015mva}, who used a different simulation code and $f(R)$ gravity solver.

\subsection{True density approach}
\label{sec:true_approach}

Mappings can also be done for haloes identified with the true density field. For these haloes, the radius, $R_{\rm 500}^{\rm true}$, would enclose an average \textit{true} density of 500 times the critical density of the Universe. For haloes in $f(R)$ gravity and GR with the same true mass, $M^{\rm GR} = M_{\rm true}^{f(R)}$, the total gravitational potential at $R_{\rm 500}^{\rm true}$ in $f(R)$ gravity would be a factor of $M_{\rm dyn}^{f(R)}/M_{\rm true}^{f(R)}$ higher than in GR (where both the dynamical and true mass are measured within $R_{\rm 500}^{\rm true}$). According to the self-similar model predictions, the gas temperature in $f(R)$ gravity is expected to be higher by the same factor:
\begin{equation}
    T^{f(R)}_{\rm gas}\left(M^{f(R)}_{\rm true}\right) = \frac{M_{\rm dyn}^{f(R)}}{M_{\rm true}^{f(R)}}T^{\rm GR}_{\rm gas}\left(M^{\rm GR}=M^{f(R)}_{\rm true}\right).
    \label{eq:temp_equiv_true}
\end{equation}
On the other hand, for haloes in $f(R)$ gravity and GR with the same true mass, the gas density profiles are expected to agree in the outer regions.

To check these assumptions, let us once again examine Fig.~\ref{fig:profiles}, this time looking at the third and fourth columns from the left, which show the stacked gas density and temperature profiles for groups in mass bins $10^{13.7}M_{\odot}<M_{\rm true}\left(<R_{500}^{\rm true}\right)<10^{14.0}M_{\odot}$ and $10^{13.0}M_{\odot}<M_{\rm true}\left(<R_{500}^{\rm true}\right)<10^{13.3}M_{\odot}$, respectively. The radial range is shown up to the mean logarithm of $R_{500}^{\rm true}$ for all profiles. Referring to the bottom two rows, which show the non-radiative and full-physics temperature profiles, it appears that the $f(R)$ gravity profiles with the $M_{\rm dyn}^{f(R)}/M_{\rm true}^{f(R)}$ rescaling (shown by the dashed curves) show reasonable agreement with GR in the outer regions. Again, the only exception is for the high-mass bin of the full-physics data, where there is a small disparity between F5 and GR. Looking at the top two rows, the $f(R)$ gravity gas density profiles agree very well with GR in the outer-most regions for both mass bins.

These results for the temperature and gas density profiles yield the following predictions for haloes in GR and $f(R)$ gravity with $M^{\rm GR}=M_{\rm true}^{f(R)}$:
\begin{equation}
\begin{aligned}
& \int_0^r {\rm d} r'4\pi r'^2\left(\rho_{\rm gas}^{f(R)}\right)^a\left(T_{\rm gas}^{f(R)}\right)^b \\
& \approx \left(\frac{M_{\rm dyn}^{f(R)}}{M_{\rm true}^{f(R)}}\right)^b\int_0^r {\rm d} r'4\pi r'^2 \left(\rho_{\rm gas}^{\rm GR}\right)^a\left(T_{\rm gas}^{\rm GR}\right)^b,
\end{aligned}
\label{eq:true_rescaling}
\end{equation}
where this time the two integrations both have upper limit $r=R^{\rm true}_{500}$. This prediction yields the following new mappings between the mass scaling relations in $f(R)$ gravity and GR:
\begin{equation}
    Y_{\rm SZ}^{f(R)}\left(M_{\rm true}^{f(R)}\right) \approx \frac{M_{\rm dyn}^{f(R)}}{M_{\rm true}^{f(R)}}Y_{\rm SZ}^{\rm GR}\left(M^{\rm GR}=M_{\rm true}^{f(R)}\right),
    \label{eq:ysz_mapping_true}
\end{equation}
\begin{equation}
    Y_{\rm X}^{f(R)}\left(M_{\rm true}^{f(R)}\right) \approx \frac{M_{\rm dyn}^{f(R)}}{M_{\rm true}^{f(R)}}Y_{\rm X}^{\rm GR}\left(M^{\rm GR}=M_{\rm true}^{f(R)}\right),
    \label{eq:yx_mapping_true}
\end{equation}
\begin{equation}
    L_{\rm X}^{f(R)}\left(M_{\rm true}^{f(R)}\right) \approx \left(\frac{M_{\rm dyn}^{f(R)}}{M_{\rm true}^{f(R)}}\right)^{1/2}L_{\rm X}^{\rm GR}\left(M^{\rm GR}=M_{\rm true}^{f(R)}\right).
    \label{eq:lx_mapping_true}
\end{equation}
For the $Y_{\rm SZ}$ and $Y_{\rm X}$ mappings, the $M_{\rm dyn}^{f(R)}/M_{\rm true}^{f(R)}$ factor comes from the dependence on the gas temperature to power one in Eqs.~(\ref{eq:ysz_obs}) and (\ref{eq:yx_obs}). On the other hand, the X-ray luminosity, given by Eq.~(\ref{eq:lx_obs}), depends on the gas temperature to power half, which means that the corresponding $f(R)$ gravity and GR scaling relations are expected to differ by factor $\left(M_{\rm dyn}^{f(R)}/M_{\rm true}^{f(R)}\right)^{1/2}$ only. In Sec.~\ref{sec:results:scaling_relations}, we show the results of our tests of these alternative predictions using both our non-radiative and full-physics simulations.

In this section, we have referred to two different definitions of the halo radius: $R_{500}^{\rm eff}$ and $R_{500}^{\rm true}$. The radius $R_{500}^{\rm eff}$ is defined in terms of the effective density field. For an unscreened halo in $f(R)$ gravity, the effective density is up to $4/3$ times greater than the true density. As such, $R_{500}^{\rm eff}$ is typically a higher radius than $R_{500}^{\rm true}$ for haloes in $f(R)$ gravity. On the other hand, the two radii are exactly the same in GR, where the effective density field is equivalent to the true one.

\section{Simulations and methods}
\label{sec:methods:scaling_relations}

In Sec.~\ref{sec:simulations:scaling_relations}, we describe the non-radiative and full-physics simulations that are used in this chapter. Then, in Sec.~\ref{sec:methods:scaling_relations:groups}, we describe how we have measured the cluster mass and four observable mass proxies from these simulations.

\subsection{Simulations}
\label{sec:simulations:scaling_relations}

The results discussed in this chapter were generated using a subset of the \textsc{shybone} simulations \citep{Arnold:2019vpg,Hernandez-Aguayo:2020kgq}. These full-physics hydrodynamical simulations, which have been run using \textsc{arepo} \citep{2010MNRAS.401..791S}, employ the IllustrisTNG galaxy formation model \citep{2017MNRAS.465.3291W,Pillepich:2017jle} and include runs for GR, HS $f(R)$ gravity and nDGP (we will only use the GR and $f(R)$ runs in this chapter). 

For every full-physics run used in this chapter, we also utilise a non-radiative counterpart which does not include cooling, star formation or stellar and black hole feedback processes.  Both the full-physics and non-radiative simulations span a comoving box of length 62$h^{-1}{\rm Mpc}$. These runs each start with $512^3$ dark matter particles and the same number of initial gas cells, and begin at redshift $z=127$. All results in this chapter are computed at $z=0$. The cosmological parameters have values ($h$, $\Omega_{\rm M}$, $\Omega_{\rm b}$, $\Omega_{\Lambda}$, $n_{\rm s}$, $\sigma_8$) $=$ ($0.6774$, $0.3089$, $0.0486$, $0.6911$, $0.9667$, $0.8159$), where $n_{\rm s}$ is the power-law index of the primordial density power spectrum. The mass resolution is set by the DM particle mass $m_{\rm DM}=1.28\times10^8h^{-1}M_{\odot}$ and an average gas cell mass of $m_{\rm gas}\approx2.5\times10^7h^{-1}M_{\odot}$. In addition to GR, the runs include the F6 and F5 HS models, all starting from identical initial conditions at $z=127$.

In the calculation of the gas temperature, we have assumed that the primordial hydrogen mass fraction has a value $X_{\rm H}=0.76$ and set the adiabatic index to $\gamma=5/3$ (for a monatomic gas). For the non-radiative simulations we assume that the gas is made up of fully ionised hydrogen and helium.

\subsection{Group catalogues}
\label{sec:methods:scaling_relations:groups}

The group catalogues were constructed using \textsc{subfind} \citep{springel2001}, which locates the gravitational potential minimum of the FOF groups using the true density field. For each group, both radii $R_{500}^{\rm true}$ and $R_{500}^{\rm eff}$ were computed around the gravitational potential minimum, enclosing, respectively, average \textit{true} and \textit{effective} densities of 500 times the critical density of the Universe. For each radius definition, the total enclosed dynamical and true masses were measured, in addition to the group observables. Quantities measured within $R_{500}^{\rm eff}$ have been used to test the scaling relation mappings from the effective density approach described in Sec.~\ref{sec:eff_approach}, while the quantities measured within $R_{500}^{\rm true}$ have been used to test the predictions of the true density approach discussed in Sec.~\ref{sec:true_approach}.

In the computation of the halo temperature, we have excluded the core regions in which the complex thermal and dynamical processes during cluster formation and evolution can lead to a significant degree of dispersion between the halo temperature profiles. We have set the core region to the radial range $r<0.15R$, where $R$ can be either $R_{500}^{\rm eff}$ or $R_{500}^{\rm true}$. This range is consistent with previous studies of cluster scaling relations \citep[e.g.,][]{Fabjan:2011,Brun:2016jtk,Truong:2016egq}.

The halo gas temperature has been computed using a mass-weighted average:
\begin{equation}
    \bar{T}_{\rm gas} = \frac{\sum_i m_{{\rm gas},i}T_i}{\sum_i m_{{\rm gas},i}},
    \label{eq:mass_weighted_temperature}
\end{equation}
where $m_{{\rm gas},i}$ and $T_i$ are, respectively, the mass and temperature of gas cell $i$. The summations have been performed over all gas cells whose positions fall within the radial range $0.15R<r<R$. The integrated SZ flux is given by:
\begin{equation}
    Y_{\rm SZ} = \frac{\sigma_{\rm T}}{m_{\rm e}c^2}\sum_i N_{{\rm e},i}T_i,
    \label{eq:ysz}
\end{equation}
where $N_{{\rm e},i}$ is the number of electrons in gas cell $i$ and the sum includes the same cells as for $\bar{T}_{\rm gas}$. The X-ray analogue of the integrated SZ flux is equal to the product of the total gas mass $M_{\rm gas}$, of all gas cells within $R_{500}$, and $\bar{T}_{\rm gas}$:
\begin{equation}
    Y_{\rm X} = M_{\rm gas}\times \bar{T}_{\rm gas}.
    \label{eq:yx}
\end{equation}
Finally, the X-ray luminosity is calculated using:
\begin{equation}
    L_{\rm X} = \sum_im_{{\rm gas},i}\rho_{{\rm gas},i}T_i^{1/2},
    \label{eq:lx}
\end{equation}
where $\rho_{{\rm gas},i}$ is the gas density of gas cell $i$ and the summation is performed over the same gas cells as for the $\bar{T}_{\rm gas}$ calculation.

\section{Results}
\label{sec:results:scaling_relations}

In Sec.~\ref{sec:results:scaling_relations:scaling_relations}, we discuss our results for the cluster scaling relations in HS $f(R)$ gravity. Then, in Sec.~\ref{sec:results:scaling_relations:mass_ratio}, we test the validity of our analytical tanh formula for the dynamical mass enhancement, given by Eq.~(\ref{eq:mdyn_enhancement}), in the presence of full physics. There we will also present an example in which we map between the GR and $f(R)$ mass scaling relations based on this approximate fitting formula, rather than the actual values of $M^{f(R)}_{\rm dyn}/M^{f(R)}_{\rm true}$ from the simulations. Finally, in Sec.~\ref{sec:results:scaling_relations:yx-t_relation}, we propose a new test of gravity using the $Y_{\rm X}$-$\bar{T}_{\rm gas}$ relation, which does not require inferences of the cluster mass.

\subsection{Scaling relations}
\label{sec:results:scaling_relations:scaling_relations}

Using our simulation data, we have tested the scaling relation mappings described in Sec.~\ref{sec:background:scaling_relations}. Due to the small box size, $62h^{-1}{\rm Mpc}$ (comoving), of our simulations, we can only examine haloes with mass $M_{500}\lesssim10^{14.5}M_{\odot}$. We show all objects with $M_{500}\geq10^{13}M_{\odot}$ (groups and clusters) in Figs.~\ref{fig:T_gas}--\ref{fig:Lx_scaling_relation}, which typically includes $\sim100$ haloes for a given model. Note that it is difficult to rigorously test our scaling relation mappings for the cluster regime ($M_{500}\gtrsim10^{14}M_{\odot}$), where there are only 5-10 haloes in the present simulations. However, for F5, groups are typically unscreened and low-mass clusters are partially screened, while for F6 low-mass groups are partially screened and higher-mass objects are completely screened; haloes with $M_{500}\gtrsim10^{14.5}M_{\odot}$ will be mostly screened for F5 and entirely screened for F6 (see, e.g., Fig.~\ref{fig:mdyn_mtrue} below and Fig.~\ref{fig:unweighted_matrix}). Therefore, while we do not have a significant number of such large cluster-sized objects, we do expect the scaling relations calibrated for GR to be valid for them.

In addition to showing data points for individual haloes, all plots include curves showing a moving average. This is computed using a moving window of fixed size equal to 10 haloes, for which the mean logarithm of the mass and the median proxy are displayed. We note that the highest-mass haloes have been included in the moving average, even though the highest mean mass is only $\sim10^{14}M_{\odot}$. We also show sub-plots with the smoothed relative difference between the $f(R)$ and GR curves, as well as the halo scatter in GR. The latter is computed by fitting a linear model to the GR data and computing the root-mean-square residuals within mass bins.

For the panels labelled `effective' in Figs.~\ref{fig:T_gas}--\ref{fig:Lx_scaling_relation}, all measurements of the mass and observables have been taken within $R_{500}^{\rm eff}$ (see Sec.~\ref{sec:eff_approach}), and the relations are plotted against the dynamical mass. This allows the effective density mappings given by Eqs.~(\ref{eq:temp_equiv_eff}) and (\ref{eq:ysz_mapping})-(\ref{eq:lx_mapping}), originally proposed by \citet{He:2015mva}, to be tested. On the other hand, an outer radius $R_{500}^{\rm true}$ (see  Sec.~\ref{sec:true_approach}) is imposed for all measurements for the data displayed in the panels labelled `true'. These are plotted against the true mass, and can be used to test the true density mappings given by Eqs.~(\ref{eq:temp_equiv_true}) and (\ref{eq:ysz_mapping_true})-(\ref{eq:lx_mapping_true}).

\subsubsection{Temperature scaling relations}
\label{sec:results:scaling_relations:tgas}

\begin{figure*}
\centering
\includegraphics[width=1.0\textwidth]{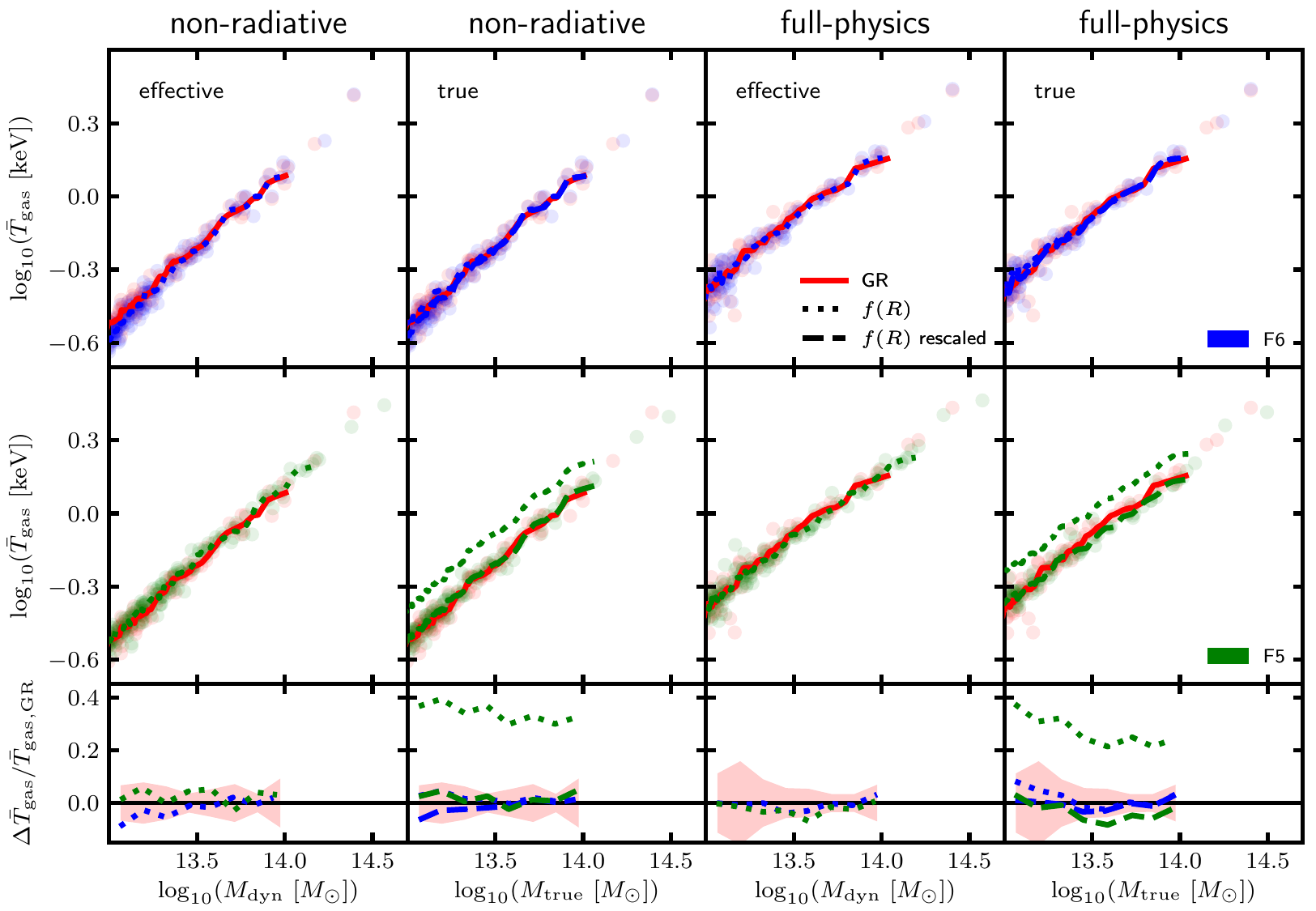}
\caption[Mass-weighted temperature as a function of mass for haloes in $f(R)$ gravity and GR, using the full-physics and non-radiative \textsc{shybone} simulations.]{Gas temperature plotted as a function of mass for FOF haloes from the non-radiative and full-physics \textsc{shybone} simulations (see Sec.~\ref{sec:simulations:scaling_relations}). The curves correspond to the median mass-weighted temperature and the mean logarithm of the mass computed within a moving window of fixed size equal to 10 haloes. Data has been included for GR (\textit{red solid lines}) together with the F6 (\textit{blue lines}) and F5 (\textit{green lines}) $f(R)$ gravity models. Rescalings to the $f(R)$ gravity temperature have been carried out as described in Secs.~\ref{sec:eff_approach} and \ref{sec:true_approach}. For the `true' density approach, the rescaled data (\textit{dashed lines}) is shown along with the unaltered data (\textit{dotted lines}). For this data, the mass corresponds to the total true mass within the radius $R_{500}^{\rm true}$, and the temperature has also been computed within this radius. For the `effective' density approach, no rescaling is necessary, the mass corresponds to the total dynamical mass within $R_{500}^{\rm eff}$, and the temperature has also been computed within this radius. Data points are displayed, with each point corresponding to a GR halo (\textit{red points}) or to a halo in F6 (\textit{blue points}) or F5 (\textit{green points}), including the rescaling for the `true' density data. \textit{Bottom row}: the smoothed relative difference between the $f(R)$ gravity and GR curves in the above plots; the red shaded regions indicate the size of the halo scatter in GR.}
\label{fig:T_gas}
\end{figure*}

The results for the gas temperature scaling relations are shown in Fig.~\ref{fig:T_gas}. The non-radiative data is displayed in the left two columns and the full-physics data is shown in the right two columns. For all models and hydrodynamical schemes, the data follows a power-law behaviour, as expected from the self-similar model. The correlation is particularly tight for the non-radiative data, with an overall scatter of 7\%. The non-radiative runs contain gas and dark matter particles, but do not feature baryonic processes (apart from basic hydrodynamics) such as radiative cooling, stellar and black hole feedback and star formation. It is therefore expected that the thermodynamical properties can be largely determined from the gravitational potential, which is observed in the results. On the other hand, there is $\sim10\%$ overall scatter in the full-physics data, and the gas temperature is typically higher. This can be explained by the inclusion of feedback mechanisms which act as an additional source of heating of the surrounding gas and cause some departures from self-similarity. These mechanisms have a stronger effect on lower-mass haloes, resulting in a particularly high ($10$-$20\%$) scatter for these objects.

For the effective density approach, Eq.~(\ref{eq:temp_equiv_eff}) is expected to hold: the temperature is predicted to be equal for haloes in GR and $f(R)$ gravity with the same dynamical mass. In Fig.~\ref{fig:T_gas}, the non-radiative and full-physics results from our effective catalogue are shown in the first and third columns from the left, respectively. For both F6 and F5, there is excellent agreement with the GR data, with typical agreement $\lesssim5\%$. This agreement for the non-radiative data backs up the findings from \citet{He:2015mva}, while the full-physics results do not show clear evidence for a departure from Eq.~(\ref{eq:temp_equiv_eff}) caused by feedback processes and cooling. These results agree with the self-similar model predictions: two haloes in $f(R)$ gravity and GR which have the same dynamical mass $M_{\rm dyn}$ (and therefore the same radius $R_{500}^{\rm eff}$) also have the same gravitational potential, $\phi = GM_{\rm dyn}/R_{500}^{\rm eff}$.

In order to test the new mappings predicted by the true density approach, the temperature of each halo in $f(R)$ gravity has been divided by the mass ratio $M_{\rm dyn}^{f(R)}/M_{\rm true}^{f(R)}$. In Fig.~\ref{fig:T_gas}, both the data for individual haloes and corresponding moving averages are shown with this rescaling applied (dashed lines), along with the moving averages for the unaltered data (dotted lines). It is expected, from Eq.~(\ref{eq:temp_equiv_true}), that the data with the rescaling should agree with GR. For the non-radiative results in the second column, this indeed appears to be the case, with an excellent agreement that is generally within just a few percent. For the full-physics data there is still reasonable $\lesssim10\%$ agreement, but the F5 temperature appears to be lower than the GR temperature for $\log_{10}\left[M_{\rm true}\left(<R_{500}^{\rm true}\right)/M_{\odot}\right]\gtrsim13.5$.

This small deviation is consistent with the full-physics temperature profiles shown in Fig.~\ref{fig:profiles}. The plots in the bottom right of that figure show the temperature profiles with the $M_{\rm dyn}^{f(R)}/M_{\rm true}^{f(R)}$ rescaling applied. The profiles are shown for the halo mass bins $10^{13.7}M_{\odot}<M_{\rm true}\left(<R_{500}^{\rm true}\right)<10^{14.0}M_{\odot}$ and $10^{13.0}M_{\odot}<M_{\rm true}\left(<R_{500}^{\rm true}\right)<10^{13.3}M_{\odot}$. 
For the high-mass full-physics profiles, the rescaled F5 profile is clearly lower than the profile in GR across most of the radial range. This can explain the lower rescaled F5 temperature observed at the high-mass end of the full-physics data. On the other hand, for the lower-mass bin with full-physics and for the non-radiative data the agreement between the rescaled $f(R)$ gravity and GR temperature profiles is very good, particularly at the outer radii which have greater overall contribution to the mass-weighted temperature. Similar agreement is shown between the $f(R)$ gravity and GR profiles for the plots in the bottom-left of Fig.~\ref{fig:profiles}. Again, for the high-mass full-physics data there is some deviation between F5 and GR, particularly in the outermost regions. But this is not as noticeable as for the profiles with the true density rescalings.

The small difference between F5 and GR is likely to be caused by a difference in the levels of feedback -- which itself is determined by the interrelations between modified gravity (including screening or the lack of it) and baryonic physics -- in these higher-mass haloes for the two models. Encouragingly, this appears to have only a small effect on the effective density data, where there is good agreement between F5 and GR for high-mass groups. However, the effect is greater for the true density data. To understand the implications that this could have on tests of gravity using the cluster regime, we will need simulations with a larger box size. This will be addressed in Chapter \ref{chapter:baryonic_fine_tuning}, where we will use a re-calibrated full-physics model to probe masses up to $M_{500}\sim10^{15}M_{\odot}$.

\subsubsection{\texorpdfstring{$Y_{\rm SZ}$}{YSZ} and \texorpdfstring{$Y_{\rm X}$}{YX} scaling relations}
\label{sec:results:scaling_relations:y_params}

\begin{figure*}
\centering
\includegraphics[width=1.0\textwidth]{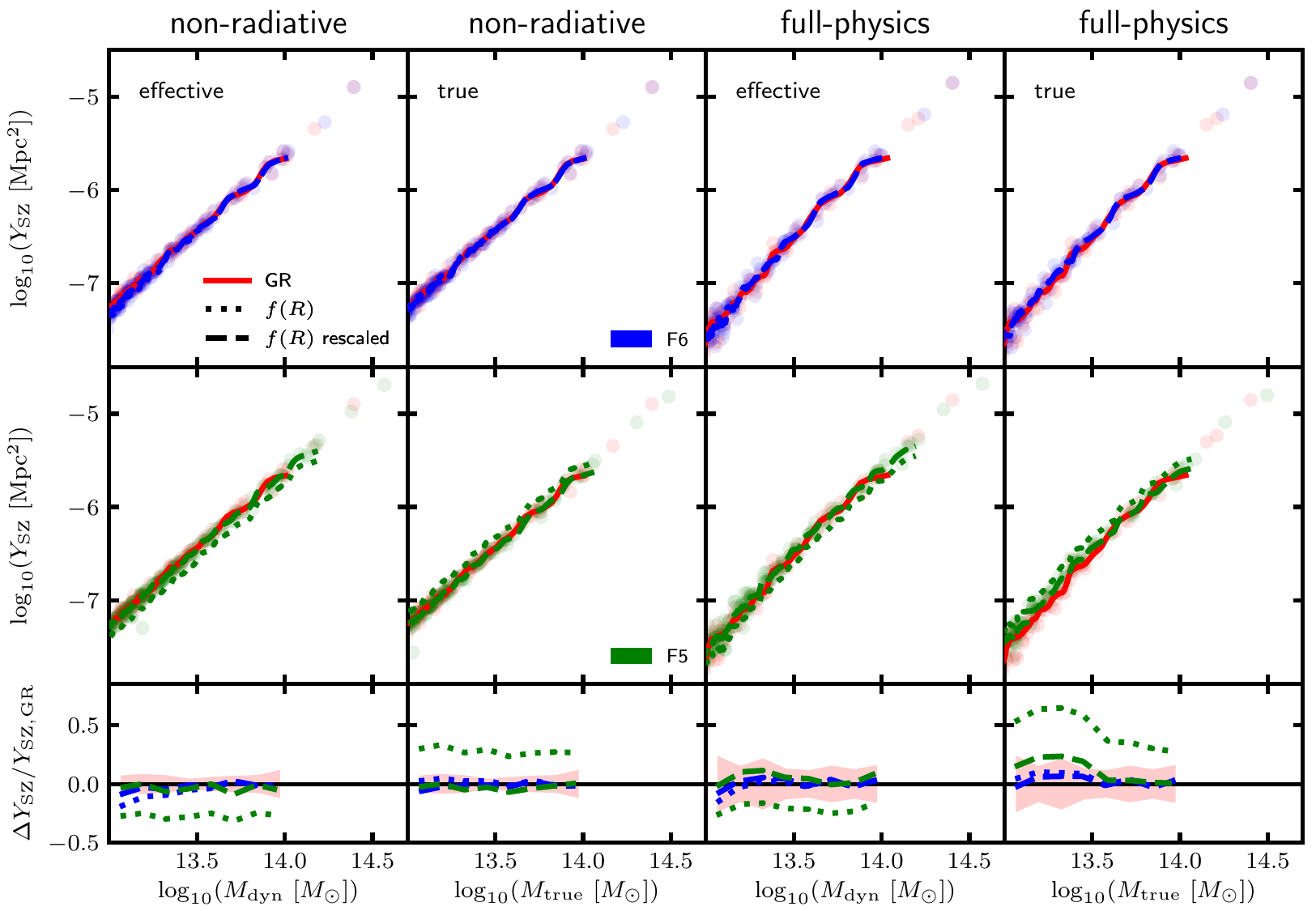}
\caption[$Y_{\rm SZ}$ parameter as a function of mass for haloes in $f(R)$ gravity and GR, using the full-physics and non-radiative \textsc{shybone} simulations.]{Compton $Y$-parameter of the SZ effect plotted as a function of mass for FOF haloes from the non-radiative and full-physics \textsc{shybone} simulations (see Sec.~\ref{sec:simulations:scaling_relations}). Apart from the $Y_{\rm SZ}$ parameter being shown rather than the mass-weighted temperature, the layout of this figure is otherwise identical to Fig.~\ref{fig:T_gas}. The rescalings of the $f(R)$ $Y_{\rm SZ}$ data are carried out as described in Sec.~\ref{sec:background:scaling_relations}.}
\label{fig:Ysz_scaling_relation}
\end{figure*}

\begin{figure*}
\centering
\includegraphics[width=1.0\textwidth]{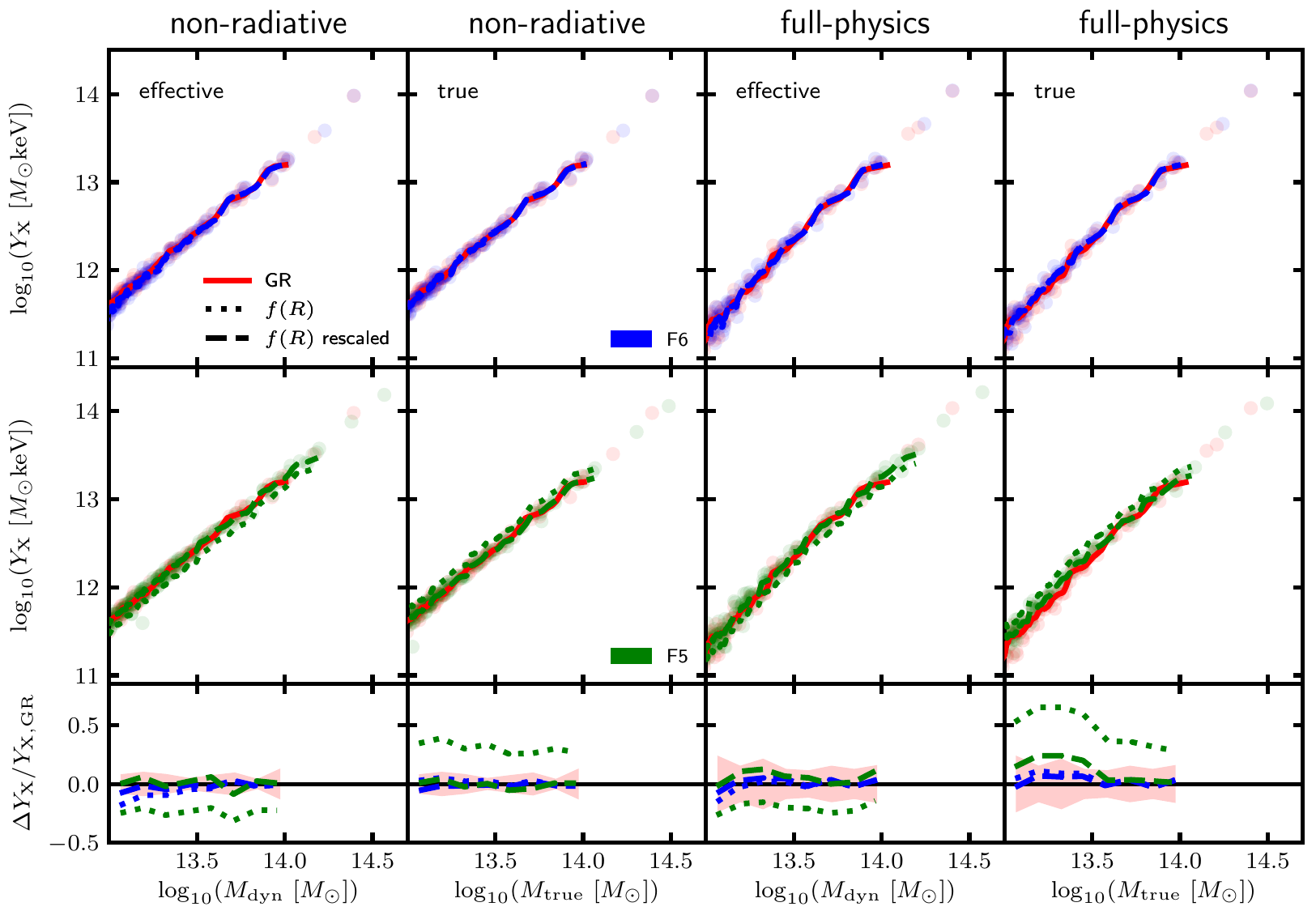}
\caption[$Y_{\rm X}$ parameter as a function of mass for haloes in $f(R)$ gravity and GR, using the full-physics and non-radiative \textsc{shybone} simulations.]{X-ray analogue of the Compton $Y$-parameter plotted as a function of mass for FOF haloes from the non-radiative and full-physics \textsc{shybone} simulations (see Sec.~\ref{sec:simulations:scaling_relations}). Apart from the $Y_{\rm X}$ parameter being shown rather than the mass-weighted temperature, the layout of this figure is otherwise identical to Fig.~\ref{fig:T_gas}. The rescalings of the $f(R)$ $Y_{\rm X}$ data are carried out as described in Sec.~\ref{sec:background:scaling_relations}.}
\label{fig:Yx_scaling_relation}
\end{figure*}

Our results for the $Y_{\rm SZ}$ and $Y_{\rm X}$ scaling relations are shown in Figs.~\ref{fig:Ysz_scaling_relation} and \ref{fig:Yx_scaling_relation}. The $Y_{\rm SZ}$ and $Y_{\rm X}$ parameters are, by definition, tightly correlated. Their results therefore follow similar patterns, and both show very tight correlations with the halo mass, with a scatter of $\sim8\%$ and $\sim19\%$ for the non-radiative and full-physics data, respectively. There are also no clear outliers in the full-physics data, unlike for the temperature and the X-ray luminosity data (see below). This is because of the competing effects of feedback processes on the gas density and gas temperature \citep{Fabjan:2011}. Comparing the non-radiative and full-physics profiles in Fig.~\ref{fig:profiles}, it can be seen that the additional processes in the full-physics runs cause haloes to have a lower gas density, particularly at the inner regions, and a higher gas temperature. This is caused by stellar and black hole feedbacks, which generate high-energy winds that heat up the surrounding gas and blow it out from the central regions. Such competing effects are approximately
cancelled out in the product of the gas density with the gas temperature, as in Eqs.~(\ref{eq:ysz_obs}) and (\ref{eq:yx_obs}). 

In order to test the mappings predicted by the effective density approach, given by Eqs.~(\ref{eq:ysz_mapping}) and (\ref{eq:yx_mapping}), the $Y_{\rm SZ}$ and $Y_{\rm X}$ values measured for $f(R)$ gravity have been multiplied by the mass ratio $M_{\rm dyn}^{f(R)}/M_{\rm true}^{f(R)}$. For the non-radiative plots in Figs.~\ref{fig:Ysz_scaling_relation} and \ref{fig:Yx_scaling_relation}, there is excellent agreement between this rescaled data and GR. There is also a strong agreement for the higher-mass full-physics data. For the mass range $13.2<\log_{10}(M_{\rm dyn}(<R_{500}^{\rm eff})M_{\odot}^{-1})<13.5$, however, there is some disparity of $\lesssim20\%$ between the rescaled F5 data and GR.

A similar level of accuracy is observed for the mappings given by Eqs.~(\ref{eq:ysz_mapping_true}) and (\ref{eq:yx_mapping_true}), which are predicted by the true density approach. To test these, $Y_{\rm SZ}$ and $Y_{\rm X}$ are divided by $M_{\rm dyn}^{f(R)}/M_{\rm true}^{f(R)}$ to generate rescaled data for F6 and F5. Again, this data shows excellent agreement, within a few percent, with GR for the non-radiative simulations and the high-mass end of the full-physics data. But for the lower-mass full-physics data there is a significant disagreement between F5 and GR of up to $\sim30\%$, which is higher than for the effective density rescalings. 

The disparities found in the low-mass full-physics data can be explained using Fig.~\ref{fig:profiles}. Looking at the full-physics profiles for the true density mass bins, it is observed that for a large portion of the inner halo regions the gas density is higher in F5 than in GR. These profiles converge at $r\approx10^{2.5}{\rm kpc}$ for both mass bins. The high-mass haloes have higher overall radius $R_{500}^{\rm true}$, which means that the profiles are converged for a large portion of the outer radii. This means the disparities at lower radii have a negligible overall contribution to the integrals for $Y_{\rm SZ}$ and $Y_{\rm X}$. But for the lower-mass haloes, the profiles are converged only for a small portion of the overall radius range, causing $Y_{\rm SZ}$ and $Y_{\rm X}$ to be greater in F5 than in GR. A similar reasoning can be used to explain the disparities for the results with the effective density rescaling, although the difference in agreement at the outer regions for each mass bin is not quite as substantial here, which explains why the full-physics data for the effective density approach shows less overall deviation between F5 and GR in Figs.~\ref{fig:Ysz_scaling_relation} and \ref{fig:Yx_scaling_relation}.

The difference between the full-physics $f(R)$ and GR scaling relations at low mass is likely to be explained by baryonic processes such as feedback which are absent in the non-radiative simulations. However, in studies of clusters, these lower-mass groups are of less interest. The strong agreement at the higher masses is therefore very encouraging for our framework to constrain $f(R)$ gravity using the high-mass end of the halo mass function.

\subsubsection{X-ray luminosity scaling relations}
\label{sec:results:scaling_relations:lx}

\begin{figure*}
\centering
\includegraphics[width=1.0\textwidth]{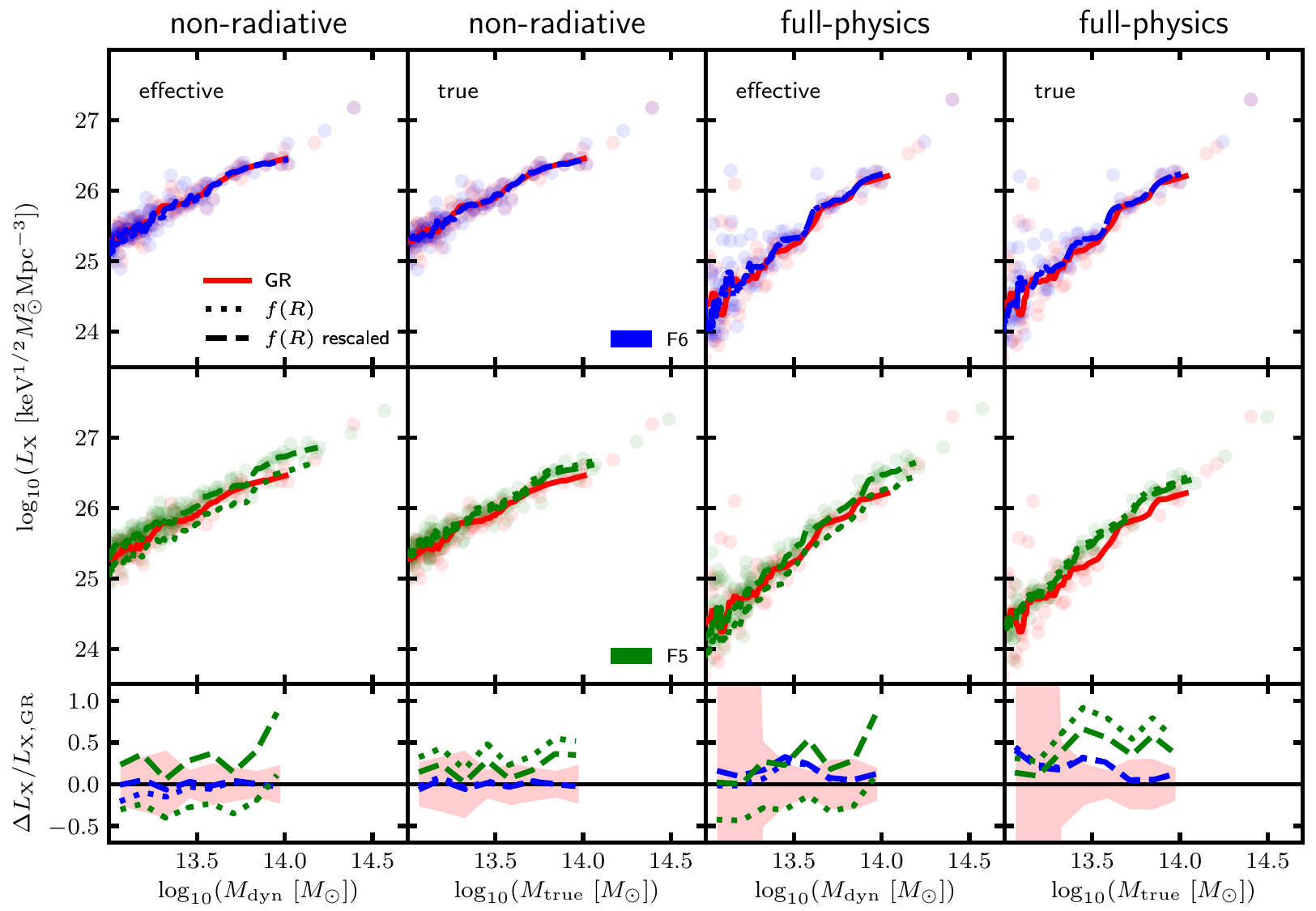}
\caption[X-ray luminosity as a function of mass for haloes in $f(R)$ gravity and GR, using the full-physics and non-radiative \textsc{shybone} simulations.]{X-ray luminosity plotted as a function of the mass for FOF haloes from the non-radiative and full-physics \textsc{shybone} simulations (see Sec.~\ref{sec:simulations:scaling_relations}). Apart from the X-ray luminosity being shown rather than the mass-weighted temperature, the layout of this figure is otherwise identical to Fig.~\ref{fig:T_gas}. The rescalings of the $f(R)$ $L_{\rm X}$ data are carried out as described in Sec.~\ref{sec:background:scaling_relations}.}
\label{fig:Lx_scaling_relation}
\end{figure*}

The results for the X-ray luminosity scaling relations are shown in Fig.~\ref{fig:Lx_scaling_relation}. Compared with the temperature, $Y_{\rm SZ}$ and $Y_{\rm X}$ data, the X-ray luminosity is much more scattered, with particularly large dispersion in lower-mass haloes and $\sim25\%$ scatter at higher masses. One explanation for this is that the X-ray luminosity, defined in Eq.~(\ref{eq:lx_obs}), depends on the gas density to power two. This means that the inner regions of the group, which have a higher gas density than the outer regions, have a greater overall contribution to the $L_{\rm X}$ integral than for the other observables discussed in this chapter. The inner halo regions are expected to be more impacted by unpredictable dynamical processes during cluster formation, including halo mergers. In particular, they are more prone to gas heating and blowing-out of gas caused by feedback mechanisms. While the competing effects of these processes on the gas density and gas temperature roughly cancel for the $Y_{\rm SZ}$ and $Y_{\rm X}$ observables, this is not the case for $L_{\rm X}$, which depends on the gas density to power two and the gas temperature to power half. This results in a number of significant outliers, as can be seen in the full-physics data of Fig.~\ref{fig:Lx_scaling_relation}. 

For the mapping defined using the effective density field, given by Eq.~(\ref{eq:lx_mapping}), it has been expected that the GR X-ray luminosity should be equal to the $f(R)$ gravity value multiplied by the factor $\left(M_{\rm dyn}^{f(R)}/M_{\rm true}^{f(R)}\right)^2$. From Fig.~\ref{fig:Lx_scaling_relation}, the rescaled data in F5 appears to be higher than in GR by $\sim30\%$ on average for both the non-radiative and the full-physics simulations. A similar level of deviation is also observed for the true density results, where Eq.~(\ref{eq:lx_mapping_true}) predicts that the GR and $f(R)$ gravity X-ray luminosity should be equal after the values in $f(R)$ gravity are divided by the factor $\left(M_{\rm dyn}^{f(R)}/M_{\rm true}^{f(R)}\right)^{1/2}$. Again, the rescaled F5 X-ray luminosity is significantly greater than in GR on average. 

As for the $Y_{\rm SZ}$ and $Y_{\rm X}$ mappings, the disparity found here can be explained by looking at the gas density profiles in Fig.~\ref{fig:profiles}. For both the non-radiative and full-physics data, the gas density in the inner halo regions is greater for F5 (with appropriate rescaling applied) than for GR. Because the inner regions have a greater contribution to the X-ray luminosity than for other proxies, as described above, this causes these differences in the inner regions to become significant overall, even for the non-radiative data for which the F5 and GR profiles are converged above a lower radius. This results in the general offset for the full range of masses as shown in Fig.~\ref{fig:Lx_scaling_relation}. As described above, the X-ray luminosity is also more strongly influenced by feedback processes, which can further increase the offset between F5 and GR if the feedback behaves differently in these two models.

Our observation that the mappings have a poorer performance for the X-ray luminosity than for the other proxies is consistent with \citet{He:2015mva}, who observed a disparity of $\sim13\%$. Unless further corrections are applied to account for the unpredictable effects of feedback in the mappings, therefore, the X-ray luminosity is unlikely to be a reliable proxy for mass determination in accurate cluster tests of $f(R)$ gravity.

\subsubsection{Further comments}

As described in Sec.~\ref{sec:methods:scaling_relations:groups}, we have excluded the core region of $r<0.15R_{500}$ when calculating the thermal properties of our simulated FOF groups. We have also experimented excluding core regions of size $r<0.1R_{500}$ and $r<0.2R_{500}$: for the scaling relations $Y_{\rm SZ}$-$M$, $Y_{\rm X}$-$M$ and $\bar{T}_{\rm gas}$-$M$, the relative differences between $f(R)$ gravity and GR are barely affected; on the other hand, the effect is larger for the $L_{\rm X}$-$M$ scaling relation, because $L_{\rm X}$ is more sensitive to the inner halo regions than the other proxies. However, the $L_{\rm X}$-$M$ relation is not ideal for reliable tests of gravity anyway, as noted above, and is mainly included in this thesis for completeness and for comparison with the other relations. Even if the entire core is included in the calculations, we have found that the effect on the $Y_{\rm SZ}$-$M$ and $Y_{\rm X}$-$M$ results is still very small, providing further confirmation that these relations can be used for reliable tests of gravity.

The scatter of the full-physics GR scaling relations shown in Figs.~\ref{fig:T_gas}--\ref{fig:Lx_scaling_relation} is typically higher than the scatter quoted in recent studies that have also used simulations which include star formation, cooling and stellar and black hole feedback. For example, we observe a root-mean-square dispersion of $\sim19\%$ for the $Y_{\rm SZ}$-$M$ and $Y_{\rm X}$-$M$ relations, while \citet{Brun:2016jtk} and \citet{Truong:2016egq} reported $\sim10\%$ and $\sim15\%$, respectively. This is likely to be caused by our restricted halo population (modified gravity simulations are much more computationally expensive than their standard gravity counterparts which limits the affordable box-size and resolution), which contains a large number of low-mass groups that are more susceptible to feedback. Our results suggest that the $\bar{T}_{\rm gas}$-$M$ relation has the lowest scatter and the $L_{\rm X}$-$M$ relation has the highest scatter, and this is consistent with the above works.

\subsection{Halo mass ratio calculation}
\label{sec:results:scaling_relations:mass_ratio}

For the results discussed in Sec.~\ref{sec:results:scaling_relations:scaling_relations}, the rescalings to the observables in $f(R)$ gravity have been computed using direct measurements of the true mass and the dynamical mass from the simulations. However, for studies of clusters using real observations, measurements of both the true mass and dynamical mass are unlikely to be available. In this case, our analytical model for the ratio of the dynamical mass to the true mass, given by Eq.~(\ref{eq:mdyn_enhancement}), can be used. 

\begin{figure*}
\centering
\includegraphics[width=1.0\textwidth]{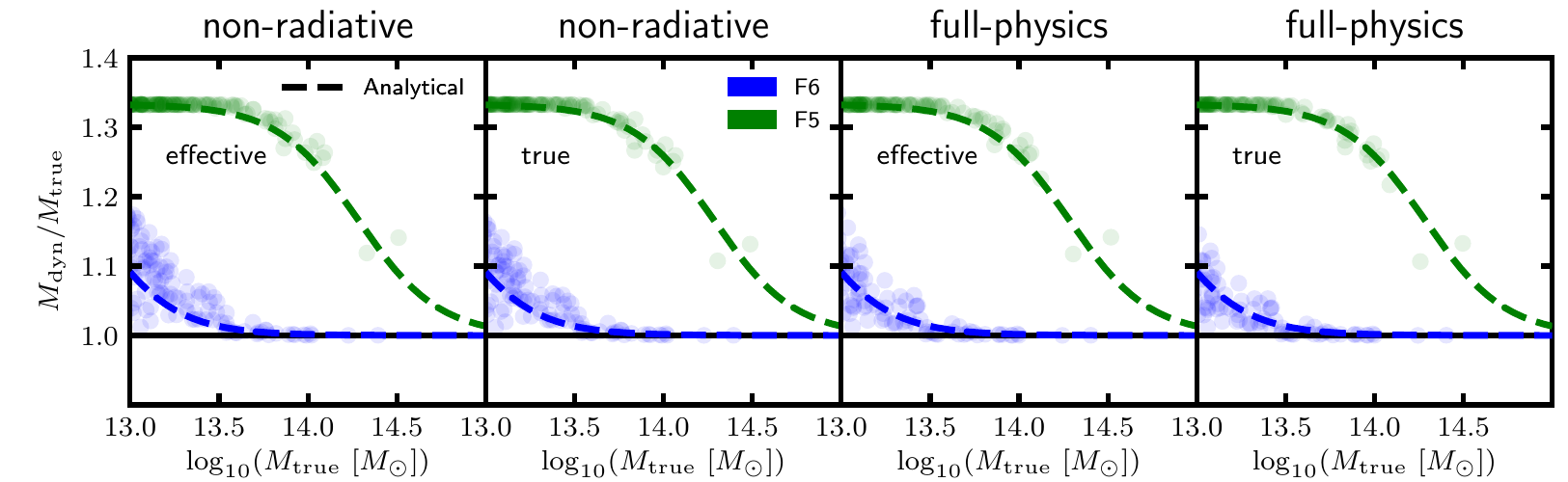}
\caption[Dynamical mass enhancement in $f(R)$ gravity as a function of the true mass for haloes from the non-radiative and full-physics \textsc{shybone} simulations.]{Ratio of the dynamical mass to the true mass of haloes plotted as a function of the true mass. The data points correspond to FOF haloes from the non-radiative and full-physics \textsc{shybone} simulations (see Sec.~\ref{sec:simulations:scaling_relations} for details). Data is included for F6 (\textit{blue}) and F5 (\textit{green}). For the data labelled `effective', the dynamical and true mass have been measured within the radius $R_{500}^{\rm eff}$ (defined in Sec.~\ref{sec:eff_approach}), while the radius $R_{500}^{\rm true}$ (defined in Sec.~\ref{sec:true_approach}) has been used for the data labelled `true'. Analytical predictions for the mass enhancement have been computed using Eq.~(\ref{eq:mdyn_enhancement}) and are shown (\textit{dashed curves}) for each model.}
\label{fig:mdyn_mtrue}
\end{figure*}

This model was calibrated using a suite of DMO simulations. To check how it performs for data that includes full physics, we have plotted the model predictions on top of actual measurements of the dynamical mass enhancement for the FOF groups in the \textsc{shybone} simulations. This is shown in Fig.~\ref{fig:mdyn_mtrue}. Data for both the effective density and true density catalogues have been included, for which the dynamical and true halo masses have been measured within $R_{500}^{\rm eff}$ and $R_{500}^{\rm true}$, respectively. We have made use of all available data with $M_{\rm true}>10^{13}M_{\odot}$, including haloes with $M_{\rm true}\sim10^{14.5}M_{\odot}$. These results indicate that there is very good agreement between the analytical predictions and the actual data for both F6 and F5, regardless of the hydrodynamical scheme that is employed. Interestingly, even though Eq.~(\ref{eq:mdyn_enhancement}) was originally calibrated using measurements of the dynamical and true mass within $R_{500}^{\rm true}$, it still performs very well for data measured within $R_{500}^{\rm eff}$, which is typically a higher radius.

\begin{figure*}
\centering
\includegraphics[width=1.0\textwidth]{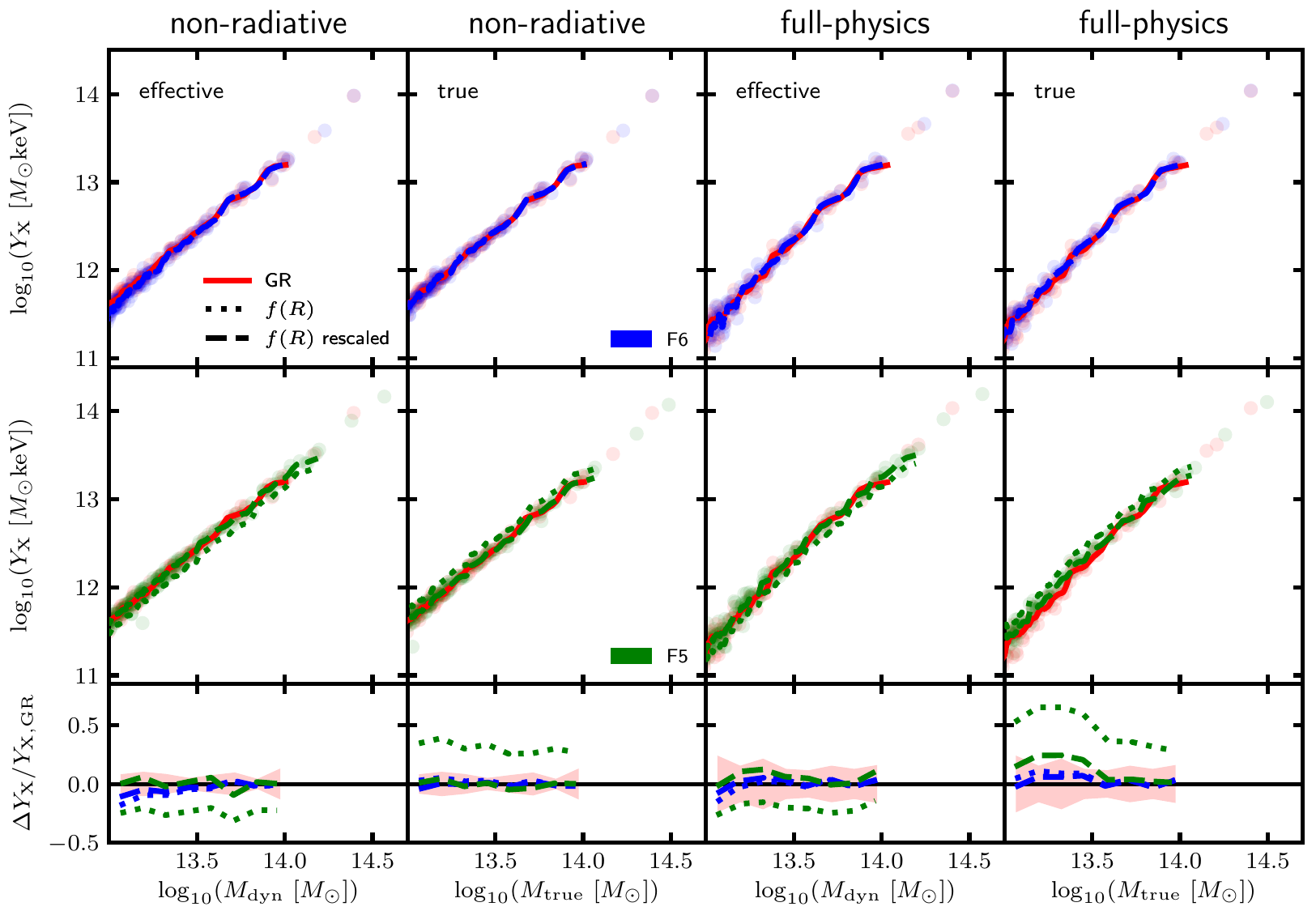}
\caption[Same as Fig.~\ref{fig:Yx_scaling_relation} but using Eq.~(\ref{eq:mdyn_enhancement}) for the mapping between the $f(R)$ and GR scaling relations.]{X-ray analogue of the Compton $Y$-parameter plotted as a function of the mass for FOF haloes from the non-radiative and full-physics \textsc{shybone} simulations (see Sec.~\ref{sec:simulations:scaling_relations}). The layout and format of this figure are identical to those of Fig.~\ref{fig:Yx_scaling_relation}. The results shown here are also mostly the same as for Fig.~\ref{fig:Yx_scaling_relation}, however the rescalings of $Y_{\rm X}$ in the $f(R)$ gravity data have been generated using the analytical tanh formula given by Eq.~(\ref{eq:mdyn_enhancement}).}
\label{fig:Yx_scaling_relation_TANH}
\end{figure*}

We have also tested the mappings of Eqs.~(\ref{eq:yx_mapping}) and (\ref{eq:yx_mapping_true}) for the $Y_{\rm X}$ parameter, with Eq.~(\ref{eq:mdyn_enhancement}) used to compute the required rescalings to the $f(R)$ gravity data. This is shown in Fig.~\ref{fig:Yx_scaling_relation_TANH}. From comparing this plot with Fig.~\ref{fig:Yx_scaling_relation}, it can be seen that there is almost no difference in the rescaled data in both figures. This confirms that Eq.~(\ref{eq:mdyn_enhancement}) can be applied to derive the mappings between GR and $f(R)$ scaling relations for, at least, the mass range $10^{13}M_{\odot}<M_{500}<10^{14}M_{\odot}$. Given the very good agreement up to $10^{14.5}M_{\odot}$ shown in Fig.~\ref{fig:mdyn_mtrue}, it is expected that our formula can be applied in the cluster regime as well. However, this should be tested more rigorously using full-physics simulations with a larger box size.

\subsection{\texorpdfstring{$Y_{\rm X}$}{YX}-temperature scaling relation}
\label{sec:results:scaling_relations:yx-t_relation}

So far, we have only considered scaling relations that can be used to infer the cluster mass, which is a vital ingredient for tests of gravity that use the cluster abundance (Fig.~\ref{fig:mg_flow_chart}). However, tests of gravity can also be conducted using the relations themselves. For example, \citet{Hammami:2016npf,DelPopolo:2019oxn} used the temperature-mass scaling relation to probe screened MG models. The cluster mass can be determined using other observations, such as weak lensing, making such scaling relations observable, and our full-physics results (Figs.~\ref{fig:T_gas}--\ref{fig:Yx_scaling_relation}) confirm that the $\bar{T}_{\rm gas}$-$M$ and $Y_{\rm X}$-$M$ (and $Y_{\rm SZ}$-$M$) relations can be used as reliable probes on group and cluster scales, with differences between GR and F5 in the range 20-50\% (when no rescaling is applied).

\begin{figure*}
\centering
\includegraphics[width=1.0\textwidth]{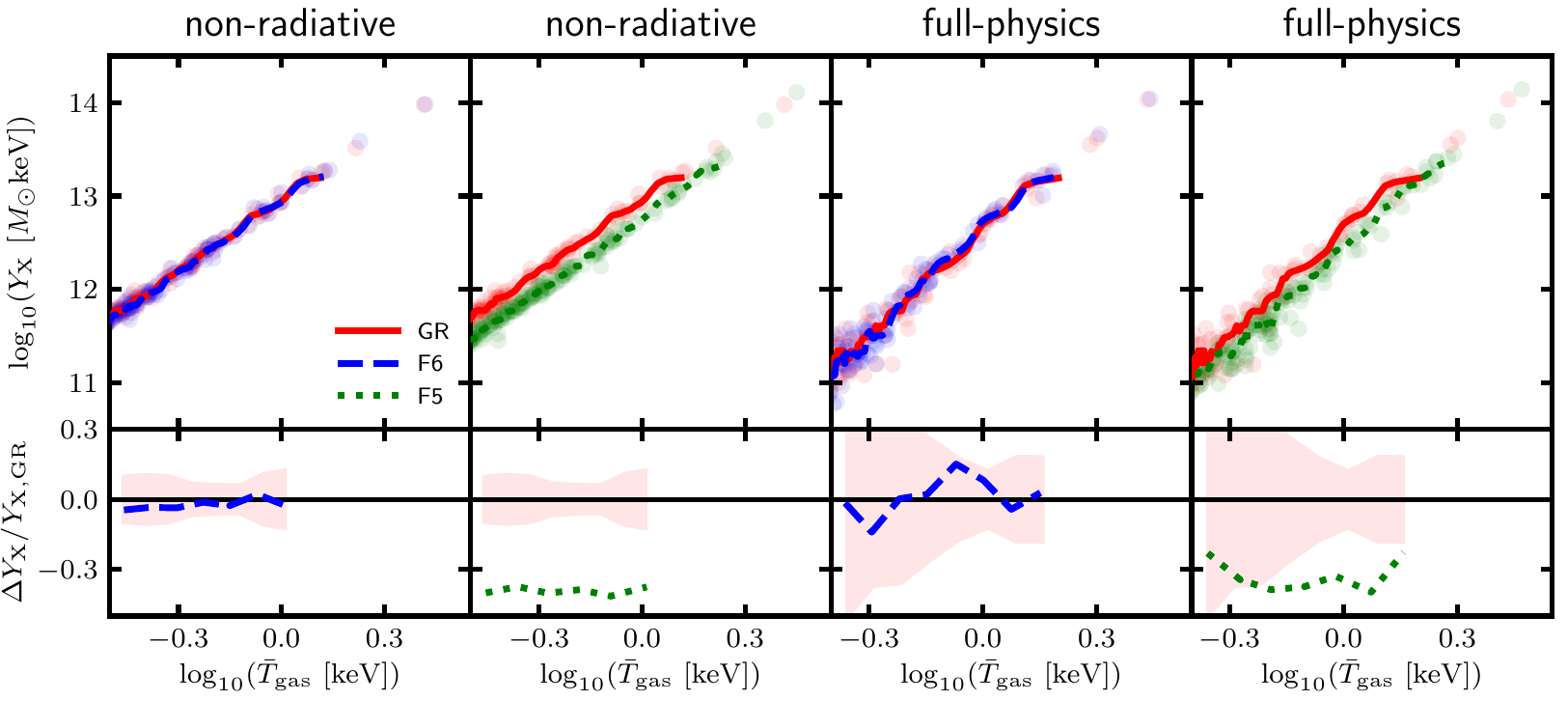}
\caption[$Y_{\rm X}$ parameter as a function of the mass-weighted temperature for haloes in $f(R)$ gravity and GR, using the full-physics and non-radiative \textsc{shybone} simulations.]{The X-ray analogue of the Compton $Y$-parameter plotted as a function of the mass-weighted temperature for FOF haloes from the non-radiative and full-physics \textsc{shybone} simulations (see Sec.~\ref{sec:simulations:scaling_relations}). The curves correspond to the median $Y_{\rm X}$ versus the mean logarithm of the temperature computed within a moving window of fixed size equal to 10 haloes. Data has been included for GR (\textit{red solid lines}) together with the F6 (\textit{blue lines}) and F5 (\textit{green lines}) $f(R)$ gravity models. $Y_{\rm X}$ and the temperature have been computed within the radius $R_{500}^{\rm true}$ (see Sec.~\ref{sec:true_approach}). Data points are displayed, with each point corresponding to a GR halo (\textit{red points}), or to a halo in F6 (\textit{blue points}) or F5 (\textit{green points}). \textit{Bottom row}: the smoothed relative difference between the $f(R)$ gravity and GR curves in the above plots; the red shaded regions indicate the size of the halo scatter in GR.}
\label{fig:Yx-T_scaling_relation}
\end{figure*}

However, scaling relations which do not involve the mass can also be modelled and used. In Fig.~\ref{fig:Yx-T_scaling_relation}, we show the scaling relation between the $Y_{\rm X}$ parameter and the gas temperature, where both observables have been computed within the radius $R_{500}^{\rm true}$. A significant relative difference of 30-40\% is observed between the GR and F5 models for both the non-radiative and full-physics data, indicating that this relation can offer a powerful test of gravity using group- and cluster-sized objects. A key advantage of using the $Y_{\rm X}$--$\bar{T}_{\rm gas}$ (or $Y_{\rm SZ}$--$\bar{T}_{\rm gas}$) scaling relation is that it does not involve measuring the cluster mass, and hence no need for mass calibrations or synergies with other observations such as weak lensing. It also has a relatively low scatter compared to, for example, the $L_{\rm X}$-$\bar{T}_{\rm gas}$ relation that was considered by \citet{arnold:2014}.

\section{Summary, Discussion and Conclusions}
\label{sec:conclusions:scaling_relations}



In this chapter, we have made use of the first full-physics simulations that have been run for both GR and $f(R)$ gravity (along with non-radiative counterparts), to study the effects of the fifth force of $f(R)$ gravity on the scaling relations between the cluster mass and four observable proxies: the gas temperature (Fig.~\ref{fig:T_gas}), the $Y_{\rm SZ}$ and $Y_{\rm X}$ parameters (Figs.~\ref{fig:Ysz_scaling_relation} and \ref{fig:Yx_scaling_relation}) and the X-ray luminosity (Fig.~\ref{fig:Lx_scaling_relation}). To understand these effects in greater detail, we have also examined the effects of both $f(R)$ gravity and full-physics on the gas density and temperature profiles (see Fig.~\ref{fig:profiles}). In doing so, we have been able to test two methods for mapping between scaling relations in $f(R)$ gravity and GR.

The first method was proposed by \citet{He:2015mva}. This proposes a set of mappings, given by Eqs.~(\ref{eq:temp_equiv_eff}) and (\ref{eq:ysz_mapping})-(\ref{eq:lx_mapping}), that can be applied to haloes whose mass and radius are measured using the effective density field (see Sec.~\ref{sec:eff_approach}). A second, new, approach is proposed in Sec.~\ref{sec:true_approach}, and predicts another set of mappings, given by Eqs.~(\ref{eq:temp_equiv_true}) and (\ref{eq:ysz_mapping_true})-(\ref{eq:lx_mapping_true}), that can be applied to haloes whose mass and radius are measured using the true density field. Both sets of mappings involve simple rescalings that depend only on the ratio of the dynamical mass to the true mass in $f(R)$ gravity. As shown by Figs.~\ref{fig:mdyn_mtrue} and \ref{fig:Yx_scaling_relation_TANH}, even with the inclusion of full-physics processes this ratio can be computed with high accuracy using our analytical tanh formula, which is given by Eq.~(\ref{eq:mdyn_enhancement}).

For the mass-weighted gas temperature and the $Y_{\rm SZ}$ and $Y_{\rm X}$ observables, we found that the F6 and F5 scaling relations, with appropriate rescaling applied (using either method discussed above), match the GR relations to within a few percent for the full mass-range tested for the non-radiative simulations. With the inclusion of full-physics effects such as feedbacks, star formation and cooling, the rescaled $Y_{\rm SZ}$ and $Y_{\rm X}$ scaling relations continue to show excellent agreement with GR for mass $M_{500}\gtrsim10^{13.5}M_{\odot}$, which includes group- and cluster-sized objects. These proxies also show relatively low scatter as a function of the cluster mass, compared with other observables. $Y_{\rm SZ}$ and $Y_{\rm X}$ are therefore likely to be suitable for accurate determination of the cluster mass in tests of $f(R)$ gravity. The mappings for the gas temperature show a very high accuracy for lower-mass objects, but show a small $\lesssim5\%$ offset between F5 and GR for higher-mass objects.

The mappings do not work as well for the X-ray luminosity $L_{\rm X}$, for which the F5 relations after rescaling are typically enhanced by $\sim30\%$ compared with GR. This is caused by the unique dependency of $L_{\rm X}$ on the gas density to power two, and the gas temperature to power half, which means that the inner halo regions have a greater contribution than for the other proxies and the competing effects of feedback on the temperature and gas density profiles are less likely to cancel out. This issue, in addition to the fact that $L_{\rm X}$ has a highly scattered correlation with the cluster mass, means that this proxy is unlikely to be suitable for cluster mass determination in tests of $f(R)$ gravity.

We also considered the $Y_{\rm X}$-$\bar{T}_{\rm gas}$ scaling relation (Fig.~\ref{fig:Yx-T_scaling_relation}), and found that this is suppressed by $30$-$40\%$ in the F5 model relative to GR. This offers a potential new and useful test of gravity with group- and cluster-sized objects which avoids the systematic uncertainties incurred from mass calibration.

We note that the box size $62 h^{-1}{\rm Mpc}$ of the simulations used in this chapter is more suited to studying galaxy-sized objects than group- or cluster-sized objects. Indeed, there are only $\sim100$ objects with $M_{500}>10^{13}M_{\odot}$ and $\sim5$-$10$ objects with $M_{500}>10^{14}M_{\odot}$ in the simulations, making it impossible to test the mappings discussed in this chapter for the most massive galaxy clusters to be observed. In Chapter \ref{chapter:baryonic_fine_tuning}, we will present a re-calibrated full-physics model which can be used to run larger simulations that can be used to reliably probe halo masses up to $M_{500}\sim10^{15}M_{\odot}$.

Our results also provide insights into the viability of extending cluster tests of gravity to the group-mass regime. An advantage of using lower-mass objects is that these objects can be unscreened (or partially screened) even for weaker $f(R)$ models, offering the potential for tighter constraints using data from ongoing and upcoming SZ and X-ray surveys \citep[e.g.,][]{erosita,Planck_SZ_cluster} which are now entering this regime. On the other hand, as we have seen above, the scatter induced by feedback mechanisms becomes more significant in group-sized haloes, which means that additional work will need to be conducted to characterise this effect and to understand its impact on model tests.

Finally, we note that our parameter $p_2$, which is used to compute the ratio of the dynamical mass to the true mass, depends only on the quantity $|\bar{f}_R|/(1+z)$, and not on the model parameters $n$ and $f_{R0}$ of HS $f(R)$ gravity. This dependence was derived by using the thin-shell model (Chapter \ref{chapter:mdyn}), which does not depend on the details of the $f(R)$ model. We therefore expect our scaling relation mappings to perform similarly for any combination of the HS $f(R)$ parameters, and potentially other chameleon-type or thin-shell-screened models. However, due to the high computational cost of running full-physics simulations of $f(R)$ gravity and other models, we do not seek to confirm this conjecture in this thesis.  
\graphicspath{{./gfx/}}

\chapter{\boldmath A self-consistent pipeline for unbiased constraints of \texorpdfstring{$f(R)$}{f(R)} gravity}
\label{chapter:constraint_pipeline}

\section{Introduction}
\label{sec:introduction_pipeline}

In Chapters \ref{chapter:mdyn}-\ref{chapter:scaling_relations}, we used a combination of DMO and full-physics simulations to model the effects of the $f(R)$ gravity fifth force on the cluster dynamical mass, the halo concentration and the observable-mass scaling relations. These models are core components of our proposed framework for $f(R)$ constraints using cluster number counts (see Fig.~\ref{fig:fr_flow_chart}): our model for the enhancement of the concentration can be used for conversions between cluster mass definitions, which is required if, for example, the theoretical predictions and observations use different spherical overdensities; and our model for the dynamical mass enhancement can be used to predict the $f(R)$ scaling relation given a GR counterpart relation, and this can be used to relate the observable mass function (${\rm d}n/{\rm d}Y$) to the theoretical mass function (${\rm d}n/{\rm d}M$). In this chapter, we will incorporate the remaining components of this framework, including the theoretical model for the HMF in $f(R)$ gravity and the MCMC sampling used to constrain $f_{R0}$, and we will test the full constraint pipeline using mock cluster catalogues generated for both GR and $f(R)$ fiducial cosmologies. 

The chapter is arranged as follows: in Sec.~\ref{sec:background_pipeline}, we provide an overview of the effects of the $f(R)$ fifth force on the cluster properties; in Sec.~\ref{sec:methods_pipeline}, we describe our MCMC constraint pipeline, including the calculation of the log-likelihood and the generation of the mocks; in Sec.~\ref{sec:results_pipeline}, we present constraints using the GR and $f(R)$ mocks; then, in Sec.~\ref{sec:bias}, we highlight potential sources of bias in our pipeline; finally, we summarise our main findings in Sec.~\ref{sec:conclusions_pipeline}.

\section{Background}
\label{sec:background_pipeline}

In this section, we summarise the main effects of $f(R)$ gravity on the properties of galaxy clusters. In Sec.~\ref{sec:background_pipeline:dynamical_mass}, we recap our model for the dynamical mass enhancement and present a new model for its scatter. Then, in Secs.~\ref{sec:background_pipeline:concentration} and \ref{sec:background_pipeline:scaling_relations}, we recap our model for the concentration enhancement and our mapping between observable-mass scaling relations in $f(R)$ gravity and GR. Finally, in Sec.~\ref{sec:background_pipeline:hmf}, we outline the modelling by \citet{Cataneo:2016iav} for the $f(R)$ enhancement of the HMF.

\subsection{Dynamical mass enhancement and its scatters}
\label{sec:background_pipeline:dynamical_mass}

In Chapter \ref{chapter:mdyn}, we presented our general formula for the ratio $\mathcal{R}$ of the dynamical mass to the true mass:
\begin{equation}
\mathcal{R} = \frac{M^{f(R)}_{\rm dyn}}{M^{f(R)}_{\rm true}} = \frac{7}{6}-\frac{1}{6}\tanh\left(p_1\left[\log_{10}\left(M^{f(R)}_{\rm true}M_{\odot}^{-1}h\right)-p_2\right]\right).
\label{eq:pipeline:mdyn_enhancement}
\end{equation}
We showed that $p_1$ is approximately constant, with best-fit value $2.21\pm0.01$, while the parameter $p_2$ closely follows the following physically motivated linear relation:
\begin{equation}
    p_2=(1.503\pm0.006)\log_{10}\left(\frac{|\bar{f}_R(z)|}{1+z}\right)+(21.64\pm0.03).
\label{eq:pipeline:p2}
\end{equation}

\begin{figure}
\centering
\includegraphics[width=0.7\textwidth]{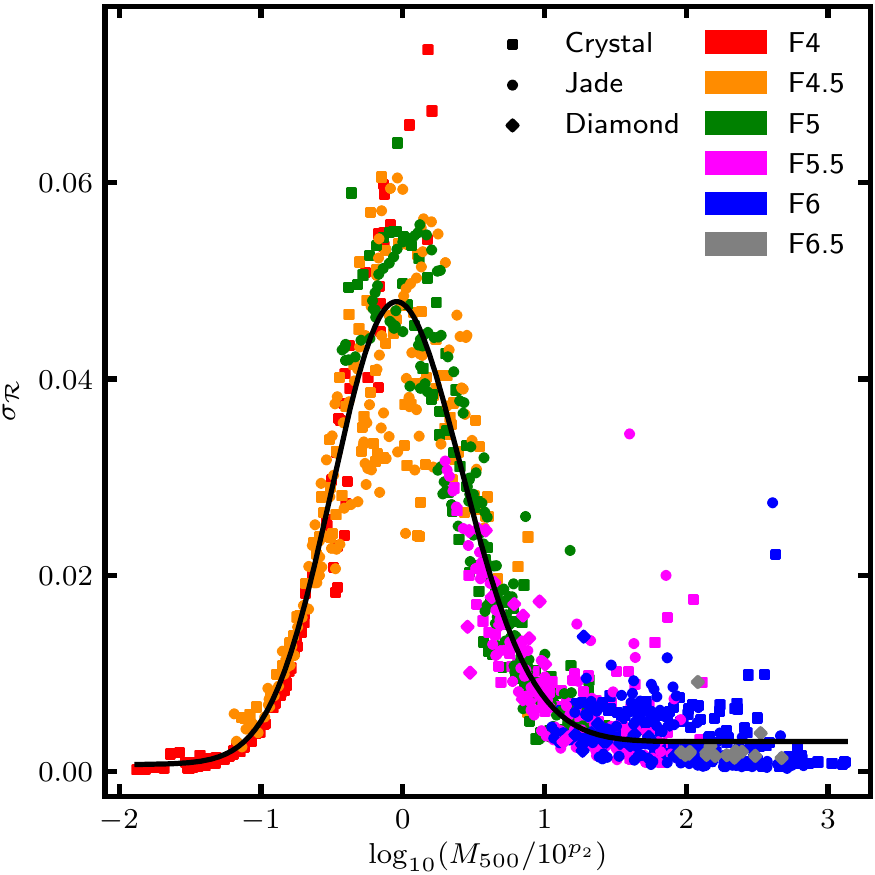}
\caption[Root-mean-square scatter of the dynamical mass enhancement in $f(R)$ gravity as a function of the rescaled halo mass $\log_{10}(M_{500}/10^{p_2})$.]{Root-mean-square scatter in the dynamical mass enhancement as a function of the rescaled halo mass $\log_{10}(M_{500}/10^{p_2})$, where $p_2$ is given by Eq.~(\ref{eq:pipeline:p2}). The data points, which correspond to mass bins spanning $10^{11}h^{-1}M_{\odot}\lesssim M_{500}\lesssim 10^{15}h^{-1}M_{\odot}$, have been generated using the DMO simulations Crystal (\textit{squares}), Jade (\textit{circles}) and Diamond (\textit{diamonds}), which are described in Chapter \ref{chapter:mdyn}. The data spans redshifts $0\leq z\leq1$ and includes present-day scalar field amplitudes $|f_{R0}|=10^{-6.5}$ (\textit{grey}), $10^{-6}$ (\textit{blue}), $10^{-5.5}$ (\textit{magenta}), $10^{-5}$ (\textit{green}), $10^{-4.5}$ (\textit{orange}) and $10^{-4}$ (\textit{red}). The solid line represents our best-fit model, which is given by Eq.~(\ref{eq:scatter_model}).}
\label{fig:rms_scatter}
\end{figure}

For this chapter, we have again used the simulation data from Chapter \ref{chapter:mdyn} (Table \ref{table:simulations}) to model the root-mean-square scatter of the dynamical mass enhancement, $\sigma_{\mathcal{R}}$. Our model is shown by the solid line in Fig.~\ref{fig:rms_scatter} (we provide a detailed description of this model and our fitting procedure in Appendix \ref{sec:appendix:pipeline:mdyn_scatter}). The rescaled mass, $\log_{10}(M_{500}M_{\odot}^{-1}h)-p_2\equiv\log_{10}(M_{500}/10^{p_2})$, is expected to take positive values for haloes that are screened and negative values for haloes that are unscreened. The scatter peaks for haloes that are partially screened, with $\log_{10}(M_{500}M_{\odot}^{-1}h)\sim p_2$, whereas it falls to roughly zero for lower and higher masses. Physically, this makes sense: at sufficiently high masses where \textit{all} haloes are screened and have $\mathcal{R}\approx1$, it follows that the scatter $\sigma_{\mathcal{R}}$ is very small, and a similar argument can be applied for haloes deep in the unscreened regime. Between these two regimes, the physics is more complicated, giving rise to greater dispersion in the chameleon screening; for example, haloes which do not have a high enough mass to be self-screened can still be environmentally screened by nearby massive haloes.

\subsection{Halo concentration}
\label{sec:background_pipeline:concentration}

In Chapter \ref{chapter:concentration}, we used the rescaled logarithmic mass $x=\log_{10}(M_{500}/10^{p_2})$ to calibrate the following formula for the enhancement of the halo concentration $c_{200}$:
\begin{equation}
\begin{split}
\log_{10}\left|\frac{c}{c_{\rm GR}}\right|_{200} = &\frac{1}{2}\left(\frac{\lambda}{\omega_{\rm s}}\phi(x')\left[1+\rm{erf}\left(\frac{\alpha x'}{\sqrt[]{2}}\right)\right]+\gamma\right)\\
&\times(1-\tanh\left(\omega_{\rm t}\left[x+\xi_{\rm t}\right]\right)),
\end{split}
\label{eq:pipeline:c_model}
\end{equation}
where $x'=(x-\xi_{\rm s})/\omega_{\rm s}$, $\phi(x')$ is the normal distribution and $\rm{erf}(\alpha x'/\sqrt{2})$ is the error function. The parameters have best-fit values $\lambda=0.55$, $\omega_{\rm s}=1.7$, $\xi_{\rm s}=-0.27$, $\alpha=-6.5$, $\gamma=-0.07$, $\omega_{\rm t}=1.3$ and $\xi_{\rm t}=0.1$.

\subsection{Observable-mass scaling relations}
\label{sec:background_pipeline:scaling_relations}

In Chapter \ref{chapter:scaling_relations}, we used the \textsc{shybone} simulations to verify a set of proposed mappings between the $f(R)$ scaling relations and their GR power-law counterparts. Here, we will only recap the `true density' rescaling for the $Y_{\rm SZ}(M)$ relation, which is used in our constraint pipeline. For haloes in $f(R)$ gravity and GR that have the same true mass, $M_{\rm true}^{f(R)}=M^{\rm GR}$, we showed:
\begin{equation}
    Y_{\rm SZ}^{f(R)}\left(M_{\rm true}^{f(R)}\right) \approx \frac{M_{\rm dyn}^{f(R)}}{M_{\rm true}^{f(R)}}Y_{\rm SZ}^{\rm GR}\left(M^{\rm GR}=M_{\rm true}^{f(R)}\right).
    \label{eq:pipeline:ysz_mapping_true}
\end{equation}
In this case, the total gravitational potential of the $f(R)$ haloes is enhanced by a factor of $M_{\rm dyn}^{f(R)}/M_{\rm true}^{f(R)}$ compared to the GR haloes. The temperature is then enhanced by the same amount, giving rise to this factor in Eq.~(\ref{eq:pipeline:ysz_mapping_true}). We showed that this mapping holds for halo masses $M_{500}\gtrsim10^{13.5}M_{\odot}$.

\subsection{Halo mass function}
\label{sec:background_pipeline:hmf}

In this section, we will outline the \citet{Cataneo:2016iav} model for the $f(R)$ enhancement of the HMF, which we have adopted for our constraint pipeline. This is computed using the \citet{Sheth:1999mn} prescription of the HMF:
\begin{equation}
    n_{\rm ST} \equiv \frac{{\rm d}n}{{\rm d}\ln M} = \frac{\bar{\rho}_{\rm M}}{M}\frac{{\rm d}\ln\nu}{{\rm d}\ln M}\nu f(\nu),
    \label{eq:st_hmf}
\end{equation}
where the multiplicity function $\nu f(\nu)$ is given by:
\begin{equation}
    \nu f(\nu) = A\sqrt{\frac{2}{\pi}a\nu^2}\left[1+(a\nu^2)^{-p}\exp\left(-\frac{a\nu^2}{2}\right)\right].
    \label{eq:multiplicity}
\end{equation}
For the parameters $A$, $a$ and $p$, \citet{Cataneo:2016iav} used the fits by \citet{Despali:2015yla}, which extend the \citet{Sheth:1999mn} HMF to be a function of generic halo overdensity $\Delta$. For the latter, \citet{Cataneo:2016iav} used value $300\Omega_{\rm M}(z)$ (i.e., here the halo mass $M$ is $M_{300{\rm m}}$). The peak height $\nu$ is given by:
\begin{equation}
    \nu = \frac{\delta_{\rm c}}{\sigma(M,z)},
    \label{eq:nu}
\end{equation}
where $\delta_{\rm c}$ is the linearly extrapolated threshold density for spherical collapse and $\sigma(M,z)$ is the linear root-mean-square fluctuation of the matter density within spheres of mass $M$ containing an average density of $\bar{\rho}_{\rm M}(z)$. The latter can be computed using the $\Lambda$CDM linear power spectrum (for both GR and $f(R)$ gravity) with the publicly available code \textsc{camb} \citep{Lewis:1999bs}. 

The $f(R)$ effects are incorporated through $\delta_{\rm c}$: in GR, this is given by:
\begin{equation}
    \delta_{\rm c}^{\rm GR}(z)\approx\frac{3}{20}(12\pi)^{\frac{2}{3}}\left[1 + 0.0123\log_{10}\Omega_{\rm M}(z)\right],
    \label{eq:delta_c_gr}
\end{equation}
while in $f(R)$ gravity it can be expressed as:
\begin{equation}
    \delta_{\rm c}^{\rm eff}(M, z)\equiv\epsilon(M, z)\times\delta_{\rm c}^{f(R)}(M, z).
    \label{eq:delta_eff}
\end{equation}
The function $\delta_{\rm c}^{f(R)}(M,z)$ is the prediction of the linearly extrapolated threshold density for spherical collapse in $f(R)$ gravity. This treats haloes and their surrounding environment as co-centred spherically symmetric top-hat overdensities (note the environment can be underdensities) which are co-evolved from an initial time to the time of collapse. This procedure, which is based on the method developed by \citet{LE2012,Lombriser:2013wta}, takes into account both the mass-dependent self-screening and the environmental screening of the fifth force. However, while giving qualitatively correct predictions, the method is unable to very accurately capture the complex nonlinear dynamics of structure formation in $f(R)$ gravity. This limitation is accounted for using the correction factor $\epsilon(M,z)$, which \citet{Cataneo:2016iav} modelled and fitted using DMO simulations. Their best-fit model can accurately reproduce the $f(R)$ enhancement of the HMF for redshifts $0.0\leq z\leq0.5$ and field strengths $10^{-6}\leq |f_{R0}|\leq10^{-4}$.

\begin{figure}
\centering
\includegraphics[width=0.6\textwidth]{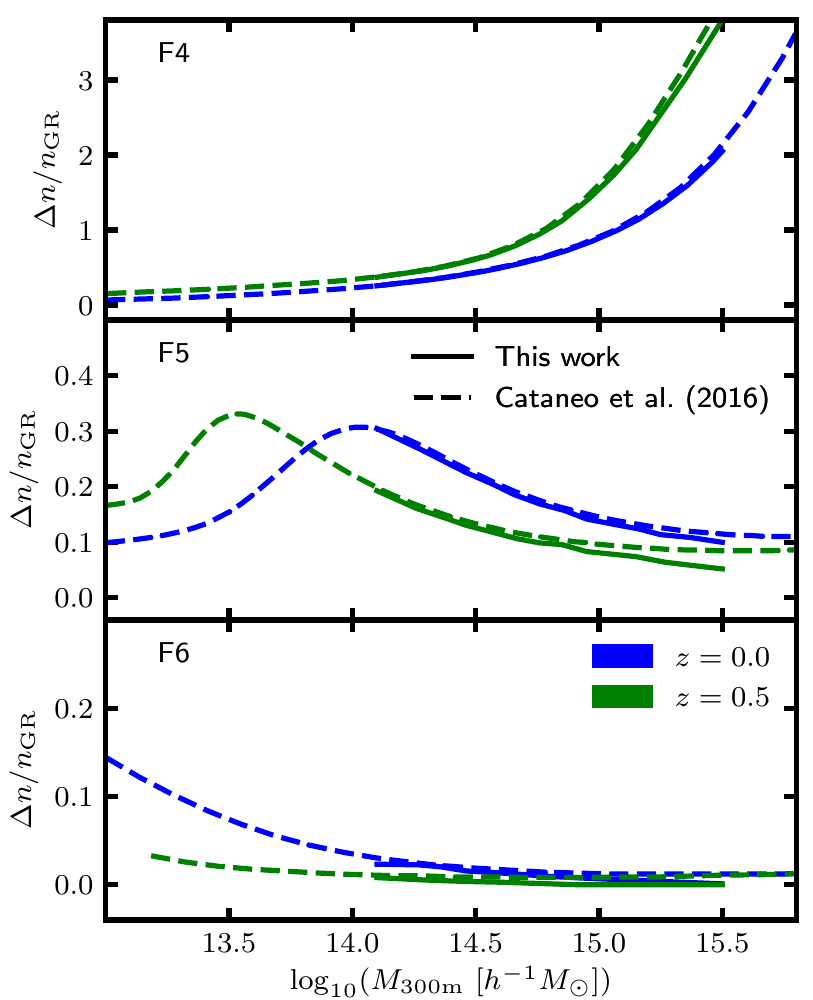}
\caption[Predicted halo mass function enhancement in $f(R)$ gravity with respect to GR.]{Halo mass function enhancement in $f(R)$ gravity with respect to GR as a function of the halo mass. The solid lines show the predictions from our 5D interpolation of $\delta_{\rm c}^{f(R)}$ (see Sec.~\ref{sec:background_pipeline:hmf}) and the dashed lines show the results from \citet{Cataneo:2016iav}. The predictions have been generated using the WMAP9 cosmological parameters and $f(R)$ models F4 (\textit{top row}), F5 (\textit{middle row}) and F6 (\textit{bottom row}), at redshifts 0 (\textit{blue lines} and 0.5 (\textit{green lines}).}
\label{fig:hmf_enhancement}
\end{figure}

For this thesis, we have evaluated $\delta_{\rm c}^{f(R)}$ on a grid of $M$, $z$, $\Omega_{\rm M}$, $\sigma_8$ and $f_{R0}$, and obtained the relation $\delta_{\rm c}^{f(R)}(M,z,\Omega_{\rm M},\sigma_8,f_{R0})$ using 5D interpolation. For a given set of cosmological and $f(R)$ parameters, we can use this to predict $\delta_{\rm c}^{f(R)}(M,z)$, which can then be used to predict $\delta_{\rm c}^{\rm eff}(M,z)$ using the model for $\epsilon(M,z)$ taken from \citet{Cataneo:2016iav}. The $f(R)$ enhancement of the HMF is given by the ratio between $n_{\rm ST}|_{f(R)}$ and $n_{\rm ST}|_{\rm GR}$, which are evaluated using $\delta_{\rm c}=\delta_{\rm c}^{\rm eff}$ and $\delta_{\rm c}=\delta_{\rm c}^{\rm GR}$, respectively.

For illustrative purposes, we show, in Fig.~\ref{fig:hmf_enhancement}, our predictions of the HMF enhancement as a function of the halo mass for F6, F5 and F4 at redshifts 0.0 and 0.5. We also show the predictions from \citet{Cataneo:2016iav} as a comparison. Both sets of predictions assume the 9-year WMAP cosmological parameter estimates \citep{2013ApJS..208...19H}. There are some small differences between the two sets of predictions, which are likely caused by subtle differences in the calculations of $\delta_{\rm c}^{f(R)}$. The largest difference is observed at $M_{\rm 300m}\gtrsim10^{15}h^{-1}M_{\odot}$ for F5 at $z=0.5$. We note that the enhancement is expected to drop to zero at high masses where haloes become completely screened, therefore the behaviour of the solid lines here 
appears to be physically reasonable. We also note that we set the enhancement to zero wherever our calculations predict a negative (unphysical) enhancement. This is the case for $M_{\rm 300m}\gtrsim10^{15}h^{-1}M_{\odot}$ for F6 at $z=0.5$.

\section{Methods}
\label{sec:methods_pipeline}

In this section, we describe the main components of our constraint pipeline, including the mass function predictions (Sec.~\ref{sec:methods_pipeline:hmf}), the observable-mass scaling relation (Sec.~\ref{sec:methods_pipeline:scaling_relation}), the mock cluster catalogues (Sec.~\ref{sec:methods_pipeline:mock}) and the MCMC sampling (Sec.~\ref{sec:methods_pipeline:likelihood}).

\subsection{Theoretical mass function}
\label{sec:methods_pipeline:hmf}

In order to make constraints using cluster number counts, it is necessary to have a parameter-dependent theoretical model for the HMF. For our pipeline, we start with a GR HMF and apply the $f(R)$ enhancement using: 
\begin{equation}
    n^{f(R)} = n^{\rm GR}\times\frac{n_{\rm ST}|_{f(R)}}{n_{\rm ST}|_{\rm GR}},
    \label{eq:fr_hmf}
\end{equation}
where the ratio is computed using the \citet{Sheth:1999mn} prescription, as described in Sec.~\ref{sec:background_pipeline:hmf}, and we have chosen the \citet{Tinker:2008ff} calibration for $n^{\rm GR}$. 

Before Eq.~(\ref{eq:fr_hmf}) can be applied, the halo mass definition must be considered. As mentioned in Sec.~\ref{sec:background_pipeline:hmf}, the model for the ratio in Eq.~(\ref{eq:fr_hmf}) was calibrated by \citet{Cataneo:2016iav} using overdensity $\Delta=300\Omega_{\rm M}(z)$; however, with the framework in Fig.~\ref{fig:fr_flow_chart}, we hope to use data from SZ and X-ray surveys, which often measure cluster properties with overdensity $500$. Therefore, it is necessary to convert the HMF between these two definitions. 

\begin{figure}
\centering
\includegraphics[width=0.7\textwidth]{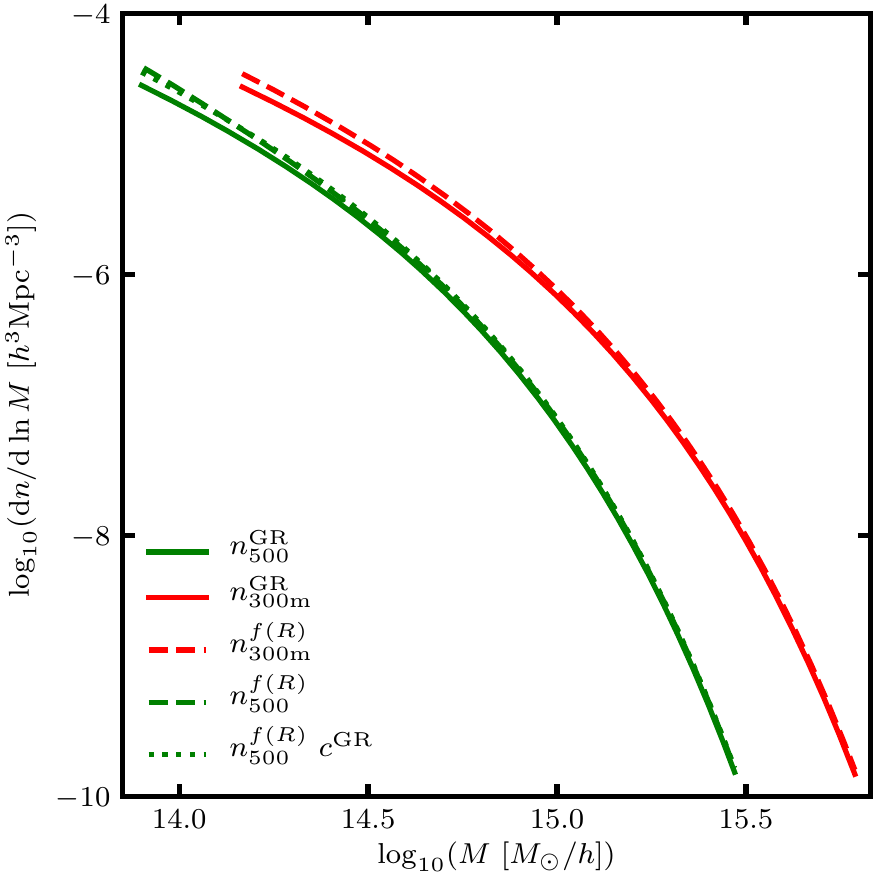}
\caption[Halo mass function in GR and F5 for spherical overdensities $500$ and $300\Omega_{\rm M}(z)$.]{Halo mass function in GR (\textit{solid lines}) and F5 (\textit{dashed lines}), with the mass defined using spherical overdensities $500$ (\textit{green lines}) and $300\Omega_{\rm M}(z)$ (\textit{dashed lines}). The mass conversions and $f(R)$ enhancement have been applied as described in Sec.~\ref{sec:methods_pipeline:hmf}; the green dotted line shows the F5 HMF prediction that results from neglecting the $f(R)$ enhancement of the halo concentration in the mass conversion $300\Omega_{\rm M}(z)\rightarrow500$.}
\label{fig:hmf_conversion}
\end{figure}

In Fig.~\ref{fig:hmf_conversion}, we show each step of the mass conversion procedure for the F5 model at $z=0$. We start with the \citet{Tinker:2008ff} HMF with overdensity $500$ ($n_{500}^{\rm GR}$), which we compute using the python package \textsc{hmf} \citep{Murray:2013qza}, and convert this to overdensity $300\Omega_{\rm M}(z)$ ($n_{\rm 300m}^{\rm GR}$) using the \citet{Duffy:2008pz} concentration-mass-redshift relation. We then apply the $f(R)$ enhancement using Eq.~(\ref{eq:fr_hmf}) to get $n_{\rm 300m}^{f(R)}$. Finally, to convert this back to overdensity $500$ ($n_{500}^{f(R)}$), we use the $f(R)$ concentration-mass-redshift relation, which is computed by applying the concentration enhancement, given by Eq.~(\ref{eq:pipeline:c_model}), to the \citet{Duffy:2008pz} relation. We also show, with the dotted green line, the prediction with the concentration enhancement neglected; the effect here is quite small, since cluster-size haloes are mostly screened in F5. For further details of the formulae used to convert the halo mass and the HMF from one mass definition to another, we refer the reader to Appendix \ref{sec:appendix:pipeline:mass_conversions}.

The final result $n_{500}^{f(R)}(M_{500})$ provides the theoretical prediction of the cluster abundance in $f(R)$ gravity. This is computed following the above steps for each set of parameter values sampled by our MCMC pipeline. We note that our mass conversions are evaluated assuming an NFW profile, which has also been used in previous cluster tests of $f(R)$ gravity \citep[e.g.,][]{PhysRevD.92.044009}. However, this may not provide an accurate description for haloes that are not dynamically relaxed and it does not account for the effects of baryons on the total mass profile. We plan to investigate the latter effect using clusters identified from the realistic full hydrodynamical simulations in $f(R)$ gravity described in Chapter \ref{chapter:baryonic_fine_tuning}. However, we remark here that the main use of the concentration-mass relation in our pipeline is to perform mass conversions as described above, and so it would not be strictly needed if a theoretical HMF for the required mass definition $M_\Delta$ ($M_{500}$ in this case) is already in place.

\subsection{Observable-mass scaling relation}
\label{sec:methods_pipeline:scaling_relation}

As discussed in Sec.~\ref{sec:background_pipeline:scaling_relations}, the $f(R)$ scaling relation can be computed by simply rescaling a GR relation using our model for the dynamical mass enhancement. For the GR relation, we adopt the power-law mapping between $Y_{\rm SZ}$ and the halo mass calibrated by the Planck Collaboration \citep{Planck_SZ_cluster}:
\begin{equation}
    E^{-\beta}(z)\left[\frac{D_{\rm A}^2(z)\bar{Y}_{500}}{10^{-4}{\rm Mpc}^2}\right] = Y_{\star}\left[\frac{h}{0.7}\right]^{-2+\alpha}\left[\frac{(1-b)M_{500}}{6\times10^{14}M_{\odot}}\right]^{\alpha},
    \label{eq:planck_ysz}
\end{equation}
where $E(z)=H(z)/H_0$ and $D_{\rm A}(z)$ is the angular diameter distance. This includes parameters $\beta$ for the $z$-evolution, $Y_{\star}$ for the normalisation and $\alpha$ for the power-law slope with respect to the mass. It also includes a bias parameter $(1-b)$ which accounts for differences between the X-ray determined masses used in the calibration, which are subject to hydrostatic equilibrium bias, and the true mass. Planck have also provided the following formula for the intrinsic lognormal scatter of the relation:
\begin{equation}
    P(\log Y_{500}) = \frac{1}{\sqrt{2\pi}\sigma_{\log Y}}\exp\left[-\frac{\log^2(Y_{500}/\bar{Y}_{500})}{2\sigma_{\log Y}^2}\right],
    \label{eq:intrinsic_scatter}
\end{equation}
where $\sigma_{\log Y}$ is a fixed spread.

We assume a fixed value of 0.8 for the hydrostatic equilibrium bias parameter, which is consistent with the range 0.7 to 1.0 adopted by Planck, and we treat $Y\equiv D_{\rm A}^2(z)Y_{\rm SZ}$ as the cluster SZ observable, rather than $Y_{\rm SZ}$. This leaves four scaling relation parameters which are allowed to vary in our MCMC sampling. We adopt the following Gaussian priors from Planck: $\log Y_{\star}=-0.19\pm0.02$, $\alpha=1.79\pm0.08$, $\beta=0.66\pm0.50$ and $\sigma_{\log Y}=0.075\pm0.010$. 

To obtain the $f(R)$ scaling relation $Y_{\rm SZ}^{f(R)}(M_{500})$ from the above $Y_{\rm SZ}^{\rm GR}(M_{500})$ relation, we rescale the right-hand side of Eq.~(\ref{eq:planck_ysz}) by the mass ratio $\mathcal{R}$, which is predicted using Eq.~(\ref{eq:pipeline:mdyn_enhancement}) with scatter given by Eq.~(\ref{eq:scatter_model}). This rescaling is based on Eq.~(\ref{eq:pipeline:ysz_mapping_true}), which means that the mass $M_{500}$ in the expressions $Y_{\rm SZ}^{f(R)}(M_{500})$ and $Y_{\rm SZ}^{\rm GR}(M_{500})$ above is the true mass; we note that, although the Planck masses were originally determined using X-ray measurements, the value $(1-b)=0.8$ assumed for the mass bias is consistent with weak lensing measurements \citep[e.g.,][]{Hoekstra:2015gda}. 

Finally, we note that the scaling relation adopted in this work is intended to be representative of general scaling relations between the mass and SZ and X-ray observables, not just the Planck $Y_{\rm SZ}(M_{500})$ relation. This justifies our decision to encapsulate $D_{\rm A}^2(z)$ in the cluster observable and to fix the hydrostatic equilibrium bias; indeed, scaling relations for other observables --- for example, the SZ significance and the $Y_{\rm X}$ parameter --- do not include the function $D_{\rm A}^2(z)$ or a bias parameter \citep[e.g.,][]{deHaan:2016qvy,Bocquet:2018ukq}. Regardless of the observable, the main purpose of this chapter is to check that our constraint pipeline can give reasonable constraints of $f_{R0}$ using a realistic scaling relation which includes both intrinsic scatter and the $f(R)$ enhancement. It would be very straightforward to adapt this pipeline for other cluster observables, or for more than one cluster observable.

\subsection{Mock catalogues}
\label{sec:methods_pipeline:mock}

We test our framework (Fig.~\ref{fig:fr_flow_chart}) using mock cluster catalogues in place of observational data. We have generated mocks for both the GR and F5 models, using fiducial cosmological parameter values based on the Planck 2018 CMB constraints \citep{Planck:2018vyg}: $(\Omega_{\rm M},\sigma_8,h,\Omega_{\rm b},n_{\rm s})=(0.3153,0.8111,0.6736,0.04931,0.9649)$. For the scaling relation parameters, we assume the central values of the Gaussian priors listed in Sec.~\ref{sec:methods_pipeline:scaling_relation}. 

To generate the mocks, we first compute the predicted count per unit mass per unit redshift:
\begin{equation}
    \frac{{\rm d}N}{{\rm d}z{\rm d}\ln M} = \frac{{\rm d}n}{{\rm d}\ln M}\times\frac{{\rm d}V_{\rm c}(z)}{{\rm d}z},
    \label{eq:count_density}
\end{equation}
where $V_{\rm c}(z)$ is the comoving volume enclosed by the survey area between redshifts 0 and $z$ and the first term is the theoretical HMF $n_{500}^{f(R)}$, which is computed as described in Sec.~\ref{sec:methods_pipeline:hmf} for the fiducial cosmology. For this work, we assume a survey area of 5000 deg$^2$ and a maximum redshift of $z=0.5$, which is the upper redshift used to calibrate the $f(R)$ enhancement of the HMF (Sec.~\ref{sec:background_pipeline:hmf}). In the future, it will be important to develop models of the $f(R)$ HMF that work for a wider redshift range, which will be applicable to real cluster survey data.

The predicted number of clusters is:
\begin{equation}
    N_{\rm tot} = \int_{0.0}^{0.5}{\rm d}z\int_{-\infty}^{\infty}{\rm d}\ln M\frac{{\rm d}N}{{\rm d}z{\rm d}\ln M}.
\end{equation}
For each mock, we randomly draw the masses and redshifts of $N_{\rm tot}$ clusters using ${\rm d}N/{\rm d}z{\rm d}\ln M$, which is effectively a probability density. For each cluster $i$, we then draw a mass ratio $\mathcal{R}_i$ using a normal distribution with mean given by Eq.~(\ref{eq:pipeline:mdyn_enhancement}) and standard deviation given by Eq.~(\ref{eq:scatter_model}). The intrinsic observable $Y'_i(=D_{\rm A}^2Y_{500,i})$ of each cluster is then drawn using the lognormal distribution given by Eq.~(\ref{eq:intrinsic_scatter}), where $\bar{Y}_{500}$ is computed using Eq.~(\ref{eq:planck_ysz}) and rescaled by $\mathcal{R}_i$. 

We assume a fixed $1\sigma$ measurement uncertainty of $10\%$. The measured observable $Y_i$ is therefore drawn from a normal distribution with mean $Y'_i$ and standard deviation $0.1Y'_i$. We note that this choice of a fixed fractional uncertainty is intended to keep our calculations simple and general (for example, a more complicated model may be specific to a particular observational survey). We have also considered $5\%$ and $20\%$ uncertainties and have found that the inferred parameter constraints do not significantly differ, suggesting that this uncertainty is not the dominant source of error in the constraint pipeline (e.g., compared to the intrinsic scatters in the cluster scaling relation or the $f(R)$ dynamical mass enhancement).

Finally, we remove all clusters for which $Y_i$ is below some observational flux limit $Y_{\rm cut}$. For the main results of this chapter, we use $Y_{\rm cut}=1.5\times10^{-5}{\rm Mpc}^2$; however, we will also discuss the effects of using cuts $10^{-5}{\rm Mpc}^2$, $2\times10^{-5}{\rm Mpc}^2$ and $2.5\times10^{-5}{\rm Mpc}^2$. For each mock, we store only the cluster redshift $z_i$ (which is assumed to have no error) and the measured observable $Y_i$.

\begin{figure}
\centering
\includegraphics[width=0.7\textwidth]{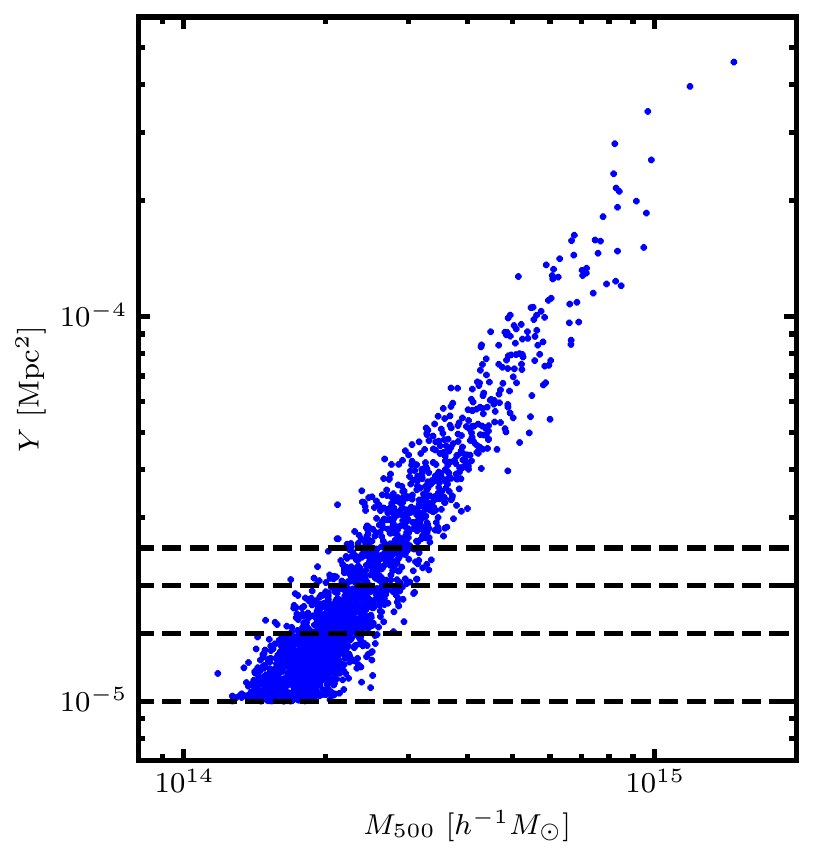}
\caption[SZ $Y$-parameter as a function of the halo mass for clusters from a GR mock catalogue.]{SZ $Y$-parameter as a function of the halo mass for clusters from a GR mock catalogue with observational flux limit $Y_{\rm cut}=10^{-5}{\rm Mpc}^2$. The dashed lines indicate the cuts $10^{-5}{\rm Mpc}^2$, $1.5\times10^{-5}{\rm Mpc}^2$, $2\times10^{-5}{\rm Mpc}^2$ and $2.5\times10^{-5}{\rm Mpc}^2$.}
\label{fig:mock_sr}
\end{figure}

An F5 mock with $Y_{\rm cut}=1.5\times10^{-5}{\rm Mpc}^2$ contains $\sim1350$ clusters. Generating a GR mock is more straightforward, since there is no need to include the $f(R)$ enhancements of the HMF or the scaling relation. In this case, there are $\sim1150$ clusters for $Y_{\rm cut}=1.5\times10^{-5}{\rm Mpc}^2$. For illustrative purposes, in Fig.~\ref{fig:mock_sr} we show the measured $Y$-parameters of the clusters as a function of the mass $M_{500}$ for a GR mock with $Y_{\rm cut}=10^{-5}{\rm Mpc}^2$. Horizontal dashed lines are included to indicate the four flux thresholds considered in this work, to give an idea of the mass range of clusters found above each.

\subsection{MCMC sampling}
\label{sec:methods_pipeline:likelihood}

For our parameter constraints, we use the unbinned Poisson likelihood \citep[e.g.,][]{Artis:2021tjj}:
\begin{equation}
    \ln\mathcal{L} = -\int {\rm d}z{\rm d}Y\frac{{\rm d}N}{{\rm d}z{\rm d}Y}(z,Y) + \sum_i\ln\frac{{\rm d}N}{{\rm d}z{\rm d}Y}(z_i,Y_i),
    \label{eq:log_likelihood}
\end{equation}
where the first term represents the predicted cluster count and the second term is a summation performed over all mock clusters. The expression ${\rm d}N/{\rm d}z{\rm d}Y$ represents the theoretical prediction of the count per unit $z$ per unit $Y$. 

Since our theoretical HMF is defined in terms of the mass $M$ rather than the observable $Y$, it is more convenient to re-express the first term with an integral over $\ln M$ \citep[e.g.,][]{deHaan:2016qvy}:
\begin{equation}
    \begin{split}
    &-\int_{0.0}^{0.5}{\rm d}z\int_{Y_{\rm cut}}^{\infty}{\rm d}Y\frac{{\rm d}N}{{\rm d}z{\rm d}Y}(z,Y)\\
    &= -\int_{0.0}^{0.5}{\rm d}z\int_{-\infty}^{\infty}{\rm d}\ln MP(Y>Y_{\rm cut}|M,z)\frac{{\rm d}N}{{\rm d}z{\rm d}\ln M}(M,z),
    \end{split}
    \label{eq:predicted_count}
\end{equation}
where ${\rm d}N/{\rm d}z{\rm d}\ln M$ can be computed using 
Eq.~(\ref{eq:count_density}), and the redshift integral is evaluated between $z=0$ and the maximum redshift $z=0.5$ of the mock. $P(Y>Y_{\rm cut}|M,z)$ represents the probability that, for a given mass and redshift, the measured $Y$-parameter exceeds the flux threshold. This depends on both the measurement uncertainty and the intrinsic log-normal scatter of $Y$:
\begin{equation}
    P(Y > Y_{\rm cut}|M,z) = \int_{-\infty}^{\infty}{\rm d}\ln Y' P(Y > Y_{\rm cut}|Y')P(Y'|M,z),
    \label{eq:cut_prob}
\end{equation}
where $P(Y > Y_{\rm cut}|Y')$ is the probability that the measured value $Y$ exceeds $Y_{\rm cut}$, given an intrinsic value $Y'$, and $P(Y'|M,z)$ is the probability density of a cluster having intrinsic value $Y'$ given that it has mass $M$ and redshift $z$. As discussed in Sec.~\ref{sec:methods_pipeline:mock}, the mocks use a fixed measurement uncertainty of $10\%$, which means that the former can be estimated using a normal distribution with mean $Y'$ and standard deviation $0.1Y'$. The probability density $P(Y'|M,z)$ is more complicated, since this depends both on the intrinsic scatter of the $Y(M)$ scaling relation and the scatter of the mass ratio $\mathcal{R}$:
\begin{equation}
    P(Y'|M,z) = \int_{1}^{4/3}{\rm d}\mathcal{R} P(Y'|\bar{Y}(M,z,\mathcal{R}))P(\mathcal{R}|M,z),
    \label{eq:ratio_integral}
\end{equation}
where $P(\mathcal{R}|M,z)$ is the probability density of a cluster having mass ratio $\mathcal{R}$ given that it has mass $M$ and redshift $z$. This is computed using a normal distribution with mean given by Eq.~(\ref{eq:pipeline:mdyn_enhancement}) and standard deviation given by Eq.~(\ref{eq:scatter_model}). The other probability density, $P(Y'|\bar{Y}(M,z,\mathcal{R}))$, is computed using Eq.~(\ref{eq:intrinsic_scatter}), with $\bar{Y}$ calculated using Eq.~(\ref{eq:planck_ysz}) and rescaled by a factor of $\mathcal{R}$. Together, Eqs.~(\ref{eq:predicted_count})-(\ref{eq:ratio_integral}) form a 4D integral, which we compute using a fixed grid in ($\ln M,z,\ln Y,\mathcal{R}$).

For the second term in Eq.~(\ref{eq:log_likelihood}), we can again re-express into a form that depends on ${\rm d}N/({\rm d}z{\rm d}\ln M)$ using:
\begin{equation}
    \begin{split}
    \frac{{\rm d}N}{{\rm d}z{\rm d}Y}(z_i,Y_i) = &\int{\rm d}\ln Y'\int{\rm d}\ln M'\\ 
    &\times P(Y_i|Y')P(Y'|M',z_i)\frac{{\rm d}N}{{\rm d}z{\rm d}\ln M'}(M',z_i),
    \end{split}
    \label{eq:second_term}
\end{equation}
where the probability density functions $P(Y_j|Y')$ and $P(Y'|M',z_i)$ represent the measurement uncertainty and intrinsic scatter, respectively. The latter is computed using Eq.~(\ref{eq:ratio_integral}), meaning that Eq.~(\ref{eq:second_term}) is really a 3D integral. We compute this for each mock cluster using a fixed grid in ($\ln Y',\ln M',\mathcal{R}$), then evaluate the sum in Eq.~(\ref{eq:log_likelihood}).

We have used the python package \textsc{emcee} \citep{2013PASP..125..306F} for the MCMC sampling. For all of the results discussed in this chapter, we have used 28 walkers each travelling 2700 steps (we discard the first 600 steps to ensure that the chains are well converged). At each step, the log-likelihood is computed for the sampled parameters as described above. In addition to the $f_{R0}$ parameter, the cosmological parameters $\Omega_{\rm M}$ and $\sigma_8$ and the four scaling relation parameters $Y_{\star}$, $\alpha$, $\beta$ and $\sigma_{\log Y}$ are sampled. For the cosmological parameters, we adopt uniform (flat) priors $\log_{10}|f_{R0}|\in[-7,-4]$ and $\sigma_8\in[0.60,0.95]$, and for $\Omega_{\rm M}$ we use either a flat prior $\Omega_{\rm M}\in[0.15,0.50]$ or a Gaussian prior $\Omega_{\rm M}=0.3153\pm0.0073$ which is based on the Planck 2018 CMB constraints \citep{Planck:2018vyg}. For the scaling relation parameters, we adopt the Gaussian priors listed in Sec.~\ref{sec:methods_pipeline:scaling_relation}.

The flat prior $[-7,-4]$ for $\log_{10}|f_{R0}|$ extends beyond the range $[-6,-4]$ used to calibrate the HMF enhancement model \citep{Cataneo:2016iav}. For sampled values in the range $-7\leq\log_{10}|f_{R0}|\leq-6$, we first calculate the HMF enhancement for $\log_{10}|f_{R0}|=-6$, then linearly interpolate between $|f_{R0}|=0$ (GR) and $|f_{R0}|=10^{-6}$ to estimate the enhancement. For example, this means that the estimated enhancement for $|f_{R0}|=10^{-7}$ would be $10\%$ of the enhancement for $|f_{R0}|=10^{-6}$. We note that, because clusters are expected to be completely screened for this range of $\log_{10}|f_{R0}|$ values, it is not necessary to use a physically accurate method here, so long as the predicted enhancement lies between GR and F6. We use a similar approach to estimate the dynamical mass enhancement for this range of $\log_{10}|f_{R0}|$, where, again, the enhancement is very close to zero anyway.

\section{Results}
\label{sec:results_pipeline}

In this section, we discuss the main results of this chapter. In Sec.~\ref{sec:results_pipeline:gr_pipeline}, we use a GR mock to check that our pipeline can give reasonable constraints of the $\Lambda$CDM and scaling relation parameters. Then, in Sec.~\ref{sec:results_pipeline:fr_pipeline}, we use our full pipeline to constrain the $f_{R0}$ parameter of $f(R)$ gravity, using a combination of GR and F5 mocks.

\subsection{GR pipeline}
\label{sec:results_pipeline:gr_pipeline}

\begin{figure*}
\centering
\includegraphics[width=1.0\textwidth]{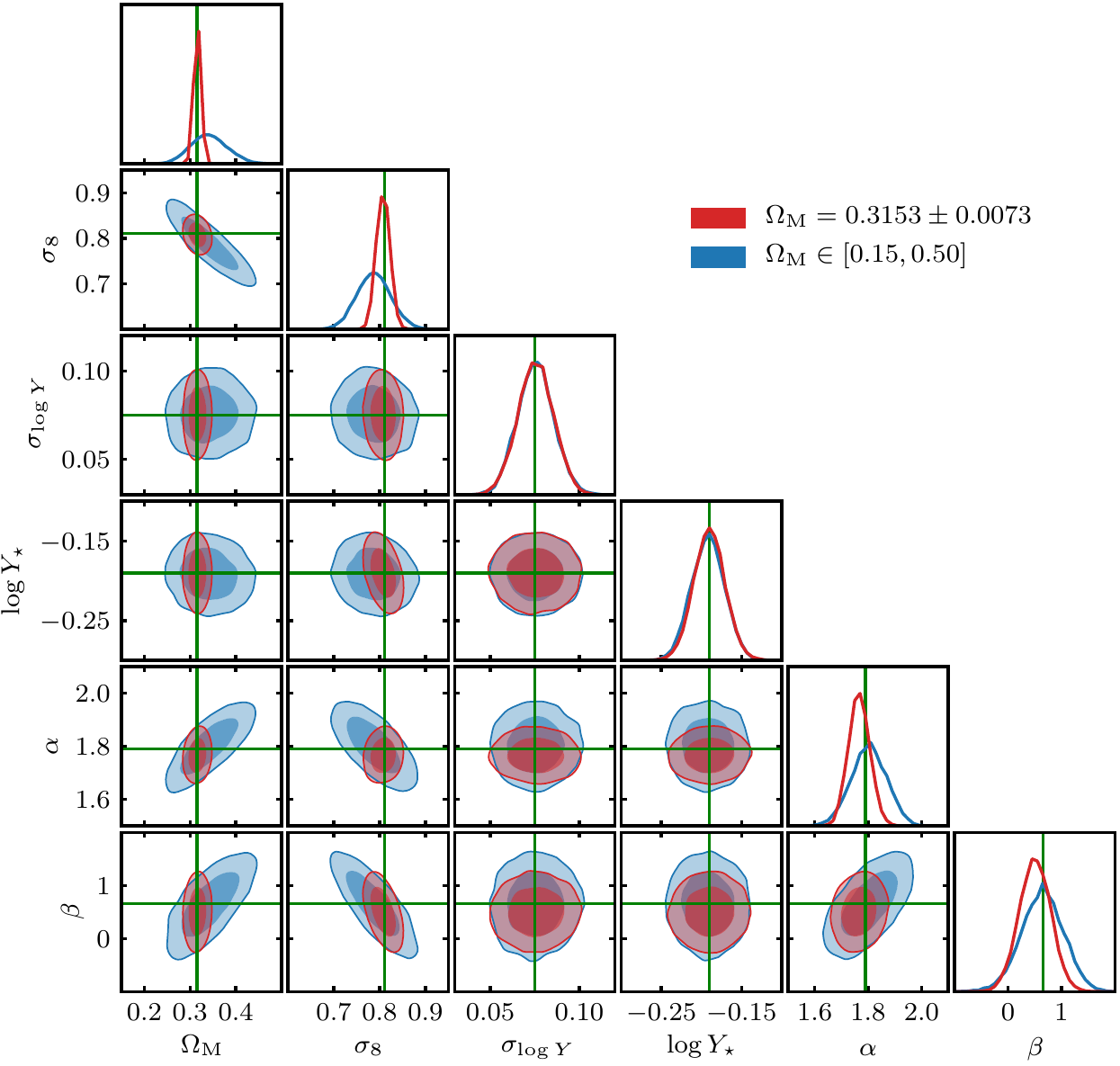}
\caption[Parameter constraints using our GR pipeline for a $\Lambda$CDM fiducial cosmology.]{Parameter constraints using our GR pipeline, which does not include $f(R)$ enhancements of the HMF and the scaling relation (see Sec.~\ref{sec:methods_pipeline}), using a GR mock with observational flux threshold $Y_{\rm cut}=1.5\times10^{-5}{\rm Mpc}^2$. The two sets of constraints are generated using a flat prior $[0.15,0.50]$ (\textit{blue}) and a Gaussian prior $0.3153\pm0.0073$ (\textit{red}) in $\Omega_{\rm M}$. The dark and light regions of the contours represent 68\% and 95\% confidences, respectively. The distributions of the sampled parameter values are shown in the top panels of each column, with the mean and standard deviation of each parameter quoted in Table~\ref{table:gr_pipeline}. The fiducial cosmological parameter values of the GR mock are indicated by the green lines.}
\label{fig:gr_pipeline}
\end{figure*}

In order to verify that our pipeline can give reasonable $\Lambda$CDM constraints and successfully account for the intrinsic scatter of the $Y(M_{500})$ relation and measurement uncertainty in the mock, we first test our `GR pipeline', where the $f(R)$ corrections to the HMF and scaling relation are excluded. We show the constraints, which have been inferred using a GR mock with $Y_{\rm cut}=1.5\times10^{-5}{\rm Mpc}^2$, in Fig.~\ref{fig:gr_pipeline}. The blue contours are obtained using the flat prior $\Omega_{\rm M}\in[0.15,0.50]$, while the red contours are obtained using the Gaussian prior $\Omega_{\rm M}=0.3153\pm0.0073$ from Planck 2018. 

\begin{sidewaystable*}
\centering

\small
\begin{tabular}{ c@{\hskip 0.5in}cc@{\hskip 0.5in}cc@{\hskip 0.5in}cc } 
 \toprule
 
  & & & \multicolumn{2}{c}{Flat $\Omega_{\rm M}$ prior} & \multicolumn{2}{c}{Gaussian $\Omega_{\rm M}$ prior} \\
 Parameter & Fiducial value & Prior & 68\% range & $\mathcal{L}_{\rm max}$ & 68\% range & $\mathcal{L}_{\rm max}$ \\

 \midrule

 $\Omega_{\rm M}$ & $0.3153$ & --- & $0.34\pm0.04$ & $0.3384$ & $0.316\pm0.007$ & $0.3175$ \\ 
 $\sigma_8$ & $0.8111$ & $[0.60,0.95]$ & $0.79\pm0.04$ & $0.7888$ & $0.808\pm0.015$ & $0.8037$ \\
 $\sigma_{\log Y}$ & $0.075$ & $0.075\pm0.010$ & $0.076\pm0.010$ & $0.076$ & $0.076\pm0.010$ & $0.073$ \\
 $\log Y_{\star}$ & $-0.19$ & $-0.19\pm0.02$ & $-0.19\pm0.02$ & $-0.19$ & $-0.19\pm0.02$ & $-0.19$\\
 $\alpha$ & $1.79$ & $1.79\pm0.08$ & $1.80\pm0.07$ & $1.80$ & $1.77\pm0.04$ & $1.76$\\
 $\beta$ & $0.66$ & $0.66\pm0.50$ & $0.6\pm0.4$ & $0.63$ & $0.5\pm0.3$ & $0.58$\\
 
 \bottomrule
 
\end{tabular}

\caption[Parameter constraints using our GR pipeline for a $\Lambda$CDM fiducial cosmology.]{Parameter constraints using our GR pipeline. The mean and standard deviation are quoted (68\% range) along with the parameter combinations giving the highest log-likelihood ($\mathcal{L}_{\rm max}$). The constraints correspond to the distributions shown in Fig.~\ref{fig:gr_pipeline}.}
\label{table:gr_pipeline}

\end{sidewaystable*}

For the flat $\Omega_{\rm M}$ prior, the contours are in good agreement with the fiducial parameter values, which are indicated by the green lines. In the top panel of each column, we show the marginalised distributions of each parameter, with the mean and standard deviation quoted in Table~\ref{table:gr_pipeline}. In Table~\ref{table:gr_pipeline}, we also show the combination of parameters that gave the highest log-likelihood during the sampling ($\mathcal{L}_{\rm max}$); these can be thought of as the `most likely' set of values. The distributions of the scaling relation parameters closely match the Gaussian priors. Meanwhile, the constraints $0.34\pm0.04$ for $\Omega_{\rm M}$ and $0.79\pm0.04$ for $\sigma_8$ -- while still within $1\sigma$ agreement -- are slightly offset from the fiducial values, and the same goes for the highest-likelihood values 0.34 and 0.79. As shown by the constraints in red, using a tighter Gaussian prior in $\Omega_{\rm M}$ results in narrower contours and constraints $\Omega_{\rm M}=0.316\pm0.007$ and $\sigma_8=0.808\pm0.015$ which match the fiducial values more closely.

The initial offset of the $\Omega_{\rm M}$ and $\sigma_8$ constraints from the fiducial values is caused by a well-known degeneracy between these two parameters: increasing either of these will boost the predicted amplitude of the HMF. Therefore, the effects of increasing (decreasing) $\Omega_{\rm M}$ and decreasing (increasing) $\sigma_8$ on the HMF can roughly cancel out. This causes the elongated shape of the blue $\Omega_{\rm M}$-$\sigma_8$ contour. 

We also observe degeneracies between $\Omega_{\rm M}$, $\sigma_8$, $\alpha$ and $\beta$. One explanation for this is that $\alpha$ and $\beta$ can also affect the predicted HMF. For example, increasing $\alpha$ (i.e., increasing the slope of the $Y(M)$ scaling relation) will cause the predicted $Y$-parameter to be reduced for clusters with $0.8M_{500}<6\times10^{14}M_{\odot}$ (since $(1-b)^{-1}6\times10^{14}M_\odot$ is the pivot mass of the power-law function in Eq.~\eqref{eq:planck_ysz}), which includes the majority of clusters in our mocks (see Fig.~\ref{fig:mock_sr}). This means that fewer clusters will be predicted to have $Y>Y_{\rm cut}$, and therefore the inferred cluster count will be lower, which can be countered by a larger $\Omega_{\rm M}$. The effects of changing $\alpha$, $\beta$, $\Omega_{\rm M}$ and $\sigma_8$ may balance out overall, giving rise to the observed degeneracies in the blue contours of Fig.~\ref{fig:gr_pipeline}. 

By adopting the tighter $\Omega_{\rm M}$ prior, these degeneracies appear to be mostly eliminated. This shows the importance of accurate independent measurements of $\Omega_{\rm M}$ in the use of galaxy cluster number counts to constrain cosmological models and parameters.

\subsection{\texorpdfstring{$f(R)$}{f(R)} pipeline}
\label{sec:results_pipeline:fr_pipeline}

\begin{figure*}
\centering
\includegraphics[width=1.0\textwidth]{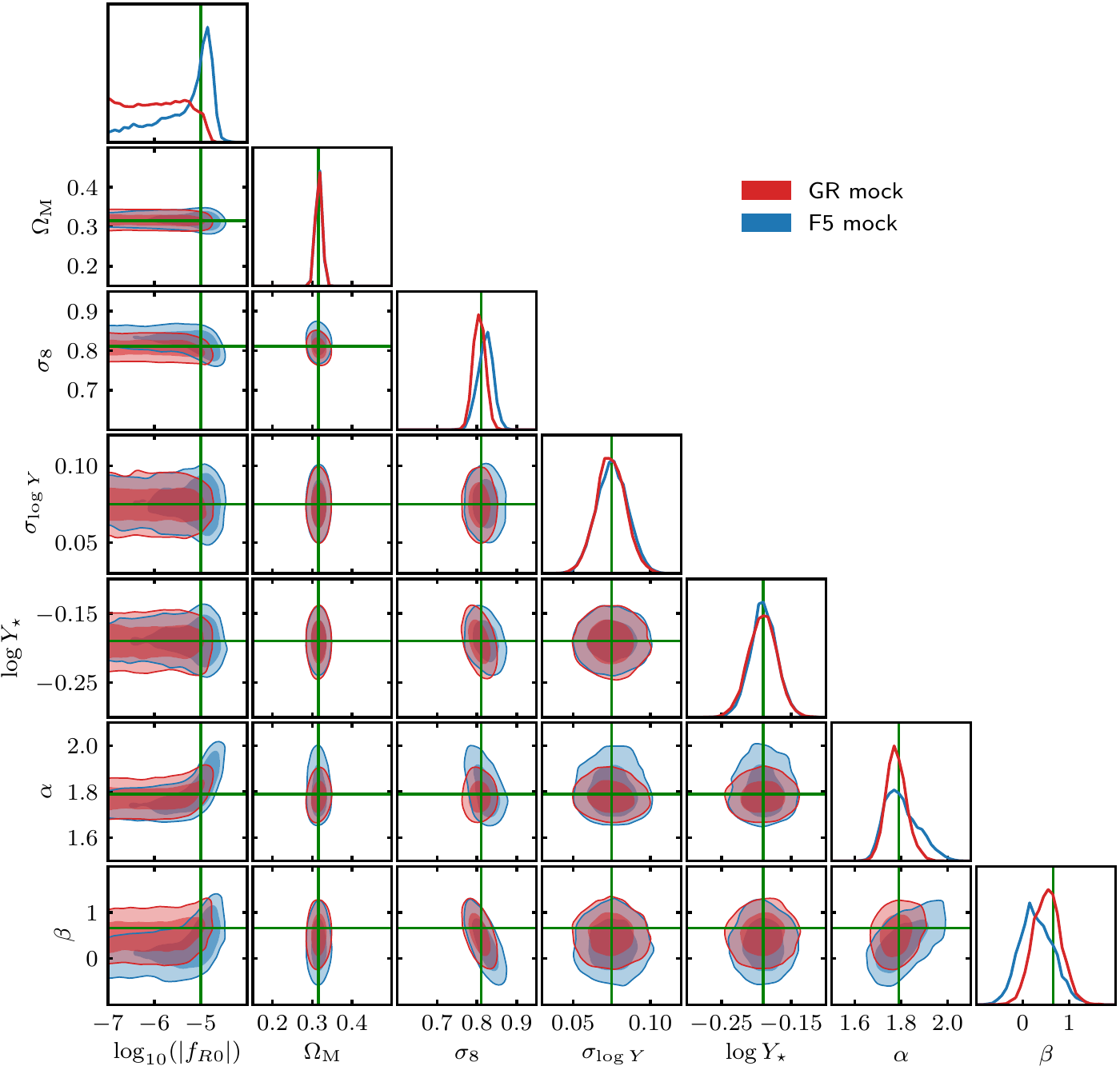}
\caption[Parameter constraints using our full $f(R)$ pipeline for both $\Lambda$CDM and F5 fiducial cosmologies.]{Parameter constraints using our full $f(R)$ pipeline, as detailed in Sec.~\ref{sec:methods_pipeline}, using GR (\textit{red}) and F5 (\textit{blue}) mocks with observational flux threshold $Y_{\rm cut}=1.5\times10^{-5}{\rm Mpc}^2$. The dark and light regions of the contours represent 68\% and 95\% confidences, respectively. The distributions of the sampled parameter values are shown in the top panels of each column, with the mean and standard deviation quoted in Table~\ref{table:fr_pipeline}. The fiducial cosmological parameter values of the mocks are indicated by the green lines, including the value $\log_{10}|f_{R0}|=-5$ for the F5 mock.}
\label{fig:full_fr_pipeline}
\end{figure*}

We now test the full $f(R)$ gravity constraint pipeline, which includes the $f(R)$ effects on the HMF and scaling relation, as described in Secs.~\ref{sec:methods_pipeline:hmf} and \ref{sec:methods_pipeline:scaling_relation}. In Fig.~\ref{fig:full_fr_pipeline}, we show the constraints inferred using GR and F5 mocks with $Y_{\rm cut}=1.5\times10^{-5}{\rm Mpc}^2$. For these results, we use the Gaussian prior of $\Omega_{\rm M}$ in order to prevent the $\Omega_{\rm M}$--$\sigma_8$ degeneracy observed in Fig.~\ref{fig:gr_pipeline}. As we will show in Sec.~\ref{sec:bias}, using a flat prior for $\Omega_{\rm M}$ can otherwise lead to biased constraints of $\log_{10}|f_{R0}|$.

\begin{sidewaystable*}
\centering

\small
\begin{tabular}{ c@{\hskip 0.5in}cc@{\hskip 0.5in}cc@{\hskip 0.5in}cc } 
 \toprule
 
  & & & \multicolumn{2}{c}{GR constraints} & \multicolumn{2}{c}{F5 constraints} \\
 Parameter & Fiducial value & Prior & 68\% range & $\mathcal{L}_{\rm max}$ & 68\% range & $\mathcal{L}_{\rm max}$ \\

 \midrule

 $\log_{10}|f_{R0}|$ & --- & $[-7,-4]$ & $\leq-5.56$ & $-6.75$ & $-5.1^{+0.3}_{-1.0}$ & $-4.92$ \\ 
 $\Omega_{\rm M}$ & $0.3153$ & $0.3153\pm0.0073$ & $0.316\pm0.007$ & $0.317$ & $0.316\pm0.008$ & $0.313$ \\ 
 $\sigma_8$ & $0.8111$ & $[0.60,0.95]$ & $0.806\pm0.015$ & $0.806$ & $0.821\pm0.019$ & $0.815$ \\
 $\sigma_{\log Y}$ & $0.075$ & $0.075\pm0.010$ & $0.075\pm0.010$ & $0.072$ & $0.075\pm0.010$ & $0.079$ \\
 $\log Y_{\star}$ & $-0.19$ & $-0.19\pm0.02$ & $-0.19\pm0.02$ & $-0.19$ & $-0.190\pm0.019$ & $-0.18$\\
 $\alpha$ & $1.79$ & $1.79\pm0.08$ & $1.78\pm0.04$ & $1.77$ & $1.80\pm0.07$ & $1.82$\\
 $\beta$ & $0.66$ & $0.66\pm0.50$ & $0.5\pm0.3$ & $0.51$ & $0.3\pm0.4$ & $0.44$\\
 
 \bottomrule
 
\end{tabular}

\caption[Parameter constraints using our full $f(R)$ pipeline for both $\Lambda$CDM and F5 fiducial cosmologies.]{Parameter constraints using our full $f(R)$ pipeline. The 68\% range columns show the mean and standard deviation for all parameters other than $\log_{10}|f_{R0}|$; for the latter, the 68\% upper bound is shown for the GR mock constraints and the median and 68-percentile is shown the F5 mock constraints. The parameter combinations giving the highest log-likelihood ($\mathcal{L}_{\rm max}$) are also shown. The constraints correspond to the distributions shown in Fig.~\ref{fig:full_fr_pipeline}.}
\label{table:fr_pipeline}

\end{sidewaystable*}

For the constraints obtained from the GR mock, which are indicated by the red contours in Fig.~\ref{fig:full_fr_pipeline}, the $\log_{10}|f_{R0}|$ posterior distribution is roughly uniform for the range $-7\leq\log_{10}|f_{R0}|\lesssim-5$ and drops to zero for $\log_{10}|f_{R0}|\gtrsim-5$. This rules out $f(R)$ models stronger than F5, whereas weaker models are difficult to distinguish from GR for this sample of clusters. We show our constraints of the parameter values in Table~\ref{table:fr_pipeline}. Since the $\log_{10}|f_{R0}|$ posterior does not follow a normal distribution, we quote an upper bound rather than the mean and standard deviation. In this case, 68\% of the sampled points have $\log_{10}|f_{R0}|\leq-5.56$. We note that this threshold may depend on the width of the $\log_{10}|f_{R0}|$ prior: for a wider prior (i.e., extending the lower bound of the prior to some value smaller than $-7$ while fixing the upper bound of the prior) and a uniform $\log_{10}|f_{R0}|$ posterior, it is reasonable to expect the 68\% upper bound to be lower. Therefore, it is perhaps more useful to look at the combination of parameter values that give the highest log-likelihood. In this case, the most likely combination has $\log_{10}|f_{R0}|=-6.75$, which is quite close to the lower bound of the prior (although we note that, given the flat posterior distribution of $\log_{10}|f_{R0}|$, the point with $\mathcal{L}_{\rm max}$ might not be much more significant than points with only slightly smaller log-likelihood values). The constraints for the other parameters are in excellent agreement with the fiducial values. Therefore, the results suggest that our pipeline can successfully constrain $f_{R0}$ using cluster samples in a GR universe.

The constraints for the F5 mock are indicated by the blue contours in Fig.~\ref{fig:full_fr_pipeline}. The $\log_{10}|f_{R0}|$ constraints appear to be in good agreement with the fiducial value $-5$, which lies within the 68\% confidence region of the contours. This region only extends down to $\log_{10}|f_{R0}|\approx-6.5$, clearly favouring $f(R)$ gravity over GR. The constraints also appear to rule out models with $\log_{10}|f_{R0}|\gtrsim-4.5$. The median and 68-percentile range of the sampled values is $\log|f_{R0}|=-5.1^{+0.3}_{-1.0}$, while the highest-likelihood parameter combination has $\log_{10}|f_{R0}|=-4.92$. Both of these results are very close to the fiducial value of $-5$. The constraints for the other parameters are again in very reasonable agreement with the fiducial values. This result suggests that our pipeline can clearly identify if the underlying universe model is F5.

Despite this promising agreement, it is interesting to note that the $\log_{10}|f_{R0}|$ posterior distribution has a long tail over the range $-7<\log_{10}|f_{R0}|<-5$. Over this range of points, $\sigma_8$ appears to have value $0.83$-$0.84$ on average, while $\alpha$ and $\beta$ have values $\sim1.75$ and $\sim0.0$ on average (see the blue contours in Fig.~\ref{fig:full_fr_pipeline}). As $\log_{10}|f_{R0}|$ is lowered, the predicted amplitude of the HMF will be reduced. The increased $\sigma_8$ can act against this, as can the lowered $\alpha$, which, as discussed in Sec.~\ref{sec:results_pipeline:gr_pipeline}, can increase the predicted cluster count for clusters with $0.8M_{500}<6\times10^{14}M_{\odot}$. The latter can also give a scaling relation that more closely matches the F5 result: this is because the scaling relation in F5 is enhanced at lower masses, which may be approximated by the constraint pipeline as a power-law with shallower slope. This degeneracy also comes into play for $\log_{10}|f_{R0}|>-5$, where $\sigma_8$ becomes slightly lower on average and $\alpha$ and $\beta$ become higher. Overall, this reduces the precision of the $\log_{10}|f_{R0}|$ constraint, and is perhaps the reason why the $\log_{10}|f_{R0}|$ posterior peaks at a value that is slightly higher than $-5$. This can also explain why the $\beta$ constraints predict a value $0.3\pm0.4$ that is slightly lower than the fiducial value 0.66. By using tighter priors in $\sigma_8$, $\alpha$ or $\beta$ it may be possible to eliminate this bias (see Sec.~\ref{sec:bias:degeneracies} for a detailed discussion).

We have also tested our pipeline using an F4.5 mock (with $\log_{10}|f_{R0}|=-4.5$), and in Appendix \ref{sec:appendix:pipeline:F4.5} we show that this model is clearly distinguished from F5.

\section{Potential biases in model constraints}
\label{sec:bias}

In Sec.~\ref{sec:results_pipeline}, we demonstrated that our framework can give very reasonable constraints of $\log_{10}|f_{R0}|$ for both GR and F5 mocks (Fig.~\ref{fig:full_fr_pipeline}). An important feature of this constraint framework (Fig.~\ref{fig:fr_flow_chart}) is the inclusion of corrections for the effects of $f(R)$ gravity on the internal cluster properties, which are expected to prevent biased constraints. In Sec.~\ref{sec:bias:pipeline}, we will assess potential sources of bias in the constraint pipeline, including an incomplete treatment of the scaling relation. Then, in Sec.~\ref{sec:bias:sample}, we will check the effects of the cluster sample, including selection criteria, on the constraints. Finally, we will demonstrate how the various parameter degeneracies can be prevented by using tighter parameter priors in Sec.~\ref{sec:bias:degeneracies}. 

For all of the figures in this section, we will only show constraints for parameters that are either biased or contribute to parameter degeneracies. Therefore, we exclude the $\log Y_{\star}$ and $\sigma_{\log Y}$ constraints, since these always match the Gaussian priors very closely (e.g., see Figs.~\ref{fig:gr_pipeline} and \ref{fig:full_fr_pipeline}). For similar reasons, we will also exclude $\Omega_{\rm M}$ constraints that have been inferred using the Gaussian prior from Planck 2018.

\subsection{Constraint pipeline}
\label{sec:bias:pipeline}

\subsubsection{Power-law scaling relation}
\label{sec:bias:pipeline:power_law_sr}

\begin{figure}
\centering
\includegraphics[width=0.7\textwidth]{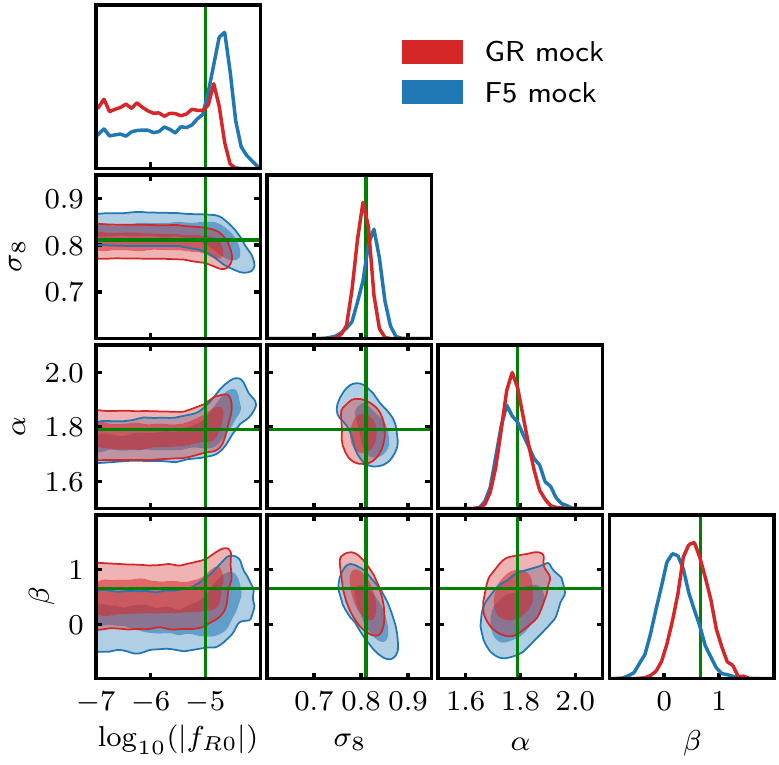}
\caption[Parameter constraints from our simplified pipeline with the $f(R)$ effects on the observable-mass scaling relation neglected.]{Parameter constraints generated using the same GR (\textit{red}) and F5 (\textit{blue}) mocks as Fig.~\ref{fig:full_fr_pipeline}, but with a simplified constraint pipeline in which the $f(R)$ effects on the observable-mass scaling relation are neglected.}
\label{fig:gr_sr}
\end{figure}

In Fig.~\ref{fig:gr_sr}, we show constraints inferred using the same GR and F5 mocks as used for Fig.~\ref{fig:full_fr_pipeline}. However, here the $f(R)$ effects on the SZ scaling relation (Eq.~(\ref{eq:planck_ysz})) have been neglected, i.e., a power-law scaling relation without $f(R)$ corrections is used in the (incomplete) $f(R)$ pipeline. 

For the GR mock constraints, shown by the red contours in Fig.~\ref{fig:gr_sr}, the $\log_{10}|f_{R0}|$ posterior appears to be uniformly distributed over the range $-7\leq\log_{10}|f_{R0}|\lesssim-4.5$. This extends beyond the range $-7\leq\log_{10}|f_{R0}|\lesssim-5$ observed using the full pipeline in Fig.~\ref{fig:full_fr_pipeline}, and the range $\log_{10}|f_{R0}|\leq-5.36$ containing 68\% of the sampled points has a higher upper bound than the range $\leq-5.56$ given in Table~\ref{table:fr_pipeline} for the full pipeline. Therefore, even though the GR mock is generated using a power-law scaling relation, it seems that neglecting the $f(R)$ effects on the scaling relation in the pipeline leads to less precise and weaker constraints of $\log_{10}|f_{R0}|$ overall. 

The F5 mock constraints, which are shown by the blue contours, still give a peaked $\log_{10}|f_{R0}|$ posterior distribution. However, there are now a greater proportion of sampled points within the range $-7\leq\log_{10}|f_{R0}|\lesssim-5$. This means that the 68\% confidence contours extend to $\log_{10}|f_{R0}|=-7$, indicating that the pipeline is unable to convincingly rule out GR. There are also a greater number of sampled points with $\log_{10}|f_{R0}|\gtrsim-4.5$; indeed, the highest-likelihood parameter combination has $\log_{10}|f_{R0}|=-4.56$, which is much higher than the fiducial value $-5$ and the value $-4.92$ when using the full pipeline. The median and 68-percentile range is $\log_{10}|f_{R0}|=-5.1^{+0.5}_{-1.2}$, which is less precise than the constraint $\log_{10}|f_{R0}|=-5.1^{+0.3}_{-1.0}$ with the full $f(R)$ pipeline. 

In summary, our constraints for the GR and F5 mocks indicate that assuming a power-law observable-mass scaling relation can lead to imprecise and biased constraints of $f(R)$ gravity. This appears to be linked to parameter degeneracies, where we again observe a lowered $\sigma_8$ and increased $\alpha$ for $\log_{10}|f_{R0}|\gtrsim-5$, and an increased $\sigma_8$ and lowered $\alpha$ and $\beta$ at $\log_{10}|f_{R0}|\lesssim-5$.

\subsubsection{Mass ratio scatter}
\label{sec:bias:pipeline:ratio_scatter}

\begin{figure}
\centering
\includegraphics[width=0.7\textwidth]{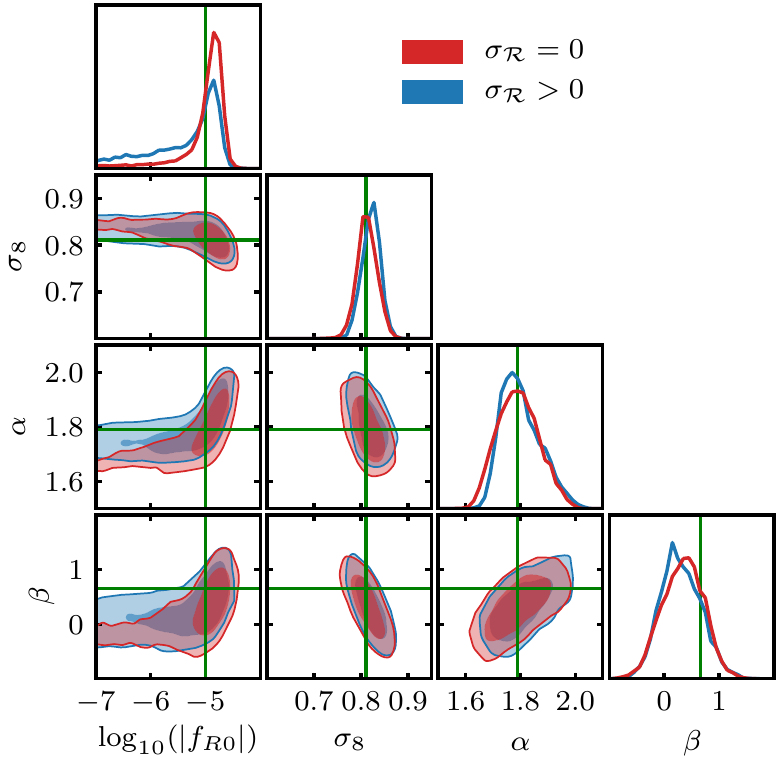}
\caption[Effect of the dynamical mass enhancement scatter on the parameter constraints from our $f(R)$ pipeline.]{Parameter constraints generated using our constraint pipeline, where the blue constraints are the same as the F5 mock constraints in Fig.~\ref{fig:full_fr_pipeline} and the red constraints are generated with the scatter of the dynamical mass enhancement set to zero in both the mock and log-likelihood.}
\label{fig:corner_ratio_scatter}
\end{figure}

For our constraints using the F5 mock in Fig.~\ref{fig:full_fr_pipeline}, we included the scatter of the dynamical mass enhancement, given by Eq.~(\ref{eq:scatter_model}), in both the mock and the log-likelihood calculation. We now consider the effect of neglecting this scatter from the mock and the likelihood. The new result is shown by the red contours in Fig.~\ref{fig:corner_ratio_scatter}, along with the previous results in blue. Without this scatter, the observable-mass scaling relation is less scattered overall; as a result, the $f(R)$ constraints are more precise, with 68-percentile range $\log_{10}|f_{R0}|=-4.89^{+0.15}_{-0.35}$ as opposed to $\log_{10}|f_{R0}|=-5.1^{+0.3}_{-1.0}$. In particular, the red 68\% contours do not feature the tail towards low $\log_{10}|f_{R0}|$. These results indicate that excluding the scatter could lead to $f(R)$ constraints with an unrealistically high precision.

\subsection{Cluster sample}
\label{sec:bias:sample}

\subsubsection{Flux threshold}
\label{sec:bias:sample:ysz_cut}

\begin{figure}
\centering
\includegraphics[width=0.7\textwidth]{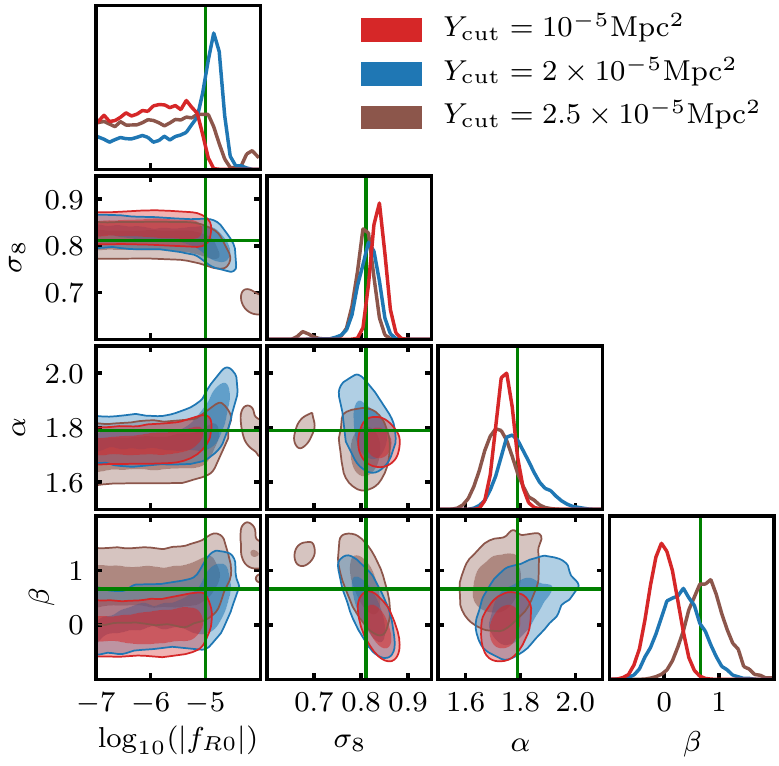}
\caption[Parameter constraints generated using our $f(R)$ pipeline for F5 mocks with different flux thresholds.]{Parameter constraints generated using our constraint pipeline, using F5 mocks with observational flux thresholds of $10^{-5}{\rm Mpc}^2$ (\textit{red}), $2\times10^{-5}{\rm Mpc}^2$ (\textit{blue}) and $2.5\times10^{-5}{\rm Mpc}^2$ (\textit{brown}).}
\label{fig:corner_ycut}
\end{figure}

In addition to the observational cut $Y_{\rm cut}=1.5\times10^{-5}{\rm Mpc}^2$ which is used in the main results of this chapter, we have also generated mocks with cuts $10^{-5}{\rm Mpc^2}$, $2\times10^{-5}{\rm Mpc^2}$ and $2.5\times10^{-5}{\rm Mpc^2}$. From Fig.~\ref{fig:mock_sr}, a cut of $10^{-5}{\rm Mpc}^2$ means that the lowest mass clusters, with $M_{500}\sim10^{14}h^{-1}M_{\odot}$, are included in the sample. In the F5 model, the HMF is more enhanced at these lower halo masses (see Fig.~\ref{fig:hmf_enhancement}), therefore it is expected that using lower-mass objects can give more precise constraints of $\log_{10}|f_{R0}|$. 

In Fig.~\ref{fig:corner_ycut}, we show constraints generated from F5 mocks with these three cuts. For $Y_{\rm cut}=2.5\times10^{-5}{\rm Mpc}^2$, the sampled $\log_{10}|f_{R0}|$ distribution is quite uniform for $-7<\log_{10}|f_{R0}|\lesssim-5$, indicating that this high-mass cluster sample cannot be used to distinguish the F5 model from weaker models, including GR. This is not surprising, given that higher-mass clusters are better-screened in F5, which means that their number count deviates from the GR prediction less strongly (see Fig.~\ref{fig:hmf_enhancement}). On the other hand, the constraints for $Y_{\rm cut}=2\times10^{-5}{\rm Mpc}^2$ clearly favour $\log_{10}|f_{R0}|$ values close to $-5$. However, the 68\% contours still extend to $\log_{10}|f_{R0}|=-7$, which is very close to GR. This is improved upon with $Y_{\rm cut}=1.5\times10^{-5}{\rm Mpc}^2$, which is able to convincingly distinguish the F5 model from GR, as we showed in Fig.~\ref{fig:full_fr_pipeline}. 

In Fig.~\ref{fig:corner_ycut}, we also show the constraints from the F5 mock with $Y_{\rm cut}=10^{-5}{\rm Mpc}^2$. Interestingly, despite containing lower-mass clusters than the other mocks, the sampled $\log_{10}|f_{R0}|$ values are approximately evenly distributed over $-7\lesssim\log_{10}|f_{R0}|\lesssim-5$. One possible reason is that this mock catalogue includes many more low-mass, unscreened, clusters, and the main constraining power comes from different objects than the previous cases. We note that for these constraints, the $\sigma_8$, $\alpha$ and $\beta$ parameters are all biased. As we have already discussed, these parameters can be varied in such a way that the predicted theoretical HMF in GR (i.e., with low $\log_{10}|f_{R0}|$) can match the F5 HMF with the fiducial cosmological parameters. Our results here show that this can cause biased constraints which appear to prefer GR over $f(R)$ gravity even though this is an F5 mock, and this seems to be more relevant for cluster samples that extend to lower masses. As we will show in Sec.~\ref{sec:bias:degeneracies}, these degeneracies can be prevented by using tighter parameter priors.

\begin{figure}
\centering
\includegraphics[width=0.7\textwidth]{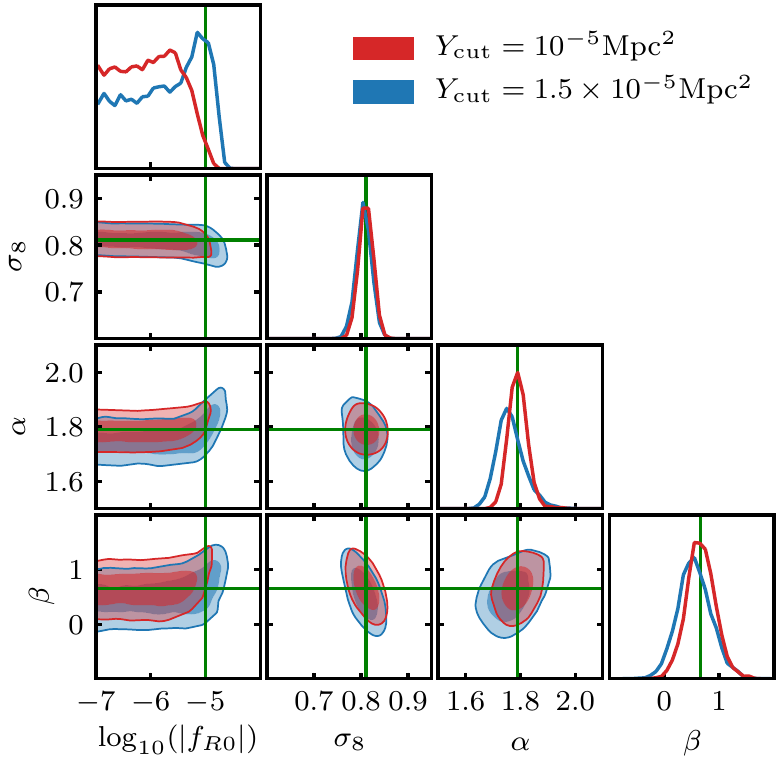}
\caption[Parameter constraints generated using our $f(R)$ pipeline for GR mocks with different flux thresholds.]{Parameter constraints generated using our constraint pipeline, using GR mocks with observational flux thresholds $10^{-5}{\rm Mpc}^2$ (\textit{red}) and $1.5\times10^{-5}{\rm Mpc}^2$ (\textit{blue}). The latter is a different realisation (generated in the same way) from the GR mock used in Fig.~\ref{fig:full_fr_pipeline}, and is included to show the potential effects of sample variance on the constraints.}
\label{fig:corner_sample_variance}
\end{figure}

We note that the biased results described above only apply to an F5 fiducial cosmology. The red contours in Fig.~\ref{fig:corner_sample_variance} show the constraints inferred using a GR mock with $Y_{\rm cut}=10^{-5}{\rm Mpc}^2$. These are consistent with GR, with 68\% of the sampled points in the range $\log_{10}|f_{R0}|\leq-5.71$, which is even more precise than the $\log_{10}\leq-5.56$ constraint from Fig.~\ref{fig:full_fr_pipeline}. Meanwhile, the constraints for $\sigma_8$, $\alpha$ and $\beta$ show an excellent match with the fiducial values. Therefore, the bias described above may not be an issue for cluster samples in a GR universe.

\subsubsection{Sample variance}
\label{sec:bias:sample:variance}

In order to check the effect of sample variance on the constraints, we have generated several GR mocks with $Y_{\rm cut}=1.5\times10^{-5}{\rm Mpc}^2$, following the method discussed in Sec.~\ref{sec:methods_pipeline:mock}. In all cases, the inferred constraints of $\log_{10}|f_{R0}|$ are consistent with GR, with the 68\% constraint contours spanning $-7\leq\log_{10}|f_{R0}|\lesssim-5$ just like the red contours in Fig.~\ref{fig:full_fr_pipeline}. 

However, we have occasionally observed peaks in the $\log_{10}|f_{R0}|$ posterior distribution close to $-5$, which are related to the degeneracies between $\log_{10}|f_{R0}|$, $\sigma_8$, $\alpha$ and $\beta$ mentioned above. An example is shown with the blue contours in Fig.~\ref{fig:corner_sample_variance}. As we have discussed, in the constraints using the F5 mock in Fig.~\ref{fig:full_fr_pipeline}, we can see a `rise' in the $\log_{10}|f_{R0}|$--$\alpha$ contour at $\log_{10}|f_{R0}|>-5$; there is a similar `rise' in the case of the blue contours in Fig.~\ref{fig:corner_sample_variance}. This is because a larger $\alpha$, which means a steeper scaling relation and hence underpredicted cluster number counts, could be compensated by a stronger gravity, so that to the pipeline, the GR mock would appear to be reasonably fitted with an $f(R)$ model with slightly larger $\alpha$. We also see a slight `drop' in the $\log_{10}|f_{R0}|$--$\sigma_8$ contour, where the lowered $\sigma_8$ can again counteract the strengthened gravity. These effects can lead to more points sampled around $\log_{10}|f_{R0}|=-5$, and because even stronger gravity is disfavoured an artificial peak is formed at $-5$. While the peak in $\log_{10}|f_{R0}|$ here is smaller than the peak observed for the F5 mock in Fig.~\ref{fig:full_fr_pipeline}, it is important to be wary that degeneracies can lead to a particular value of $\log_{10}|f_{R0}|$ being favoured even for a GR fiducial cosmology. Like the other sources of bias discussed in this chapter, this issue can be eliminated by using tighter priors, as we will show in the next section.

\subsection{Tighter priors}
\label{sec:bias:degeneracies}

\begin{figure*}
\centering
\includegraphics[width=0.9\textwidth]{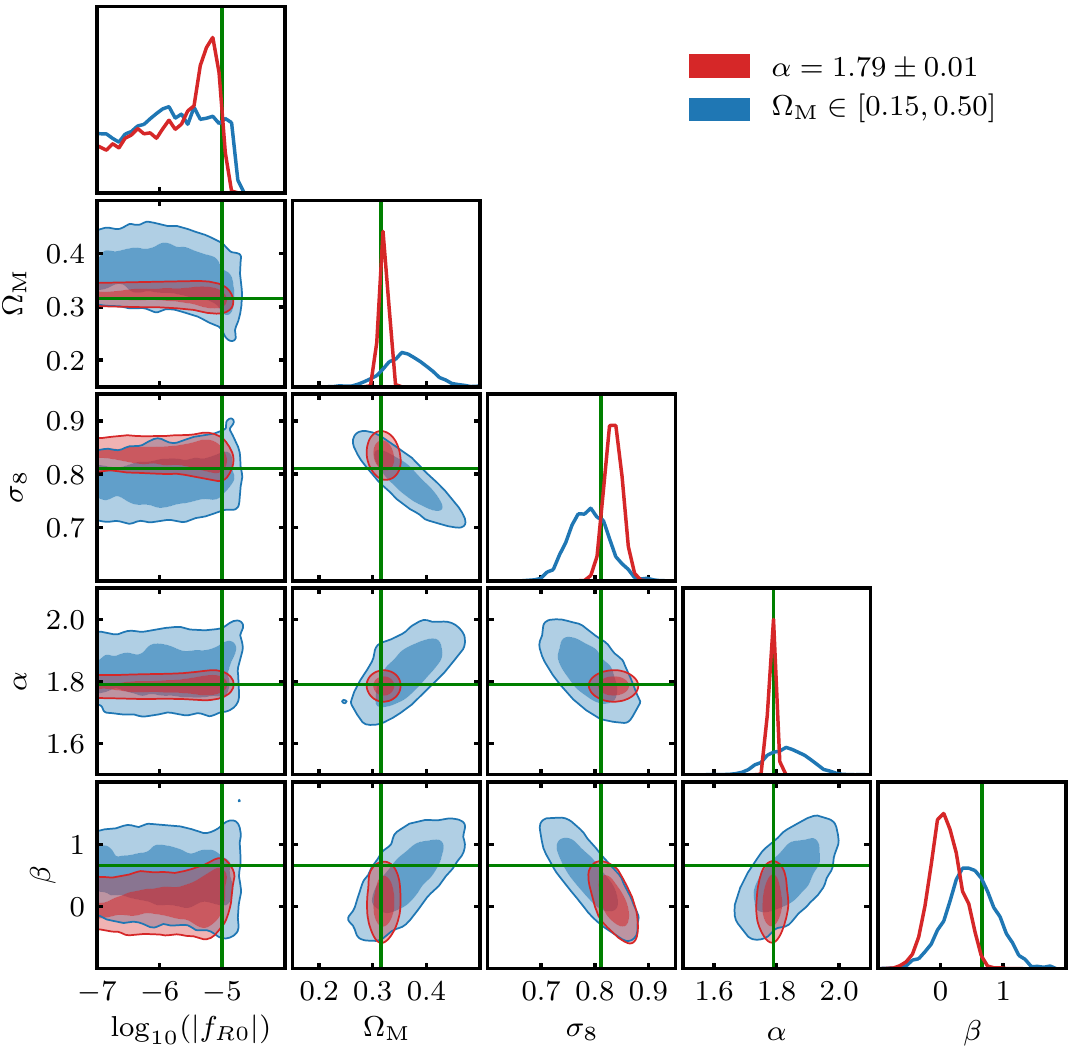}
\caption[Parameter constraints generated by our $f(R)$ pipeline using different priors for $\alpha$ and $\Omega_{\rm M}$.]{Parameter constraints generated by our constraint pipeline using: an F5 mock with flux threshold $Y_{\rm cut}=10^{-5}{\rm Mpc}^2$ and a tight Gaussian prior $1.79\pm0.01$ for $\alpha$ (\textit{red}); and an F5 mock with flux threshold $Y_{\rm cut}=1.5\times10^{-5}{\rm Mpc}^2$ and a flat prior $[0.15,0.50]$ for $\Omega_{\rm M}$ (blue).}
\label{fig:corner_tight_priors}
\end{figure*}

\begin{figure*}
\centering
\includegraphics[width=1.0\textwidth]{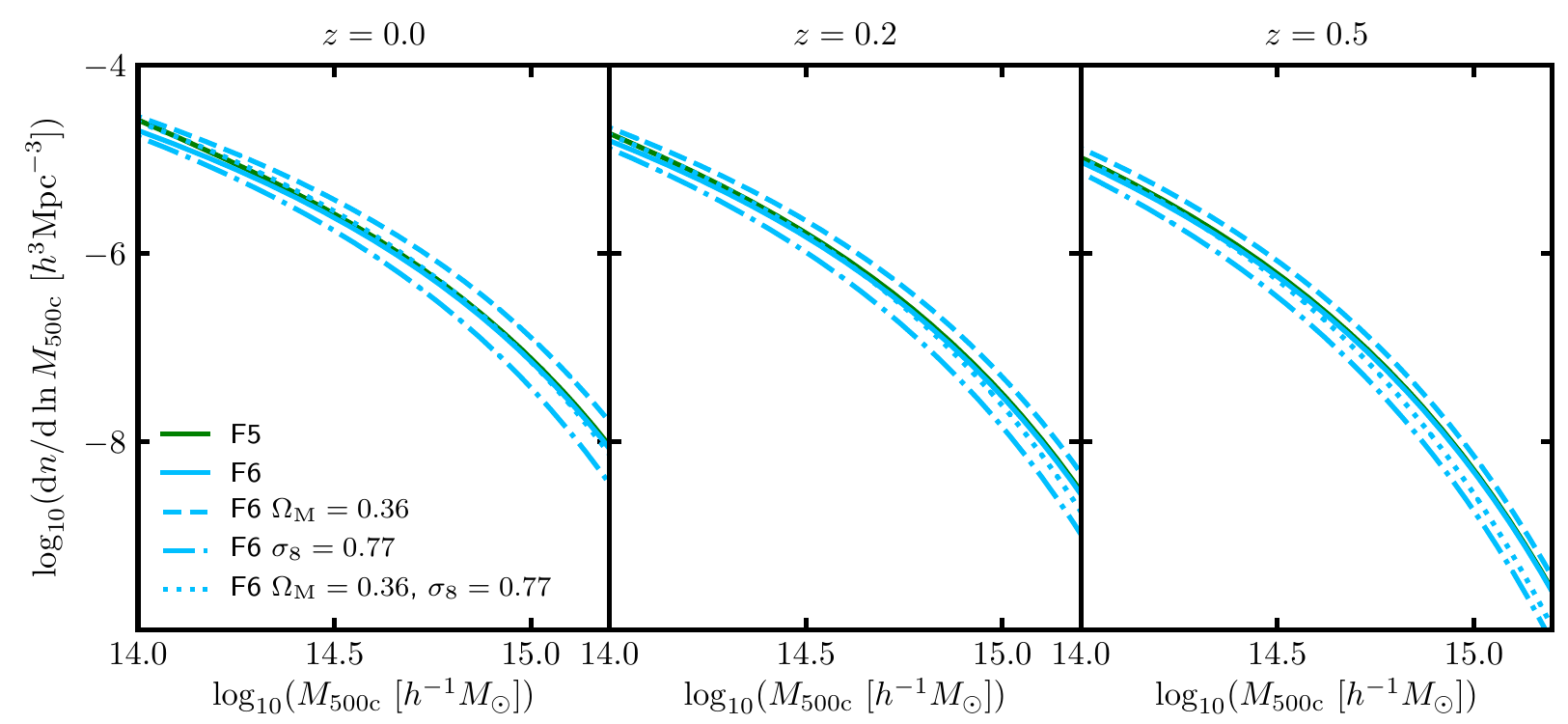}
\caption[Predictions of the halo mass function in F6 and F5 for different combinations of values of $\Omega_{\rm M}$ and $\sigma_8$.]{Predictions of the HMF in F6 (\textit{light blue}) and F5 (\textit{dark green}) at redshifts 0.0, 0.2 and 0.5. We show predictions with the fiducial parameter values $\Omega_{\rm M}=0.3153$ and $\sigma_8=0.8111$ (\textit{solid lines}), an increased $\Omega_{\rm M}$ (\textit{dashed line}), a reduced $\sigma_8$ (\textit{dash-dotted line}) and both an increased $\Omega_{\rm M}$ and reduced $\sigma_8$ (\textit{dotted line}). This figure illustrates not only the well-known degeneracy between $\Omega_{\rm M}$ and $\sigma_8$ in determining the HMF, but also their degeneracy with $f_{R0}$: by tuning the values of these two parameters, an F6 model can closely mimic the HMF of an F5 model; note that the latter degeneracy may be broken by looking at multiple redshifts or by having more precise knowledge of $\Omega_{\rm M}$ and/or $\sigma_8$.}
\label{fig:hmf_degeneracy}
\end{figure*}

For many of the results discussed in this chapter, we have observed degeneracies between $\log_{10}|f_{R0}|$, $\Omega_{\rm M}$, $\sigma_8$, $\alpha$ and $\beta$. Together, these parameters can vary such that the theoretical GR HMF is consistent with the F5 mocks, or similarly the theoretical F5 HMF can be made consistent with the GR mocks. For our main results with the $f(R)$ constraint pipeline (Fig.~\ref{fig:full_fr_pipeline}), we have been using a tight Gaussian $\Omega_{\rm M}$ prior from Planck 2018. In Fig.~\ref{fig:corner_tight_priors}, the blue constraints have been generated using the same F5 mock as the blue constraints in Fig.~\ref{fig:full_fr_pipeline}; however, here a flat prior [0.15,0.50] has been adopted for $\Omega_{\rm M}$. This gives rise to the degeneracy between $\Omega_{\rm M}$ and $\sigma_8$ (observed earlier in Fig.~\ref{fig:gr_pipeline}), which leads to a uniform distribution in $\log_{10}|f_{R0}|$. Fig.~\ref{fig:hmf_degeneracy} provides an illustration of this degeneracy: here, the HMF prediction for F6 with increased $\Omega_{\rm M}$ and reduced $\sigma_8$ closely resembles the F5 prediction, particularly at lower masses which dominate the mock cluster samples. By using the tight $\Omega_{\rm M}=0.3153\pm0.0073$ prior for our main results in Sec.~\ref{sec:results_pipeline:fr_pipeline}, we have prevented this issue. The tight prior on $\Omega_{\rm M}$ can potentially be replaced by combining cluster number counts with other cosmological probes that are sensitive to $\Omega_{\rm M}$, such as the CMB.

We have also shown that there is a degeneracy between $\sigma_8$ and the SZ scaling relation parameters $\alpha$ and $\beta$. Although we have used Gaussian priors for the latter, they can still vary enough to cause biased constraints. In Sec.~\ref{sec:bias:sample:ysz_cut}, we found that this degeneracy caused the $\log_{10}|f_{R0}|$ constraints using the F5 mock with $Y_{\rm cut}=10^{-5}{\rm Mpc}^2$ to resemble GR (see Fig.~\ref{fig:corner_ycut}). In Fig.~\ref{fig:corner_tight_priors}, the red contours show the $\log_{10}|f_{R0}|$ constraints for the same mock, but this time using a tighter $\alpha$ prior of $1.79\pm0.01$. The $\log_{10}|f_{R0}|$ posterior distribution now peaks close to the fiducial value $-5$, though the constraints on $\sigma_8$ and $\beta$ are similarly biased as before. In this case, as in Sec.~\ref{sec:bias:sample:ysz_cut}, the constrained $\beta$ value is lower, which means less time evolution; because the time evolution is normalised at $z=0$, this implies that, for a given cluster mass $M_{500}$, the measured observable $Y$ at $z>0$ is smaller than the true value, and so fewer detectable clusters would be predicted. This is compensated by a larger $\sigma_8$ (actually a similar degeneracy can be observed in the GR case, see the $\sigma_8$--$\beta$ contour in Fig.~\ref{fig:gr_pipeline}), but one side effect is that smaller $\log_{10}|f_{R0}|$ values are more likely to be allowed, leading to a uniform posterior distribution in Fig~\ref{fig:corner_ycut}, which is alleviated in Fig.~\ref{fig:corner_tight_priors} with the tighter prior on $\alpha$ but nevertheless not completely eliminated. Looking at the red contours in the left column of Fig.~\ref{fig:corner_tight_priors}, we can see that at $\log_{10}|f_{R0}|\approx-5$, $\beta$ and $\sigma_8$ both match their correct values, which suggests that if we can tighten the prior on either $\sigma_8$ or $\beta$, the constraint on $\log_{10}|f_{R0}|$ can be further improved. 

Therefore, a conclusion from this discussion is that, with better knowledge of the scaling relation parameters, it is possible to reduce the effect of these degeneracies. However, we note that it may be difficult to constrain the scaling relation parameters with even greater precision. In this case, the degeneracies could be prevented by using a synergy with weak lensing data, which can estimate the cluster mass with higher precision. Even if this data is only available for a subset of the clusters, it can still be incorporated in the log-likelihood \citep[e.g.,][]{Bocquet:2018ukq}.

\section{Summary, Discussion and Conclusions}
\label{sec:conclusions_pipeline}


In this chapter, we have combined all of our models for the effects of $f(R)$ gravity on cluster properties into an MCMC pipeline for constraining the amplitude of the present-day background scalar field, $|f_{R0}|$. We have adopted the model from \citet{Cataneo:2016iav} for the $f(R)$ enhancement of the HMF, and used this, along with our model for the enhancement of the halo concentration, to produce a model-dependent prediction of the cluster number counts (Sec.~\ref{sec:methods_pipeline:hmf}). We have also used our model for the enhancement of the dynamical mass in $f(R)$ gravity to convert a GR power-law observable-mass scaling relation, which is based on the Planck $Y_{\rm SZ}(M_{500})$ relation \citep{Planck_SZ_cluster}, into a form consistent with $f(R)$ gravity, where the fifth force enhances the relation at sufficiently low masses (Sec.~\ref{sec:methods_pipeline:scaling_relation}). These models are all incorporated in our log-likelihood (Sec.~\ref{sec:methods_pipeline:likelihood}), which we have used to infer parameter constraints using a set of mock cluster catalogues (Sec.~\ref{sec:methods_pipeline:mock}).

Using a combination of GR and F5 mocks, we have shown that our pipeline is able to give reasonable parameter constraints that are consistent with the fiducial cosmology (Figs.~\ref{fig:gr_pipeline} and \ref{fig:full_fr_pipeline}). For the GR mock, the constraints conclusively rule out $f(R)$ models with $\log_{10}|f_{R0}|\gtrsim-5$ and favour values in the range $-7\leq\log_{10}|f_{R0}|\lesssim-5$ where $-7$ is the lowest value considered by our MCMC sampling. Meanwhile, the constraints inferred using the F5 mock favour values close to the fiducial value of $-5$, with 68\% range $-5.1^{+0.3}_{-1.0}$ and a `most likely' value of $-4.92$. We have also shown that the constraints inferred from both mocks can be imprecise and biased if the $f(R)$ enhancement of the scaling relation is not accounted for (Fig.~\ref{fig:gr_sr}). Therefore, this should be properly modelled in future tests of $f(R)$ gravity in order to prevent biased constraints. This will become particularly relevant as cluster catalogues start to enter the galaxy group regime \citep[e.g.,][]{Pillepich:2018sin,2021Univ....7..139L}, where more objects can be unscreened in $f(R)$ gravity.

Throughout this chapter, the main obstacle to precise and unbiased constraints has stemmed from degeneracies between $f_{R0}$, $\Omega_{\rm M}$, $\sigma_8$ and the scaling relation parameters $\alpha$ and $\beta$, all of which can influence the predicted cluster count. We have shown that the degeneracies can be prevented by using a tighter Gaussian prior for $\Omega_{\rm M}$ and by having better knowledge of the scaling relation parameters (Fig.~\ref{fig:corner_tight_priors}). The latter can potentially be achieved by including lensing data for a subset of the clusters. If wide or flat parameter priors are used, this may give rise to biased constraints of $\log_{10}|f_{R0}|$. For example, we have found that the parameter degeneracies can have a more significant effect for cluster samples that extend to lower masses (Sec.~\ref{sec:bias:sample:ysz_cut}).

Our constraint pipeline can be improved in a couple of ways. First, while the HMF model of \citet{Cataneo:2016iav} is accurate, it only covers the redshift range $[0,0.5]$. An extended model that works for a larger redshift range, as well as for wider ranges of other cosmological parameters (not restricted to the $\Omega_{\rm M}$ and $\sigma_8$ parameters as we have focused on here), would be very useful. Calibrating this model for spherical overdensity $\Delta=500$ would also mean that conversions between halo mass definitions would no longer be required. Second, the MCMC pipeline should be extended so that independent cluster data, such as weak lensing, can be included in the model constraint. Once these tasks are completed, this pipeline can be used to constrain $f(R)$ gravity using observations. It is also straightforward to extend our framework to other gravity models; we have already started to do this for the nDGP model (Chapter \ref{chapter:DGP_clusters}). 
\graphicspath{{./gfx/}}

\chapter{Cluster and halo properties in DGP gravity}
\label{chapter:DGP_clusters}

\section{Introduction}
\label{sec:introduction:dgp}

So far in this thesis, we have calibrated models for the effects of the $f(R)$ gravity fifth force on the dynamical mass (Chapter \ref{chapter:mdyn}), the halo concentration (Chapter \ref{chapter:concentration}) and the observable-mass scaling relations (Chapter \ref{chapter:scaling_relations}) of clusters, and we have used these to create a robust MCMC constraint pipeline that uses cluster number counts to probe gravity (Chapter \ref{chapter:constraint_pipeline}). However, our general framework for cluster tests of gravity (Fig.~\ref{fig:mg_flow_chart}) is intended to be easily extended to other MG models. In this chapter, we will study cluster and halo properties in nDGP. As discussed in Chapter \ref{chapter:intro}, this model gives rise to departures from GR above a particular `cross-over' scale, resulting in a fifth force which enhances the total strength of gravity (at smaller scales, the fifth force is screened out by the Vainshtein screening mechanism). As with the fifth force in HS $f(R)$ gravity, the fifth force in nDGP could alter cluster properties such as the temperature and density profile. If these are not taken into account in cluster constraints, then cluster mass measurements could become biased. We will address this by studying four models of nDGP, which exhibit different strengths of the fifth force, using a combination of DMO and full-physics simulations that cover a wide range of resolutions and box sizes. This allows us to study and model the effects of the nDGP fifth force on the halo concentration and observable-mass scaling relations. By combining our DMO simulations, we also examine the halo abundance over a continuous mass range extending from Milky Way galaxy-sized to large cluster-sized haloes.

This chapter is organised as follows: in Sec.~\ref{sec:methods:dgp}, we describe the nDGP simulations used in the analyses of this chapter and the method for calculating the halo properties; our main results are presented and discussed in Sec.~\ref{sec:results:dgp}; and, in Sec.~\ref{sec:conclusions:dgp}, we give the main conclusions and discuss the significance of our results.

\section{Simulations and methods}
\label{sec:methods:dgp}

Since Eq.~\eqref{eq:DGP_scalar_field} for the scalar field is highly nonlinear, the fifth force in the nDGP model can display a wide spectrum of behaviours, depending on time, scale and the mass of the objects being considered. Therefore, numerical simulations are essential for predicting its cosmological properties and implications accurately. For earlier works that make use of nDGP simulations, see, e.g., \citet{Chan:2009ew,Schmidt:2009sg,Khoury:2009tk,Li:2013nua,Falck:2014jwa,Falck:2015rsa}. We describe the DMO and full-physics simulations used in this chapter in Sec.~\ref{sec:methods:dgp:simulations}. Then, in Sec.~\ref{sec:methods:dgp:measurements}, we describe our methods for computing the thermal properties and concentration of our haloes.

\subsection{Simulations}
\label{sec:methods:dgp:simulations}

Our simulations were run using \textsc{arepo} \citep{2010MNRAS.401..791S}. One of these is the first cosmological simulation to simultaneously incorporate both full baryonic physics (implemented using the IllustrisTNG model) and nDGP\footnote{We note that the IllustrisTNG model was tuned using standard gravity simulations. However, the differences between the GR and nDGP predictions for the stellar and gas properties of galaxies are generally small compared to typical observational scatters \citep[see, e.g., Fig.~8 of][]{Hernandez-Aguayo:2020kgq}, making a full retuning of the TNG parameters for the nDGP model unnecessary.}. This simulation, which is part of the \textsc{shybone} simulation suite \citep[see][]{Arnold:2019vpg,Hernandez-Aguayo:2020kgq}, has box size $62h^{-1}{\rm Mpc}$ and consists of $512^3$ dark matter particles, with mass $1.28\times10^8h^{-1}M_{\odot}$, and (initially) the same number of Voronoi gas cells, which have mass $\sim2.5\times10^7h^{-1}M_{\odot}$ on average. We also have four DMO $N$-body simulations, whose specifications are provided in Table \ref{table:simulations:dgp}. These have box sizes $62h^{-1}{\rm Mpc}$, $200h^{-1}{\rm Mpc}$, $500h^{-1}{\rm Mpc}$ and $1000h^{-1}{\rm Mpc}$. Throughout this chapter, we will refer to these as L62, L200, L500 and L1000, respectively. These span a wide range of mass resolutions -- from $1.52\times10^8h^{-1}M_{\odot}$ in L62 to $7.98\times10^{10}h^{-1}M_{\odot}$ in L1000 -- allowing us to study haloes spanning, continuously, the mass range $\sim10^{11}h^{-1}M_{\odot}$ to $\sim10^{15}h^{-1}M_{\odot}$.

\begin{table*}
\centering

\small
\begin{tabular}{ c@{\hskip 0.4in}cccc } 
 \toprule
 
 Specifications & \multicolumn{4}{c}{Simulations} \\
 and models & L62 & L200 & L500 & L1000 \\

 \midrule

 box size / $h^{-1}$Mpc & 62 & 200 & 500 & 1000 \\ 
 particle number & $512^3$ & $1024^3$ & $1024^3$ & $1024^3$ \\ 
 particle mass / $h^{-1}M_{\odot}$ & $1.52\times10^8$ & $6.39\times10^8$ & $9.98\times10^9$ & $7.98\times10^{10}$ \\
 nDGP models & N1,N5 & N0.5,N1,N2,N5 & N0.5,N1,N2,N5 & N0.5,N1,N2,N5\\
 
 \bottomrule
 
\end{tabular}

\caption[Specifications of the four dark-matter-only simulations used to study cluster and halo properties in nDGP.]{Specifications of the four DMO simulations that are used, along with the full-physics \textsc{shybone} simulations, to study cluster and halo properties in nDGP. These are labelled L62, L200, L500 and L1000, according to their box size. The simulations have all been run for GR in addition to the nDGP models listed, where N0.5, N1, N2 and N5 correspond to $H_0r_{\rm c}=0.5, 1, 2, 5$, respectively.}
\label{table:simulations:dgp}

\end{table*}

The simulations have all been run with cosmological parameters $(h,\Omega_{\rm M},\Omega_{\rm b},\sigma_8,n_{\rm s})$=$(0.6774,0.3089,0.0486,0.8159,0.9667)$. All simulations include runs with N5 and N1, in addition to GR. The L200, L500 and L1000 simulations also feature runs with N2 and N0.5, allowing us to thoroughly explore the effects of different strengths of the fifth force on halo properties. The simulations all begin at redshift $z=127$. For this study, we use 12 particle snapshots from each simulation which span the redshift range $0\leq z\leq3$.

\begin{figure*}
\centering
\includegraphics[width=1.0\textwidth]{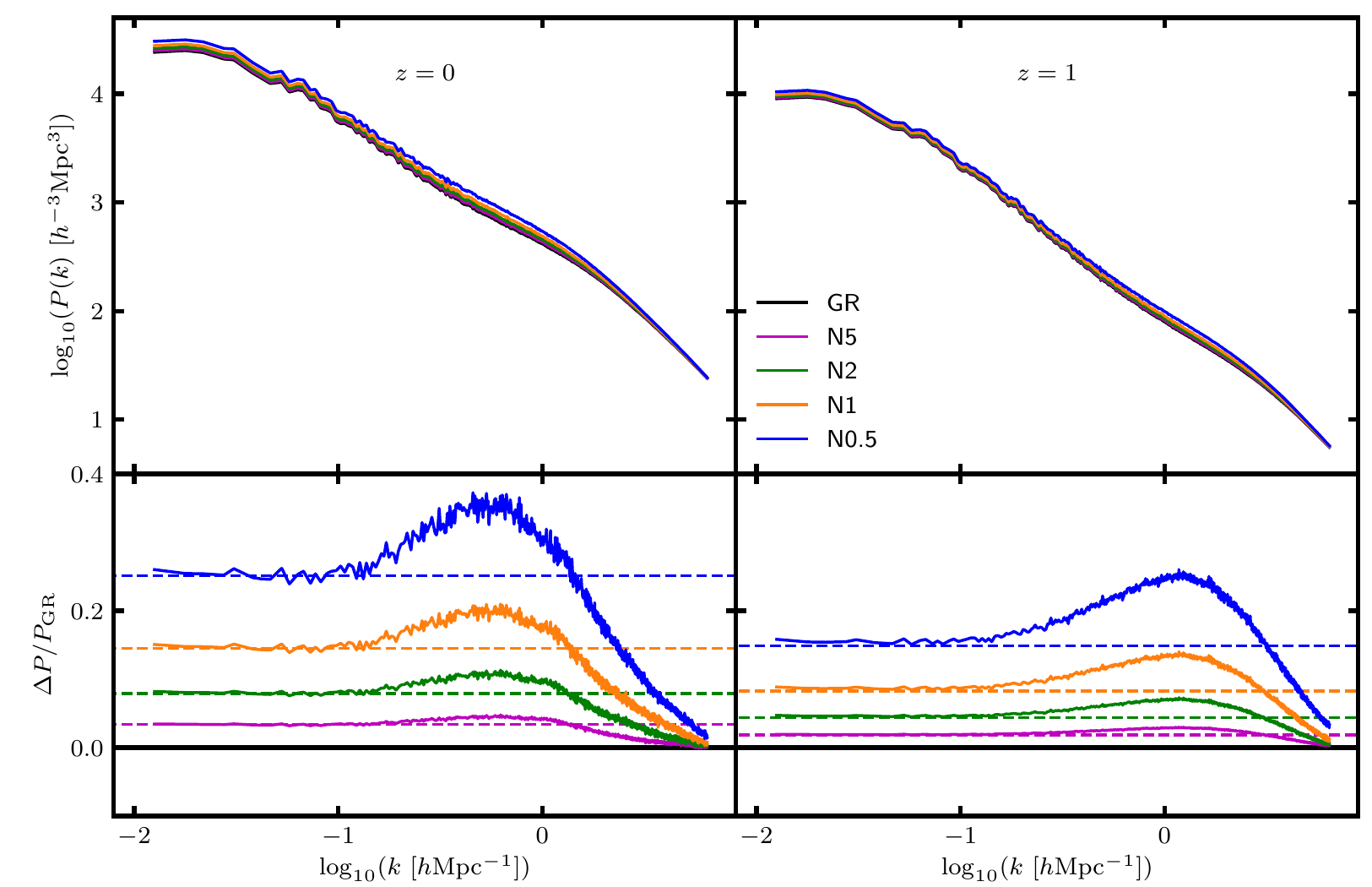}
\caption[Matter power spectra in GR and four nDGP models, generated using the L500 \textsc{arepo} simulation.]{Matter power spectrum (\textit{top row}) and its relative difference in nDGP with respect to GR (\textit{bottom row}), as a function of the wavenumber at redshifts $0$ and $1$. The data has been generated using our dark-matter-only L500 simulation (see Table \ref{table:simulations:dgp}), which has been run for GR (\textit{black}) and the nDGP models N5 (\textit{magenta}), N2 (\textit{green}), N1 (\textit{orange}) and N0.5 (\textit{blue}). The dashed lines in the bottom row show the linear theory predictions of the relative difference.}
\label{fig:Pk}
\end{figure*}

For completeness and as a first check, we show here the matter power spectra of the simulated nDGP models; because this is not the primary focus, we shall only discuss the result briefly. The upper panels of Fig.~\ref{fig:Pk} show the matter power spectra generated using the $z=0$ and $z=1$ snapshots of L500 (similar results can be found for the L200 and L1000 boxes). The relative differences between the nDGP and GR spectra are shown in the lower panels, where we have also included the predictions from linear theory (dashed lines). On large scales ($k\lesssim0.1h{\rm Mpc}^{-1}$), the observed relative differences closely match the linear predictions; here, the fifth force enhances the power by $\sim25\%$ in N0.5 and by a few percent in N5 at $z=0$. The enhancement is even greater at intermediate scales, where the N0.5 power is enhanced by up to $\sim35\%$. This is a consequence of mode-coupling at these scales. At smaller scales ($k\gtrsim1h{\rm Mpc}^{-1}$), which correspond to halo scales, the Vainshtein screening of the fifth force suppresses the power spectrum enhancement. These results are consistent with previous works \citep[e.g.,][]{2009PhRvD..80l3003S,2010PhRvD..81f3005S,Winther:2015wla}. While the trends are similar at $z=0$ and $z=1$, the nDGP enhancement is smaller for the latter due to the fifth force being weaker at earlier times.

\subsection{Halo catalogues}
\label{sec:methods:dgp:measurements}

At each simulation snapshot, we have generated halo catalogues using the \textsc{subfind} code \citep{springel2001}. For each halo, we have measured the mass-weighted gas temperature and the $Y_{\rm SZ}$ and $Y_{\rm X}$ parameters using Eqs.~(\ref{eq:mass_weighted_temperature})-(\ref{eq:yx}), again excluding gas cells within the radial range $r<0.15R_{500}$. The halo concentration is measured using full fitting of the NFW profile to the halo density, as described in Sec.~\ref{sec:c_measurement}.

\section{Results}
\label{sec:results:dgp}

In Sec.~\ref{sec:results:dgp:scaling_relations}, we present our results for the observable-mass scalings using our full-physics simulations. Then, in Sec.~\ref{sec:results:dgp:concentration}, we study and model the concentration-mass-redshift relation in nDGP. Finally, in Sec.~\ref{sec:results:dgp:hmf}, we examine the HMF in nDGP.

\subsection{Observable-mass scaling relations}
\label{sec:results:dgp:scaling_relations}

\begin{figure}
\centering
\includegraphics[width=0.6\columnwidth]{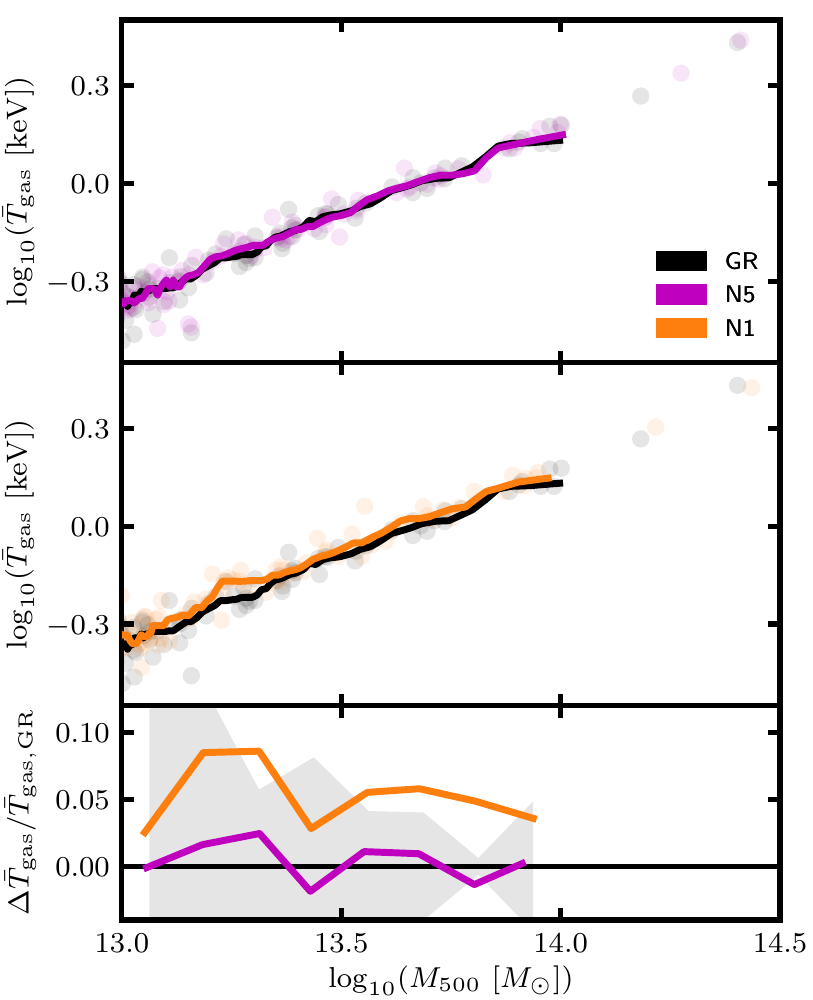}
\caption[Mass-weighted temperature as a function of mass for haloes in nDGP and GR, using the full-physics \textsc{shybone} simulations.]{Gas temperature as a function of mass for haloes from our full-physics \textsc{shybone} simulations (see Sec.~\ref{sec:methods:dgp:simulations}) at $z=0$. Data is included for GR (\textit{black}) and the nDGP models N5 (\textit{magenta}) and N1 (\textit{orange}). The data points correspond to individual haloes. The lines show the median temperature and mean logarithm of the mass which have been computed using a moving window. \textit{Bottom panel}: relative difference between the median temperatures in nDGP and GR; the grey shaded region shows the size of the GR halo scatter.}
\label{fig:tgas}
\end{figure}

\begin{figure}
\centering
\includegraphics[width=0.6\columnwidth]{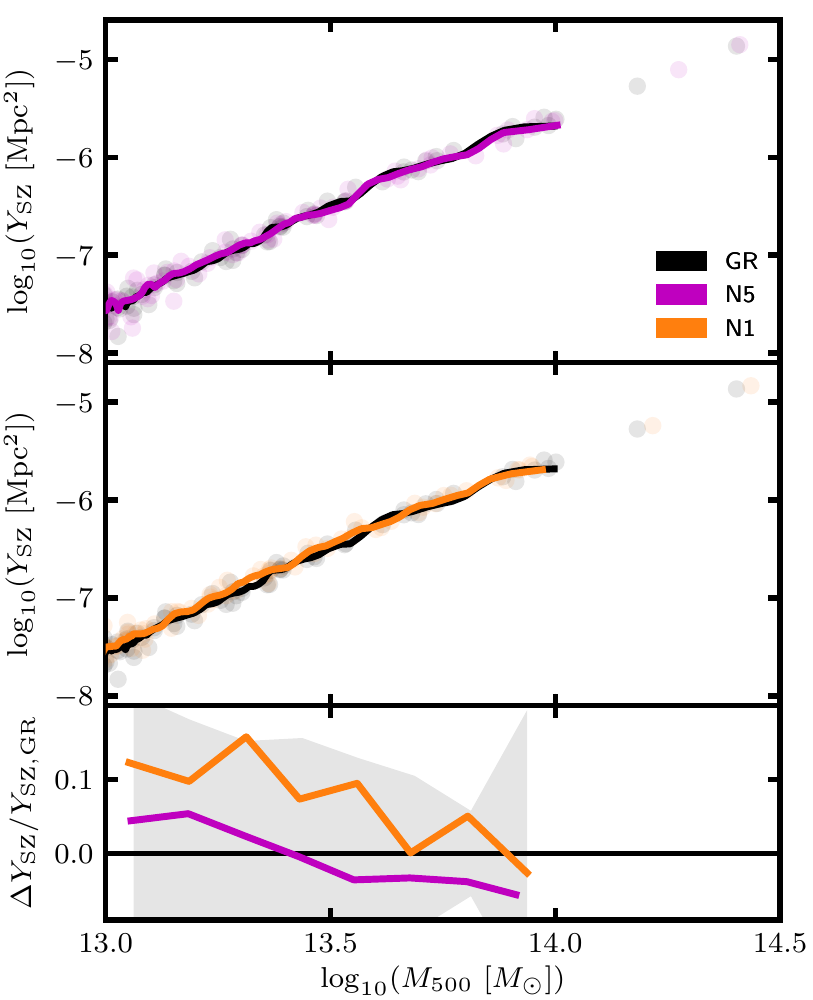}
\caption[$Y_{\rm SZ}$ parameter as a function of mass for haloes in nDGP and GR, using the full-physics \textsc{shybone} simulations.]{SZ Compton $Y$-parameter as a function of mass for haloes from our full-physics \textsc{shybone} simulations (see Sec.~\ref{sec:methods:dgp:simulations}) at $z=0$. Data is included for GR (\textit{black}) and the nDGP models N5 (\textit{magenta}) and N1 (\textit{orange}). The data points correspond to individual haloes. The lines show the median $Y$-parameter and mean logarithm of the mass which have been computed using a moving window. \textit{Bottom panel}: relative difference between the median $Y$-parameters in nDGP and GR; the grey shaded region shows the size of the GR halo scatter.}
\label{fig:ysz}
\end{figure}

\begin{figure}
\centering
\includegraphics[width=0.6\columnwidth]{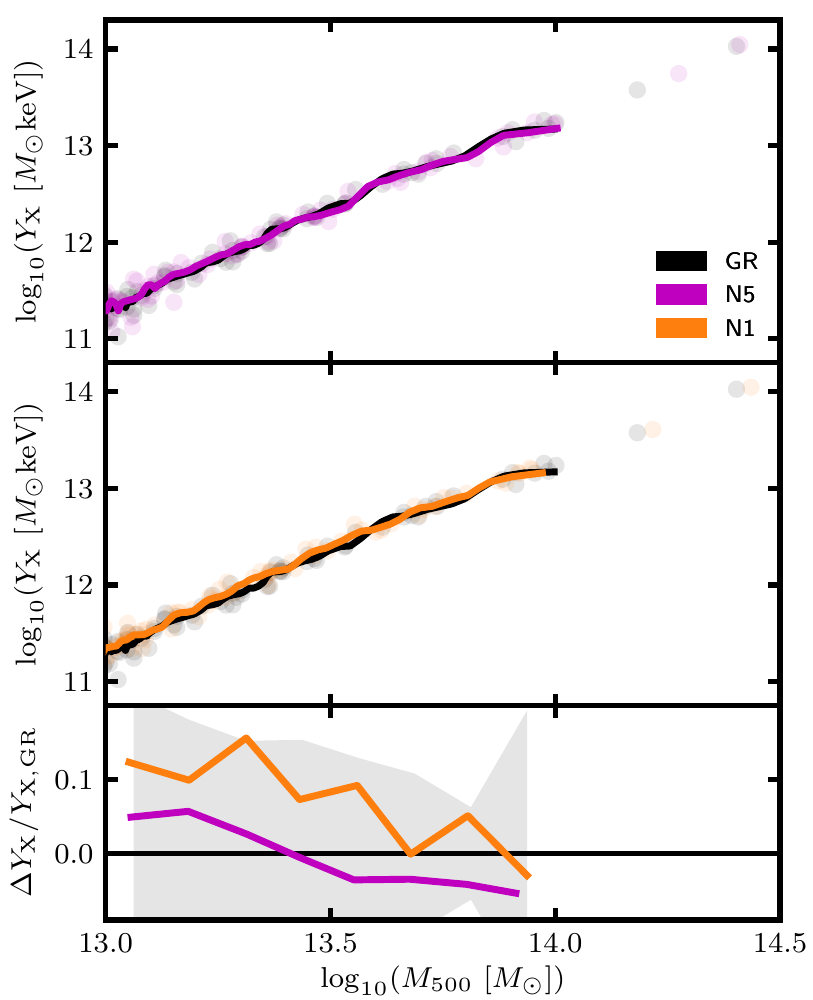}
\caption[$Y_{\rm X}$ parameter as a function of mass for haloes in nDGP and GR, using the full-physics \textsc{shybone} simulations.]{X-ray analogue of the Compton $Y$-parameter as a function of mass for haloes from our full-physics \textsc{shybone} simulations (see Sec.~\ref{sec:methods:dgp:simulations}) at $z=0$. Data is included for GR (\textit{black}) and the nDGP models N5 (\textit{magenta}) and N1 (\textit{orange}). The data points correspond to individual haloes. The lines show the median $Y$-parameter and mean logarithm of the mass which have been computed using a moving window. \textit{Bottom panel}: relative difference between the median $Y$-parameters in nDGP and GR; the grey shaded region shows the size of the GR halo scatter.}
\label{fig:yx}
\end{figure}

In Figs.~\ref{fig:tgas}-\ref{fig:yx}, we plot the mass-weighted gas temperature and the $Y_{\rm SZ}$ and $Y_{\rm X}$ parameters against the halo mass $M_{500}$. In addition to showing individual data points for each halo in the mass range $M_{500}>10^{13}M_{\odot}$, we also plot lines showing the median observable as a function of the mean logarithm of the mass. These averages have been computed using a moving window with a fixed size of 10 haloes. This approach, which is consistent with our study of the observable-mass scaling relations using the $f(R)$ \textsc{shybone} simulations (Chapter \ref{chapter:scaling_relations}), is preferred over using a set of fixed-width bins, which would contain much fewer haloes at high mass than at low mass. The moving averages make use of all haloes with mass $M_{500}>10^{13}M_{\odot}$, including cluster-sized haloes with $M_{500}\gtrsim10^{14}M_{\odot}$. We note, however, that because there are only a few haloes with this mass (owing to the small box size of the full-physics simulations), the highest mean mass of the moving average is only $\sim10^{14}M_{\odot}$. The lower panels of Figs.~\ref{fig:tgas}-\ref{fig:yx} show the relative differences between the observable medians in nDGP and GR. These are smoothed by computing the mean relative difference within 8 mass bins. We also show the root-mean-square halo scatter in GR for each of these bins (grey shaded regions). 

The $\bar{T}_{\rm gas}$-$M$ relation is shown in the top two panels of Fig.~\ref{fig:tgas}. Both the GR and nDGP data follow a power-law relation as a function of the mass. From the lower panel of Fig.~\ref{fig:tgas}, we see that the median temperature in N5 agrees very closely with GR, typically within a couple of percent. This is consistent with the fact that the fifth force has a very small amplitude in this model (see the discussion below Eq.~\eqref{eq:DGP_beta}). However, the temperature in N1 is enhanced by about 5\% relative to GR on average. This result is quite surprising: using the same full-physics simulations, \citet{Hernandez-Aguayo:2020kgq} found that the N1 fifth force reaches just 2\%-3\% of the strength of the Newtonian force at the radius $R_{500}$ for galaxy group-sized haloes and is even more efficiently screened at smaller radii. Therefore, the total gravitational potential at radius $R_{500}$, within which we have calculated the gas temperature, is expected to be just a few percent deeper than the Newtonian potential. 
We would therefore expect the temperature to be enhanced by just a few percent rather than the 5\% that we observe. However, we note that, in nDGP, gravity is enhanced at the outer halo regions even at redshift $z=2$ \citep[see, for example, Fig.~7 of][]{Hernandez-Aguayo:2020kgq}. Therefore, between $0<z<2$, gas at the outer halo radii will undergo a gravitational acceleration in nDGP that is enhanced compared to GR. Consequently, it will have a higher speed than in GR as it reaches smaller radii where it gets shock-heated. The fact that this happens over a long period of time can potentially explain how the gas temperature is enhanced by as much as 5\% within $R_{500}$. 

\begin{figure*}
\centering
\includegraphics[width=0.82\textwidth]{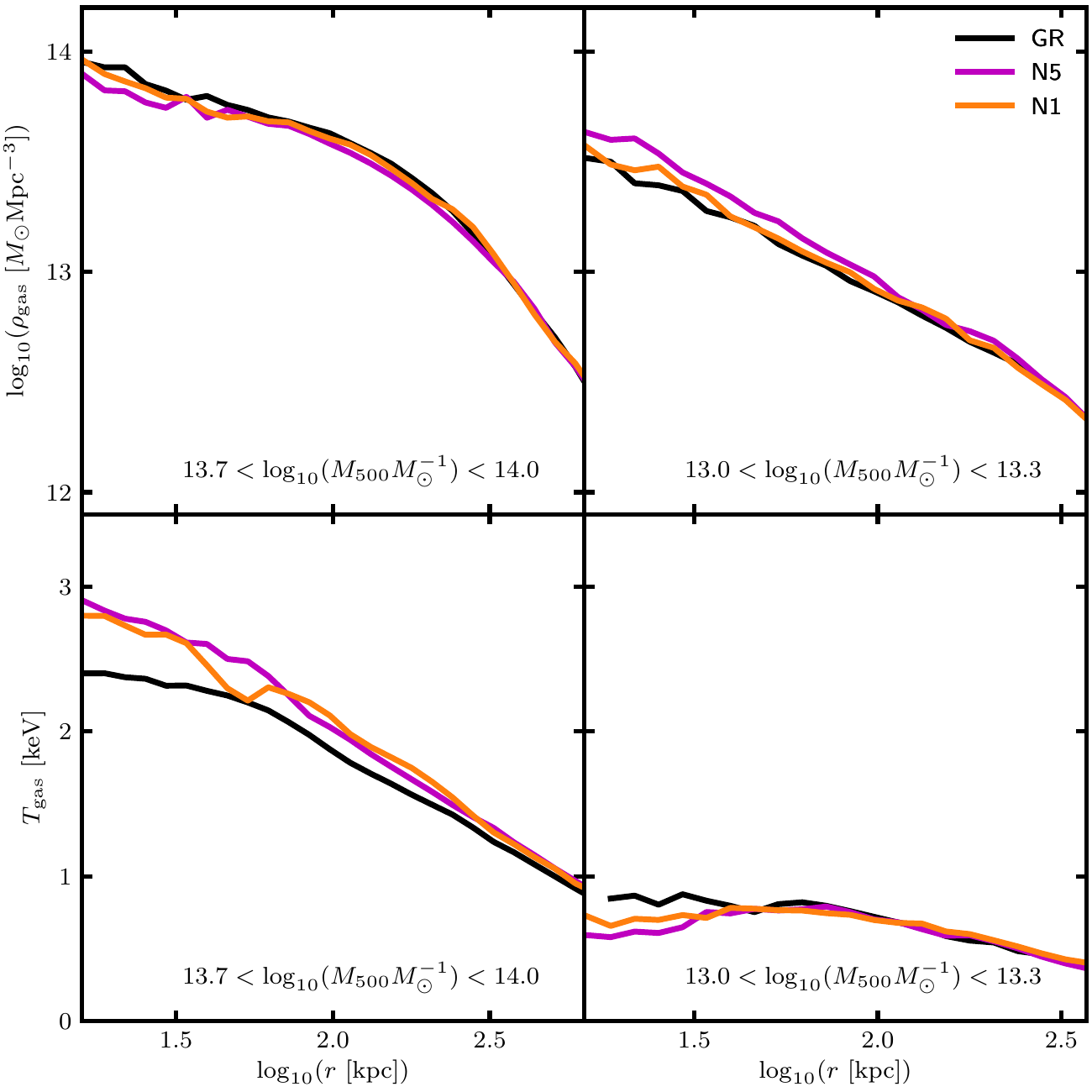}
\caption[Median gas density and temperature profiles of haloes in nDGP and GR, using the full-physics \textsc{shybone} simulations.]{Median gas density (\textit{top row}) and temperature (\textit{bottom row}) profiles of haloes from our full-physics \textsc{shybone} simulations (see Sec.~\ref{sec:methods:dgp:simulations}) at $z=0$. Data is shown for GR (\textit{black}) and the nDGP models N5 (\textit{magenta}) and N1 (\textit{orange}). The two mass bins used to measure the median profiles are annotated. The maximum radius shown for each column corresponds to $R_{500}$.}
\label{fig:profiles:dgp}
\end{figure*}

Our results for the $Y_{\rm SZ}$-$M$ and $Y_{\rm X}$-$M$ scaling relations are shown in Figs.~\ref{fig:ysz} and \ref{fig:yx}, respectively. The $Y_{\rm SZ}$ and $Y_{\rm X}$ parameters are closely related to each other, and so the results appear similar for both: the enhancement of the $Y$-parameters in the N1 model ranges from zero at high masses to 10\%-15\% at low masses, while in N5 it ranges between a 5\% suppression at high masses and 5\% enhancement at low masses. The low-mass enhancement in N1 can in part be explained by the enhanced temperature seen in Fig.~\ref{fig:tgas}. Even for N5, the temperature appears to be enhanced on average for masses $M_{500}\lesssim10^{13.4}h^{-1}M_{\odot}$, so this can also partly explain the $\sim5\%$ enhancement of the $Y$-parameters at these masses. The $Y$-parameters are also correlated with the gas density. In the top row of Fig.~\ref{fig:profiles:dgp}, we show the median gas density profiles for haloes from two mass bins (annotated). For the low-mass bin, both the N5 and N1 gas profiles appear to be enhanced, on average, with respect to GR, while for the high-mass bin the profiles appear to be suppressed. This can help explain why the $Y$-parameters are enhanced in nDGP at lower masses and closer to GR or suppressed at higher masses.

The physical origin of these effects on the gas density is not entirely clear. They could be related to the complex interrelations between the nDGP fifth force and baryonic processes such as cooling and feedback. For example, if the fifth force leads to a larger amount of feedback, this would heat up and blow out surrounding gas. This would be consistent with the results shown for the gas density and temperature profiles in the high-mass bin in Fig.~\ref{fig:profiles:dgp}, where the gas density is suppressed and the temperature is enhanced in nDGP compared to GR. The opposite trend is present in the low-mass bin, which would be consistent with a lowering of feedback efficiency in nDGP compared to GR. Another possibility is that the enhancement of the gas speeds due to the fifth force leads to differences in the density profiles between nDGP and GR. This effect can be inherited from times before the gas falls into haloes and is screened from the fifth force. Haloes of different mass will experience this effect to a different extent as larger haloes are formed from matter and gas further afield.

In the lower panels of Fig.~\ref{fig:profiles:dgp} we show the halo gas temperature profiles. For the higher mass bin, the profiles in N5 and N1 are both enhanced compared to GR. For N1, this is consistent with the result for the mass-weighted temperature discussed above; however, for N5, the enhancement relative to GR appears to contradict Fig.~\ref{fig:tgas}. This is actually related to the difference in binning: while the median mass-weighted temperature has been computed using a moving window containing a fixed number of haloes, the temperature profile is computed within a single wide bin. The mean mass of this bin is actually higher in N5 than in GR, indicating that this bin contains a greater number of high-mass haloes in N5 which also have a higher temperature. This supports our decision to use a moving average in Figs.~\ref{fig:tgas}-\ref{fig:yx}, which avoids the issues that arise from having a fixed set of bins for each model. For the lower mass bin, the nDGP temperature profiles are suppressed for radii $r\lesssim 100{\rm kpc}$ and the N1 profile is just slightly enhanced at higher radii. We note that, because there are more particles at the outer radii, which cover a larger volume, these regions have a greater overall contribution to the mass-weighted temperature, which can explain why the latter is enhanced in N1 even though the temperature profile is suppressed at lower radii compared to GR. And, as described above, the difference in binning can make it difficult to directly compare Figs.~\ref{fig:tgas} and \ref{fig:profiles:dgp}.

We finally note that, due to the small box size of our full-physics simulations, we can only rigorously study the scaling relations for halo masses corresponding to galaxy groups. A larger box will be required to rigorously probe the interplay between the fifth force and baryonic physics in galaxy clusters. Galaxy groups, particularly low-mass groups, are typically more susceptible to feedback than cluster-sized objects. This is why, in Fig.~\ref{fig:tgas}, the scatter in the GR halo temperature is above 10\% for low-mass groups and less than 5\% for high-mass groups. It will therefore be interesting to see how the nDGP scaling relations compare to GR at these larger masses, where the unpredictable effects from feedback are not as significant. In Chapter \ref{chapter:baryonic_fine_tuning}, we will present a retuned baryonic physics model which we plan to use for running large-box full-physics simulations of the nDGP model in the future.

\subsection{Concentration-mass relation}
\label{sec:results:dgp:concentration}

In Sec.~\ref{sec:results:dgp:c_DMO}, we discuss the concentration results from our DMO simulations (dashed lines in Figs.~\ref{fig:c_M}-\ref{fig:density}). Then, in Sec.~\ref{sec:results:dgp:c_FP}, we summarise the results from full-physics simulations (solid lines in Figs.~\ref{fig:c_M}-\ref{fig:density}), including the effect of baryons on the model differences. Finally, in Sec.~\ref{sec:results:dgp:concentration_model}, we present a general model for the concentration-mass relation in nDGP.

\subsubsection{Dark-matter-only concentration}
\label{sec:results:dgp:c_DMO}

In order to study the concentration over a wide and continuous halo mass range, we have combined the data from our DMO simulations into a single catalogue. In order to avoid resolution issues with the concentration measurement, we exclude haloes which have fewer than 2000 particles (within the radius $R_{200}$) and we leave out L1000 due to its low mass resolution. The resulting catalogue consists of haloes spanning masses $3.04\times10^{11}h^{-1}M_{\odot}\lesssim M_{200}\lesssim10^{15}h^{-1}M_{\odot}$. We note that, because L62 has not been run for N2 and N0.5, the data for these models only extends down to mass $1.278\times10^{12}h^{-1}M_{\odot}$ ($\equiv2000$ particles from L200). Throughout this section, we will only refer to the results from this combined catalogue; however, in Appendix \ref{sec:appendix:dgp_clusters:consistency}, we also compare the concentration predictions from each of our DMO simulations, including L1000.

\begin{figure*}
\centering
\includegraphics[width=1.0\textwidth]{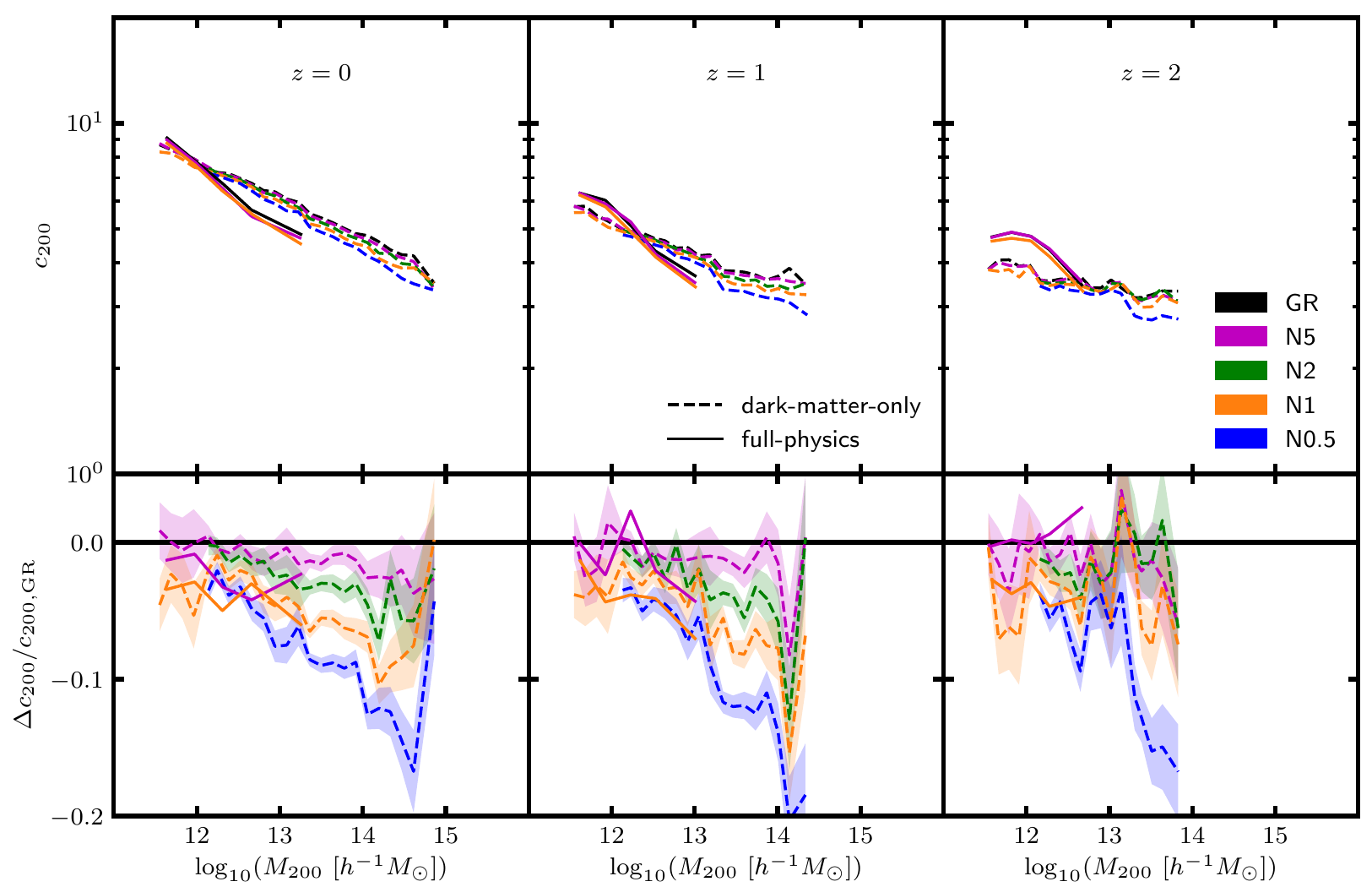}
\caption[Halo concentration as a function of mass for haloes in GR and four nDGP models at redshifts 0, 1 and 2.]{Median halo concentration (\textit{top row}) and its relative difference with respect to GR (\textit{bottom row}), as a function of the mean logarithm of the halo mass at redshifts $0$, $1$ and $2$. The data is generated using the dark-matter-only simulations L62, L200 and L500 (\textit{dashed lines}) and our full-physics simulation (\textit{solid lines}), the specifications of which are given in Table \ref{table:simulations:dgp}. These have been run for GR (\textit{black}) and the nDGP models N5 (\textit{magenta}), N2 (\textit{green}), N1 (\textit{orange}) and N0.5 (\textit{blue}). The shaded regions in the lower panels show the $1\sigma$ uncertainty in the relative difference.}
\label{fig:c_M}
\end{figure*}

The top row of Fig.~\ref{fig:c_M} shows the median concentration as a function of mass for redshifts $0$, $1$ and $2$ (from left to right). The median has been computed using mass bins containing a minimum of 100 GR haloes each: the bins all have equal width in logarithmic mass apart from the highest-mass bin, which is wide enough to enclose the 100 highest-mass haloes. The same set of bins is used for each gravity model. As expected from literature \citep[e.g.,][]{Duffy:2008pz}, the median concentration appears to follow a descending power-law relation with the mass. This behaviour arises due to the hierarchical nature of structure formation: higher-mass haloes form at later times when the background density is lower. Therefore, the concentration of these haloes is also typically lower.

The bottom row of Fig.~\ref{fig:c_M} shows the relative difference between the nDGP and GR median concentrations. The shaded region shows the $1\sigma$ error. To calculate this, the standard error of the mean (equal to the standard deviation divided by the square root of the halo count) is computed for each mass bin for GR and nDGP, and then combined in quadrature. We note that, although the nDGP and GR simulations are started from the same initial conditions, the differing gravitational forces affect the trajectories of the simulation particles, which end up at different positions with different velocities, essentially losing much of the memory of their initial states. Therefore, the concentration measurements of each model can be treated as independent, so that the errors may be combined as described. Our results show that the nDGP fifth force causes the concentration to be reduced, since particles experience the fifth force and hence have enhanced velocities before they fall into haloes, so that after entering the haloes their higher kinetic energy makes it harder for them to settle towards the central regions. The effect is greater for models which have a stronger fifth force, so the concentration suppression is highest in N0.5 ($\sim${$10\%$} on average) and lowest in N5 (at percent level). At $z=0$ and $z=1$, the suppression is greater at higher mass. This appears to be the case for N0.5 at $z=2$ as well, but not for weaker models, where the suppression appears to have a much weaker dependence on the halo mass.

\begin{figure*}
\centering
\includegraphics[width=1.0\textwidth]{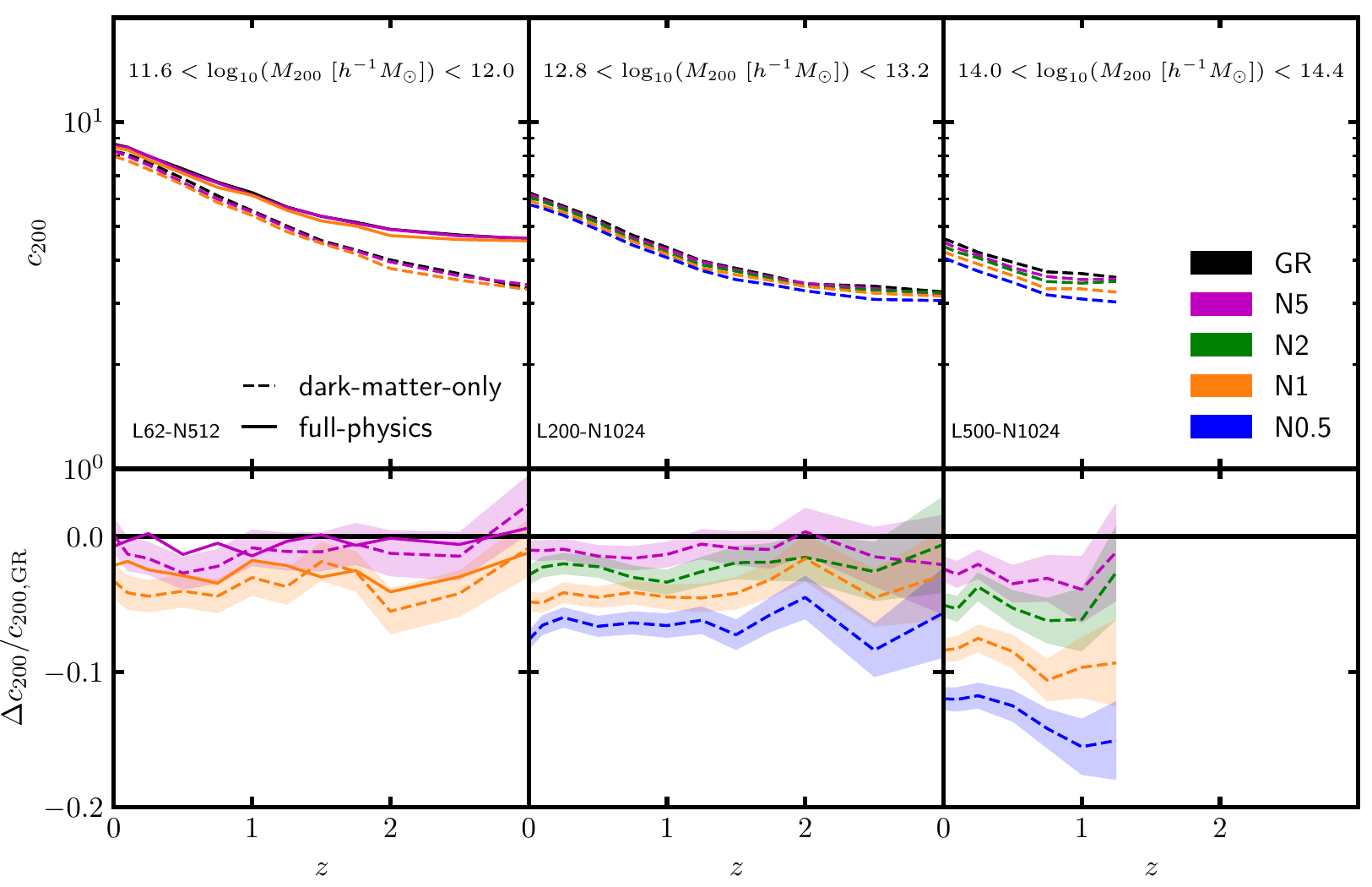}
\caption[Halo concentration as a function of redshift for haloes in GR and four nDGP models, for three mass bins.]{Median halo concentration (\textit{top row}) and its relative difference with respect to GR (\textit{bottom row}), as a function of redshift for three mass bins. The data is generated using the dark-matter-only simulations L62, L200 and L500 (\textit{dashed lines}) and the full-physics simulation  (\textit{solid lines}), the specifications of which are given in Table \ref{table:simulations:dgp}. These have been run for GR (\textit{black}) and the nDGP models N5 (\textit{magenta}), N2 (\textit{green}), N1 (\textit{orange}) and N0.5 (\textit{blue}). The shaded regions in the lower panels show the $1\sigma$ uncertainty in the relative difference.}
\label{fig:c_z}
\end{figure*}

To complement these results, we show the median concentration, computed within three mass bins, as a function of redshift in Fig.~\ref{fig:c_z}. The lower-mass bin, $10^{11.6}h^{-1}M_{\odot}<M_{200}<10^{12}h^{-1}M_{\odot}$, corresponds to galaxy-sized haloes: here, we use haloes from L62, for which we again note that only the GR, N5 and N1 models are available. For the middle-mass bin, $10^{12.8}h^{-1}M_{\odot}<M_{200}<10^{13.2}h^{-1}M_{\odot}$, we use haloes from L200. For both of these bins, the nDGP suppression of the concentrations appears to be approximately constant over the redshift range $0\leq z\leq3$, ranging from a couple of percent at most in N5 to about $7\%$ in N0.5.

The higher-mass bin, $10^{14}h^{-1}M_{\odot}<M_{200}<10^{14.4}h^{-1}M_{\odot}$, shown in Fig.~\ref{fig:c_z} corresponds to cluster-sized objects; for this, we use haloes from L500. Because clusters typically form at later times, this bin consists of fewer than 100 haloes for redshifts $z\gtrsim1.25$, and we therefore exclude these redshifts from the figure. The suppression of the concentration in nDGP is greater for this bin than for the lower-mass bins, reaching $\sim15\%$ in N0.5. This is consistent with the results of Fig.~\ref{fig:c_M}. As for the other bins, the suppression does not appear to evolve with redshift in N5, N2 and N1. However, for N0.5, the suppression is slightly greater at $z=1$ ($\sim15\%$), than at $z=0$ ($\sim12\%$). We note that the error is also greater at high redshift due to the reduced number of objects, so these results alone do not provide compelling evidence of a redshift evolution of the concentration suppression.

\begin{figure*}
\centering
\includegraphics[width=0.82\textwidth]{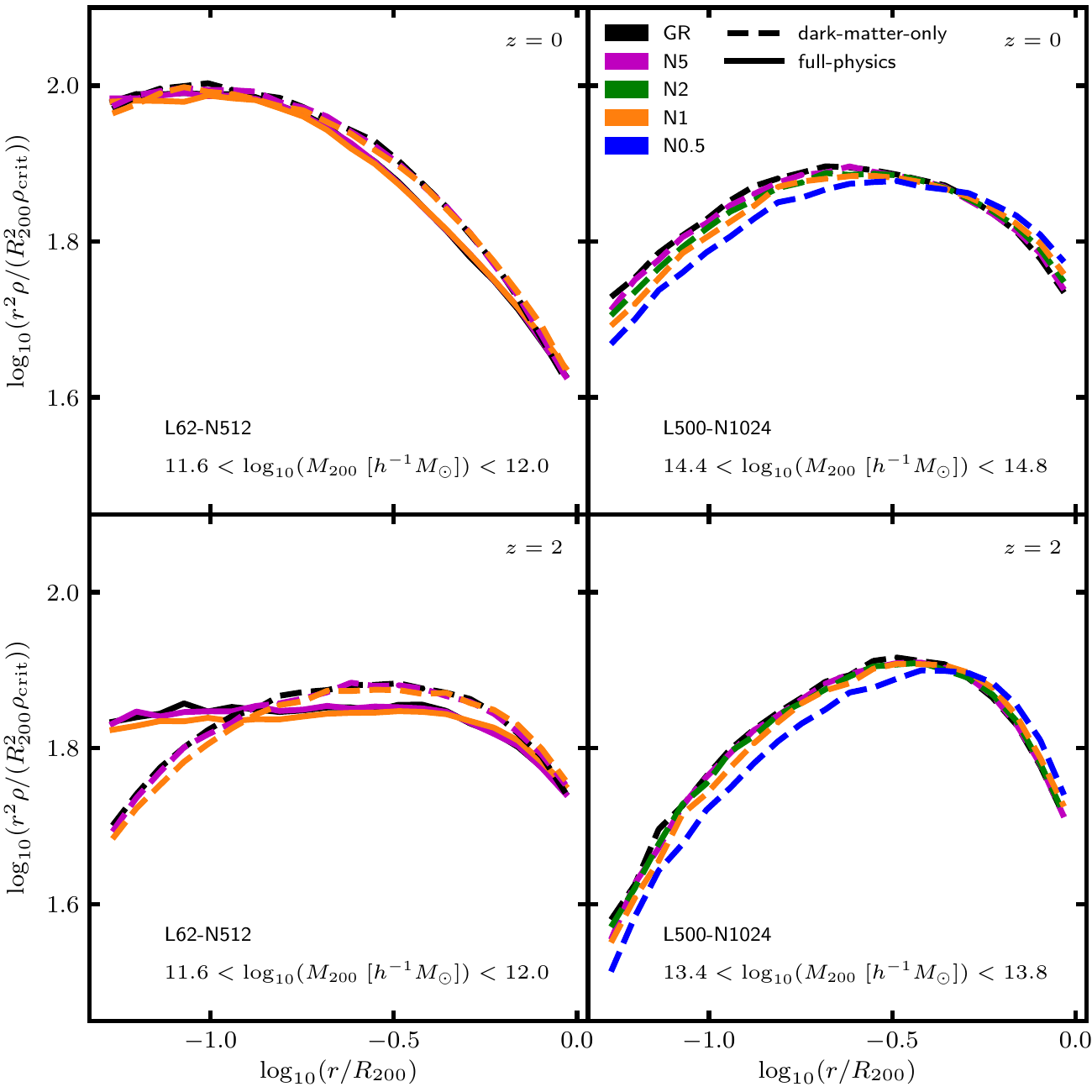}
\caption[Median density profiles of haloes in GR and four nDGP models for a sample of mass bins and redshifts, generated using full-physics and dark-matter-only simulations.]{Median density profiles of haloes from the \textsc{arepo} simulations L62 (\textit{left column}) and L500 (\textit{right column}) at redshifts $0$ (\textit{top row}) and $2$ (\textit{bottom row}). Data from both the full-physics (\textit{solid lines}) and dark-matter-only (\textit{dashed lines}) counterparts of L62 are shown. The L500 simulation includes runs for GR (\textit{black}) and the nDGP models N5 (\textit{magenta}), N2 (\textit{green}), N1 (\textit{orange}) and N0.5 (\textit{blue}), while the L62 simulation includes GR, N5 and N1 only. The mass bins used to measure the median density are annotated.}
\label{fig:density}
\end{figure*}

To help make sense of these results, in Fig.~\ref{fig:density} we show the median density profiles of haloes from a few mass bins at redshifts 0 and 2. These have been computed by measuring the median density, in radial bins spanning $0.05R_{200}$ to $R_{200}$, using the binned haloes. The density has been scaled by $r^2$ so that the profiles peak at the scale radius, $R_{\rm s}$. This means that the concentration, $c_{200}=R_{200}/R_{\rm s}$, can effectively be read off from the peak radius: a higher (lower) peak radius corresponds to a lower (higher) concentration. In the left column of Fig.~\ref{fig:density}, we show the median profile for haloes from L62 in the mass bin $10^{11.6}h^{-1}M_{\odot}<M_{200}<10^{12}h^{-1}M_{\odot}$. In the right column, we use haloes from L500 within mass bins $10^{14.4}h^{-1}M_{\odot}<M_{200}<10^{14.8}h^{-1}M_{\odot}$ and $10^{13.4}h^{-1}M_{\odot}<M_{200}<10^{13.8}h^{-1}M_{\odot}$ at redshifts 0 and 2, respectively. We use a lower mass for the $z=2$ profile due to the limited number of haloes at higher masses.

For the higher mass bins -- where we have seen that there is a greater suppression of the concentration in nDGP models -- a clear trend is present: at the outer (inner) regions of haloes, the density is greater (lower) in nDGP than in GR. As mentioned above, this is related to the nature of the Vainshtein screening in nDGP, which suppresses the fifth force on small scales or distances. This means that the fifth force is stronger at large scales, which correspond to the outer regions of these haloes and regions further away from the halo-formation sites. This causes orbiting dark matter particles to undergo an enhanced gravitational acceleration at these regions and have higher kinetic energy, which prevents them from relaxing and settling into lower-radius orbits where the fifth force is suppressed. This causes $r^2\rho(r)$ to peak at a higher radius in the nDGP models than in GR, resulting in a suppressed concentration. The effect is greatest in N0.5.

For the lower mass bins, we have seen in Figs.~\ref{fig:c_M} and \ref{fig:c_z} that the effect of the fifth force is not as strong. This is consistent with the low-mass density profiles in Fig.~\ref{fig:density}, where the nDGP profiles are closer to GR. However, the density is still slightly reduced at the inner regions and increased at the outer regions, and so the concentration is still suppressed. The reason that the effect is not as strong at low mass is again due to the nature of the Vainshtein screening: lower-mass haloes have a smaller spatial extent, therefore the small-scale suppression of the fifth force is more substantial throughout the range $r<R_{200}$. In addition, smaller haloes generally form at higher redshifts, so that the particles inside them have spent less time outside the haloes and are therefore less affected by the fifth force; this is because, once these particles enter haloes, the fifth force is strongly suppressed.

\subsubsection{Full-physics concentration}
\label{sec:results:dgp:c_FP}

In Figs.~\ref{fig:c_M}-\ref{fig:density}, we have also included data from our full-physics simulations, which are represented with solid lines. Because these data are only available for the $62h^{-1}{\rm Mpc}$ box, the data only extends to low-mass galaxy clusters (although, we note that the mean logarithmic mass of the rightmost bin shown in Fig.~\ref{fig:c_M} is only slightly above $10^{13}h^{-1}M_{\odot}$). Nevertheless, by comparing this to the data from the combined DMO data, we can get an idea of how the results differ when gas and processes such as star formation and feedbacks are included.

From the solid lines in Figs.~\ref{fig:c_M} and \ref{fig:c_z}, we see that the full-physics concentration is typically greater at lower masses and reduced at higher masses. The full-physics simulations include a gaseous component which, unlike dark matter, is affected by turbulence. This causes the gas cells to slow down and settle at the inner regions of haloes. Also, at the centre of a halo, we are likely to see stellar particles concentrate. This means that the total halo density is enhanced in the inner regions, which is consistent with the stacked density profiles of the full-physics simulations in Fig.~\ref{fig:density}. According to these results, the rescaled density profile becomes approximately flat at the inner regions, corresponding to a $\rho(r)\propto r^{-2}$ power-law. This clearly deviates from the NFW profile, which follows an $r^{-1}$ power law in these regions. Because the concentration is a parameter of the NFW profile, we still have to fit Eq.~(\ref{eq:nfw_fit}) in order to measure this. Doing so produces a value that is either higher or lower than for DMO haloes with the same mass.

Despite the difference in the absolute concentrations, the suppression of the concentration in nDGP appears to have a similar magnitude in the full-physics and DMO simulations, according to the bottom-left panel of Fig.~\ref{fig:c_z}, for galaxy-sized haloes in the redshift range $0\leq z\leq3$. In Fig.~\ref{fig:c_M}, the dashed and solid lines in the lower panels also appear to have a similar magnitude; however, we are unable to rigorously test this for masses $M_{200}\gtrsim10^{13}h^{-1}M_{\odot}$, which would require full-physics simulations of nDGP that have a much larger box size. Such simulations are highly expensive, and are therefore left for future work.

\subsubsection{Modelling the concentration in nDGP}
\label{sec:results:dgp:concentration_model}

\begin{figure*}
\centering
\includegraphics[width=1.0\textwidth]{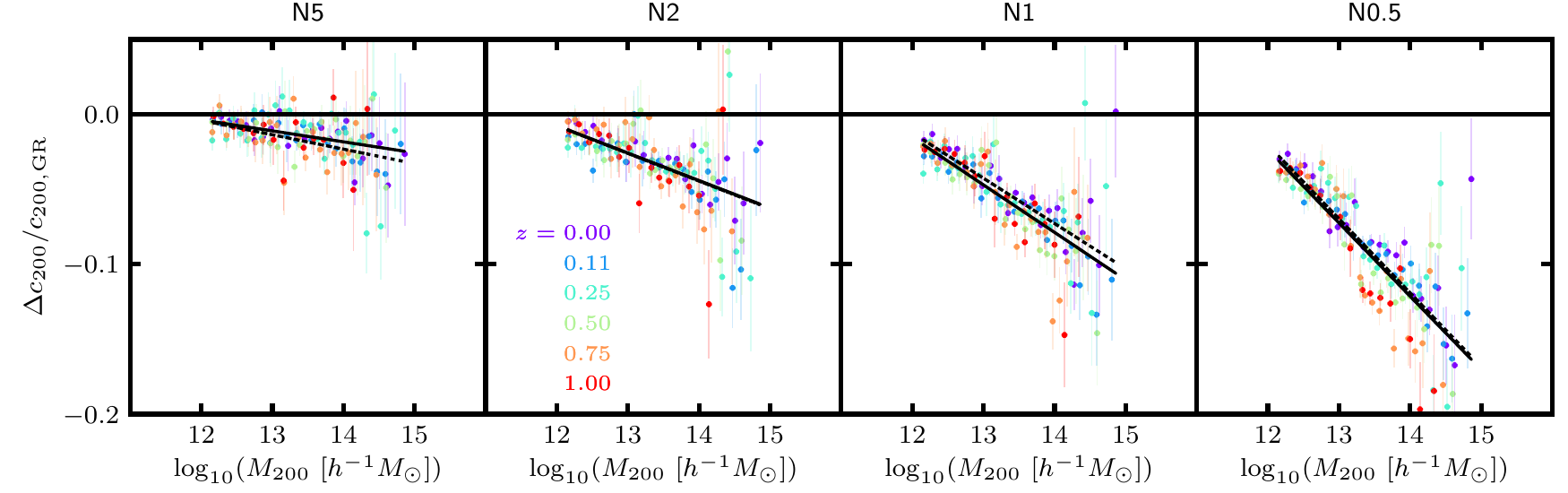}
\caption[Relative difference between the median halo concentrations in nDGP and GR as a function of mass, plotted along with the predictions from our general model.]{Relative difference between the median halo concentration in nDGP with respect to GR, as a function of the mean logarithm of the halo mass. Binned data are shown for all snapshots with redshift $z\leq1$, where the redshift is represented by colour. The data is generated using the dark-matter-only simulations L62, L200 and L500, the specifications of which are given in Table \ref{table:simulations:dgp}. These have been run for GR and the nDGP models N5, N2, N1 and N0.5 (shown from left to right). The error bars indicate the $1\sigma$ uncertainties. The solid lines represent the best-fit linear relations for each panel, while the dashed lines show the predictions from our general model, which is given by Eq.~(\ref{eq:c_model}). For all models, and across a halo mass range of four orders of magnitude, the fitting function gives a percent-level agreement with the simulation measurement of the concentration decrement at $0\leq z\leq1$.}
\label{fig:linear_fit}
\end{figure*}

From Figs.~\ref{fig:c_M} and \ref{fig:c_z}, it appears that the suppression of the DMO halo concentration in nDGP grows with mass and is approximately constant as a function of redshift. In Fig.~\ref{fig:linear_fit}, we show the binned relative difference data from our combined DMO simulation data for all snapshots at $z\leq1$. The data appears to follow a linear trend as a function of the mass, therefore we can model this using:
\begin{equation}
    \Delta c/c_{\rm GR} = A - B\log_{10}(M_{200}M_{\odot}^{-1}h),
    \label{eq:linear_fit}
\end{equation}
where $A$ and $B$ are parameters representing the amplitude and slope of the relation, respectively. This does not include any dependence on redshift. For the N0.5 data, there is a clear $z$-dependence, with low-$z$ (blue) data having a smaller suppression than high-$z$ (red) data; however, the suppression in different snapshots is still quite close, and there does not appear to be any $z$-evolution for the other, more realistic, models of nDGP.

\begin{figure*}
\centering
\includegraphics[width=1.0\textwidth]{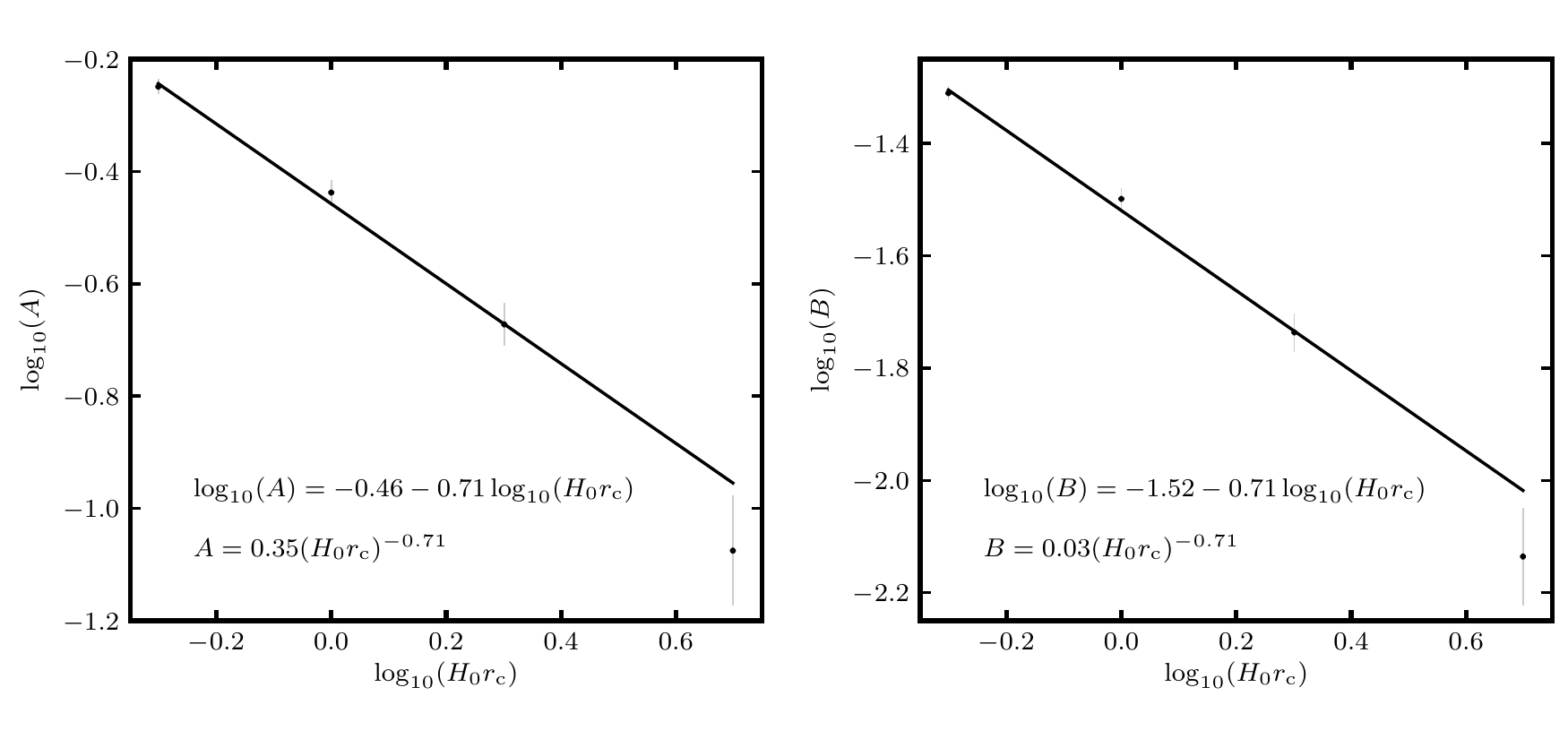}
\caption[Best-fit values of the parameters $A$ and $B$ of Eq.~(\ref{eq:linear_fit}) as a function of the $H_0r_{\rm c}$ parameter of nDGP.]{Best-fit values of the parameters $A$ and $B$ of Eq.~(\ref{eq:linear_fit}) as a function of the logarithm of $H_0r_{\rm c}$, where $r_{\rm c}$ is the cross-over scale of nDGP gravity. The best-fit values of the data points have been computing by fitting Eq.~(\ref{eq:linear_fit}) to the data shown in the four panels of Fig.~\ref{fig:linear_fit}, which correspond to the models N5, N2, N1 and N0.5. The error bars represent the $1\sigma$ uncertainties in the data, obtained from the weighted least squares fits. The solid lines show best-fit power-law fits of the four data points, Eq.~(\ref{eq:power_laws}), which are annotated.}
\label{fig:power_law}
\end{figure*}

The solid lines in Fig.~\ref{fig:linear_fit} are the best-fit relations for each model. These are created by using weighted least squares to fit Eq.~(\ref{eq:linear_fit}) to the data points, where points with large (small) error bars are given smaller (larger) weighting. In Fig.~\ref{fig:power_law}, we show the best-fit values of $A$ and $B$ as a function of the $H_0r_{\rm c}$ parameter which characterises the nDGP models. Both $A$ and $B$ appear to be well-described by a power-law relation. Using weighted least squares to fit the four data points, we obtain the following best-fit relations:
\begin{equation}
    \begin{split}
    &A = (0.35\pm0.01)(H_0r_{\rm c})^{-0.71\pm0.05};\\
    &B = (0.0302\pm0.0008)(H_0r_{\rm c})^{-0.71\pm0.05}.
    \end{split}
    \label{eq:power_laws}
\end{equation}
Interestingly, the relations both have power-law slope $-0.71\pm0.05$. They can therefore be combined with Eq.~(\ref{eq:linear_fit}) to form the following simple relation:
\begin{equation}
    \begin{split}
    \frac{\Delta c}{c_{\rm GR}} = &[(0.35\pm0.01) - (0.0302\pm0.0008)\log_{10}(M_{200}M_{\odot}^{-1}h)]\\
    &\times(H_0r_{\rm c})^{-0.71\pm0.05}.
    \end{split}
    \label{eq:c_model}
\end{equation}
This 
can be used to predict the suppression of the concentration in nDGP, as a function of the halo mass $M_{200}$ and model parameter $H_0r_{\rm c}$. The dashed lines in Fig.~\ref{fig:linear_fit} show the model predictions for our four nDGP models. The agreement with the data is generally very good for the full mass range, $10^{12}h^{-1}M_{\odot}\lesssim M_{200} \lesssim 10^{15}h^{-1}M_{\odot}$, of our simulation data. The agreement is particularly good for weaker models, where it appears to match $z=0$ and $z=1$ data equally well. For N0.5, which is our strongest model, our relation appears to slightly underestimate the concentration suppression for high-redshift data; however, the overall level of agreement is still very good, considering that the model is able to give reasonable predictions for such a wide range of nDGP models and masses.

\begin{figure*}
\centering
\includegraphics[width=1.0\textwidth]{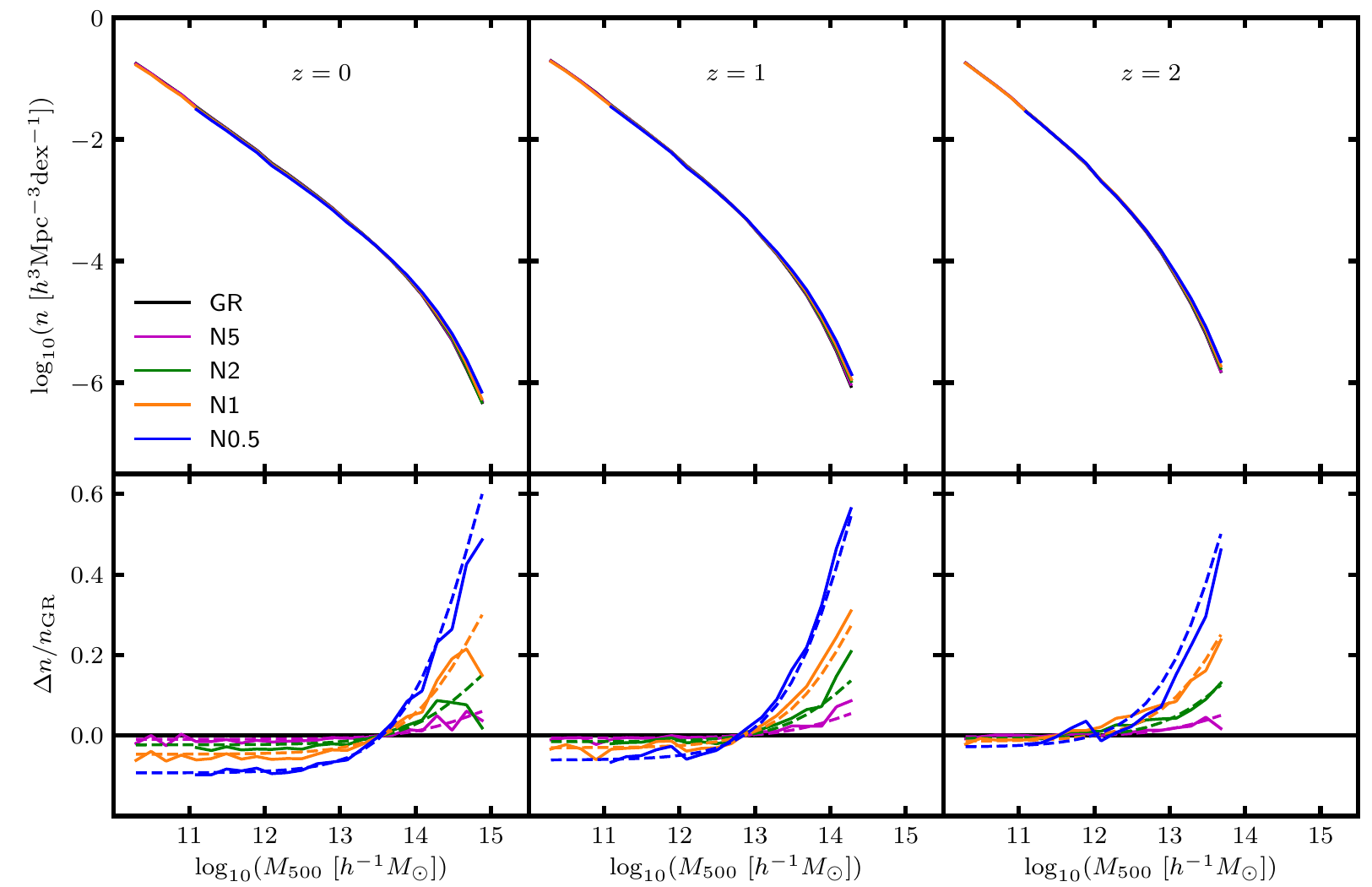}
\caption[Halo mass function and its relative difference in nDGP with respect to GR at redshifts $0$, $1$ and $2$, plotted along with our best-fit model.]{Halo mass function (\textit{top row}) and its relative difference in nDGP with respect to GR (\textit{bottom row}), as a function of the mean logarithm of the halo mass at redshifts $0$, $1$ and $2$. The data is generated using our \textsc{arepo} dark-matter-only simulations, the specifications of which are given in Table \ref{table:simulations:dgp}. These have been run for GR (\textit{black}) and the nDGP models N5 (\textit{magenta}), N2 (\textit{green}), N1 (\textit{orange}) and N0.5 (\textit{blue}). The dashed lines show the predictions from our general fitting model, which is given by Eqs.~(\ref{eq:hmf}, \ref{eq:hmf_params}).}
\label{fig:hmf_combined}
\end{figure*}

\subsection{Halo mass function}
\label{sec:results:dgp:hmf}

The HMF does not have a strict mass resolution requirement like the concentration, therefore we use all haloes which have at least 100 particles within the radius $R_{500}$. We again combine the halo data from our DMO simulations, and the relaxed resolution requirement means that we can now also include L1000 haloes. The HMF is computed using mass bins with equal logarithmic width 0.2. The halo count in each bin is divided by the total volume from all contributing simulations: for example, the volume is $62^3h^{-3}{\rm Mpc}^3$ for the lowest-mass bins where only the L62 box has sufficient resolution, and $(62^3+200^3+500^3+1000^3)h^{-3}{\rm Mpc}^3$ for the highest-mass bins where all simulations have sufficient resolution. In Appendix \ref{sec:appendix:dgp_clusters:consistency}, we also assess the consistency of our DMO simulations by comparing the HMF predictions at different resolutions.

The binned HMF is shown in Fig.~\ref{fig:hmf_combined} for redshifts 0, 1 and 2, where only mass bins containing at least 100 haloes are displayed. We note that, because our highest-resolution simulation L62 has been run for N5 and N1 only, the data for these models extends to lower masses than the other models. The relative difference between the nDGP and GR results is shown in the lower panels. For all three redshifts, the HMF is significantly enhanced in nDGP relative to GR at high mass: for N0.5, the HMF is enhanced by up to $60\%$, while for N5 the enhancement is less than $10\%$. On the other hand, the HMF is suppressed at lower masses in nDGP, by up to $\sim10\%$ in N0.5 and a couple of percent in N5. The threshold mass above which the HMF is enhanced and below which it is suppressed is higher at lower redshifts, with values $\sim10^{13.5}h^{-1}M_{\odot}$ at $z=0$ and $\sim10^{12}h^{-1}M_{\odot}$ at $z=2$. The low-mass suppression of the HMF also decreases with redshift.

These results can again be explained by the behaviour of the fifth force, which enhances the overall strength of gravity on large scales, accelerating the formation of high-mass haloes so that there is a greater abundance of these objects in nDGP compared to GR at a given time. On the other hand, the abundance of low-mass haloes, which undergo an increased number mergers, is reduced. The mass threshold between HMF enhancement and HMF suppression is reduced at higher redshifts, which is likely simply because the masses of a given population of haloes are lower at earlier times. 

Structure formation is sped up by a greater extent in models which feature a stronger fifth force, so the effects described above are greater for N0.5 than for weaker models. The enhancement of the HMF is greatest at the high-mass end. Therefore, by using observations of high-mass galaxy clusters from ongoing and upcoming galaxy surveys \citep[e.g.,][]{desi,euclid,lsst}, it will be possible to make powerful constraints of nDGP. However, any tests of the nDGP model of this kind may be affected by the cluster observable-mass scaling relations discussed earlier, and this should be investigated in a future work.

From the lower panels of Fig.~\ref{fig:hmf_combined}, it appears that, for any model, i.e., for a given choice of $H_0r_{\rm c}$, the HMF enhancement has a constant shape, but shifts downwards and towards larger $M_{500}$ as one goes to lower redshifts. Therefore, it can be well-described by the following model:
\begin{equation}
    \frac{\Delta n}{n_{\rm GR}} = A(H_0r_{\rm c})\left[\tanh\left(\log_{10}(M_{500}M_{\odot}^{-1}h) - B(z)\right) + C(z)\right].
    \label{eq:hmf}
\end{equation}
We use a portion of a $\tanh$ function to represent the mass-dependent shape, which is level at low mass and rises steeply at high mass. We also include the following parameters: $A(H_0r_{\rm c})$ controls the amplitude, which depends on the model parameter $H_0r_{\rm c}$; $B(z)$ represents the $z$-dependent shift along the mass axis; and $C(z)$ represents the $z$-dependent shift along the $\Delta n/n_{\rm GR}$ axis. By adopting simple linear models for each of these parameters, and by combining the data from all simulation snapshots in the range $0\leq z\leq2$, we have used unweighted least squares to obtain the following best-fit results:
\begin{equation}
    \begin{split}
    &A(H_0r_{\rm c}) = (0.342\pm0.014){\left(H_0r_{\rm c}\right)^{-1}},\\
    &B(z) = (14.87\pm0.03) - (0.481\pm0.010)z,\\
    &C(z) = (0.864\pm0.008) + (0.047\pm0.005)z.
    \end{split}
    \label{eq:hmf_params}
\end{equation}
The predictions of this calibrated model are indicated by the dashed lines in Fig.~\ref{fig:hmf_combined}. The agreement with the simulation data is excellent for all models for the mass ranges shown, which span $4$--$5$ decades depending on redshift. At $z=0$ and $z=1$, apart from the highest mass bin where data is noisy, the agreement between the fitting function and simulation measurements is within $\sim3\%$; at $z=2$, the agreement is within $\sim3\%$ for all but the strongest model ($H_0r_{\rm c}=0.5$) where we still have a $\lesssim5\%$ accuracy. In the limit $H_0r_{\rm c}\rightarrow\infty$, where nDGP becomes GR, our model predicts a relative difference of zero as expected. However, we note that our model will predict a constant relative difference if extrapolated to higher masses. This behaviour may not be physically accurate, but the high halo masses are beyond the dynamical range of our simulations and so we cannot test this reliably. Therefore, the model in Eqs.~(\ref{eq:hmf}, \ref{eq:hmf_params}) should only be used for the mass range $10^{11}h^{-1}M_{\odot}\lesssim M_{500}\lesssim M_{\rm max}(z)$, where $M_{\rm max}(z)$ is the maximum mass used to calibrate the above model at a given redshift. The latter can be estimated using the relation:
\begin{equation}
    \log_{10}\left(M_{\rm max}M_{\odot}^{-1}h\right) = 14.81 - 0.54z,
\end{equation}
which we have calibrated using snapshots in the range $0\leq z\leq2$. 

In this section, we have focused on the mass definition $M_{500}$, which is commonly used in cluster number counts studies \citep[e.g.,][]{Planck_SZ_cluster}. For completeness, we also present, in Appendix \ref{sec:appendix:dgp_clusters:hmf}, results and modelling for mass definition $M_{200}$.

\section{Summary, Discussion and Conclusions}
\label{sec:conclusions:dgp}


In this chapter, we have extended our general framework for cluster tests of gravity (Fig.~\ref{fig:mg_flow_chart}) to the popular nDGP model, in which a fifth force is able to act over sufficiently large scales.

Using the first cosmological simulations that simultaneously incorporate full baryonic physics and the nDGP model, we have studied the observable-mass scaling relations for three mass proxies (see Sec.~\ref{sec:results:dgp:scaling_relations}). For groups and clusters in the mass range $M_{500}\lesssim10^{14.5}M_{\odot}$, our results show that for the N1 model, the $\bar{T}_{\rm gas}(M)$ relation is enhanced by about 5\% with respect to GR, while the $Y_{\rm SZ}(M)$ and $Y_{\rm X}(M)$ relations are both enhanced by 10\%-15\% at low masses but more closely match the GR relations at high masses. For N5, which is much weaker than N1, the $\bar{T}_{\rm gas}(M)$ relation closely resembles the GR relation, while the $Y_{\rm SZ}(M)$ and $Y_{\rm X}(M)$ relations are enhanced by up to 5\% at low mass and suppressed by up to 5\% at high mass. These deviations from GR could be related to the effect of the fifth force on gas velocities during cluster formation, and they also hint at an interplay between the fifth force and stellar and black hole feedback.

Using a suite of DMO $N$-body simulations, which cover a wide range of resolutions and box sizes, we have found that, in nDGP, the concentration is typically suppressed relative to GR, varying from a few percent in N5 to up to $\sim15\%$ in N0.5 (see Sec.~\ref{sec:results:dgp:concentration}). Using stacked density profiles at different mass bins, we have shown that this behaviour is caused by a reduced (increased) density at the inner (outer) halo regions. Including full baryonic physics significantly affects the concentration-mass relation; however, our results show that, for masses $M_{200}\lesssim10^{13}h^{-1}M_{\odot}$, the model differences between nDGP and GR still have a similar magnitude compared to the DMO simulations.

By combining the data from our $z\leq1$ simulation snapshots, we have calibrated a general model, given by Eq.~(\ref{eq:c_model}), which is able to accurately predict the suppression of the halo concentration with respect to the GR results as a function of the halo mass and the $H_0r_{\rm c}$ parameter of nDGP over ranges $10^{12}h^{-1}M_{\odot}\lesssim M_{200}\lesssim 10^{15}h^{-1}M_{\odot}$ and 0.5-5, respectively. This model can be included in our MCMC pipeline for converting between mass definitions in case, for example, the theoretical predictions and observables are defined with respect to different spherical overdensities. Our model can also be used, along with the HMF, to predict the nonlinear matter power spectrum, which can also be used to constrain gravity.

We have also used our DMO simulations to study the HMF over the mass range $1.52\times10^{10}h^{-1}M_{\odot}\leq M_{500}\lesssim 10^{15}h^{-1}M_{\odot}$ at redshifts 0, 1 and 2 (see Sec.~\ref{sec:results:dgp:hmf}). Our results (Fig.~\ref{fig:hmf_combined}), indicate that the nDGP HMF is enhanced at high masses (by up to $\sim60\%$ in N0.5) and suppressed at low masses (by $\sim10\%$ in N0.5) compared to GR. These results indicate the potential constraining power from using the observed mass function to probe the $H_0r_{\rm c}$ parameter of nDGP. By combining the data from our $z\leq2$ snapshots, we have calibrated a general model, given by Eq.~(\ref{eq:hmf}), which can accurately reproduce the HMF enhancement as a function of the halo mass, redshift and $H_0r_{\rm c}$ parameter. This model can be used for theoretical predictions of the nDGP HMF (using a parameter-dependent GR calibration) in our MCMC pipeline.

In Chapter \ref{chapter:scaling_relations}, we showed that a model for the $f(R)$ dynamical mass enhancement can be used to predict observable-mass scaling relations in $f(R)$ gravity using their GR counterparts. Such a model in nDGP could similarly be useful to help understand the enhancements of the temperature and SZ and X-ray $Y$-parameters observed in this work. This is left to a future study. For now, though, we note that the scaling relations in nDGP still appear to follow power-law relations as a function of the mass: the $\bar{T}_{\rm gas}(M)$ relation in N1 can be related to the GR relation by a simple rescaling of the amplitude, whereas the $Y_{\rm SZ}(M)$ and $Y_{\rm X}(M)$ relations appear to have shallower slopes in N5 and N1 than in GR. Therefore, in our future MCMC pipeline for obtaining constraints of nDGP, we can still assume the GR power-law form of the scaling relations by allowing the parameters controlling the amplitude and slope to vary along with the cosmological and nDGP parameters \citep[e.g.,][]{deHaan:2016qvy,Bocquet:2018ukq}. 

Although our simulations have only been run for a single choice of cosmological parameters, we expect that our models for the enhancements of the halo concentration and HMF will have a reasonable accuracy for other (not too exotic) parameter values. The gravitational force enhancement in nDGP, given by $\left[1 + 1/(3\beta)\right]$, has only a weak dependence on $\Omega_{\rm M}$: for the N1 model ($\Omega_{\rm rc}=0.25$), the force enhancement varies within a very small range (roughly $12.1\%-12.6\%$) for $\Omega_{\rm M} \in [0.25,0.35]$ at the present day, and the range of variation is even smaller at higher redshifts. Therefore, for now we assume that the effects of the cosmological parameters on the concentration and HMF are approximately cancelled out in the ratios $\Delta c/c_{\rm GR}$ and $\Delta n/n_{\rm GR}$. However, we will revisit this in a future work, using a large number of nDGP simulations that are currently being run for different combinations of cosmological parameters, before these models are used in tests of gravity using observational data.

Finally, we note that, because the \textsc{shybone} simulations have a small box size ($62h^{-1}{\rm Mpc}$), it is difficult to robustly model the observable-mass scaling relations for cluster-sized objects ($M_{500}\gtrsim10^{14}M_{\odot}$). It would therefore be useful to revisit this study using full-physics nDGP simulations with a larger box. We have been fine-tuning a new baryonic model which can allow TNG-like simulations to be run at a much lower resolution, making it possible to run large simulations with reduced computational cost. We will present this model in Chapter \ref{chapter:baryonic_fine_tuning}, in which we will also revisit our $f(R)$ scaling relation results using much larger simulations. 
\graphicspath{{./gfx/}}

\chapter{The impact of modified gravity on the Sunyaev-Zel'dovich effect}
\label{chapter:sz_power_spectrum}

\section{Introduction}
\label{sec:introduction:sz}

The SZ effect is caused by the inverse Compton scattering of CMB photons off of high-energy electrons within ionised gas. The effect is made up of two measurable components: a thermal (tSZ) component which arises due to the random thermal motions of the electrons; and a (much smaller) kinematic (kSZ) component resulting from the bulk motion of the gas relative to the CMB rest frame \citep[e.g.,][]{1972CoASP...4..173S,1980ARA&A..18..537S}. The tSZ and kSZ signals are both highly correlated with the presence of large-scale structures such as groups and clusters of galaxies. Their power spectra are therefore extremely sensitive to the values of cosmological parameters which affect the growth of large-scale structure, offering the possibility of probing a wide range of cosmological models, including MG theories in which the strength of gravity is enhanced.

A number of works have made use of the tSZ power spectrum to constrain cosmological parameters including 
$\Omega_{\rm M}$, $\sigma_8$, the dark energy equation of state parameter and the neutrino mass \citep[e.g.,][]{Horowitz:2016dwk,Hurier:2017jgi,Bolliet:2017lha,Salvati:2017rsn}. Meanwhile, as the precision of measurements of the kSZ power continues to improve, a number of works have identified this as a promising probe for future constraints of dark energy and MG theories \citep[e.g.,][]{Ma:2013taq,Bianchini:2015iaa,Roncarelli:2018kud}. The wealth of high-quality observational data coming from current and upcoming surveys \citep[e.g.,][]{Sievers:2013ica,Aghanim:2015eva,George:2014oba,Reichardt:2020jrr,Ade:2018sbj,Abazajian:2016yjj} for both the tSZ and kSZ effects make it an exciting time for this growing area. 

In addition to their sensitivity to cosmology, the tSZ and kSZ power spectra are also highly sensitive to non-gravitational processes, such as star formation, cooling and stellar and black hole feedback, which can alter the thermal state of the intra-cluster medium \citep[e.g.,][]{McCarthy:2013qva,Park:2017amo}. Without a careful consideration of these processes, which are still not fully understood, this could pose a barrier to making reliable constraints. Incorporating full-physics baryonic models in numerical simulations, along with the cosmological model of interest, is now a vital step in order to understand the potential sensitivity of the constraints to baryonic physics.

In this chapter, we study the effects of $f(R)$ gravity and nDGP on the tSZ and kSZ power spectra. These are expected to be altered by the effects of the fifth force on the abundance and peculiar motion of large-scale structures, and on the temperature of the intra-cluster gas via the enhancement of the halo gravitational potential. We make use of the full-physics \textsc{shybone} simulations, which were also used in Chapters \ref{chapter:scaling_relations} and \ref{chapter:DGP_clusters} to study the observable-mass scaling relations in $f(R)$ gravity and nDGP, respectively. We measure the power spectra using mock maps of the tSZ and kSZ effects, which are generated using the simulation data. We also study non-radiative simulations (using the same cosmological parameters and initial conditions), allowing us to single out fifth force and baryonic feedback effects.

The chapter is structured as follows: in Sec.~\ref{sec:methods:sz}, we describe the simulations used in this chapter and our methods for predicting the SZ power spectra; our main results are presented in Sec.~\ref{sec:results:sz}; and, finally, we present a summary of our findings and their significance in Sec.~\ref{sec:conclusions:sz}.

\section{Simulations and methods}
\label{sec:methods:sz}

In Sec.~\ref{sec:methods:sz:simulations}, we describe the simulations used in this chapter. Then, in Sec.~\ref{sec:methods:sz:maps}, we present our procedure for generating SZ maps from the simulation data.

\subsection{Simulations}
\label{sec:methods:sz:simulations}

The results discussed in this chapter have been produced using the {\sc shybone} simulations \citep{Arnold:2019vpg,Hernandez-Aguayo:2020kgq}, which have already been described in Chapters \ref{chapter:scaling_relations} and \ref{chapter:DGP_clusters}. In addition to these full-physics simulations, we have also used non-radiative counterparts for the $f(R)$ model, using identical initial conditions and cosmological parameters. We have not run non-radiative counterparts for the nDGP model since these are computationally expensive to perform and there is already a lot of information provided by the existing simulations. Particle data has been saved at various snapshots: the $f(R)$ data consists of 46 snapshots between $z=3$ and $z=0$, whereas the nDGP data includes 100 snapshots between $z=20$ and $z=0$.

\subsection{SZ maps}
\label{sec:methods:sz:maps}

In the generation of each SZ map, a light cone is first constructed using our simulation snapshots with $z\leq3$. We use a field of view of $1^{\circ}\times1^{\circ}$ for the light cone, which is aligned along a specified direction from an imaginary observer placed at the centre of the simulation box at $z=0$. The box is repeated along this direction and, at a given distance from the observer, the snapshot that is closest to the corresponding redshift is used. Each snapshot is randomly rotated and shifted in order to reduce statistical correlations caused by the repetition of the box.

The $1^{\circ}\times1^{\circ}$ field of view is split into a $512\times 512$ grid of pixels, and an imaginary light ray is fired along the central axis of each pixel from $z=3$ to the observer. For each gas cell, an effective size, $s$, is defined, which can be used to determine whether it intersects with the light ray. By approximating the gas cells as spherical, the radius, $r_{\rm cell}$, of a gas cell can be estimated using:
\begin{equation}
    r_{\rm cell} = 2.5\left(\frac{3V_{\rm cell}}{4\pi}\right)^{\frac{1}{3}},
\end{equation}
where $V_{\rm cell}$ is the volume of the gas cell. This quantity is similar to the smoothing radius in smoothed-particle hydrodynamics, with the factor $2.5$ used to smooth the gas distribution. However, in the mock SZ map, the minimum length scale that is resolved (at a given distance) is the pixel side-length, $r_{\rm pixel}$. The effective size of a gas cell is therefore set as follows:
\begin{equation}
    s = 
    \begin{cases}
        r_{\rm pixel} & \text{if } r_{\rm cell} < r_{\rm pixel}.\\
        r_{\rm cell} & \text{if } r_{\rm cell} \geq r_{\rm pixel}.
    \end{cases}
\end{equation}
A gas cell contributes to the SZ signal of a pixel if the distance between its centre of mass and the light ray is smaller than $s$.

The tSZ effect is quantified by the Compton $y$-parameter, which can be computed via an integral of the electron pressure along the line of sight as follows:
\begin{equation}
    y = \frac{\sigma_{\rm T}}{m_{\rm e}c^2}\int n_{\rm e}T_{\rm gas}{\rm d}l.
    \label{eq:actual_ysz}
\end{equation}
This is evaluated for each pixel $ij$ via a summation over all gas cells that intersect with the light ray:
\begin{equation}
    y_{ij} = \frac{\sigma_{\rm T}}{m_{\rm e}c^2}\sum_{\alpha}p_{\alpha}w_{\alpha,ij},
    \label{eq:pixel_ysz}
\end{equation}
where $w_{\alpha,ij}$ is a normalised smoothing kernel. The quantity $p_{\alpha}$ is given by:
\begin{equation}
    p_{\alpha} = \frac{N_{\rm e,\alpha}}{s_{\alpha}^2}T_{\alpha},
\end{equation}
where $N_{\rm e,\alpha}$, $s_{\alpha}$ and $T_{\alpha}$ are the electron number count, the effective size and the temperature of gas cell $\alpha$, respectively. Note that we have not accounted for the relativistic SZ (rSZ) effect in our calculations. The rSZ effect can induce a significant bias in the measurement of the $y$-parameter for the most massive clusters \citep[see, e.g.,][]{Erler:2017dok}. However, the effect is much smaller for lower-mass objects which have a lower gas temperature. Since our simulations contain only galaxy groups and low-mass clusters ($M_{500}\lesssim10^{14.5}M_{\odot}$), we expect that including the rSZ effect would have a modest impact on our tSZ power spectrum results. In particular, we expect the effect on the model differences to be very small, but this is something that should be tested in the future with large simulations that contain a fair sample of cluster-sized objects.

The kSZ effect is quantified by the $b$-parameter:
\begin{equation}
    b = \sigma_{\rm T}\int \frac{n_{\rm e}v_{\rm r}}{c}{\rm d}l,
    \label{eq:actual_bsz}
\end{equation}
where $v_{\rm r}$ is the radial component of the gas peculiar velocity and $b$ is positive (negative) for gas that is moving away from (towards) the observer. The $b$-parameter is equivalent to the CMB temperature fluctuation due to the kSZ effect: $b = -\Delta T/T$. This is evaluated for each pixel as follows:
\begin{equation}
    b_{ij} = \frac{\sigma_{\rm T}}{c}\sum_{\alpha}q_{\alpha}w_{\alpha,ij}.
    \label{eq:pixel_bsz}
\end{equation}
The quantity $q_{\alpha}$ is given by:
\begin{equation}
    q_{\alpha} = \frac{N_{\rm e,\alpha}}{s_{\alpha}^2}v_{\rm r,\alpha},
\end{equation}
where $v_{\rm r,\alpha}$ is the radial component of the peculiar velocity of gas cell $\alpha$.

\begin{figure*}
\centering
\includegraphics[width=1.0\textwidth]{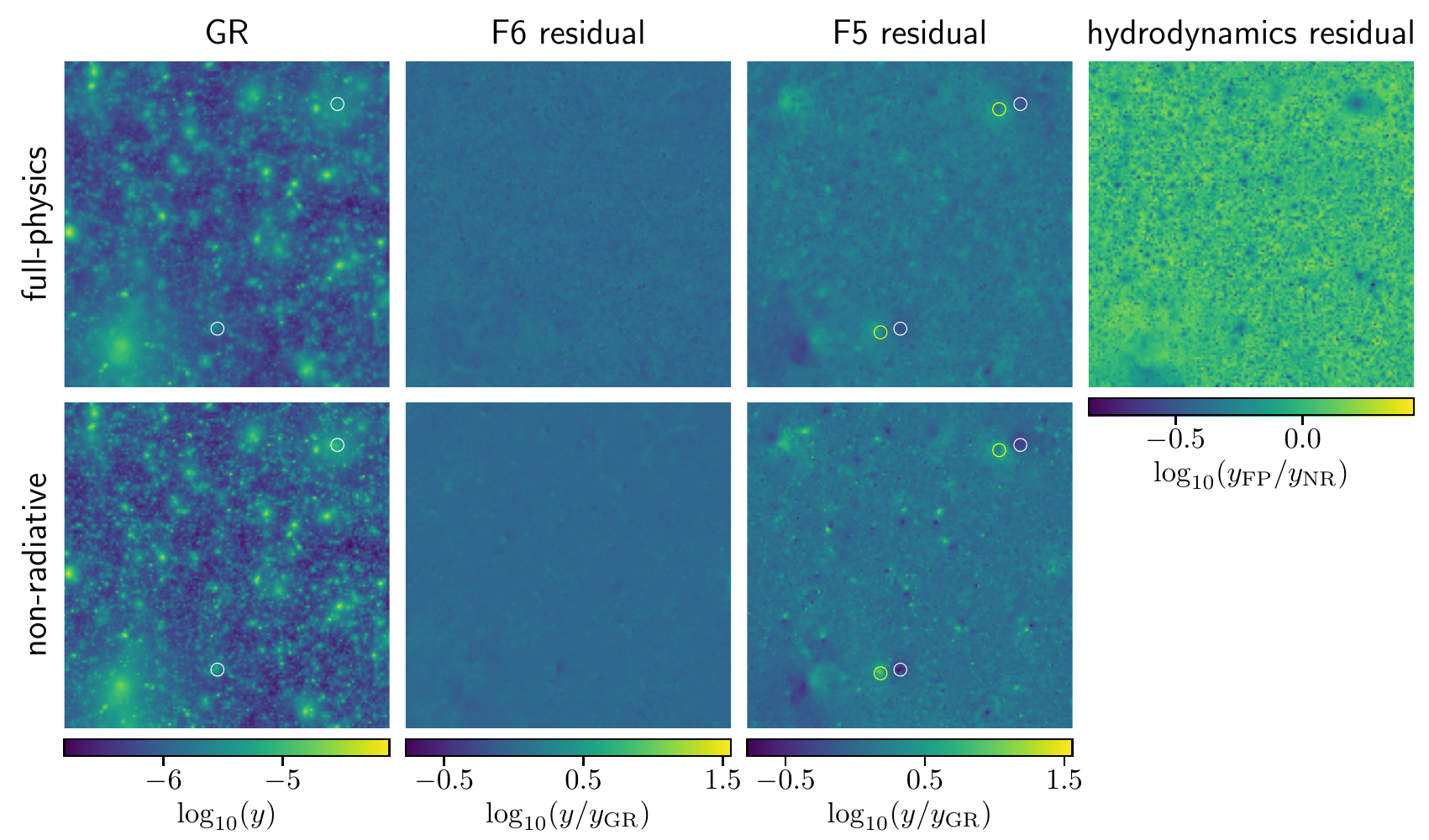}
\caption[Maps of the thermal SZ effect in GR and its relative enhancement in $f(R)$ gravity, generated using full-physics and non-radiative simulations.]{Maps of the thermal SZ effect in GR (\textit{first column}), its relative enhancement in F6 (\textit{second column}) and F5 (\textit{third column}) with respect to GR, and the relative difference between the full-physics and non-radiative GR maps (\textit{fourth column}). The maps have a side length of $1^\circ$ and a $512\times512$-pixel resolution, and have been constructed from the {\sc shybone} simulations (see Sec.~\ref{sec:methods:sz}). Both the full-physics (\textit{top row}) and non-radiative (\textit{bottom row}) runs are shown. The $y$-parameter is computed for each pixel using Eq.~(\ref{eq:pixel_ysz}). The rings indicate two haloes whose positions are shifted in F5 (\textit{yellow}) relative to GR (\textit{white}).}
\label{fig:thermal_sz}
\end{figure*}

\begin{figure*}
\centering
\includegraphics[width=1.0\textwidth]{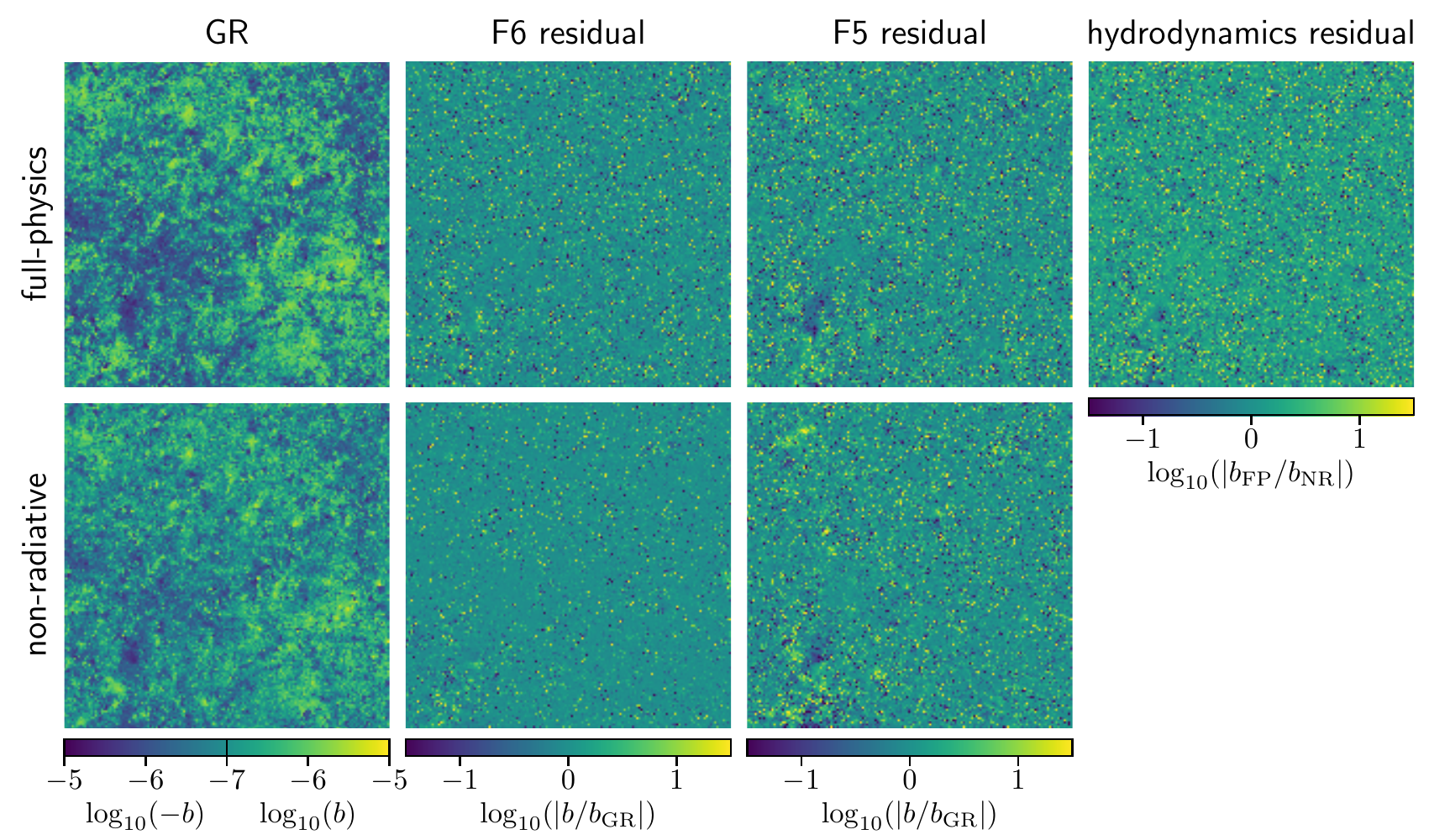}
\caption[Maps of the kinetic SZ effect in GR and its relative enhancement in $f(R)$ gravity, generated using full-physics and non-radiative simulations.]{Maps of the kinetic SZ effect in GR (\textit{first column}), its absolute relative enhancement in F6 (\textit{second column}) and F5 (\textit{third column}) with respect to GR, and the absolute relative difference between the full-physics and non-radiative GR maps (\textit{fourth column}). The maps have a side length of $1^\circ$ and a $512\times512$-pixel resolution, and have been constructed from the {\sc shybone} simulations (see Sec.~\ref{sec:methods:sz}). Both the full-physics (\textit{top row}) and non-radiative (\textit{bottom row}) runs are shown. The $b$-parameter is computed for each pixel using Eq.~(\ref{eq:pixel_bsz}).}
\label{fig:kinetic_sz}
\end{figure*}

We have generated 14 independent light cones, each aligned along a unique direction. The same set of directions has been used to construct the maps for each gravity model and for both the full-physics and non-radiative data. This means that for any two maps aligned in the same direction, the only differences are caused by the contrasting gravity models and hydrodynamics schemes. The tSZ and kSZ maps corresponding to one of the light cones, generated using the $f(R)$ simulations, are shown in Figs.~\ref{fig:thermal_sz} and \ref{fig:kinetic_sz}, respectively. For each figure, the GR maps are shown in the left column, with the map from the full-physics run in the top row and the map for the non-radiative simulation in the bottom row.

The bright yellow peaks in the tSZ maps, which correspond to a high $y$-parameter, trace hot gas within groups and clusters of galaxies. These peaks are found at the same positions in both the full-physics and non-radiative maps. However, the addition of feedback mechanisms, which create winds that heat up and blow gas out of haloes, cause the peaks to appear more diffuse in the full-physics map. The kSZ map is made up of dark and bright regions, which correspond to negative and positive values of the $b$-parameter, respectively.

Rather than the absolute maps of F6 and F5, which are visually very similar to the GR maps, we display residual maps to indicate the main differences. These are shown in the second and third columns of Figs.~\ref{fig:thermal_sz} and \ref{fig:kinetic_sz}. The tSZ residuals represent the enhancement of the $f(R)$ $y$-parameter with respect to GR for each pixel. The F6 residuals are quite close to zero across the field of view, owing to the efficient screening of the fifth force in galaxy groups and clusters for this model. However, for the F5 model, for which the fifth force is more prominent, the residuals appear more complex. Pairs of bright and dark regions, two of which are indicated by rings placed in Fig.~\ref{fig:thermal_sz}, are visible throughout the images. These are caused by the shift of halo positions in F5 compared to GR, with each dark (bright) region corresponding to the position in GR (F5). While this in itself does not provide useful information about the effect of the fifth force on the tSZ effect, we note that at the extremes the positive residuals ($\log_{10}(y/y_{\rm GR})\approx1.5$) are greater in magnitude than the negative residuals ($\log_{10}(y/y_{\rm GR})\approx-0.8$), indicating that the tSZ effect is strengthened on average in F5 compared to GR.

For the kSZ signal, the $f(R)$ gravity residuals correspond to the enhancement of the absolute value of the $b$-parameter with respect to GR. A higher value of $b$ indicates that gas is moving faster with respect to the CMB rest-frame. Many individual pixels gain much higher and much lower values of $b$, seemingly at random, across the field of view. This is caused by the effect of the fifth force on the motion of the gas. Pairs of bright and dark regions are also just visible in the F5 residual map, again corresponding to the relative shifts in halo position with respect to GR.

In the rightmost columns of Figs.~\ref{fig:thermal_sz} and \ref{fig:kinetic_sz}, we show the relative difference between the full-physics and non-radiative GR maps. The tSZ results indicate that within haloes the tSZ signal is suppressed (dark blue regions) by up to 86\% and boosted outside haloes (bright yellow regions) by up to 173\%. This is caused by the ejection of gas from haloes by feedback mechanisms, causing the electron pressure to be lowered within haloes and raised outside haloes. For the kSZ results, as for the $f(R)$ gravity residuals, the value of $b$ is increased and reduced seemingly at random, owing to the unpredictable effects of the full-physics processes on the motion of the gas.

\begin{figure*}
\centering
\includegraphics[width=0.85\textwidth]{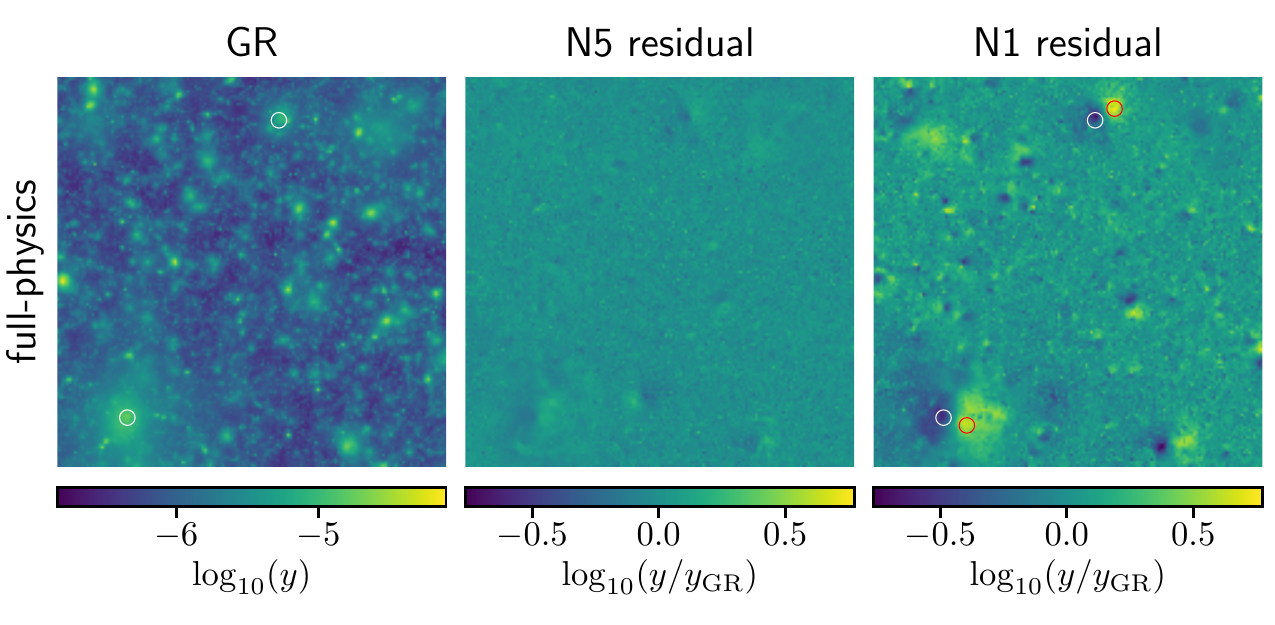}
\caption[Map of the thermal SZ effect in GR and its relative enhancement in nDGP, generated using full-physics simulations.]{Map of the thermal SZ effect in GR (\textit{left column}) and its relative enhancement in N5 (\textit{middle column}) and N1 (\textit{right column}). The maps have a side-length of $1^\circ$ and a $512\times512$-pixel resolution, and have been constructed from the {\sc shybone} simulations (see Sec.~\ref{sec:methods:sz}). The $y$-parameter is computed for each pixel using Eq.~(\ref{eq:pixel_ysz}). The rings indicate two haloes whose positions are shifted in N1 (\textit{red}) relative to GR (\textit{white}).}
\label{fig:nDGP_thermal_sz}
\end{figure*}

The nDGP tSZ maps for the same light cone are shown in Fig.~\ref{fig:nDGP_thermal_sz}, where recall that we do not have non-radiative runs. Again, the fifth force causes a shift in halo positions with respect to GR, and this is clearly visible for both nDGP models. The effect is greater in the N1 model, which is a stronger modification of GR than N5. We do not show the kSZ maps for nDGP, since these appear similar to the $f(R)$ maps and do not offer extra information.

\section{Results}
\label{sec:results:sz}

This section gives the main results of this chapter. In Sec.~\ref{sec:results:sz:profiles}, we analyse the effects of baryonic processes and the fifth force on the stacked electron pressure profiles of FOF groups from our simulations. Then, in Sec.~\ref{sec:results:sz:power}, we discuss the effects on the tSZ and kSZ angular power spectra. Finally, in Sec.~\ref{sec:results:sz:transverse_momentum}, we examine the effects on the power spectrum of the transverse component of the electron momentum field, which is closely related to the kSZ angular power spectrum.

\subsection{Electron pressure profiles}
\label{sec:results:sz:profiles}

\begin{figure*}
\centering
\includegraphics[width=1.0\textwidth]{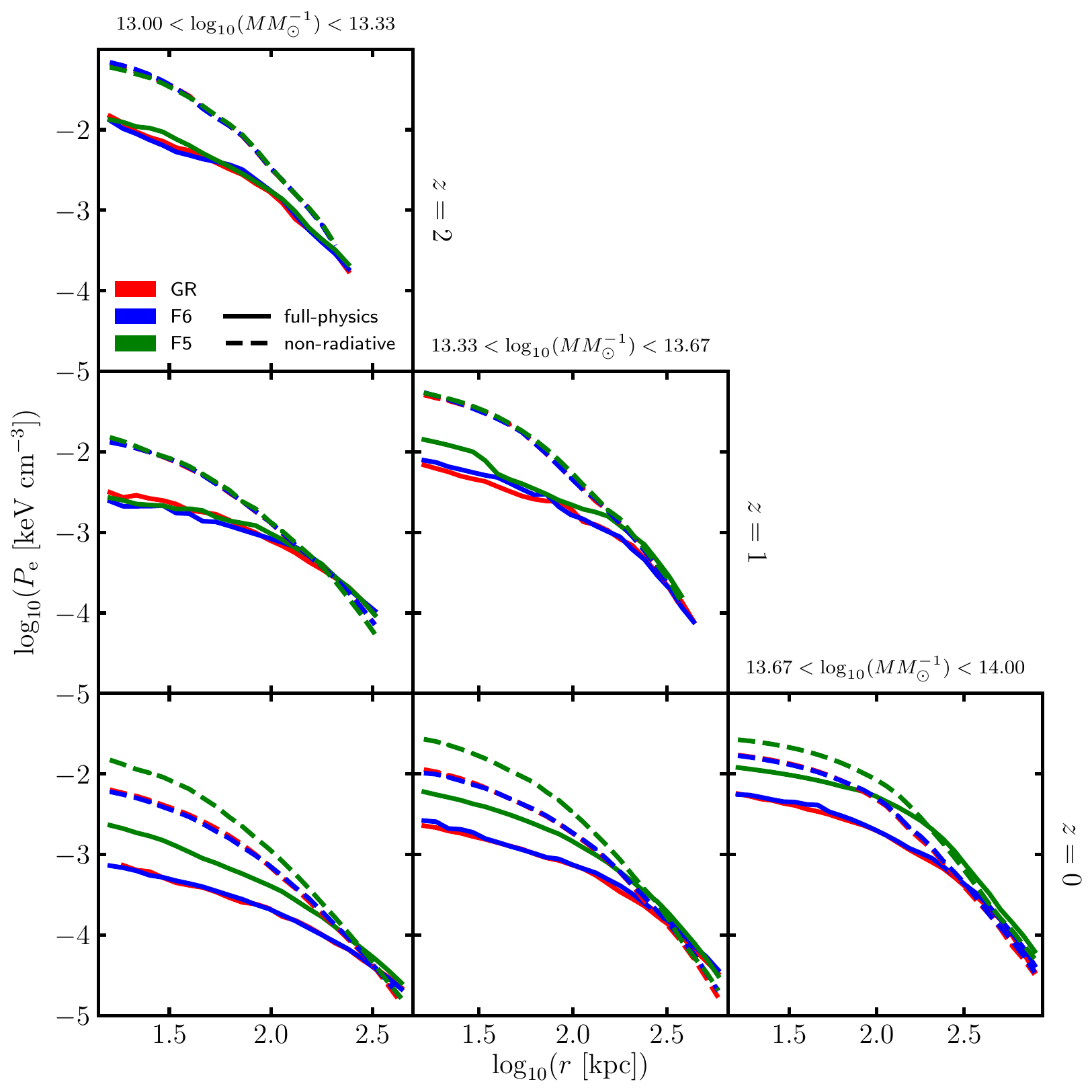}
\caption[Median electron pressure profiles for haloes in $f(R)$ gravity and GR for a selection of mass bins and redshifts, generated using full-physics and non-radiative simulations.]{Stacked electron pressure profiles for haloes from three mass bins in the range $10^{13}M_{\odot} < M_{500} < 10^{14}M_{\odot}$ and redshifts 0, 1 and 2. The haloes have been identified from the {\sc shybone} simulations (see Sec.~\ref{sec:methods:sz}), and have been generated for the GR (\textit{red}), F6 (\textit{blue}) and F5 (\textit{green}) gravity models, and for both the full-physics (\textit{solid lines}) and non-radiative (\textit{dashed lines}) hydrodynamics schemes.}
\label{fig:pressure}
\end{figure*}

\begin{figure*}
\centering
\includegraphics[width=1.0\textwidth]{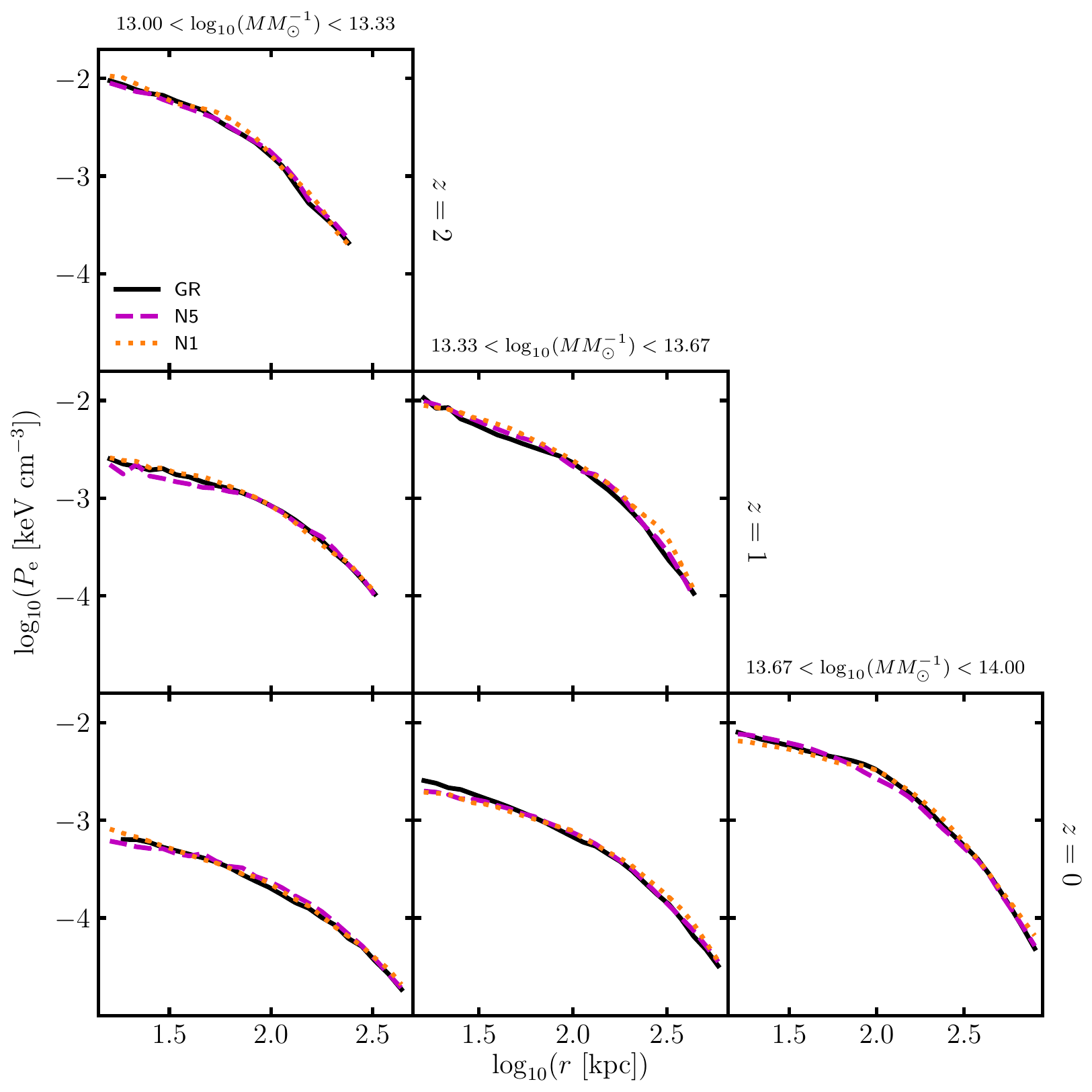}
\caption[Median electron pressure profiles for haloes in nDGP and GR for a selection of mass bins and redshifts, generated using full-physics simulations.]{Stacked electron pressure profiles for haloes from three mass bins in the range $10^{13}M_{\odot} < M_{500} < 10^{14}M_{\odot}$ and redshifts 0, 1 and 2. The haloes have been identified from the full-physics {\sc shybone} simulations (see Sec.~\ref{sec:methods:sz}), and have been generated for the GR (\textit{black solid}), N5 (\textit{magenta dashed}) and N1 (\textit{orange dotted}) gravity models.}
\label{fig:nDGP_pressure}
\end{figure*}

We show the stacked electron pressure profiles at $z=0$, $z=1$ and $z=2$ in Fig.~\ref{fig:pressure} and Fig.~\ref{fig:nDGP_pressure} for $f(R)$ gravity and nDGP, respectively. Three equally spaced logarithmic mass bins, which span the range $13 < \log_{10}(M_{500}M_{\odot}^{-1}) < 14$, are considered. The volume-weighted electron pressure is measured in radial bins for each halo. This is given by the following:
\begin{equation}
    \bar{P}_{\rm e} = \frac{\sum_iP_{{\rm e},i}V_i}{\sum_iV_i},
\end{equation}
where $P_{{\rm e},i}$ and $V_i$ are the electron pressure and volume of gas cell $i$, and the summations are evaluated over all gas cells whose centres of mass are within the current bin. The median profile is measured for each radial bin using the haloes enclosed in each mass bin, and is displayed in the figures. Because of the limitations in the size of the halo population at higher masses, only the lowest-mass bin is shown at $z=2$ and the highest-mass bin is 
not shown for $z=1$. For the highest-mass bin shown at each redshift, because the halo number is relatively small, some haloes are also excluded from each model to ensure that the same halo population is used in all models. Any small difference in population could otherwise have a significant effect in these bins, which contain only $\sim10$ haloes each. This consideration is not required for the other bins, which have $\gtrsim25$ haloes each. We are also unable to include data at $M_{500}>10^{14}M_{\odot}$, for which there are only a few haloes for each model.

For haloes in F5 at sufficiently low redshift, we find that the fifth force boosts the electron pressure. This is caused by the increase in the temperature of the intra-cluster gas, which results from the deepened gravitational potential well (see Chapter \ref{chapter:scaling_relations}). This indicates that the tSZ signal from individual haloes is expected to be significantly enhanced in F5. The magnitude of the background scalar field, $|f_R|$, increases with time, and as a result the chameleon mechanism is more efficient at screening the fifth force at earlier times. This explains why the enhancement of the pressure in F5 vanishes for $z\gtrsim1$. On the other hand, the background scalar field in the F6 model is $10$ times weaker than in F5, and as a result the fifth force is efficiently screened within group- and cluster-sized haloes even at $z=0$.

Meanwhile, Fig.~\ref{fig:nDGP_pressure} shows much smaller differences between the electron pressure profiles in GR and nDGP than in Fig.~\ref{fig:pressure}, especially at lower $z$ ($z\lesssim1$). This is because the Vainshtein mechanism is much more efficient than the chameleon mechanism at screening out the fifth force within haloes at low redshifts --- for the latter, depending on the value of $|f_{R0}|$ in the two $f(R)$ models studied here, group-sized objects could be partially or completely unscreened at low $z$, while for the former the screening efficiency is similar for haloes of different masses \citep[see, e.g., Fig.~8 of][]{Hernandez-Aguayo:2020kgq}, including the ones as small as $\sim10^{11.7}h^{-1}M_\odot$, with the fifth force always being strongly suppressed in the inner regions of haloes, at all redshifts.

By comparing the non-radiative and full-physics data in Fig.~\ref{fig:pressure}, we can see that the additional baryonic processes that are present in the latter act to suppress the pressure at the inner halo regions. This can be caused by, for example, the blowing out of gas by black hole feedback which lowers the density of electrons. Note that, while the electron pressure profiles differ significantly between the full-physics and non-radiative runs, the relative enhancement of F5 with respect to GR seems to be consistent in both cases.

\subsection{tSZ and kSZ power spectra}
\label{sec:results:sz:power}

We have used our SZ maps (see Sec.~\ref{sec:methods:sz:maps}) to generate the tSZ and kSZ angular power spectra for the $f(R)$ and nDGP models. The power has been measured for each of the 14 maps in bins of the angular wavenumber $l$. For each bin, the mean power and the mean relative difference in the power between gravity models and hydrodynamics schemes have been measured.

\begin{figure*}
\centering
\includegraphics[width=1.0\textwidth]{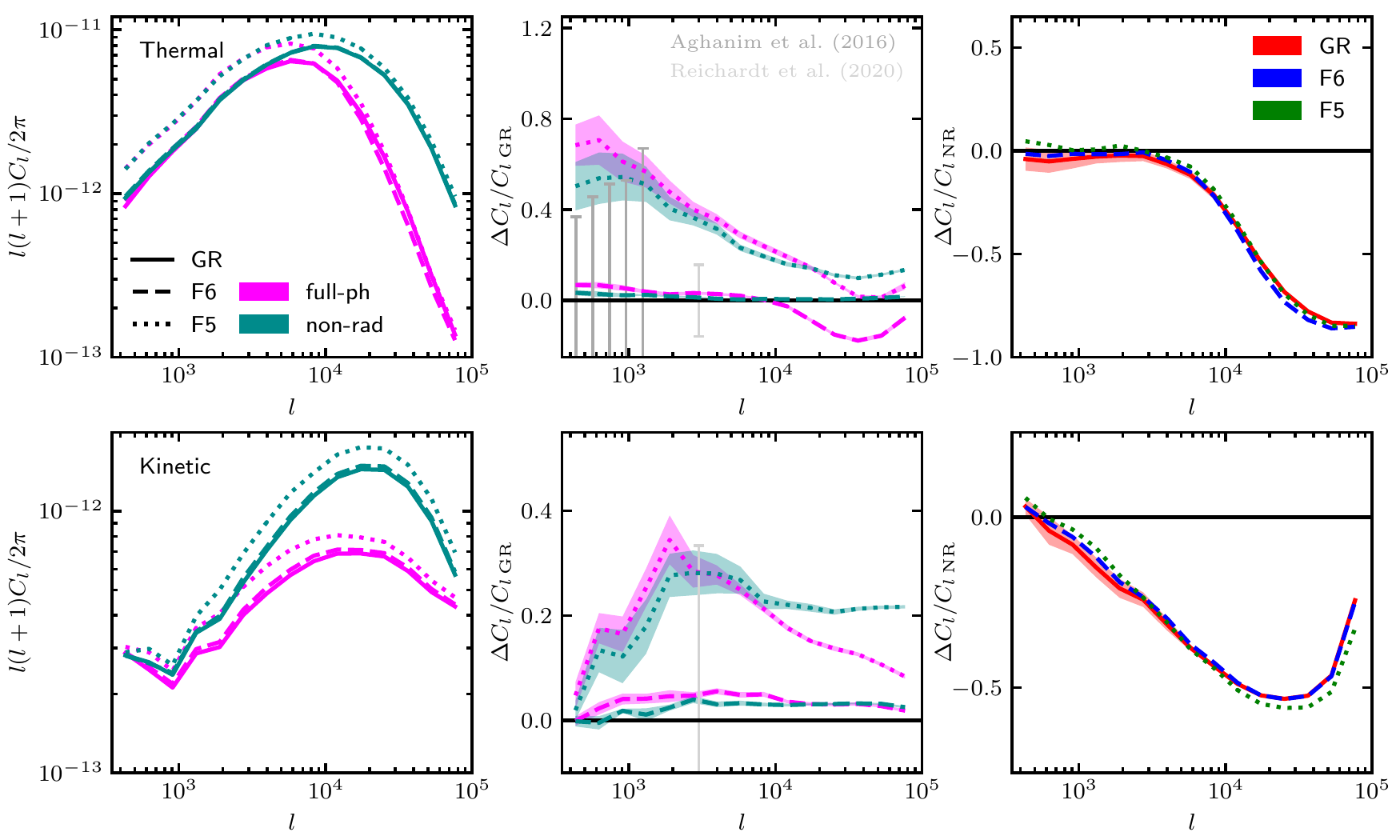}
\caption[Angular power spectra of the thermal and kinetic SZ effects in $f(R)$ gravity and GR, generated using full-physics and non-radiative simulations.]{Angular power spectra and their relative differences plotted as a function of the angular wavenumber. The data has been generated from maps of the thermal (\textit{top row}) and kinetic (\textit{bottom row}) SZ signals, which have been created using the {\sc shybone} simulations (see Sec.~\ref{sec:methods:sz}). \textit{Left column}: mean angular power spectrum plotted for GR (\textit{solid lines}), F6 (\textit{dashed lines}) and F5 (\textit{dotted lines}), including data from the full-physics (\textit{magenta}) and non-radiative (\textit{cyan}) simulations. \textit{Middle column}: mean relative enhancement of the F6 (\textit{dashed lines}) and F5 (\textit{dotted lines}) angular power spectra with respect to GR, plotted for the full-physics (\textit{magenta}) and non-radiative (\textit{cyan}) simulations. The standard error of the mean is indicated by the shaded regions. The error bars indicate the precision of the latest observations from the Planck \citep{Aghanim:2015eva} and SPT \citep{Reichardt:2020jrr} collaborations. \textit{Right column}: mean relative enhancement of the full-physics angular power spectra with respect to the non-radiative data, plotted for GR (\textit{red}), F6 (\textit{blue}) and F5 (\textit{green}). For clarity, the standard error is shown for GR only.}
\label{fig:power_spectra}
\end{figure*}

From the $f(R)$ gravity results, shown in Fig.~\ref{fig:power_spectra}, we find that the fifth force and the extra baryonic processes that are found in the full-physics simulations have different effects: the middle column shows that, for the non-radiative data, the tSZ and kSZ power spectra are both enhanced in $f(R)$ gravity relative to GR; and the right column shows that the power is suppressed in the full-physics runs relative to the non-radiative runs, particularly at smaller scales. The latter is consistent with literature:  \citet{McCarthy:2013qva} showed that, at scales $l\gtrsim1000$, the tSZ power is suppressed by the ejection of gas by black hole feedback; and \citet{Park:2017amo} found that the kSZ power is suppressed by both the locking away of free electrons in stars, black holes and neutral gas (at all scales), and the ejection of gas through black hole feedback (at smaller scales). For our data, this suppression by baryonic processes occurs at $l\gtrsim3000$ for the tSZ power and at $l\gtrsim500$ for the kSZ power. The shape and amplitude of this suppression is very similar for each gravity model, as shown in the right column: the tSZ power is suppressed by up to $\sim85\%$ and the kSZ power is suppressed by up to $\sim50\%$.

With the extra baryonic processes of cooling, star formation and feedback absent, the tSZ power is enhanced by the fifth force on all scales. The enhancement is greater in F5 than in F6, with peaks of $\sim50\%$ and a few percent, respectively, at $l<1000$. However, due to the relatively small size of the fields of view in our light cones, we cannot measure the angular power spectra at $l\lesssim500$, and so it is unclear what the asymptotic behaviour at large angular scales is, for which future works with larger simulations are needed. For the kSZ power, a roughly constant enhancement is observed (of $\sim22\%$ for F5 and $\sim3\%$ for F6) at scales $l\gtrsim3000$, with a downturn at larger scales ($l\simeq2000$). The presence of the fifth force speeds up the formation of large-scale structures, boosting the abundance and peculiar velocity of groups and clusters of galaxies and, in turn, the tSZ and kSZ power spectra. In addition to this, the electron pressure profiles of haloes at a given mass are also enhanced, as discussed in Sec.~\ref{sec:results:sz:profiles}, which could further boost the tSZ signal and tSZ power spectrum at small angular scales (the relation between the latter and halo electron pressure profiles, however, is more complicated, as we will discuss toward the end of this subsection).

The enhancement of the kSZ power by $\sim22\%$ in F5 is higher than predicted by \citet{Bianchini:2015iaa} and \citet{Roncarelli:2018kud}, who estimated an enhancement of about $15\%$ for the same model using analytical predictions and hydrodynamical simulations, respectively. We remark that our results use only the redshift range $z\leq3$ while these works used redshifts up to 9.9 and 15, including the epoch of reionisation which can have a substantial contribution to the total kSZ power. The fifth force is expected to be screened for $z\gtrsim3$, which can explain why the kSZ signal (an integral over the redshift range) shows less deviation from GR in these works. Our smaller redshift range $z\leq3$ also explains why the amplitude of the kSZ power in Fig.~\ref{fig:power_spectra} is lower than is typically predicted in literature \citep[e.g.,][]{Park:2017amo}.

The {\sc shybone} simulations are the first to simultaneously compute the fifth force of $f(R)$ gravity while incorporating full baryonic physics. The interplay between these two competing mechanisms in the full-physics simulations is therefore of particular interest. According to the middle column of Fig.~\ref{fig:power_spectra}, the extra processes in the full-physics simulations have a non-negligible effect on the relative difference between $f(R)$ gravity and GR. For the tSZ power, a suppression of the $f(R)$ enhancement is observed at very small scales ($l\gtrsim10000$), such that the F5 power is brought close to the GR power, and the F6 power becomes $\sim20\%$ lower than GR. For the kSZ power, the F5 enhancement is again suppressed at these scales, while there appears to be little change for F6. 

We note that these results are likely to be sensitive to the choice of full-physics parameters implemented by {\sc shybone}. Given that feedback is not fully understood theoretically or from observations, there is a non-negligible uncertainty in the results at small scales. In order to avoid potentially biased results, constraints should instead be made using large scales where the details of baryonic processes are not as prominent. For the tSZ power, these scales correspond to $l\lesssim3000$, although we note that even this range could be sensitive to the full-physics parameters. Our simulations predict enhancements of $\sim40$-$70\%$ in F5 and less than $10\%$ in F6, relative to GR, at these scales. For the kSZ power spectrum, again star formation, feedback and cooling appear to have a non-negligible effect at all of the scales tested here. However, the model differences between $f(R)$ gravity and GR do not differ significantly in the non-radiative and full-physics simulations for scales $l\lesssim10^4$. In this scale range, we observe relative differences of up to $\sim35\%$ and $\sim5\%$ between GR and the F5 and F6 models, respectively. Note that the non-radiative runs could be considered as an extreme case of the hydrodynamics scheme, with the most interesting physical processes neglected, and for this reason we expect that slight variations of the baryonic model should produce milder differences from the IllustrisTNG model than what is observed between the full-physics and non-radiative curves here. Also note that, due to the small box size, the full-physics runs used in this thesis could suffer from significant sample variance, e.g., due to a few large haloes experiencing unusually strong feedback in one model and not another; again, having a large simulation box in the future will help to address this question.

\begin{figure*}
\centering
\includegraphics[width=0.8\textwidth]{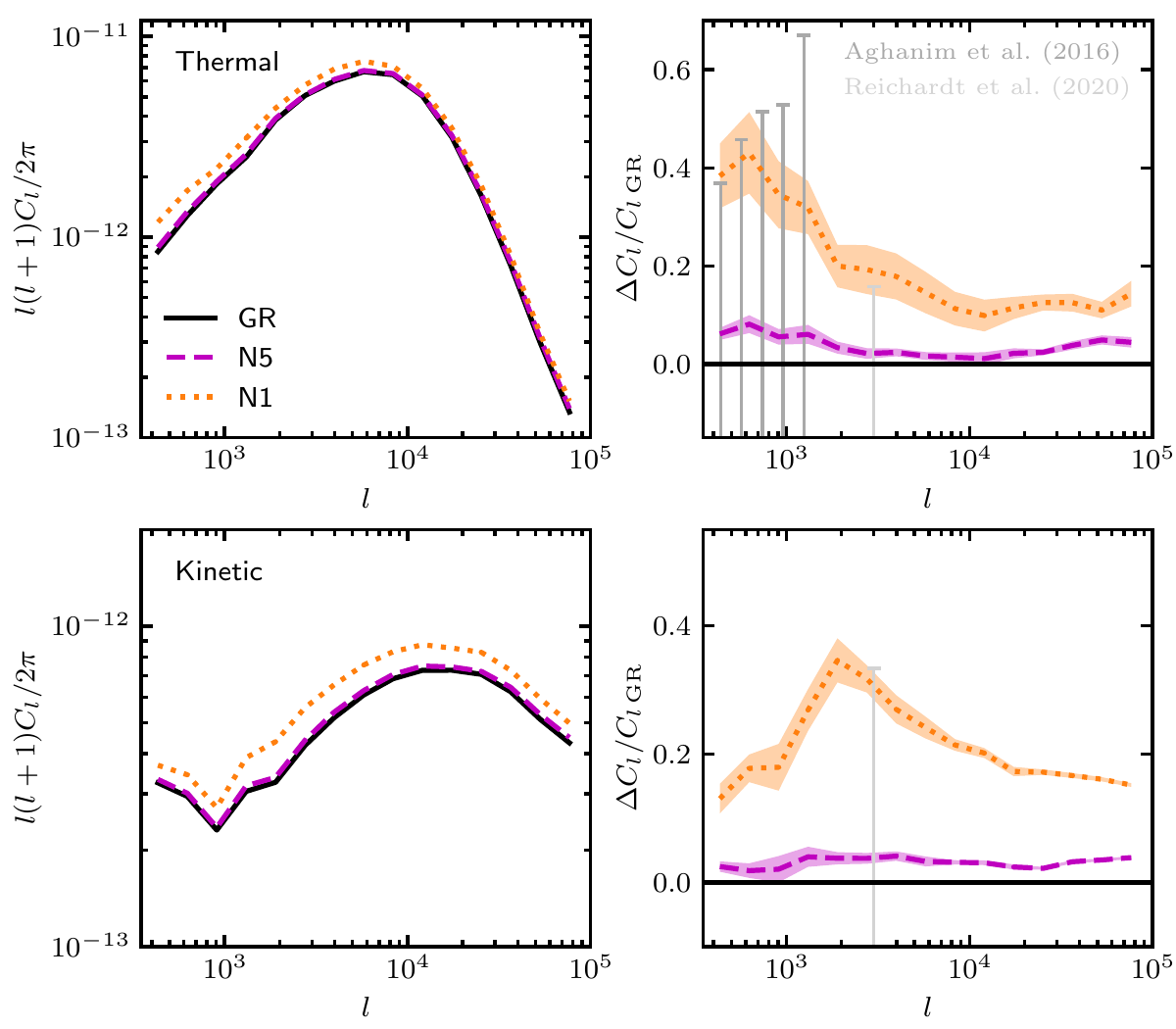}
\caption[Angular power spectra of the thermal and kinetic SZ effects in nDGP and GR, generated using full-physics simulations.]{Angular power spectra and their relative differences plotted as a function of the angular wavenumber. The data has been generated from maps of the thermal (\textit{top row}) and kinetic (\textit{bottom row}) SZ signals, which have been created using the {\sc shybone} simulations (see Sec.~\ref{sec:methods:sz}). \textit{Left column}: mean angular power spectrum plotted for GR (\textit{solid lines}), N5 (\textit{dashed lines}) and N1 (\textit{dotted lines}). \textit{Right column}: mean relative enhancement of the N5 and N1 angular power spectra with respect to GR. The standard error of the mean is indicated by the shaded regions. The error bars indicate the precision of the latest observations from the Planck \citep{Aghanim:2015eva} and SPT \citep{Reichardt:2020jrr} collaborations.}
\label{fig:nDGP_power_spectra}
\end{figure*}

The tSZ and kSZ power spectra for the nDGP model (for full-physics only), are shown in Fig.~\ref{fig:nDGP_power_spectra}. As for the $f(R)$ model, the fifth force of nDGP enhances the power on all probed scales: the tSZ power is enhanced by up to $\sim40\%$ in N1 and less than $10\%$ in N5; and the kSZ power is enhanced by up to $\sim35\%$ in N1 and $\sim5\%$ in N5. However, given the absence of a non-radiative simulation for the nDGP model, we note that it is possible that these differences could be sensitive to baryonic physics, as in $f(R)$ gravity. 

Interestingly, the tSZ power spectrum at high $l$ is significantly enhanced in N1, even though the pressure profiles (Fig.~\ref{fig:nDGP_pressure}) do not appear to show a clear deviation from GR. There are a few reasons why this can happen. First of all, the tSZ power receives contributions from outside haloes as well as from within. The fourth column of Fig.~\ref{fig:thermal_sz} indicates that outside haloes the tSZ signal can be boosted by the ejection of gas by feedback. The presence of the fifth force is expected to result in the feedback being triggered earlier, which can cause the power to be enhanced relative to GR at angular scales corresponding roughly to halo sizes\footnote{The fifth force also enhances matter clustering on large scales overall, and this is expected to be reflected in the clustering of free electrons.}. Secondly, smaller angular scales receive a greater contribution from higher redshifts \citep[see, e.g.,][]{McCarthy:2013qva}. In F5, the fifth force is efficiently screened for $z\gtrsim1$, but in N1 it can still reach a few percent of the strength of the Newtonian force at the radius $R_{200}$ and $\sim8\%$ outside it at $z\sim2$ \citep[e.g.,][]{Hernandez-Aguayo:2020kgq}. This means that the SZ power at high $l$ can be enhanced by a greater amount in N1 than in F5. In fact, for the mass bin shown at $z=2$, the N1 pressure profile is enhanced by $\sim9\%$ with respect to GR, and we also find that the nDGP pressure is enhanced in lower-mass bins which are not shown in Fig.~\ref{fig:nDGP_pressure}. We have further verified (though not shown here) that at $z\gtrsim1.5$, the 3D electron pressure power spectrum is significantly enhanced even at high $k$ values well inside the 1-halo regime.

The results discussed in this section indicate that the tSZ and kSZ power have the potential to effectively probe $f(R)$ gravity and nDGP at large scales. To demonstrate this, we have included error bars in Figs.~\ref{fig:power_spectra} and \ref{fig:nDGP_power_spectra} to indicate the uncertainties of the latest tSZ and kSZ observations from the Planck \citep{Aghanim:2015eva} and SPT \citep[SPT,][]{Reichardt:2020jrr} collaborations. The $\sim16\%$ precision of the tSZ measurement by SPT is sufficient to distinguish the F5 model from GR at $l=3000$, while the Planck measurements have sufficient precision to distinguish F5 at large angular scales ($l\lesssim500$). The $33\%$ precision of the kSZ measurement by SPT has a similar magnitude to the relative enhancements of the F5 and N1 models with respect to GR, indicating that more precise measurements from future surveys will be capable of ruling out these models. However, in order to avoid bias from uncertain baryonic physics, it will be necessary to use a range of full-physics parameters to confirm that reliable constraints can be achieved at these angular scales. It will also be important to revisit this study using simulations with a greater box size that can accurately probe the tSZ power up to angular scales $l\sim100$, where the precision of the Planck measurements is particularly high \citep{Aghanim:2015eva}. Finally, understanding the degeneracies between MG and variations in other cosmological parameters is also critical in order to have unbiased constraints.

Before finishing this section, let us note that, despite the qualitative difference in their respective screening mechanisms -- Vainshtein screening is always efficient inside dark matter haloes while the same cannot be said about the chameleon mechanism (cf.~Figs.~\ref{fig:pressure} and \ref{fig:nDGP_pressure}) -- the enhancements of both the tSZ and kSZ power spectra are very similar in these two models.

\begin{figure*}
\centering
\includegraphics[width=1.0\textwidth]{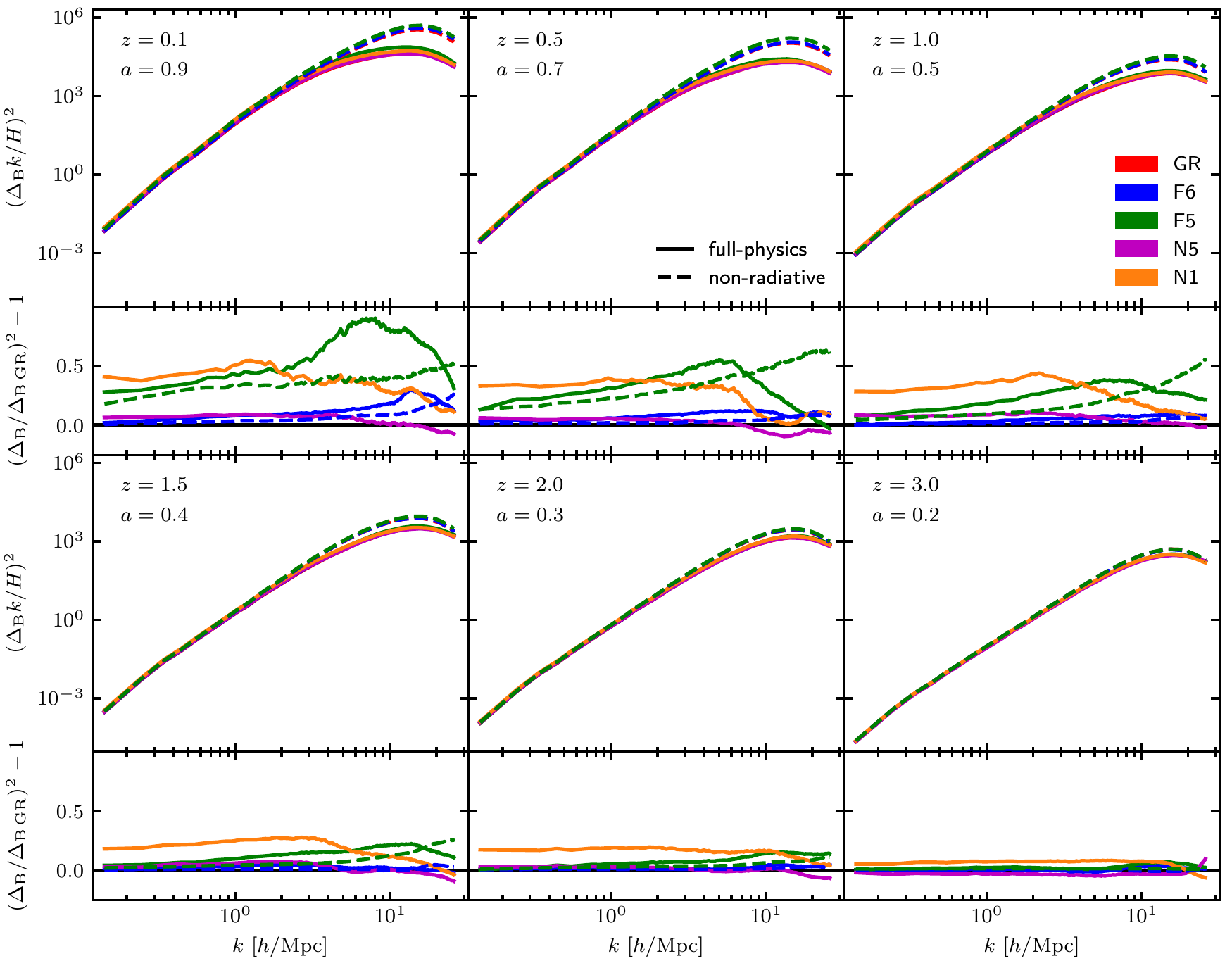}
\caption[Power spectrum of the transverse component of the electron momentum for $f(R)$ gravity and GR, generated using full-physics and non-radiative simulations.]{Power spectrum of the transverse component of the electron momentum plotted against the wavenumber. The data has been generated for six different redshifts (annotated) using the {\sc shybone} simulations (see Sec.~\ref{sec:methods:sz:simulations}) for both the non-radiative (\textit{dashed lines}) and full-physics (\textit{solid lines}) hydrodynamics schemes. In addition to GR (\textit{red lines}), data is plotted for F6 (\textit{blue lines}), F5 (\textit{green lines}), N5 (\textit{magenta lines}) and N1 (\textit{orange lines}). The lower sub-panels show the relative enhancement of the MG (F6, F5, N5 and N1) power spectra with respect to GR.}
\label{fig:transverse_momentum}
\end{figure*}

\begin{figure}
\centering
\includegraphics[width=0.7\columnwidth]{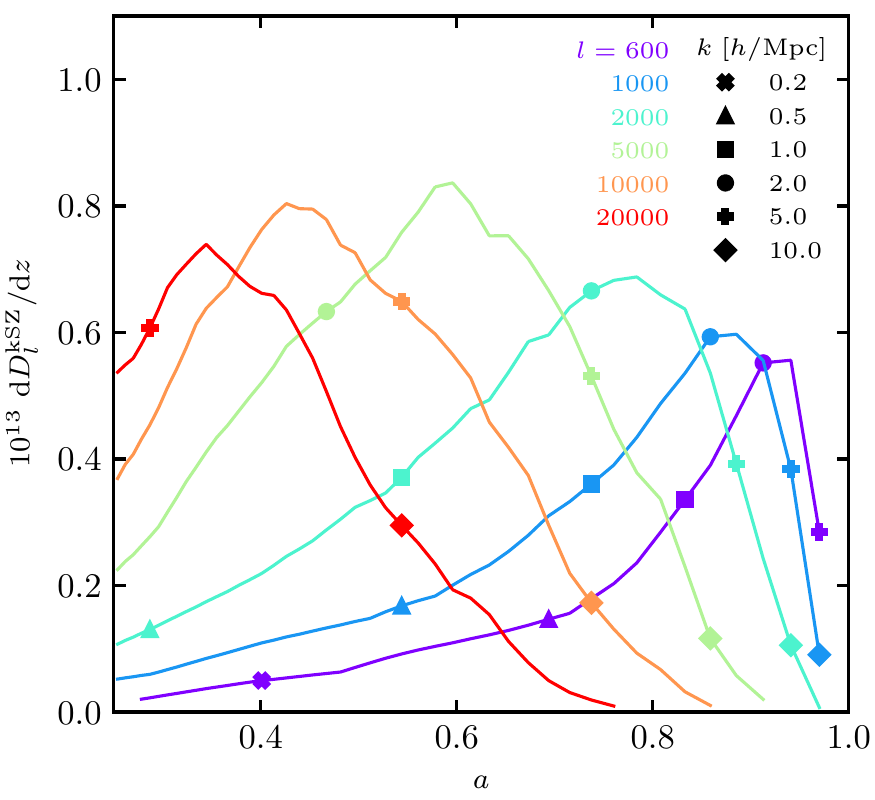}
\caption[Derivative of the GR kSZ angular power spectrum as a function of the cosmological scale factor for different values of the angular wavenumber.]{Derivative of the GR kSZ angular power spectrum as a function of the cosmological scale factor $a$ for six different values of the angular wavenumber $l$. The values are computed from the full-physics \textsc{shybone} simulation (see Sec.~\ref{sec:methods:sz:simulations}) using the Limber approximation (Eq.~\ref{eq:limber}). The scales that are spanned by each $l$ value are indicated by markers which represent unique values of the wavenumber $k$.}
\label{fig:ksz_contribution}
\end{figure}

\subsection{Transverse momentum power spectrum}
\label{sec:results:sz:transverse_momentum}

In order to understand the (similar) effects of $f(R)$ gravity and nDGP on the kSZ power in more detail, we have measured the power spectrum of the transverse component of the electron momentum field which, in the small-angle limit, can be related to the kSZ angular power spectrum using the Limber approximation \citep[e.g.,][]{2012ApJ...756...15S}:
\begin{equation}
\begin{aligned}
    & C_l^{\rm kSZ} = \frac{8\pi^2}{(2l+1)^3}\left(\frac{\sigma_{\rm T}\bar{\rho}_{\rm gas,0}}{\mu_{\rm e}m_{\rm p}}\right)^2 \\
    & \times \int_0^{z_{\rm re}}\frac{{\rm d}z}{c}(1+z)^4\chi^2\Delta_{\rm B}^2(k,z)e^{-2\tau(z)}\frac{x(z)}{H(z)},
\end{aligned}
\label{eq:limber}
\end{equation}
where $\bar{\rho}_{\rm gas,0}$ is the present-day mean background gas density, $\mu_{\rm e}m_{\rm p}$ is the mean gas mass per electron, $z_{\rm re}$ is the redshift at the epoch of reionisation, $\chi$ is the fraction of electrons that are ionised, $k=l/x$ is the wavenumber, $x(z)=\int_0^z(c{\rm d}z'/H(z'))$ is the comoving distance at redshift $z$, and the optical depth, $\tau$, is given by:
\begin{equation}
    \tau(z) = \sigma_{\rm T}c\int_0^z {\rm d}z'\frac{\bar{n}_{\rm e}(z')}{(1+z')H(z')}.
\end{equation}
We have computed the transverse momentum power, $\Delta_{\rm B}^2(k,z)$, using the electron momentum field $\boldsymbol{q}$ for a sample of snapshots from our simulations. This is defined $\boldsymbol{q} = \boldsymbol{v}(1 + \delta) = \boldsymbol{v}(n_{\rm e}/\bar{n}_{\rm e})$, where $\boldsymbol{v}$ is the velocity field of the gas. The power spectrum of the transverse momentum component, $\boldsymbol{q}_{\perp}$, is related to the power spectrum of the curl of the momentum field, $\boldsymbol{\nabla}\times\boldsymbol{q}$, by $P_{\boldsymbol{q_{\perp}}}=P_{\boldsymbol{\nabla}\times\boldsymbol{q}}/k^2$, and can be converted to the more commonly used definition $\Delta_{\rm B}^2 = P_{\boldsymbol{q_{\perp}}}k^3/(2\pi^2)$.

In Fig.~\ref{fig:transverse_momentum}, we show the dimensionless quantity $(\Delta_{\rm B}k/H)^2$ at six different redshifts for all gravity models and hydrodynamics schemes. In the lower sub-plots at each redshift, we show the MG enhancements of $\Delta_{\rm B}^2$ with respect to GR. For the non-radiative data, we see that the $f(R)$ enhancement is always increasing from large to small scales. The lowered enhancement at large scales is caused by the limited range of the fifth force, which is set by the Compton wavelength (Eq.~\eqref{eq:compton_wavelength}). For the full-physics data, the $f(R)$ enhancement follows a similar pattern at large scales but drops off for smaller scales ($k\gtrsim5h/{\rm Mpc}$) where baryonic processes are particularly prominent. For the nDGP data, the enhancement is roughly constant at large scales, since the fifth force in this model is long-range in the linear regime. There is again a suppression at small scales, which is likely to be caused by Vainshtein screening but could also be related to baryonic physics (we do not have non-radiative simulations to confirm the latter). For both $f(R)$ gravity and nDGP, the enhancement vanishes at higher redshifts where the amplitude of the scalar field is lower and the fifth force is screened out.

We also show, in Fig.~\ref{fig:ksz_contribution}, the derivative ${\rm d}D_l^{\rm kSZ}/{\rm d}z$, where 
\begin{equation}
D_l^{\rm kSZ} = l(l+1)C_l^{\rm kSZ}/(2\pi).
\end{equation}
This has been computed using the integrand and pre-factors in Eq.~(\ref{eq:limber}), and indicates the cosmic times and range of $k$-modes that have the greatest contribution to the kSZ angular power spectrum for different angular scales $l$. The enhancement of the kSZ power in F5 and N1 is observed to peak at $l=2000$ in Figs.~\ref{fig:power_spectra} and \ref{fig:nDGP_power_spectra}. From Fig.~\ref{fig:ksz_contribution}, we see that $D_{l=2000}^{\rm kSZ}$ receives a significant contribution from times $0.4 \lesssim a \lesssim 0.9$ and scales $1h/{\rm Mpc}\lesssim k \lesssim5h/{\rm Mpc}$. At these scales, the enhancement of $\Delta_{\rm B}^2$ peaks for N1 and has a similar magnitude for F5. The enhancements in these models span $\sim30\%$-$60\%$ over these scales and times, which is consistent with the peak enhancement of $C_{l}^{\rm kSZ}$. Going to larger angular scales ($600 < l < 1000$), $D_l^{\rm kSZ}$ is affected by lower $k$-modes (down to $\sim0.2h/{\rm Mpc}$) and lower redshifts. This then probes the larger scales (in Fig.~\ref{fig:transverse_momentum}) where the F5 fifth force is suppressed and the N1 enhancement levels off. This is consistent with Figs.~\ref{fig:power_spectra} and \ref{fig:nDGP_power_spectra}, where the kSZ power appears to be suppressed by a greater amount in F5 than in N1. Interestingly, this also implies that the enhancement of $C_l^{\rm kSZ}$ could be constant at angular scales larger than those available from our mock SZ maps. At smaller angular scales ($10000 \lesssim l \lesssim20000$), $D_l^{\rm kSZ}$ receives a significant contribution from high-$k$ modes ($2h/{\rm Mpc}\lesssim k\lesssim10h/{\rm Mpc}$), where Vainshtein screening suppresses the nDGP fifth force and the $f(R)$ fifth force is suppressed by baryonic processes (for the full-physics runs). In addition to this, $D_l^{\rm kSZ}$ is probing earlier times $0.25\leq a\lesssim 0.6$, where the scalar field amplitude is reduced in both models. This is therefore consistent with the lowered enhancement of $C_l^{\rm kSZ}$ at these angular scales. Note that the non-radiative $f(R)$ runs produce a higher kSZ power at high $l$ than the full-physics runs, which also agrees with the observation in Fig.~\ref{fig:transverse_momentum} that at large $k$ and $a\lesssim0.6$ the former has a larger transverse-momentum power spectrum.

The amplitude of $\Delta_{\rm B}k/H$ in Fig.~\ref{fig:transverse_momentum} appears to agree reasonably well with literature results \citep[e.g.,][]{Zhang:2003nr,2012ApJ...756...15S,Bianchini:2015iaa}, although it is slightly lower at large scales. We note that this is likely because the relatively small size $62h^{-1}{\rm Mpc}$ of our simulation box misses off longer-wavelength modes. It will therefore be useful to revisit this study with a larger box. The inclusion of longer-wavelength modes is expected to further suppress the F5 enhancement of $\Delta_{\rm B}^2$ at low-$k$ and to have little effect on the N1 enhancement.

We also note that for the entire $l$ range studied in Figs.~\ref{fig:power_spectra} and \ref{fig:nDGP_power_spectra} the kSZ power spectrum is dominated by modes with $k\gtrsim0.2h/{\rm Mpc}$ in the transverse-momentum power spectrum. From Fig.~\ref{fig:transverse_momentum}, we can see that in this regime galaxy formation has a non-negligible impact on $\Delta_{\rm B}$, which means that uncertainties in the subgrid physics can be an important theoretical systematic effect in using the kSZ power to test gravity models. Using kSZ data at $l<600$ may help reduce this effect, but the current simulation size does not allow a study of that range of $l$.

\section{Summary, Discussion and Conclusions}
\label{sec:conclusions:sz}

Over the past couple of decades, great advances have been made in the measurement of the secondary anisotropies of the CMB caused by the SZ effect, including its thermal component and even its much smaller kinematic component. The angular power spectrum of the tSZ effect has been increasingly adopted as a probe of cosmological parameters that influence the growth of large-scale structures. Also, as observations of the kSZ power spectrum continue to improve, the latter has been identified as another potentially powerful probe of cosmology. The next generation of ground-based observatories \citep{Ade:2018sbj,Abazajian:2016yjj} look set to revolutionise the constraining power of these probes.

In this chapter, we have looked at the viability of using the angular power spectra of the tSZ and kSZ effects as large-scale probes of HS $f(R)$ gravity and nDGP, which are representative of a wide-range of MG theories which exhibit the chameleon and Vainshtein screening mechanisms, respectively. In order to do so, we have made use of the {\sc shybone} simulations (cf.~Sec.~\ref{sec:methods:sz:simulations}), which are the first cosmological simulations that simultaneously incorporate full-physics plus HS $f(R)$ gravity \citep{Arnold:2019vpg} and nDGP \citep{Hernandez-Aguayo:2020kgq}. The simulations employ the IllustrisTNG full-physics model, which incorporates calibrated sub-resolution recipes for star formation and cooling as well as stellar and black hole feedback and allows realistic galaxy populations to be produced in hydrodynamical simulations.

Using these simulations, we have generated mock maps of the tSZ and kSZ signals (Sec.~\ref{sec:methods:sz:maps}), and used these maps to measure the angular power spectra. Our results (Figs.~\ref{fig:power_spectra} and \ref{fig:nDGP_power_spectra}) indicate that the fifth force, present in $f(R)$ gravity and nDGP, and the subgrid baryonic physics have different effects on the tSZ and kSZ power spectra. The former enhances the power on all scales probed by our maps ($500\lesssim{l}\lesssim8\times10^4$) by boosting the abundance and peculiar velocity of large-scale structures (e.g., dark matter haloes and free electrons inside them), while the latter brings about a suppression on scales $l\gtrsim3000$ for the tSZ effect and on all tested scales for the kSZ effect. Even with both of these effects present, we find that the power can be significantly enhanced in $f(R)$ gravity and nDGP: by up to $60\%$ for the tSZ effect and $35\%$ for the kSZ effect for the F5 and N1 models; and by $5\%$-$10\%$ for F6 and N5, which correspond to relatively weak modifications of GR. In addition, we have computed the power spectrum of the transverse component of the electron momentum field (Sec.~\ref{sec:results:sz:transverse_momentum}), which is closely related to the kSZ angular power spectrum. In particular, we show in Fig.~\ref{fig:ksz_contribution} that at angular sizes $l\geq600$ the kSZ signal is dominantly contributed by $k$-modes in the transverse-momentum power spectrum which are in the non-linear regime, and which are affected strongly by MG. The $k$-modes in the linear regime may contribute more to smaller $l$, but at least for $f(R)$ gravity the impact of MG at those $l$ values will be much less significant due to the finite range of the fifth force, as we can already see in Fig.~\ref{fig:power_spectra}.

We find that the relative difference between the MG models and GR is significantly affected by the additional baryonic processes that act in the full-physics simulations. Given that these processes are still relatively less well-constrained, this adds to the uncertainty in our theoretical predictions of the kSZ angular power spectra on small angular scales, e.g., $l>600$. Therefore, further work should be carried out using a range of full-physics parameters to precisely identify the scales on which constraints can be reliably made before the tSZ and kSZ power are used to probe $f(R)$ gravity and nDGP.

Finally, we note that the reason we are unable to study larger scales is the relatively small box size of the \textsc{shybone} simulations. In Chapter \ref{chapter:baryonic_fine_tuning}, we will present a re-calibrated full-physics model that can be used to run larger simulations, which can potentially be used to analyse larger scales in a future work. 
\graphicspath{{./gfx/}}

\chapter{Realistic galaxy formation simulations to study clusters in modified gravity}
\label{chapter:baryonic_fine_tuning}

\section{Introduction}
\label{sec:introduction_fine_tuning}

In Chapters \ref{chapter:scaling_relations} and \ref{chapter:DGP_clusters}, we used full-physics simulations to study the observable-mass scaling relations of galaxy groups and clusters in $f(R)$ gravity and nDGP. These simulations, which are from the \textsc{shybone} suite \citep{Arnold:2019vpg,Hernandez-Aguayo:2020kgq}, have a box size of $62h^{-1}{\rm Mpc}$ (`L62') and a high mass resolution. This is more suited for studying galaxies than galaxy clusters; indeed, these simulations contain just $\sim$100 galaxy group-sized objects and only 5-10 cluster-sized objects, with no objects above mass $M_{500}\sim10^{14.5}M_{\odot}$. The L62 predictions of the cluster scaling relations may therefore suffer from poor statistics and be potentially subjected to a significant influence by sample variance. Unfortunately, high-resolution simulations which incorporate both screened modified gravity and full baryonic physics are very expensive to run for larger cosmological volumes, which has made it difficult to study the interplay between baryonic physics and the fifth force at higher masses.

The \textsc{shybone} simulations make use of the IllustrisTNG baryonic physics model \citep{2017MNRAS.465.3291W,Pillepich:2017jle}, which can be used to generate galaxy populations whose stellar and gaseous properties closely match observations. However, it has been shown that a sufficiently high mass resolution is required to achieve this level of agreement \citep[e.g.,][]{Pillepich:2017jle}. For example, a lowered mass resolution means that the gas cells will have larger volumes, resulting in a smoothed out density field 
which can miss out the high-density peaks where star formation would be highest. The L62 \textsc{shybone} simulations, with $512^3$ gas cells, have $\sim15$ times lower mass resolution than that used to calibrate the TNG model ($25h^{-1}{\rm Mpc}$ box with the same number of gas cells). The gas and stellar properties of haloes from the L62 simulations still give a reasonable agreement with observational data, however the lowered resolution means that there is less star formation, resulting in the amplitudes of the stellar mass fraction, the stellar mass function and the star formation rate density being reduced compared to the fiducial TNG results \citep[for example, see Fig.~4 in][]{Arnold:2019vpg}.

For this work, running simulations with a substantial number of galaxy clusters (with masses $10^{14}M_{\odot}\lesssim M_{500}\lesssim10^{15.5}M_{\odot}$) would require a box size of at least $\sim300h^{-1}{\rm Mpc}$, which necessitates going to even lower resolutions. In order to make this possible without losing the good agreement with observational data, we have retuned the parameters of the IllustrisTNG model, including parameters which control the density threshold for star formation and the energy released by the stellar and black hole feedback mechanisms. This retuning was a significant undertaking which involved running over 200 simulations with a reduced box size, and our new model can be used to run low-resolution full-physics simulations for both standard gravity and modified gravity scenarios. We will present the model and describe the simulations used to tune it in this chapter. We have used this model to run GR and $f(R)$ gravity simulations with a significantly increased box size of $301.75h^{-1}{\rm Mpc}$, and we will present the predictions for the observable-mass scaling relations over an extended mass range $10^{13}M_{\odot}\leq M_{500}\lesssim10^{15}M_{\odot}$.

In Sec.~\ref{sec:methods_fine_tuning}, we provide a detailed description of the baryonic physics retuning and the large-box simulations, including the agreement with galaxy observations. We then present our results for the observable-mass scaling relations in Sec.~\ref{sec:results_fine_tuning}. Finally, we provide a summary of this chapter in Sec.~\ref{sec:summary}.

\section{Simulations and methods}
\label{sec:methods_fine_tuning}

In Sec.~\ref{sec:methods_fine_tuning:fine_tuning}, we discuss our retuning of the IllustrisTNG model for low-resolution simulations. Our large-box simulations, which are used for the main results of this chapter, are presented in Sec.~\ref{sec:methods_fine_tuning:L302_simulations}.

\subsection{Baryonic physics fine-tuning}
\label{sec:methods_fine_tuning:fine_tuning}

To recalibrate the baryonic physics at our desired mass resolution, we have run a large number of realisations, using \textsc{arepo} \citep{2010MNRAS.401..791S}, which have a small box size of $68h^{-1}{\rm Mpc}$ (`L68-N256'). These have $256^3$ dark matter particles with mass $1.35\times10^9h^{-1}M_{\odot}$ and, initially, the same number of gas cells with mass $\sim2.6\times10^{8}h^{-1}M_{\odot}$ on average. The calibration was carried out using GR (although, as we will show, the retuned model works equally well for F5) with the same cosmological parameter values as the \textsc{shybone} simulations. The runs were all started from the same set of initial conditions at redshift $z=127$. These have been generated using the \textsc{N-GenIC} code \citep[e.g.,][]{Springel:2005nw}, which uses the Zel’dovich approximation to displace an initially homogeneous particle distribution and obtain an initial density field with a prescribed linear power spectrum. Each of the input particles is then split into a dark matter particle and a gas cell, with the ratio of masses set by the values of the cosmic density parameters $\Omega_{\rm M}$ and $\Omega_{\rm b}$.

\begin{figure*}
\centering
\includegraphics[width=0.90\textwidth]{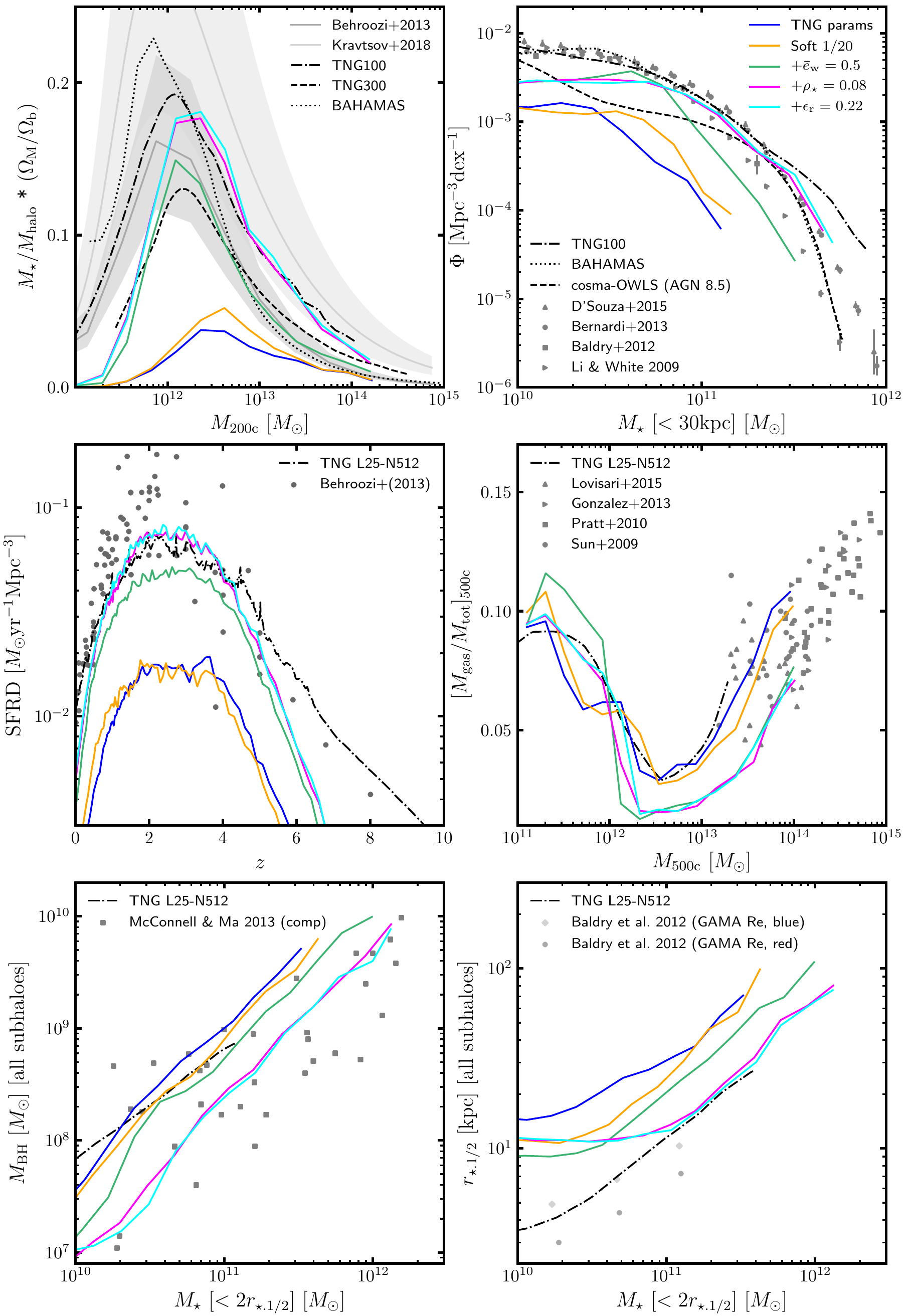}
\caption[Stellar, gas and black hole properties of haloes in a sample of the L68-N256 baryonic calibration runs.]{Stellar, gas and black hole properties in a sample of the L68-N256 calibration runs (\textit{coloured solid lines}). The properties are (\textit{clockwise from top-left}): stellar mass fraction; stellar mass function; gas mass fraction; stellar half-mass radius; black hole mass; star formation rate density. A selection of results from previous literature are shown as a comparison. See Sec.~\ref{sec:methods_fine_tuning:fine_tuning} for further details.}
\label{fig:L68_calibration}
\end{figure*}

We used the \textsc{subfind} code to locate FOF groups and the bound substructures of each group. By adjusting the baryonic physics of these calibration runs, we have aimed for reasonable agreement with observational data and empirical constraints for the six galaxy properties shown in Fig.~\ref{fig:L68_calibration}, which were also used to calibrate the IllustrisTNG model \citep{Pillepich:2017jle}. These are: the stellar mass fraction (FOF groups), with empirical constraints from \citet{2013ApJ...770...57B} and \citet{Kravtsov:2014sra}; the stellar mass function (subhaloes), with observations from \citet{2015MNRAS.454.4027D}, \citet{Bernardi:2013mqa}, \citet{2012MNRAS.421..621B} and \citet{2009MNRAS.398.2177L}; the star formation rate density (SFRD) as a function of redshift, with observations from \citet{2013ApJ...770...57B}; the gas mass fraction (FOF groups), with observations from \citet{Lovisari:2014pka}, \citet{Gonzalez:2013awy}, \citet{2010A&A...511A..85P} and \citet{Sun:2008eh}; the black hole mass versus the stellar mass (subhaloes), with the compilation of observations from \citet{2013ApJ...764..184M}; and the galaxy size versus the stellar mass (subhaloes), with observational data from \citet{2012MNRAS.421..621B}. The results for a selection of our calibration runs are represented by the coloured solid lines. Apart from the SFRD, which is a direct output of the simulations, these lines are generated using mass-binning of either FOF groups or subhaloes (see the parentheses above). The black lines show predictions from the TNG simulations \citep[e.g.,][]{Nelson:2017cxy,Springel:2017tpz,Marinacci:2017wew,Pillepich:2017fcc,2018MNRAS.477.1206N} as well as the BAHAMAS and cosmo-OWLS simulations \citep{McCarthy:2016mry,Brun:2013yva}.

The dark blue line in Fig.~\ref{fig:L68_calibration} shows the predictions using the fiducial TNG model at our lowered resolution. Star formation is significantly reduced at this resolution compared to the fiducial TNG resolution, which is used by the `TNG L25-N256' simulation (`TNG100' has a similar resolution, while `TNG300' has $\sim$5 times lower resolution). Consequently, the stellar mass fraction, the stellar mass function and the SFRD are significantly lower. The primary objective of our retuning is therefore to achieve a greater amount of star formation in order to obtain a closer match with the observational data. Our changes are described in the sections below, and the effects of these changes are shown in Fig.~\ref{fig:L68_calibration}. We note that the calibration runs discussed in this section are only a very small subset of the $\sim$200 simulations which were run for this calibration study: we provide further details on these simulations and the calibration procedure in Appendix \ref{appendix:baryonic_fine_tuning}.

\subsubsection{Gravitational softening}

In low-resolution simulations, where the gas cells have higher masses, there is a larger risk of two-body heating: this occurs when two particles come close together and incur a significant gravitational boost, which can raise the internal energy and subsequently the temperature of the gas. We have therefore increased the gravitational softening length to $1/20$ times the mean inter-particle separation, which is about twice the length used for the \textsc{shybone} simulations. The gravitational force is dampened when gas cells come within this distance, preventing extreme interactions. This change alone causes an overall reduction of the gas temperature in our simulations, which results in more cool gas that is capable of forming stars: see the orange lines in Fig.~\ref{fig:L68_calibration}, which have a greater amplitude than the dark blue lines for the stellar mass fraction and stellar mass function. 

We also considered fractions of 1/30 and 1/10 for the softening length. For higher masses (e.g., $M_{200}\gtrsim10^{12}M_{\odot}$), we observed that using a larger softening length results in greater star formation (for the reasons discussed above). However, we were unable to significantly boost star formation at lower masses; in fact, we observed that a large softening length (for example, a fraction 1/10 of the mean inter-particle separation) can even lead to less star formation at low masses. A potential effect of using a larger softening is that the gravitational potential well of haloes effectively becomes shallower. For low-mass haloes, where the gravitational potential well is already shallower than for high-mass haloes, this could potentially lead to a lower gas density (e.g., the gas is now less gravitationally bound) which in turn could reduce star formation. This is a motivation for using the fraction 1/20, for which we never observed the above effect, rather than using larger fractions.

It is evident from Fig.~\ref{fig:L68_calibration} that, while it can increase star formation, changing the gravitational softening length alone is not enough to produce stellar contents that match observational data.

\subsubsection{Stellar feedback}
\label{sec:fine_tuning:stellar_feedback}

In the TNG model, a portion of the gas mass in star-forming gas cells is converted into wind particles which are launched in random directions \citep{Pillepich:2017jle}. For a star-forming gas cell with metallicity $Z$, the available wind energy is:
\begin{equation}
    e_{\rm w} = \bar{e}_{\rm w}\left[f_{{\rm w},Z}+\frac{1 - f_{{\rm w},Z}}{1 + (Z/Z_{\rm w,ref})^{\gamma_{{\rm w},Z}}}\right]\times N_{\rm SNII}E_{\rm SN11,51}10^{51}{\rm erg}M_{\odot}^{-1},
    \label{eq:wind_energy}
\end{equation}
where $\bar{e}_{\rm w}$ is a dimensionless free parameter, $E_{\rm SNII,51}$ is the available energy from core-collapse supernovae in units of $10^{51}{\rm erg}$, $N_{\rm SNII}$ is the number of supernovae per stellar mass that is formed, and $f_{{\rm w},Z}$, $Z_{\rm w,ref}$ and $\gamma_{{\rm w},Z}$ are additional parameters of the model. A wind particle will eventually donate its thermal energy (along with its mass, momentum, and metal content) to a gas cell that is outside its local dense inter-stellar medium. This heats the gas and subsequently reduces the efficiency of star formation (gas must be sufficiently cool in order to form stars). 

Star formation efficiency is already reduced by our lowered gas cell resolution, therefore reducing the thermal heating of the gas by wind feedback can help to rectify this. We have achieved this in our retuning of the model by reducing the value of $\bar{e}_{\rm w}$ from the fiducial TNG value 3.6 to 0.5, which lowers the energy of the winds. As can be seen from the green lines in Fig.~\ref{fig:L68_calibration}, this change significantly boosts star formation over a wide range of masses. The stellar mass fraction now has a reasonable amplitude for $M_{200}\gtrsim10^{12}M_{\odot}$, while the amplitudes of the SFRD and stellar mass function are much closer to the observational data.

We have also tried varying the wind speed. This is given by \citep{Pillepich:2017jle}:
\begin{equation}
    v_{\rm w} = {\rm max}\left[\kappa_{\rm w}\sigma_{\rm DM}\left(\frac{H_0}{H(z)}\right)^{1/3},v_{\rm w,min}\right],
    \label{eq:wind_speed}
\end{equation}
where $\kappa_{\rm w}$ is a dimensionless factor, $\sigma_{\rm DM}$ is the local one-dimensional velocity dispersion of the dark matter particles and $v_{\rm w,min}$ is the minimum wind velocity allowed in the model. For our calibration runs, we tried reducing the $\kappa_{\rm w}$ and $v_{\rm w,min}$ parameters. This reduces the speed of the wind particles, which now take longer to transfer the thermal energy to the surrounding gas. Gas is therefore heated up at a slower rate, resulting in an increased amount of star formation. We found that reducing these parameters has a similar effect to reducing the $\bar{e}_{\rm w}$ parameter, with star formation boosted over a wide mass range. However, we could find no clear advantage in varying the wind speed parameters instead of $\bar{e}_{\rm w}$, or in varying all three of these parameters in combination. For simplicity, we therefore decided to adjust the stellar feedback using only the $\bar{e}_{\rm w}$ parameter.

\subsubsection{Star formation model}
\label{sec:fine_tuning:star_formation_model}

The star formation rate in IllustrisTNG is computed for gas cells using the \citet{Springel:2002uv} model. Stars can only be formed by gas cells which exceed a particular density threshold, which is approximately $n_{\rm H}\approx0.1{\rm cm}^{-3}$. We will refer to the threshold gas density as $\rho_{\star}$ in this work. At our reduced resolution, gas cells have a larger volume and therefore a smoothed density which can miss out high-density peaks in galaxies. In order to account for this, we have reduced $\rho_{\star}$ from $\approx0.1{\rm cm^{-3}}$ to a fixed value of $0.08{\rm cm^{-3}}$, allowing gas cells with lower density to form stars. 

The effect of making this change, in addition to the changes listed above, is shown by the magenta lines in Fig.~\ref{fig:L68_calibration}. This further boosts the stellar mass fraction and SFRD, which are now both in good agreement with the TNG100 results for $M_{200}\gtrsim10^{12}M_{\odot}$ and $z\lesssim5$, respectively, and there is also now a good agreement with the \citet{2015MNRAS.454.4027D}, \citet{Bernardi:2013mqa} and \citet{2012MNRAS.421..621B} observations of the stellar mass function for $M_{\star}\gtrsim10^{11}M_{\odot}$.

\subsubsection{Black hole feedback}
\label{sec:fine_tuning:bh_feedback}

The TNG model employs two types of black hole feedback, depending on the accretion state of the central supermassive black hole \citep{2017MNRAS.465.3291W}: in the low accretion state, a kinetic feedback model is employed which produces black hole-driven winds; and in the high accretion state, a thermal feedback model is employed which heats up the surrounding gas. The rate of accretion $\dot{M}$ is set by the Eddington-limited Bondi accretion:
\begin{equation}
\begin{split}
    &\dot{M}_{\rm Bondi}=\frac{4\pi G^2M_{\rm BH}^2\rho}{c_{\rm s}^3},\\
    &\dot{M}_{\rm Edd}=\frac{4\pi GM_{\rm BH}m_{\rm p}}{\epsilon_{\rm r}\sigma_{\rm T}c},\\
    &\dot{M}={\rm min}\left(\dot{M}_{\rm Bondi},\dot{M}_{\rm Edd}\right),
\end{split}
\label{eq:accretion_rate}
\end{equation}
where $M_{\rm BH}$ is the black hole mass, $\rho$ represents the ambient density around the black hole, $c_{\rm s}$ represents the ambient sound speed and $\epsilon_{\rm r}$ is the black hole radiative efficiency. The feedback mode depends on whether or not the ratio $\dot{M}/\dot{M}_{\rm Edd}$ exceeds the following threshold:
\begin{equation}
    \chi = {\rm min}\left[\chi_0\left(\frac{M_{\rm BH}}{10^8M_{\odot}}\right)^{\beta},0.1\right],
\end{equation}
where $\chi_0$ and $\beta$ are parameters. If $\dot{M}/\dot{M}_{\rm Edd}>\chi$, the resulting thermal feedback will inject thermal energy into the surrounding gas at a rate $\dot{E}_{\rm therm}=\epsilon_{\rm f,high}\epsilon_{\rm r}\dot{M}c^2$, where $\epsilon_{\rm f,high}$ is another parameter; and if $\dot{M}/\dot{M}_{\rm Edd}<\chi$, the resulting kinetic feedback will inject energy into the surroundings at a rate $\dot{E}_{\rm kin}=\epsilon_{\rm f,kin}\dot{M}c^2$, where the factor $\epsilon_{\rm f,kin}$ depends on the ambient density $\rho$. Both of these feedback modes will reduce the efficiency of star formation in the surrounding gas, either by blowing gas out, so that less gas will exceed the density threshold for star formation, or by heating up gas which, as for stellar feedback, reduces the amount of cool gas capable of forming stars. As discussed above, the star formation efficiency is already reduced by our lowered gas resolution; reducing the overall effect of black hole feedback on star formation therefore provides another means of rectifying this.

For our retuning of the black hole feedback, we have increased $\epsilon_{\rm r}$ from the fiducial TNG value 0.2 to 0.22. The effect of this change on the overall energy release is quite complex: the energy injected by thermal feedback will be boosted, unless $\dot{M}=\dot{M}_{\rm Edd}$ (i.e., $\dot{M}_{\rm Bondi}>\dot{M}_{\rm Edd}$) in which case the $\epsilon_{\rm r}$ factors will cancel and there will be no effect; on the other hand, from Eq.~(\ref{eq:accretion_rate}) we see that $\dot{M}_{\rm Edd}$ is lowered if $\epsilon_{\rm r}$ is increased, and subsequently the ratio $\dot{M}/\dot{M}_{\rm Edd}$ will be greater and there will then be less kinetic feedback. From this discussion, increasing $\epsilon_{\rm r}$ is therefore expected to increase the heating of the gas by thermal feedback and reduce the blowing out of gas by kinetic feedback: two effects which would have competing impacts on the star formation efficiency. For our calibration runs, we have observed that increasing $\epsilon_{\rm r}$ to 0.22 boosts the amount of star formation. Therefore, it seems that the reduced blowing out of gas by kinetic feedback has the dominant effect here.

The result of making this final adjustment to the baryonic physics model is shown by the cyan lines in Fig.~\ref{fig:L68_calibration}. The stellar mass fraction and stellar mass function are both slightly boosted for high-mass haloes. From the upper-right panel of Fig.~\ref{fig:L68_calibration}, our model now appears to slightly overshoot the observed stellar mass function at higher masses; this is actually a consequence of sample variance which results from using a small box-size. As we will show in Sec.~\ref{sec:methods_fine_tuning:L302_simulations}, the agreement is very good for the much larger $301.75h^{-1}{\rm Mpc}$ box size. The change to $\epsilon_{\rm r}$ also brings the galaxy size relation into closer agreement with the TNG L25-N512 runs, while the good agreement with observations for the black hole mass relation, the gas mass fraction and the SFRD is unaffected.

We also considered the minimum halo mass for black hole seeding. Central black holes are only found in haloes with mass above this threshold. Increasing the threshold means that, at a given time, black holes will have been growing for a shorter period of time and will consequently have a lower mass. This results in lower accretion and therefore reduces the energy released through feedback. We found that this can significantly boost star formation in higher-mass haloes (which contain larger black holes and are therefore more susceptible to black hole feedback) but has very little effect on the stellar content of low-mass haloes. We found no clear advantage to vary this in addition to the other parameters varied in this work, therefore we adjusted the black hole feedback using only the $\epsilon_{\rm r}$ parameter.

\subsubsection{Summary and further comments}

In summary, our retuned baryonic model uses updated parameter values $\rho_{\star}=0.08{\rm cm^{-3}}$, $\bar{e}_{\rm w}=0.5$ and $\epsilon_{\rm r}=0.22$, in addition to a larger gravitational softening length, to get sufficient star formation. 

While our retuning of the baryonic physics has significantly boosted star formation across the full mass range shown in Fig.~\ref{fig:L68_calibration}, it is still unable to give sufficient star formation at lower masses compared to observational data. Therefore, the stellar mass fraction and stellar mass function are both underestimated at the low-mass end, and the SFRD is underestimated at redshifts $z\gtrsim5$ (where there are only low-mass haloes). We attempted to rectify this by using even lower values of $\rho_{\rm gas}$ and reduced stellar and black hole feedback, but found that this offered little improvement overall. We even tried switching off feedback entirely, by setting the stellar wind energy to zero ($\bar{e}_{\rm w}=0$) and by preventing the seeding of black holes: while this resulted in a huge amount of star formation at masses $M_{200}\gtrsim10^{12}M_{\odot}$, there was still insufficient star formation at masses $M_{200}\lesssim10^{11.5}M_{\odot}$ to match observations. Therefore, the only way to have sufficient star formation across the full mass range appears to be by increasing the mass resolution. Interestingly, the BAHAMAS simulations \citep{McCarthy:2016mry} are able to achieve sufficient star formation for the full mass range (see the dotted lines in the top panels of Fig.~\ref{fig:L68_calibration}) despite having $\sim3\times$ lower mass resolution than our simulations. The BAHAMAS simulations were run using the \textsc{gadget-3} code \citep{Springel:2005mi}, which uses SPH rather than the Voronoi mesh. Perhaps the contrasting treatments of the gas by the two codes can explain the different levels of star formation at these lowered resolutions. One possible way to further boost star formation in low-mass haloes is by having a halo mass dependency for some of the baryonic parameters, but this approach is beyond the scope of this work. On the other hand, we note that our low-resolution simulations are designed primarily for studying galaxy groups and clusters ($M_{500}\gtrsim10^{13}M_{\odot}$), for which the predictions of our model appear to be very reasonable.

While retuning these parameters, we came across a number of degeneracies. For example, as discussed above in Sec.~\ref{sec:fine_tuning:stellar_feedback}, we found that the stellar-induced wind feedback can be lowered by reducing the speed of the winds rather than the wind energy. And, in our final model, we could have instead used a slightly increased $\rho_{\star}$ (e.g., $\rho_{\star}=0.09{\rm cm^{-3}}$) and reduced $\epsilon_{\rm r}$ (e.g., $\epsilon_{\rm r}=0.18$) to achieve similar results. Therefore, we note that different combinations of parameter values could have been used to achieve a similar level of agreement with the observational data.

\subsection{Large-box simulation}
\label{sec:methods_fine_tuning:L302_simulations}

Our full simulation (`L302-N1136'), which has been run using the retuned baryonic model at the same mass resolution as the L68-N256 calibration runs, has a box size of $301.75h^{-1}{\rm Mpc}$ and contains $1136^3$ dark matter particles and (initially) the same number of gas cells. The simulation has been run for both GR and F5. 

\begin{figure*}
\centering
\includegraphics[width=0.95\textwidth]{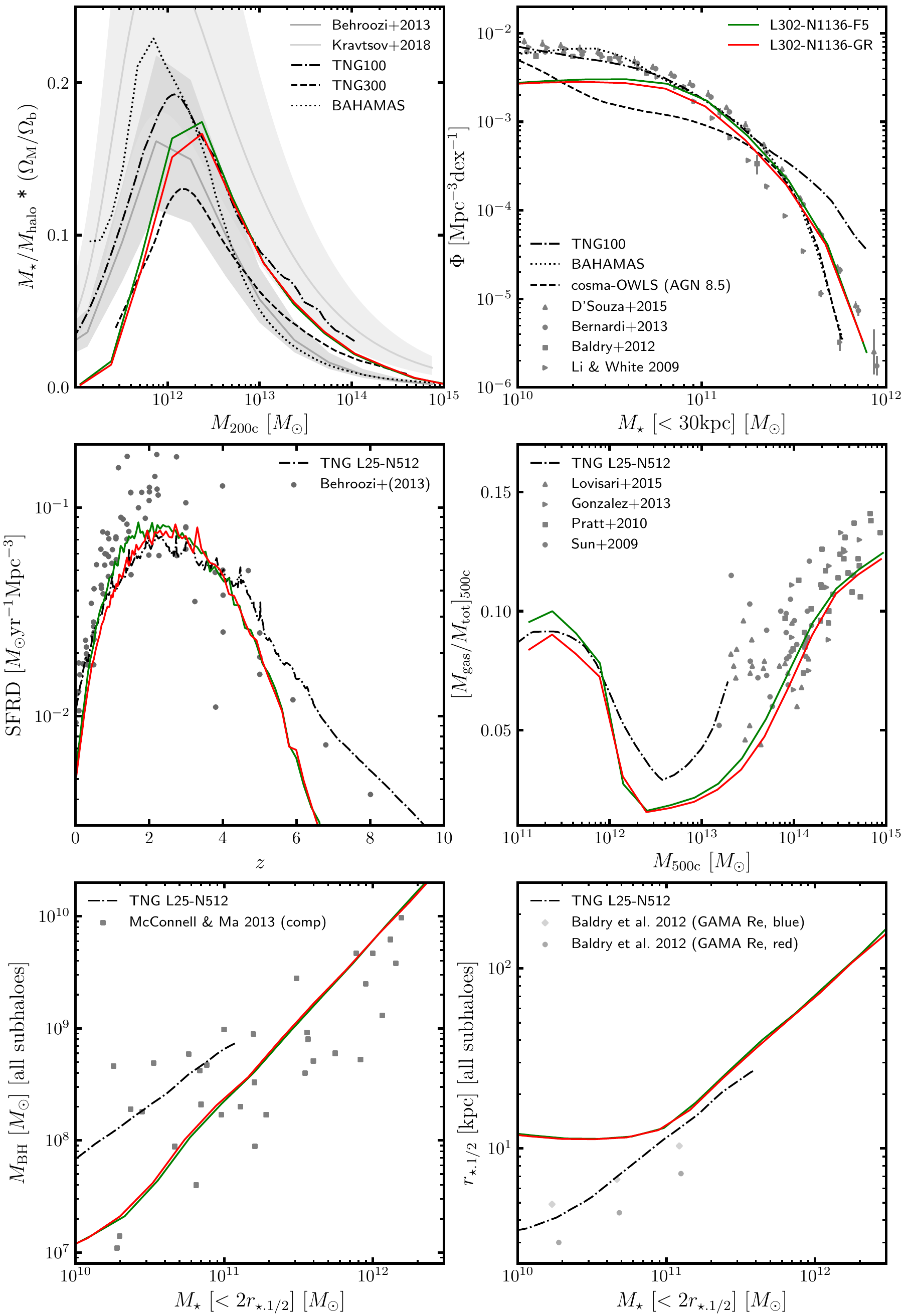}
\caption[Stellar, gas and black hole properties for F5 and GR haloes in the L302-N1136 full-physics simulations.]{Stellar, gas and black hole properties of haloes in the L302-N1136 simulations for GR (\textit{red lines}) and F5 (\textit{green lines}). Apart from the coloured lines, the contents of this figure are identical to Fig.~\ref{fig:L68_calibration}.}
\label{fig:L302_observables}
\end{figure*}

The red lines in Fig.~\ref{fig:L302_observables} show the GR predictions of the six observables used to calibrate the baryonic model. The results are slightly different compared to the cyan lines in Fig.~\ref{fig:L68_calibration}, which use the same baryonic model: the predicted amplitudes of the stellar mass fraction and stellar mass function are slightly lower, which actually improves the high-mass agreement with observations of the latter; and the amplitude of the galaxy size relation is greater for $10^{11}M_{\odot}\lesssim M_{\star}\lesssim3\times10^{11}M_{\odot}$, leading to slightly worse agreement with the TNG L25-N512 predictions at these masses. These effects are likely to be a consequence of using a much larger box size, which is less susceptible to sample variance. The L302-N1136 simulation also extends to higher masses ($M_{500}\sim10^{15}M_{\odot}$ and $M_{\star}\sim10^{12}M_{\odot}$) than the L68-N256 runs. At these masses, the agreement with the observational data in Fig.~\ref{fig:L302_observables} looks excellent. 

The predictions for the F5 model, shown by the green lines in Fig.~\ref{fig:L302_observables}, agree with the GR predictions for the galaxy size and black hole mass relations; however, the amplitudes of the other four observables are slightly boosted in F5 compared to GR. The SFRD is boosted for redshifts $0.5\lesssim z\lesssim3$: this is consistent with the results for the \textsc{shybone} simulations \citep{Arnold:2019vpg}. The stellar mass fraction and stellar mass function are boosted at $M_{200}\sim10^{12}M_{\odot}$ and $M_{\star}\sim10^{11}M_{\odot}$, respectively, and the gas mass fraction is slightly enhanced for masses $M_{500}\gtrsim10^{13}M_{\odot}$. The F5 predictions are still in excellent agreement with the observations, therefore it is not necessary to carry out a separate retuning of the baryonic physics for this model.

\section{Results}
\label{sec:results_fine_tuning}

Using the L302-N1136 simulations (see Sec.~\ref{sec:methods_fine_tuning}), we have measured the observable-mass scaling relations in GR and F5 for FOF groups in the mass range $10^{13}M_{\odot}\leq M_{500}\lesssim10^{15}M_{\odot}$. In Sec.~\ref{sec:results_fine_tuning:scaling_relations}, we will discuss the relations for the mass-weighted gas temperature $\bar{T}_{\rm gas}$, the SZ $Y$-parameter $Y_{\rm SZ}$, the X-ray analogue of the $Y$-parameter $Y_{\rm X}$ and the X-ray luminosity $L_{\rm X}$, which have all been computed in the same way as in Chapter~\ref{chapter:scaling_relations}. We again exclude gas cells found in the core region, defined by the radial range $r<0.15R_{500}$, and we will test the `true density' mappings (see Chapter \ref{chapter:scaling_relations}) between the F5 and GR relations for redshifts $0\leq z\leq1$. We will also discuss scaling relations which don't involve the cluster mass in Sec.~\ref{sec:results_fine_tuning:observable_relations}; these can potentially be used to test gravity using galaxy groups and clusters with no requirement to measure or infer the mass.

\subsection{Observable-mass scaling relations}
\label{sec:results_fine_tuning:scaling_relations}

\begin{figure*}
\centering
\includegraphics[width=1.0\textwidth]{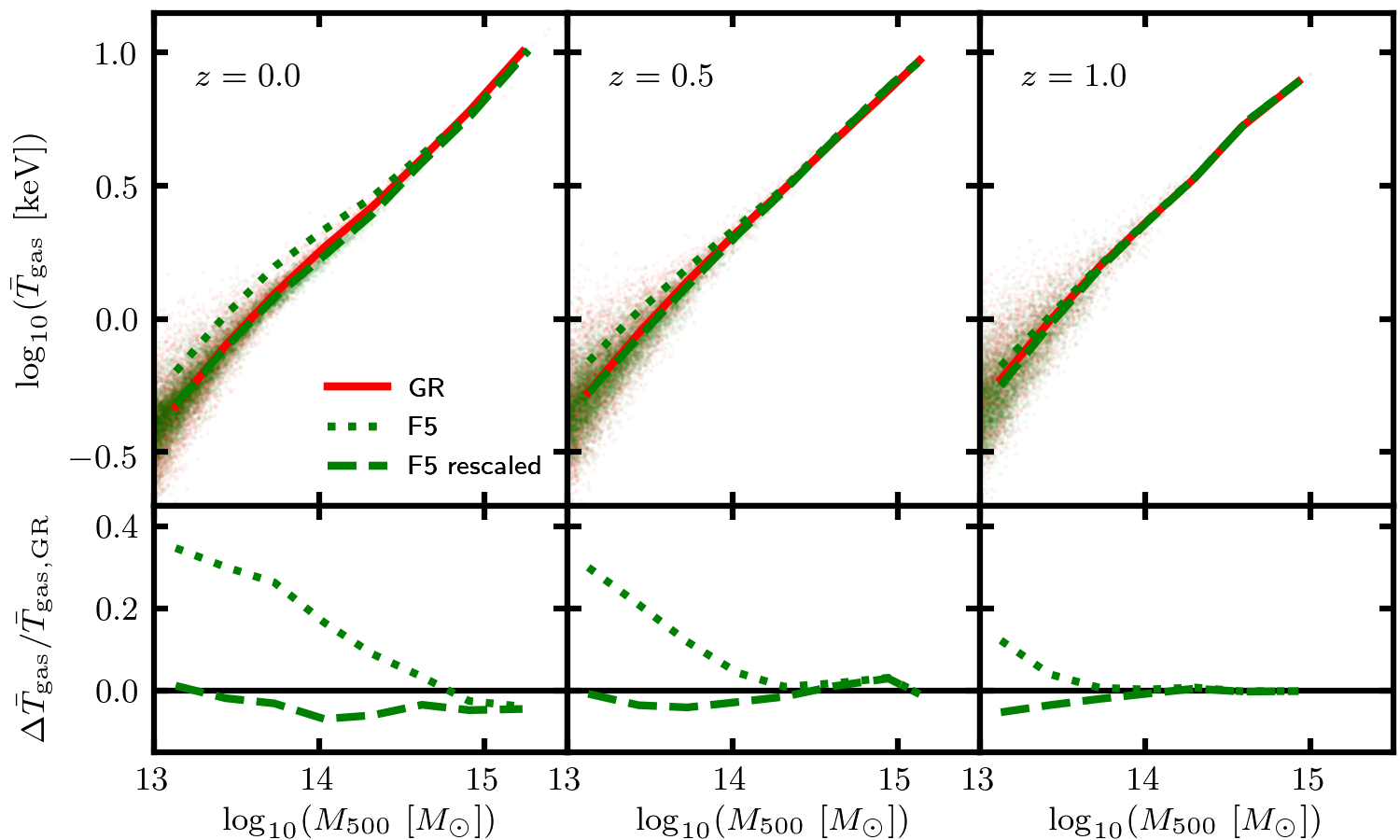}
\caption[Mass-weighted temperature as a function of mass for F5 and GR haloes in the L302 full-physics simulations.]{Gas temperature as a function of the halo mass for the full-physics L302 simulation (see Sec.~\ref{sec:methods_fine_tuning}) at redshifts 0, 0.5 and 1. The curves correspond to the median temperature and the mean logarithm of the mass computed within mass bins. Data has been included for GR (\textit{red solid lines}) and F5 (\textit{green lines}). A rescaling to the F5 temperature has been carried out as described in Sec.~\ref{sec:results_fine_tuning:scaling_relations}. Data points are displayed, with each point corresponding to a GR halo (\textit{red points}) or to a halo in F5 (\textit{green points}), including the rescaling. \textit{Bottom row}: the relative difference between the F5 and GR curves in the above plots.}
\label{fig:L302_tgas}
\end{figure*}

The top rows of Figs.~\ref{fig:L302_tgas}-\ref{fig:L302_lx} show the F5 and GR scaling relations for redshifts 0, 0.5 and 1, with data points representing individual haloes. At $z=0$, there are $\sim8000$ GR haloes with $M_{500}>10^{13}M_{\odot}$, including $\sim500$ clusters with $M_{500}>10^{14}M_{\odot}$. This is a significant improvement on the L62 \textsc{shybone} simulations, which only had $\sim100$ haloes with $M_{500}>10^{13}M_{\odot}$ at $z=0$. The curves in the top rows of the figures show the median observable as a function of the mean logarithmic mass computed within mass bins; the `true density' rescalings of the F5 relation, which are computed using our analytical tanh formula (Eq.~(\ref{eq:mdyn_enhancement})), are indicated by the dashed lines. We use eight mass bins with constant logarithmic width over the range $10^{13}M_{\odot}\leq M_{500}\leq 10^{15.4}M_{\odot}$. All bins are shown regardless of the halo count. Although there may only be a few haloes in the highest-mass bins, we note that these correspond to high-mass clusters for which scatter in the scaling relations, especially in the model difference between F5 and GR, is expected to be very low. The relative difference between the F5 and GR binned data is shown in the lower panels of the figures.

\subsubsection{Temperature scaling relation}

The results for the $\bar{T}_{\rm gas}(M)$ scaling relation are shown in Fig.~\ref{fig:L302_tgas}. The GR data appears to follow the well-known power-law behaviour for cluster-sized objects; however, the relation appears to curve at lower masses, where processes such as feedback can cause additional gas heating and break the power-law scaling. In F5, haloes are mostly screened from the fifth force for masses $M_{500}\gtrsim10^{14.5}M_{\odot}$ at $z=0$, $M_{500}\gtrsim10^{14}M_{\odot}$ at $z=0.5$ and $M_{500}\gtrsim10^{13.5}M_{\odot}$ at $z=1$, and here the F5 temperature closely follows the GR temperature. At lower masses, the F5 temperature becomes significantly enhanced, as the total gravitational potential of the halo is raised by the fifth force. 

Our rescaling of the F5 data, which we recall involves dividing the temperature by the ratio of the dynamical mass to the true mass, can successfully account for this offset at lower masses, restoring $<7\%$ agreement with the GR relation. However, the rescaled F5 relation now slightly underestimates the GR relation on average. We note that at $z=0$, this offset appears to be roughly constant for cluster-sized masses; therefore, as long as the GR scaling relation parameters are allowed to vary in MCMC sampling (which can account for small differences in the amplitude), this rescaling is still expected to work well in our constraint pipeline presented in Chapter \ref{chapter:constraint_pipeline}.

\subsubsection{\texorpdfstring{$Y_{\rm SZ}$}{YSZ} and \texorpdfstring{$Y_{\rm X}$}{YX} scaling relations}

\begin{figure*}
\centering
\includegraphics[width=1.0\textwidth]{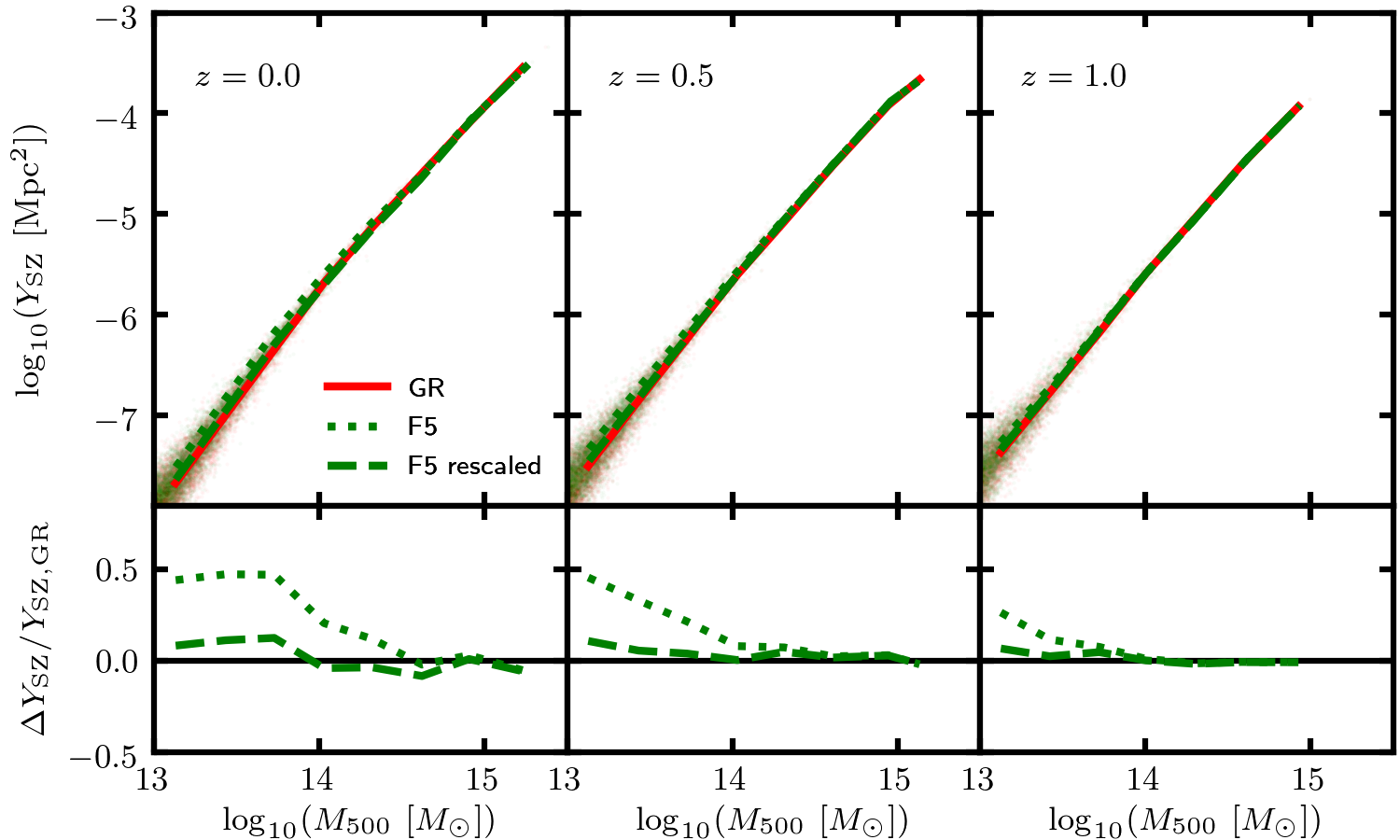}
\caption[$Y_{\rm SZ}$ parameter as a function of mass for F5 and GR haloes in the L302 full-physics simulations.]{SZ Compton $Y$-parameter as a function of the halo mass for the full-physics L302 simulation (see Sec.~\ref{sec:methods_fine_tuning}) at redshifts 0, 0.5 and 1. Apart from the observable, this figure has the same layout as Fig.~\ref{fig:L302_tgas}.}
\label{fig:L302_ysz}
\end{figure*}

\begin{figure*}
\centering
\includegraphics[width=1.0\textwidth]{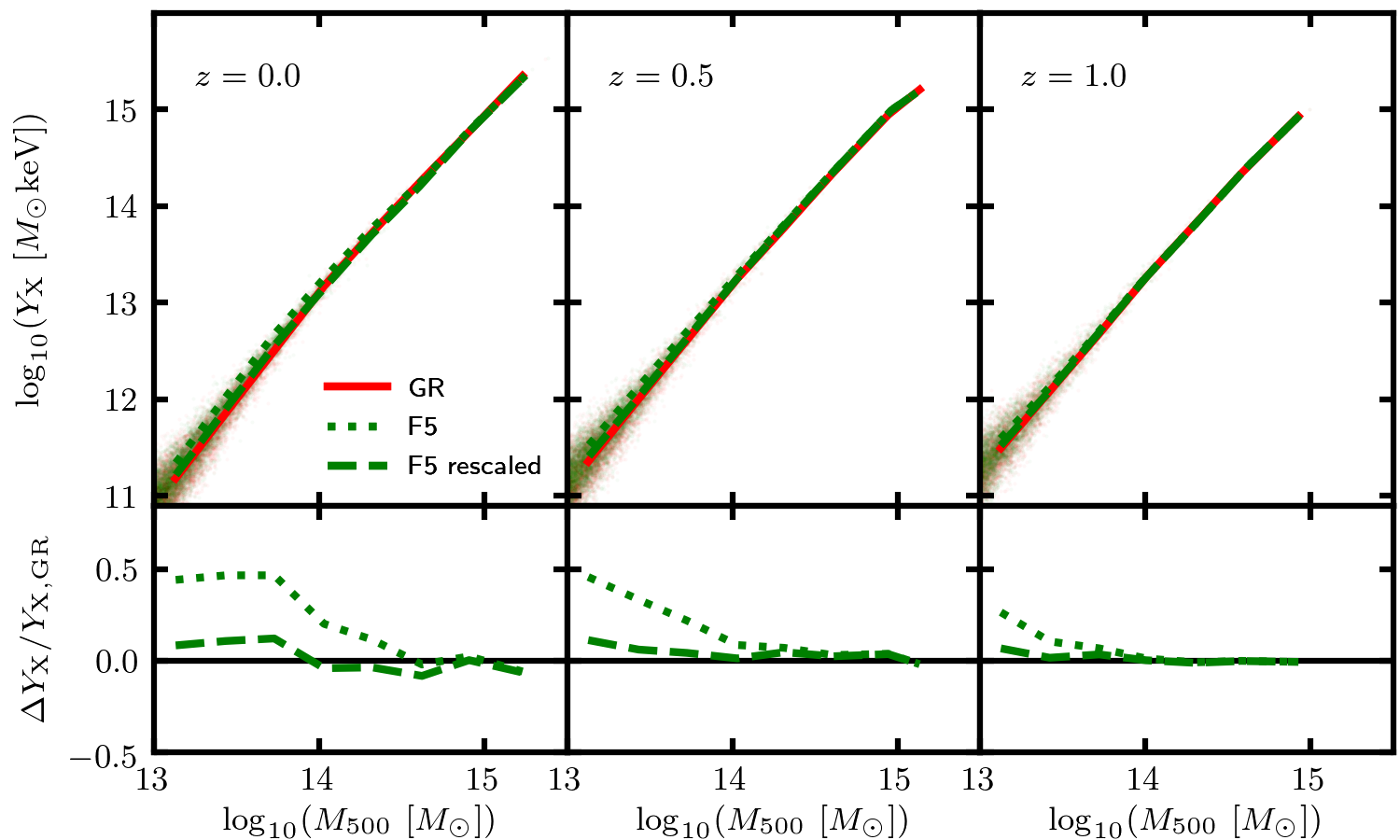}
\caption[$Y_{\rm X}$ parameter as a function of mass for F5 and GR haloes in the L302 full-physics simulations.]{X-ray analogue of the $Y$-parameter as a function of the halo mass for the full-physics L302 simulation (see Sec.~\ref{sec:methods_fine_tuning}) at redshifts 0, 0.5 and 1. Apart from the observable, this figure has the same layout as Fig.~\ref{fig:L302_tgas}.}
\label{fig:L302_yx}
\end{figure*}

The $Y_{\rm SZ}(M)$ and $Y_{\rm X}(M)$ relations are shown in Figs.~\ref{fig:L302_ysz} and \ref{fig:L302_yx}, respectively. The GR relation appears to follow a weakly broken power-law, with a slightly steeper slope for group-sized haloes ($M_{500}\lesssim10^{14}M_{\odot}$) than for cluster-sized haloes ($M_{500}\gtrsim10^{14}M_{\odot}$). Again the low-mass behaviour can be explained by feedback, which, in addition to heating up gas, also blows gas out from the inner regions which in turn can lower the $Y$ values. For example, in Chapter~\ref{chapter:scaling_relations}, we observed that the $Y$-parameter was lower in the full-physics simulations than in the non-radiative simulations, which did not include feedback. 

For lower (unscreened) masses, we observe an enhancement of the F5 relations by up to $\sim50\%$ compared to GR. This is mostly corrected by our rescaling, after which the agreement is within $\sim12\%$ for group-sized haloes and is within a few percent on average for cluster-sized objects. This is 
positive news for our constraint pipeline in Chapter \ref{chapter:constraint_pipeline}, which used this rescaling to model the $Y_{\rm SZ}(M)$ relation for clusters in the redshift range $0<z<0.5$.

\subsubsection{X-ray luminosity scaling relation}

\begin{figure*}
\centering
\includegraphics[width=1.0\textwidth]{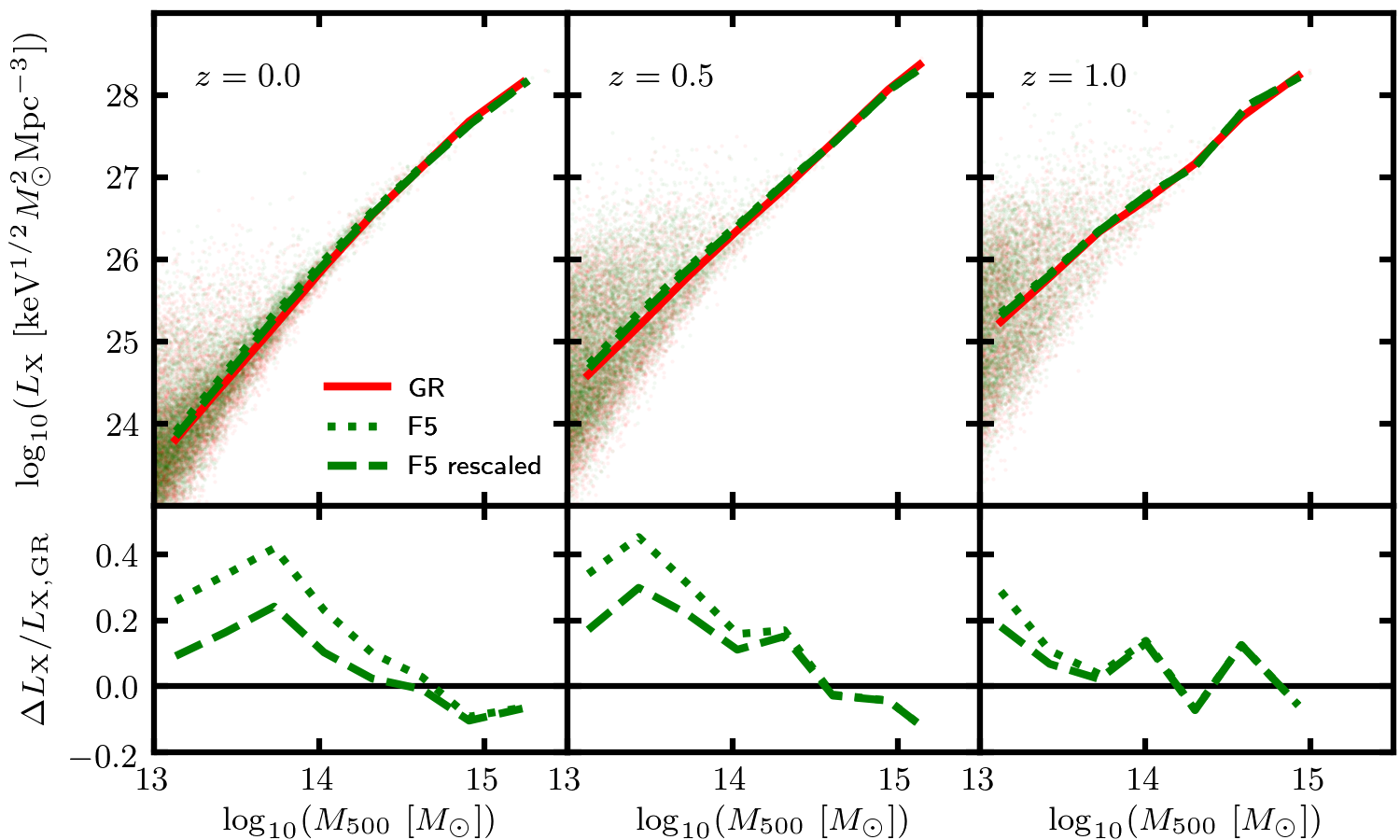}
\caption[X-ray luminosity as a function of mass for F5 and GR haloes in the L302 full-physics simulations.]{X-ray luminosity as a function of the halo mass for the full-physics L302 simulation (see Sec.~\ref{sec:methods_fine_tuning}) at redshifts 0, 0.5 and 1. Apart from the observable, this figure has the same layout as Fig.~\ref{fig:L302_tgas}.}
\label{fig:L302_lx}
\end{figure*}

The $L_{\rm X}(M)$ relation is shown in Fig.~\ref{fig:L302_lx}. As we showed in Chapter \ref{chapter:scaling_relations}, our rescalings were unable to accurately account for the difference between GR and $f(R)$ gravity. That study was carried out primarily for group-sized haloes, and for these new results the rescaling is again unsuccessful for the mass range $10^{13}M_{\odot}\lesssim M_{500}\lesssim10^{14}M_{\odot}$. The X-ray luminosity varies as $T_{\rm gas}^{1/2}\rho_{\rm gas}^2$. For the `true density' rescaling, which is applied here, it is assumed that the gas temperature is enhanced by the fifth force while the gas density is unchanged. This may be the case in non-radiative simulations, however in full-physics simulations it is not necessarily true. For example, it is likely that there are different levels of feedback in F5 and GR. A greater amount of feedback in one model would result in the blowing out of gas and subsequent lowering of the gas density. This is expected to have a much greater effect on $L_{\rm X}$, which varies as $\rho_{\rm gas}^2$, than on the other observables considered in this work. For the $Y$-parameters, which vary as $T_{\rm gas}\rho_{\rm gas}$, the effects of feedback on the gas density and the temperature can roughly balance out \citep[e.g.,][]{Fabjan:2011}, allowing our rescaling to perform better for these observables as we saw in Figs.~\ref{fig:L302_ysz} and \ref{fig:L302_yx}. 

While the above is particularly problematic for galaxy groups, which are more susceptible to feedback, our rescaling appears to work reasonably well for cluster-sized haloes in Fig.~\ref{fig:L302_lx}, where the rescaling brings the agreement to within 10\% at $z=0$. However, the $L_{\rm X}(M)$ relation is also highly scattered compared to the other relations considered in this work. For example, the agreement between F5 and GR has a large $\sim20\%$ fluctuation at $z=1$ for $M_{500}>10^{14}M_{\odot}$, even though clusters are completely screened at this redshift.

Based on this discussion, the $\bar{T}_{\rm gas}(M)$, $Y_{\rm SZ}(M)$ and $Y_{\rm X}(M)$ relations are more suitable than the $L_{\rm X}(M)$ relation for tests of gravity that involve the cluster mass.

\subsection{\texorpdfstring{$Y_{\rm X}$}{YX}-temperature and \texorpdfstring{$L_{\rm X}$}{LX}-temperature relations}
\label{sec:results_fine_tuning:observable_relations}

\begin{figure*}
\centering
\includegraphics[width=1.0\textwidth]{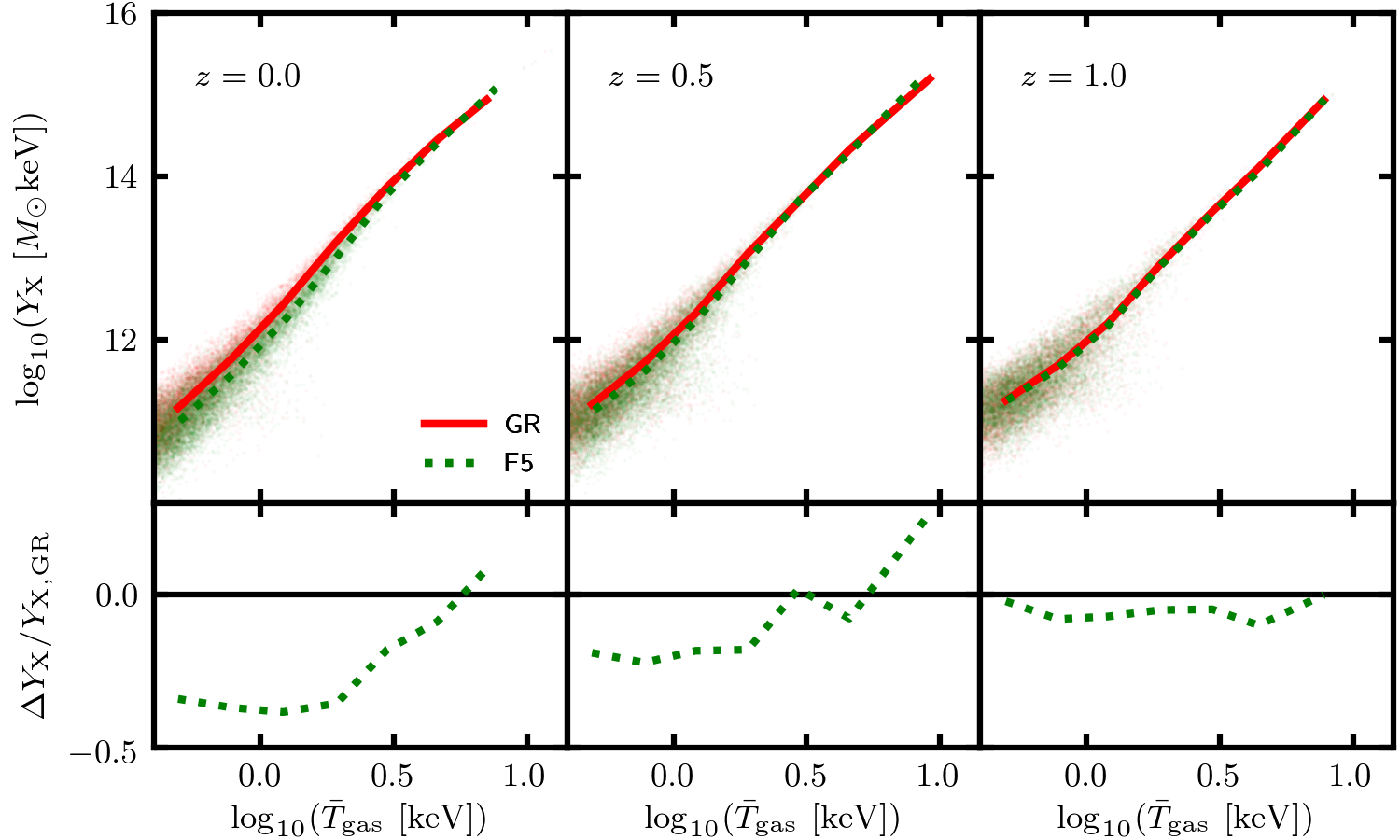}
\caption[$Y_{\rm X}$ parameter as a function of temperature for F5 and GR haloes in the L302 full-physics simulations.]{X-ray analogue of the $Y$-parameter as a function of gas temperature for haloes from the full-physics L302 simulation (see Sec.~\ref{sec:methods_fine_tuning}) at redshifts 0, 0.5 and 1. The curves correspond to the median luminosity and the mean logarithm of the temperature computed within temperature bins. Data has been included for GR (\textit{red solid lines}) and F5 (\textit{green dotted lines}). Data points are displayed, with each point corresponding to a GR halo (\textit{red points}) or to a halo in F5 (\textit{green points}). \textit{Bottom row}: the relative difference between the F5 and GR curves in the above plots.}
\label{fig:L302_yx_tgas}
\end{figure*}

\begin{figure*}
\centering
\includegraphics[width=1.0\textwidth]{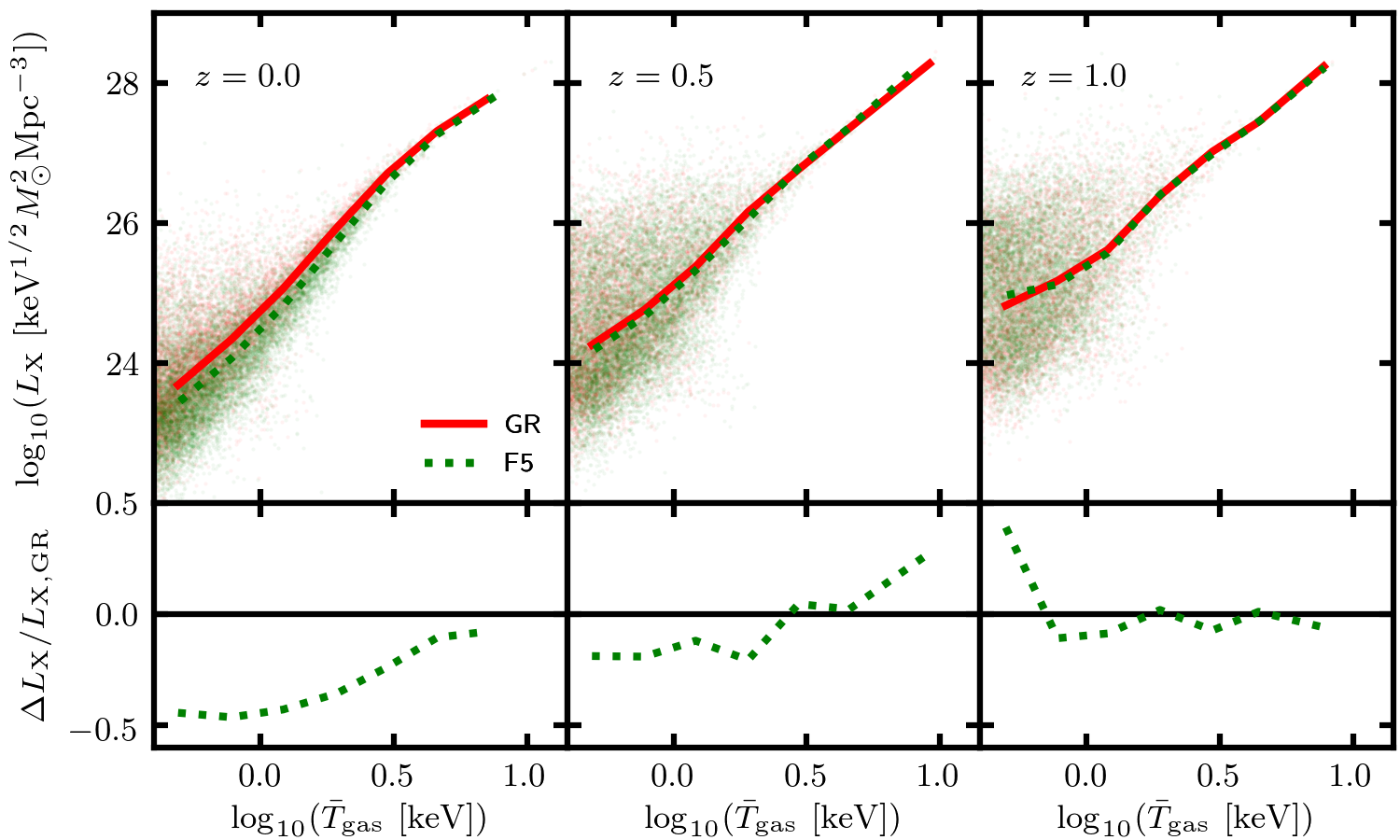}
\caption[X-ray luminosity as a function of temperature for F5 and GR haloes in the L302 full-physics simulations.]{X-ray luminosity as a function of gas temperature for haloes from the full-physics L302 simulation (see Sec.~\ref{sec:methods_fine_tuning}) at redshifts 0, 0.5 and 1. Apart from the observable used in the vertical axis, this figure has the same format as Fig.~\ref{fig:L302_yx_tgas}.}
\label{fig:L302_lx_tgas}
\end{figure*}

In Figs.~\ref{fig:L302_yx_tgas} and \ref{fig:L302_lx_tgas}, we show the $Y_{\rm X}(\bar{T}_{\rm gas})$ and $L_{\rm X}(\bar{T}_{\rm gas})$ relations, respectively, at redshifts 0, 0.5 and 1. The curves show the median $Y_{\rm X}$-parameter and mean logarithmic temperature computed within seven temperature bins, with logarithmic width 0.2, spanning the range $10^{-0.4}{\rm keV}\leq\bar{T}_{\rm gas}\leq10^1{\rm keV}$. 

Haloes in F5 and GR with the same temperature are expected to have a similar dynamical mass; in this case, the F5 haloes would have a lower true mass than the GR haloes, and therefore a lower gas density (e.g., recall the `effective density' rescalings in Chapter \ref{chapter:scaling_relations}). This explains why, for $Y_{\rm X}(\bar{T}_{\rm gas})$, the amplitude of the F5 relation is suppressed by up to $\sim40\%$ compared to GR, while for $L_{\rm X}(\bar{T}_{\rm gas})$ the F5 relation is suppressed by up to $\sim45\%$ (the differences may also be partly due to differences in the levels of feedback in the two models). For both relations, the difference is greater for lower redshifts and lower temperatures, where more haloes are unscreened. 

Neither of these relations involve the cluster mass. Therefore, these could potentially be used to test gravity using galaxy groups and clusters without the risk of bias from mass measurements. This demonstrates that, besides their abundances inferred from observables such as $Y_{\rm SZ}$ and $Y_{\rm X}$, the combination of different internal observational properties for a population of galaxy clusters or groups can also offer useful, possibly complementary, constraints on the theory of gravity.

\section{Summary}
\label{sec:summary}

Running large-box cosmological simulations which simultaneously incorporate screened modified gravity and full baryonic physics can be computationally expensive, necessitating the use of lower mass resolutions so that the calculations can involve fewer particles. However, this means that the gas density field is smoothed, resulting in high-density peaks being lost and consequently an overall reduction in star formation. This can result in poor agreement with observations of the stellar and gaseous properties of galaxies.

In this chapter, we have retuned the IllustrisTNG baryonic model so that it can be used to run full-physics simulations at a much lower resolution while still retaining a high level of agreement with galaxy observations. Calibrated using runs with a box size of $68h^{-1}{\rm Mpc}$ and, initially, $256^3$ gas cells, our model uses updated values for the following parameters (Sec.~\ref{sec:methods_fine_tuning:fine_tuning}): the threshold gas density for star formation, $\rho_{\star}$, is reduced from $\approx0.1{\rm cm^{-3}}$ to $0.08{\rm cm^{-3}}$; the parameter $\bar{e}_{\rm w}$ which controls the energy released by the stellar-driven wind feedback is reduced from 3.6 to 0.5; and the black hole radiative efficiency $\epsilon_{\rm r}$ is increased from 0.2 to 0.22. In addition to these changes, we have also increased the gravitational softening to a factor 1/20 of the mean interparticle separation. By reducing the heating and blowing out of gas by feedback and two-body interactions, and lowering the threshold density of star formation, these changes boost the amount of star formation at our lowered resolution, resulting in good agreement with observations of galaxy properties including the stellar mass fraction, the stellar mass function, the SFRD and the gas mass fraction (Fig.~\ref{fig:L68_calibration}).

Using our retuned model, we have run GR and F5 simulations with a box size 301.75$h^{-1}{\rm Mpc}$ (Sec.~\ref{sec:methods_fine_tuning:L302_simulations}). The predictions of stellar and gaseous properties in both gravity models show a very good match with galaxy observations, particularly for group- and cluster-sized masses (Fig.~\ref{fig:L302_observables}), which shows that for F5 it is not necessary to further retune the baryonic parameters. Using these simulations, we have studied, for redshifts $0\leq z\leq1$ and masses $10^{13}M_{\odot}\leq M_{500}\lesssim10^{15}M_{\odot}$, the scaling relations between the cluster mass and four observable mass proxies (Sec.~\ref{sec:results_fine_tuning:scaling_relations}): the SZ Compton $Y$-parameter $Y_{\rm SZ}$ and its X-ray analogue $Y_{\rm X}$, the mass-weighted gas temperature $\bar{T}_{\rm gas}$, and the X-ray luminosity $L_{\rm X}$.

For the $Y_{\rm SZ}(M)$ and $Y_{\rm X}(M)$ relations, our mapping between the F5 and GR relations, which involves dividing the F5 $Y$-parameter by the ratio of the dynamical mass to the true mass, is accurate to within $\sim12\%$ for galaxy groups and just a few percent for galaxy clusters. This validates our method for accounting for the effect of the fifth force on the $Y_{\rm SZ}(M)$ relation, which is currently used in our $f(R)$ constraint pipeline (Chapter \ref{chapter:scaling_relations}). For the $\bar{T}_{\rm gas}(M)$ relation, the same rescaling is again reasonable, with $\lesssim7\%$ accuracy for the full range of masses. Our rescaling does not work as well for the $L_{\rm X}(M)$ relation, which is likely due to the greater susceptibility of the X-ray luminosity to feedback processes. 

We have also shown (Sec.~\ref{sec:results_fine_tuning:observable_relations}) that the $Y_{\rm X}$-temperature and $L_{\rm X}$-temperature scaling relations can differ in F5 and GR by up to $45\%$. These relations could potentially be used for large-scale tests of gravity that do not involve measuring the cluster mass, and hence not only eliminating one potential source of uncertainty but also including additional information in the model constraints.

By running large-box full-physics simulations for a range of $f(R)$ gravity field strengths, it will be possible to test our models for the enhancements of the dynamical mass (Chapter~\ref{chapter:mdyn}) and the halo concentration (Chapter~\ref{chapter:concentration}) in the presence of full baryonic physics over a wide mass range. Our baryonic model can also potentially be used to run large full-physics simulations for other classes of modified gravity and dark energy models, e.g., the nDGP model using the MG solvers implemented in the \textsc{arepo} code, since it is likely that a recalibration of the baryonic parameters will not be necessary unless the model studied is extreme and differs strongly from the current best-fit $\Lambda$CDM (but in that case the model is likely to have already been ruled out by other observations). The application to the nDGP model will make it possible to validate our models for the nDGP enhancements of the halo concentration and the HMF in addition to extending our results for the observable-mass scaling relations to higher masses (Chapter~\ref{chapter:DGP_clusters}). Finally, in our study of the thermal and kinetic SZ angular power spectra in $f(R)$ gravity and nDGP (Chapter~\ref{chapter:sz_power_spectrum}), we were unable to study larger angular scales ($l\lesssim500$), again due to the relatively small box size of the \textsc{shybone} simulations: this can potentially be rectified by using these larger simulations. These possibilities will be explored in future works.

The ability to run large realistic galaxy and cluster formation simulations for beyond-$\Lambda$CDM models will prove highly beneficial for research in this field: not only will this endow us with numerical tools to predict observables, such as cluster properties, that cannot be studied using DMO simulations, but the hydrodynamical simulations enabled by such a tool can be used to quantify the impacts of baryons on various other observables, such as weak lensing and galaxy clustering. The lack of such a quantitative assessment would either restrict the amount of data that can be reliably used in model tests, or lead to biased constraints on models and parameters. 
\graphicspath{{./gfx/}}

\chapter{Conclusions and outlook}
\label{chapter:conclusions}

\section{Thesis summary}

\subsection{Modelling the dynamical mass enhancement in \texorpdfstring{$f(R)$}{f(R)} gravity}

In Chapter \ref{chapter:mdyn}, we found a simple model to describe the relationship between the dynamical mass and lensing mass of dark matter haloes in the Hu-Sawicki $f(R)$ model. As shown by the solid line fits of Fig.~\ref{fig:unweighted_matrix}, the $\tanh$ fitting formula of Eq.~(\ref{eq:mdyn_enhancement}) has generally shown excellent agreement with \textsc{ahf} halo data, for $z<1$, from three \textsc{ecosmog} DMO simulations, which are summarised in Table \ref{table:simulations}. By taking advantage of the variety of resolutions offered by these simulations, and using $\Lambda$CDM simulations to produce approximate data for field strengths not covered by the $f(R)$ gravity simulations, the validity of Eq.~(\ref{eq:mdyn_enhancement}) has been probed vigorously across a wide and continuous range of field values that cover $10^{-6.5}<|f_{R0}|<10^{-4}$ within $z<1$.

In addition to this, we used a simple thin-shell model (Sec.~\ref{thin_shell_modelling}) to predict the behaviours of free parameters $p_1$ and $p_2$ in Eq.~(\ref{eq:mdyn_enhancement}), which characterise the inverse width and the central logarithmic mass of the tanh-like transition respectively. The predictions, which neglect the effects of environmental screening due to nearby dark matter haloes, are given by Eqs.~(\ref{p_1}, \ref{p_2}). Using a stringent criterion to exclude unreliable snapshots in the fitting, the result for $p_2$, shown in Fig.~\ref{fig:unweighted_p_2}, is $p_2=(1.503\pm0.006)\log_{10}\left(\frac{|\bar{f}_R|}{1+z}\right)+(21.64\pm0.03)$. The slope value of $1.503\pm0.006$ shows excellent agreement with the prediction of $1.5$ by Eq.~(\ref{p_2}), and the data of Fig.~\ref{fig:unweighted_p_2} shows a clear linear trend as predicted. As shown by Fig.~\ref{fig:unweighted_p_1}, the $p_1$ data is more scattered, but given the size of the one standard deviation error bars, the constant trend predicted by Eq.~(\ref{p_1}) is not unreasonable, resulting in $p_1=(2.21\pm0.01)$. As shown by the dashed line fits of Fig.~\ref{fig:unweighted_matrix}, these results for $p_1$ and $p_2$ show good agreement with the simulation data across the full range of field values and redshifts. We also repeated the analysis using a different approach to utilise the errors in the simulation data, and the results, shown in Appendix \ref{appendix:mdyn:weighted_fitting}, also agree with the thin-shell model prediction very well. In Appendix \ref{appendix:mdyn:consistency} we further argue that the results in this chapter apply to models with different cosmological parameters such as $\sigma_8$ and $\Omega_{\rm M}$.

A generic fitting function for the relationship between the dynamical and lensing masses of dark matter haloes is an essential ingredient of the new framework, proposed in this thesis, to carry out cosmological tests of gravity in an unbiased way. Taking Eq.~(\ref{Li and He}) as an example, our general formula for the dynamical mass enhancement allows us to incorporate this particular effect of $f(R)$ gravity into galaxy cluster scaling relations in a self-consistent way. A key benefit of a fitting function is that it allows a continuous search through the model parameter space without having to run full simulations for every parameter point sampled in MCMC. The results will also be useful for other cluster tests of gravity that employ the difference between dynamical and lensing masses, such as by comparing cluster dynamical and lensing mass profiles, or by looking at measured cluster gas fractions.

The results presented in this chapter indicate that a simple model sometimes works surprisingly well despite the greatly simplified treatment of the complicated nonlinear physics of (modified) gravity. This has become a common theme in this thesis, and fitting simple models to the predictions from numerical simulations is an approach that we have repeated when studying the halo concentration in $f(R)$ gravity and the cluster and halo properties in nDGP.

Although we made a very specific choice of $f(R)$ gravity in this chapter, the theoretical model and the procedure we followed to calibrate it are expected to be applicable to general chameleon gravity theories \citep[e.g.,][]{Gronke:2015,Gronke:2016}. As discussed briefly in Appendix \ref{appendix:mdyn:consistency}, in other $f(R)$ models the transition between screened and unscreened regimes can be different from the \citet{Hu:2007nk} model with $n=1$, which may cause the exact fitted values of $p_i$ to differ from what we presented in the above. Therefore, other $f(R)$ models may require a re-calibration based on simulations. However, given that all $f(R)$ models are phenomenological, it is perhaps more sensible to focus on a representative example, such as that by \citet{Hu:2007nk}, to make precise observational constraints. The pipeline and methodology can then be applied to any other models following general parameterisation schemes \citep[e.g.,][]{Brax:2012gr,Brax:2011aw,Lombriser:2016zfz}, which are useful for capturing the essential features of large classes of models using a few parameters. Should a preferred one emerge, the conclusion for the Hu-Sawicki model can serve as a rough guideline as to what level future cluster observations can constrain scalar-tensor-type screened theories. For this reason we decided not to explore other forms of $f(R)$ in this thesis.

\subsection{Universal model for the halo concentration in \texorpdfstring{$f(R)$}{f(R)} gravity}

In Chapter \ref{chapter:concentration}, we calibrated a model for the enhancement of the halo concentration in HS $f(R)$ gravity with $n=1$ using a suite of simulations that are summarised by Table~\ref{table:simulations}. The model is shown in Fig.~\ref{fig:skewtanh_fit}, and is given by Eq.~(\ref{eq:skewtanh}) with the parameter values listed in Table \ref{table:fitting}. It has been defined in terms of a useful rescaling of the halo mass, $M_{500}/10^{p_2}$, such that the data from three different $f(R)$ gravity models can satisfy a universal description. These models have $\log_{10}(|f_{R0}|)=(-4,-5,-6)$, and the fitting was carried out using data from all simulation snapshots with $\log_{10}\left(|\bar{f}_R|/(1+z)\right)\leq-4.5$. This universal description was shown to have very good agreement with simulations for $M_{500}/10^{p_2}$ covering nearly 7 orders of magnitude, and covering five decades of the halo mass.

Our model has been tested by comparing its predictions of the enhancement of the concentration with an arbitrarily chosen set of snapshots from our simulations, as shown by the lines plotted in Figs.~\ref{fig:arepo_matrix} and \ref{fig:ecosmog_matrix}. These predictions show excellent agreement with the data for all snapshots, apart from the Crystal snapshots with $\log_{10}\left(|\bar{f}_R|/(1+z)\right)>-4.5$. This is not surprising given that this data was not used in the fit of the model. Having a general model that works for $\log_{10}\left(|\bar{f}_R|/(1+z)\right)\leq-4.5$ will prove very useful, particularly given that an analytical theoretical modelling was not available.

The data of Fig.~\ref{fig:skewtanh_fit} shows that in the unscreened regime the enhancement of the concentration reaches a distinct peak as a function of the halo mass, but drops to negative values at lower mass, where the $f(R)$ concentration is less than the GR concentration. As shown by Fig.~\ref{fig:stacked_profiles}, such negative enhancement occurs because the innermost regions of the haloes are less dense in $f(R)$ gravity than in GR. This could be caused by the velocity gained by particles in haloes, which makes it difficult for them to settle into orbits at the central regions of the halo. Meanwhile in the screened regime of Fig.~\ref{fig:skewtanh_fit} there is a small dip in the concentration. Fig.~\ref{fig:stacked_profiles} suggests that this is caused by the halo being only partially screened, so that outer particles are moved further towards the centre of the halo while the inner regions remain screened. The density profile is therefore unaffected at the innermost regions but is greater at intermediate radii. Therefore the scale radius becomes greater, and fitting an NFW profile would then result in an estimate for the concentration that is lower in $f(R)$ gravity than in GR. All of these effects are incorporated by the fitted model of Eq.~(\ref{eq:skewtanh}).

We also carried out some further investigations which can be useful for future studies of the concentration in $f(R)$ gravity, and in other similar modified gravity theories. Firstly, in addition to applying a direct NFW profile fitting to each of the haloes to measure the concentration, two simplified approaches were also used, namely the methods that are used by \cite{Prada:2011jf} and \cite{Springel:2008cc}. The resulting enhancement of the concentration from using these two methods (shown in Fig.~\ref{fig:3_panel}) shows a difference from direct NFW fitting. This is due to the effects of $f(R)$ gravity on the internal density profile, which means that the choice of regions of the halo to use in measuring the concentration becomes important. The method used by \cite{Springel:2008cc} only requires the mass enclosed by the orbital radius with the maximum circular velocity. Being found at the inner regions of a halo, which become more dense as the halo becomes unscreened, this results in the concentration being overestimated by up to 26\%. From this, we conclude that only the direct NFW fitting should be used in $f(R)$ studies. Secondly, we looked at the validity of the NFW profile fitting in $f(R)$ gravity and found that, as shown by the bottom-left panel of Fig.~\ref{fig:4_panel}, for most haloes the $\chi^2$ measure for the fit is within 20\% of the GR measure, and for some haloes the fit is even better. Therefore the systematic effects caused by fitting the NFW profile in $f(R)$ gravity are unlikely to have a significant effect on the scatter of the concentration measure.

\subsection{Observable-mass scaling relations in \texorpdfstring{$f(R)$}{f(R)} gravity}

In Chapter \ref{chapter:scaling_relations}, we made use of the first full-physics simulations that have been run for both GR and $f(R)$ gravity (along with non-radiative counterparts), to study the effects of the fifth force of $f(R)$ gravity on the scaling relations between the cluster mass and four observable proxies: the gas temperature (Fig.~\ref{fig:T_gas}), the $Y_{\rm SZ}$ and $Y_{\rm X}$ parameters (Figs.~\ref{fig:Ysz_scaling_relation} and \ref{fig:Yx_scaling_relation}) and the X-ray luminosity (Fig.~\ref{fig:Lx_scaling_relation}). To understand these effects in greater detail, we have also examined the effects of both $f(R)$ gravity and full-physics on the gas density and temperature profiles (see Fig.~\ref{fig:profiles}). In doing so, we have been able to test two methods for mapping between scaling relations in $f(R)$ gravity and GR.

The first method was proposed by \citet{He:2015mva}. This proposes a set of mappings, given by Eqs.~(\ref{eq:temp_equiv_eff}) and (\ref{eq:ysz_mapping})-(\ref{eq:lx_mapping}), that can be applied to haloes whose mass and radius are measured using the effective density field (see Sec.~\ref{sec:eff_approach}). A second, new, approach is proposed in Sec.~\ref{sec:true_approach}, and predicts another set of mappings, given by Eqs.~(\ref{eq:temp_equiv_true}) and (\ref{eq:ysz_mapping_true})-(\ref{eq:lx_mapping_true}), that can be applied to haloes whose mass and radius are measured using the true density field. Both sets of mappings involve simple rescalings that depend only on the ratio of the dynamical mass to the true mass in $f(R)$ gravity. As shown by Figs.~\ref{fig:mdyn_mtrue} and \ref{fig:Yx_scaling_relation_TANH}, even with the inclusion of full-physics processes this ratio can be computed with high accuracy using our analytical tanh formula, which is given by Eq.~(\ref{eq:mdyn_enhancement}).

For the mass-weighted gas temperature and the $Y_{\rm SZ}$ and $Y_{\rm X}$ observables, we found that the F6 and F5 scaling relations, with appropriate rescaling applied (using either method discussed above), match the GR relations to within a few percent for the full mass-range tested for the non-radiative simulations. With the inclusion of full-physics effects such as feedbacks, star formation and cooling, the rescaled $Y_{\rm SZ}$ and $Y_{\rm X}$ scaling relations continue to show excellent agreement with GR for mass $M_{500}\gtrsim10^{13.5}M_{\odot}$, which includes group- and cluster-sized objects. These proxies also show relatively low scatter as a function of the cluster mass, compared with other observables. $Y_{\rm SZ}$ and $Y_{\rm X}$ are therefore likely to be suitable for accurate determination of the cluster mass in tests of $f(R)$ gravity. The mappings for the gas temperature show a very high accuracy for lower-mass objects, but show a small $\lesssim5\%$ offset between F5 and GR for higher-mass objects.

The mappings do not work as well for the X-ray luminosity $L_{\rm X}$, for which the F5 relations after rescaling are typically enhanced by $\sim30\%$ compared with GR. This is caused by the unique dependency of $L_{\rm X}$ on the gas density to power two, and the gas temperature to power half, which means that the inner halo regions have a greater contribution than for the other proxies and the competing effects of feedback on the temperature and gas density profiles are less likely to cancel out. This issue, in addition to the fact that $L_{\rm X}$ has a highly scattered correlation with the cluster mass, means that this proxy is unlikely to be suitable for cluster mass determination in tests of $f(R)$ gravity.

We also considered the $Y_{\rm X}$-$\bar{T}_{\rm gas}$ scaling relation (Fig.~\ref{fig:Yx-T_scaling_relation}), and found that this is suppressed by $30$-$40\%$ in the F5 model relative to GR. This offers a potential new and useful test of gravity with group- and cluster-sized objects which avoids the systematic uncertainties incurred from mass calibration.

Our results also provide insights into the viability of extending cluster tests of gravity to the group-mass regime. An advantage of using lower-mass objects is that these objects can be unscreened (or partially screened) even for weaker $f(R)$ models, offering the potential for tighter constraints using data from ongoing and upcoming SZ and X-ray surveys \citep[e.g.,][]{erosita,Planck_SZ_cluster} which are now entering this regime. On the other hand, as we have seen above, the scatter induced by feedback mechanisms becomes more significant in group-sized haloes, which means that additional work will need to be conducted to characterise this effect and to understand its impact on model tests.

Finally, we note that our parameter $p_2$, which is used to compute the ratio of the dynamical mass to the true mass, depends only on the quantity $|f_R|/(1+z)$, and not on the model parameters $n$ and $f_{R0}$ of HS $f(R)$ gravity. This dependence was originally derived by using the thin-shell model (Chapter \ref{chapter:mdyn}), which does not depend on the details of the $f(R)$ model. We therefore expect our scaling relation mappings to perform similarly for any combination of the HS $f(R)$ parameters, and potentially other chameleon-type or thin-shell-screened models. However, due to the high computational cost of running full-physics simulations of $f(R)$ gravity and other models, we do not seek to confirm this conjecture in this thesis.

\subsection{Constraint pipeline for unbiased \texorpdfstring{$f(R)$}{f(R)} cluster constraints}

In Chapter \ref{chapter:constraint_pipeline}, we combined all of our models for the effects of $f(R)$ gravity on cluster properties into an MCMC pipeline for constraining the amplitude of the present-day background scalar field, $|f_{R0}|$, using cluster number counts. We have adopted the model from \citet{Cataneo:2016iav} for the $f(R)$ enhancement of the HMF, and used this, along with our model for the enhancement of the halo concentration, to produce a model-dependent prediction of the cluster number counts (Sec.~\ref{sec:methods_pipeline:hmf}). We have also used our model for the enhancement of the dynamical mass in $f(R)$ gravity to convert a GR power-law observable-mass scaling relation, which is based on the Planck $Y_{\rm SZ}(M_{500})$ relation \citep{Planck_SZ_cluster}, into a form consistent with $f(R)$ gravity, where the fifth force enhances the relation at sufficiently low masses (Sec.~\ref{sec:methods_pipeline:scaling_relation}). These models are all incorporated in our log-likelihood (Sec.~\ref{sec:methods_pipeline:likelihood}), which we have used to infer parameter constraints using a set of mock cluster catalogues (Sec.~\ref{sec:methods_pipeline:mock}).

Using a combination of GR and F5 mocks, we have shown that our pipeline is able to give reasonable parameter constraints that are consistent with the fiducial cosmology (Figs.~\ref{fig:gr_pipeline} and \ref{fig:full_fr_pipeline}). For the GR mock, the constraints conclusively rule out $f(R)$ models with $\log_{10}|f_{R0}|\gtrsim-5$ and favour values in the range $-7\leq\log_{10}|f_{R0}|\lesssim-5$ where $-7$ is the lowest value considered by our MCMC sampling. Meanwhile, the constraints inferred using the F5 mock favour values close to the fiducial value of $-5$, with 68\% range $-5.1^{+0.3}_{-1.0}$ and a `most likely' value of $-4.92$. We have also shown that the constraints inferred from both mocks can be imprecise and biased if the $f(R)$ enhancement of the scaling relation is not accounted for (Fig.~\ref{fig:gr_sr}). Therefore, this should be properly modelled in future tests of $f(R)$ gravity in order to prevent biased constraints. This will become particularly relevant as cluster catalogues start to enter the galaxy group regime \citep[e.g.,][]{Pillepich:2018sin,2021Univ....7..139L}, where more objects can be unscreened in $f(R)$ gravity.

Throughout Chapter \ref{chapter:constraint_pipeline}, the main obstacle to precise and unbiased constraints stemmed from degeneracies between $f_{R0}$, $\Omega_{\rm M}$, $\sigma_8$ and the scaling relation parameters $\alpha$ and $\beta$, all of which can influence the predicted cluster count. We have shown that the degeneracies can be prevented by using a tighter Gaussian prior for $\Omega_{\rm M}$ and by having better knowledge of the scaling relation parameters (Fig.~\ref{fig:corner_tight_priors}). The latter can potentially be achieved by including lensing data for a subset of the clusters. If wide or flat parameter priors are used, this may give rise to biased constraints of $\log_{10}|f_{R0}|$. For example, we have found that the parameter degeneracies can have a more significant effect for cluster samples that extend to lower masses (Sec.~\ref{sec:bias:sample:ysz_cut}).

\subsection{Cluster and halo properties in nDGP}

In Chapter \ref{chapter:DGP_clusters}, we extended our framework to the popular nDGP model, in which a fifth force is able to act over sufficiently large scales. 

Using the first cosmological simulations that simultaneously incorporate full baryonic physics and the nDGP model, we studied the observable-mass scaling relations for the same three mass proxies as studied in Chapter \ref{chapter:scaling_relations} (see Sec.~\ref{sec:results:dgp:scaling_relations}). For groups and clusters in the mass range $M_{500}\lesssim10^{14.5}M_{\odot}$, our results show that for the N1 model, the $\bar{T}_{\rm gas}(M)$ relation is enhanced by about 5\% with respect to GR, while the $Y_{\rm SZ}(M)$ and $Y_{\rm X}(M)$ relations are both enhanced by 10\%-15\% at low masses but more closely match the GR relations at high masses. For N5, which is much weaker than N1, the $\bar{T}_{\rm gas}(M)$ relation closely resembles the GR relation, while the $Y_{\rm SZ}(M)$ and $Y_{\rm X}(M)$ relations are enhanced by up to 5\% at low mass and suppressed by up to 5\% at high mass. These deviations from GR could be related to the effect of the fifth force on gas velocities during cluster formation, and they also hint at an interplay between the fifth force and stellar and black hole feedback.

Using a suite of DMO $N$-body simulations, which cover a wide range of resolutions and box sizes, we found that, in nDGP, the concentration is typically suppressed relative to GR, varying from a few percent in N5 to up to $\sim15\%$ in N0.5 (see Sec.~\ref{sec:results:dgp:concentration}). Using stacked density profiles at different mass bins, we have shown that this behaviour is caused by a reduced (increased) density at the inner (outer) halo regions. Including full baryonic physics significantly affects the concentration-mass relation; however, our results show that, for masses $M_{200}\lesssim10^{13}h^{-1}M_{\odot}$, the model differences between nDGP and GR still have a similar magnitude compared to the DMO simulations.

By combining the data from our $z\leq1$ simulation snapshots, we calibrated a general model, given by Eq.~(\ref{eq:c_model}), which is able to accurately predict the suppression of the halo concentration with respect to the GR results as a function of the halo mass and the $H_0r_{\rm c}$ parameter of nDGP over ranges $10^{12}h^{-1}M_{\odot}\lesssim M_{200}\lesssim 10^{15}h^{-1}M_{\odot}$ and 0.5-5, respectively. This model can be included in our MCMC pipeline for converting between mass definitions in case, for example, the theoretical predictions and observables are defined with respect to different spherical overdensities. Our model can also be used, along with the HMF, to predict the nonlinear matter power spectrum, which can also be used to constrain gravity.

We also used our DMO simulations to study the HMF over the mass range $1.52\times10^{10}h^{-1}M_{\odot}\leq M_{500}\lesssim 10^{15}h^{-1}M_{\odot}$ at redshifts 0, 1 and 2 (see Sec.~\ref{sec:results:dgp:hmf}). Our results (Fig.~\ref{fig:hmf_combined}), indicate that the nDGP HMF is enhanced at high masses (by up to $\sim60\%$ in N0.5) and suppressed at low masses (by $\sim10\%$ in N0.5) compared to GR. These results indicate the potential constraining power from using the observed mass function to probe the $H_0r_{\rm c}$ parameter of nDGP. By combining the data from our $z\leq2$ snapshots, we have calibrated a general model, given by Eq.~(\ref{eq:hmf}), which can accurately reproduce the HMF enhancement as a function of the halo mass, redshift and $H_0r_{\rm c}$ parameter. This model can be used for theoretical predictions of the nDGP HMF (using a parameter-dependent GR calibration) in our MCMC pipeline.

In Chapter \ref{chapter:scaling_relations}, we showed that a model for the $f(R)$ dynamical mass enhancement can be used to predict observable-mass scaling relations in $f(R)$ gravity using their GR counterparts. Such a model in nDGP could similarly be useful to help understand the enhancements of the temperature and SZ and X-ray $Y$-parameters observed in Chapter \ref{chapter:DGP_clusters}. For now, though, we note that the scaling relations in nDGP still appear to follow power-law relations as a function of the mass: the $\bar{T}_{\rm gas}(M)$ relation in N1 can be related to the GR relation by a simple rescaling of the amplitude, whereas the $Y_{\rm SZ}(M)$ and $Y_{\rm X}(M)$ relations appear to have shallower slopes in N5 and N1 than in GR. Therefore, in our future MCMC pipeline for obtaining constraints of nDGP, we can still assume the GR power-law form of the scaling relations by allowing the parameters controlling the amplitude and slope to vary along with the cosmological and nDGP parameters \citep[e.g.,][]{deHaan:2016qvy,Bocquet:2018ukq}. 

Although our simulations have only been run for a single choice of cosmological parameters, we expect that our models for the enhancements of the halo concentration and HMF will have a reasonable accuracy for other (not too exotic) parameter values. The gravitational force enhancement in nDGP, given by $\left[1 + 1/(3\beta)\right]$, has only a weak dependence on $\Omega_{\rm M}$: for the N1 model ($\Omega_{\rm rc}=0.25$), the force enhancement varies within a very small range (roughly $12.1\%-12.6\%$) for $\Omega_{\rm M} \in [0.25,0.35]$ at the present day, and the range of variation is even smaller at higher redshifts. Therefore, for now we assume that the effects of the cosmological parameters on the concentration and HMF are approximately cancelled out in the ratios $\Delta c/c_{\rm GR}$ and $\Delta n/n_{\rm GR}$. However, we will revisit this in a future work, using a large number of nDGP simulations that are currently being run for different combinations of cosmological parameters, before these models are used in tests of gravity using observational data.

\subsection{Sunyaev-Zel'dovich effect in \texorpdfstring{$f(R)$}{f(R)} gravity and nDGP}

Over the past couple of decades, great advances have been made in the measurement of the secondary anisotropies of the CMB caused by the SZ effect, including its thermal component and even its much smaller kinematic component. The angular power spectrum of the tSZ effect has been increasingly adopted as a probe of cosmological parameters that influence the growth of large-scale structures. Also, as observations of the kSZ power spectrum continue to improve, the latter has been identified as another potentially powerful probe of cosmology. The next generation of ground-based observatories \citep{Ade:2018sbj,Abazajian:2016yjj} look set to revolutionise the constraining power of these probes.

In Chapter \ref{chapter:sz_power_spectrum}, we used the \textsc{shybone} simulations to look at the viability of using the angular power spectra of the tSZ and kSZ effects as large-scale probes of HS $f(R)$ gravity and nDGP. We generated mock maps of the tSZ and kSZ signals (Sec.~\ref{sec:methods:sz:maps}), and used these maps to measure the angular power spectra. Our results (Figs.~\ref{fig:power_spectra} and \ref{fig:nDGP_power_spectra}) indicate that the fifth force, present in $f(R)$ gravity and nDGP, and the subgrid baryonic physics have different effects on the tSZ and kSZ power spectra. The former enhances the power on all scales probed by our maps ($500\lesssim{l}\lesssim8\times10^4$) by boosting the abundance and peculiar velocity of large-scale structures (e.g., dark matter haloes and free electrons inside them), while the latter brings about a suppression on scales $l\gtrsim3000$ for the tSZ effect and on all tested scales for the kSZ effect. Even with both of these effects present, we find that the power can be significantly enhanced in $f(R)$ gravity and nDGP: by up to $60\%$ for the tSZ effect and $35\%$ for the kSZ effect for the F5 and N1 models; and by $5\%$-$10\%$ for F6 and N5, which correspond to relatively weak modifications of GR. 

In addition, we computed the power spectrum of the transverse component of the electron momentum field (Sec.~\ref{sec:results:sz:transverse_momentum}), which is closely related to the kSZ angular power spectrum. In particular, we showed (Fig.~\ref{fig:ksz_contribution}) that at angular sizes $l\geq600$ the kSZ signal is dominantly contributed by $k$-modes in the transverse-momentum power spectrum which are in the non-linear regime, and which are affected strongly by MG. The $k$-modes in the linear regime may contribute more to smaller $l$, but at least for $f(R)$ gravity the impact of MG at those $l$ values will be much less significant due to the finite range of the fifth force, as we can already see in Fig.~\ref{fig:power_spectra}.

We found that the relative difference between the MG models and GR is significantly affected by the additional baryonic processes that act in the full-physics simulations. Given that these processes are still relatively less well-constrained, this adds to the uncertainty in our theoretical predictions of the kSZ angular power spectra on small angular scales, e.g., $l>600$. Therefore, further work should be carried out using a range of full-physics parameters to precisely identify the scales on which constraints can be reliably made before the tSZ and kSZ power are used to probe $f(R)$ gravity and nDGP.

\subsection{Realistic simulations to study clusters in modified gravity}

Running large-box cosmological simulations which simultaneously incorporate screened modified gravity and full baryonic physics can be computationally expensive, necessitating the use of lower mass resolutions so that the calculations can involve fewer particles. However, this means that the gas density field is smoothed, resulting in high-density peaks being lost and consequently an overall reduction in star formation. This can result in poor agreement with observations of the stellar and gaseous properties of galaxies.

In Chapter \ref{chapter:baryonic_fine_tuning}, we retuned the IllustrisTNG baryonic model so that it can be used to run full-physics simulations at a much lower resolution while still retaining a high level of agreement with galaxy observations. Calibrated using runs with a box size of $68h^{-1}{\rm Mpc}$ and, initially, $256^3$ gas cells, our model uses updated values for the following parameters (Sec.~\ref{sec:methods_fine_tuning:fine_tuning}): the threshold gas density for star formation, $\rho_{\star}$, is reduced from $\approx0.1$ to 0.08; the parameter $\bar{e}_{\rm w}$ which controls the energy released by the stellar-driven wind feedback is reduced from 3.6 to 0.5; and the black hole radiative efficiency $\epsilon_{\rm r}$ is increased from 0.2 to 0.22. In addition to these changes, we have also increased the gravitational softening to a factor 1/20 of the mean interparticle separation. By reducing the heating and blowing out of gas by feedback and two-body interactions, and lowering the threshold density of star formation, these changes boost the amount of star formation at our lowered resolution, resulting in good agreement with observations of galaxy properties including the stellar mass fraction, the stellar mass function, the SFRD and the gas mass fraction (Fig.~\ref{fig:L68_calibration}).

Using our retuned model, we have run GR and F5 simulations with a box size 301.75 $h^{-1}{\rm Mpc}$ (Sec.~\ref{sec:methods_fine_tuning:L302_simulations}). The predictions of stellar and gaseous properties in both gravity models show a very good match with galaxy observations, particularly for group- and cluster-sized masses (Fig.~\ref{fig:L302_observables}), which shows that for F5 it is not necessary to further retune the baryonic parameters. Using these simulations, we have studied, for redshifts $0\leq z\leq1$ and masses $10^{13}M_{\odot}\leq M_{500}\lesssim10^{15}M_{\odot}$, the scaling relations between the cluster mass and four observable mass proxies (Sec.~\ref{sec:results_fine_tuning:scaling_relations}): the SZ Compton $Y$-parameter $Y_{\rm SZ}$ and its X-ray analogue $Y_{\rm X}$, the mass-weighted gas temperature $\bar{T}_{\rm gas}$, and the X-ray luminosity $L_{\rm X}$. 

For the $Y_{\rm SZ}(M)$ and $Y_{\rm X}(M)$ relations, our mapping between the F5 and GR relations, which involves dividing the F5 $Y$-parameter by the ratio of the dynamical mass to the true mass, is accurate to within $\sim12\%$ for galaxy groups and just a few percent for galaxy clusters. This validates our method for accounting for the effect of the fifth force on the $Y_{\rm SZ}(M)$ relation, which is currently used in our $f(R)$ constraint pipeline (Chapter \ref{chapter:scaling_relations}). For the $\bar{T}_{\rm gas}(M)$ relation, the same rescaling is again reasonable, with $\lesssim7\%$ accuracy for the full range of masses. Our rescaling does not work as well for the $L_{\rm X}(M)$ relation, which is likely due to the greater susceptibility of the X-ray luminosity to feedback processes. 

We have also shown (Sec.~\ref{sec:results_fine_tuning:observable_relations}) that the $Y_{\rm X}$-temperature and $L_{\rm X}$-temperature scaling relations can differ in F5 and GR by up to $45\%$. These relations could potentially be used for large-scale tests of gravity that do not involve measuring the cluster mass, and hence not only eliminating one potential source of uncertainty but also including additional information in the model constraints.

\section{Outlook and future work}

The results in this thesis show that the $p_2$ parameter, defined in Chapter \ref{chapter:mdyn}, can be very useful in the description and modelling of complicated effects in $f(R)$ gravity. In addition to its relatively simple one-parameter definition, it can allow the combining of data generated by simulations run for different cosmological parameters, as $p_2$ encapsulates the values of $\Omega_{\rm M}$ and $\Omega_{\Lambda}$. Indeed, the data for the concentration enhancement from \textsc{arepo} and Diamond F6 shows excellent agreement (see Fig.~\ref{fig:skewtanh_fit}), even though these two simulations were run for different cosmological parameters and using very different codes. It will be interesting to see where else $p_2$ can be used in $f(R)$ studies. Of particular interest would be to see how it can simplify the modelling of the HMF. The enhancement of the HMF in $f(R)$ gravity peaks at a particular halo mass which depends on the strength of the scalar field. A stronger scalar field allows higher-mass haloes to be unscreened, and therefore results in an enhancement of the HMF at a higher mass. At the very least, the mass of the peak enhancement of the HMF can be expected to be strongly correlated to $p_2$. The enhancement of the matter power spectrum could also be investigated via a similar treatment. 

Our $f(R)$ constraint pipeline (Chapter \ref{chapter:constraint_pipeline}) can be improved in a couple of ways. First, while the HMF model of \citet{Cataneo:2016iav} is accurate, it only covers the redshift range $[0,0.5]$. An extended model that works for a larger redshift range, as well as for wider ranges of other cosmological parameters (not restricted to the $\Omega_{\rm M}$ and $\sigma_8$ parameters as we have focused on here), would be very useful. Calibrating this model for spherical overdensity $\Delta=500$ would also mean that conversions between the halo mass definitions $M_{500}$ and $M_{\rm 300m}$ would no longer be required. Second, the MCMC pipeline should be extended so that independent cluster data, such as weak lensing, can be included in the model constraint. This can improve the precision of the mass calibration and potentially prevent degeneracies between the MG and scaling relation parameters. Once these tasks are completed, this pipeline can be used to constrain $f(R)$ gravity using observations. It is also straightforward to extend our framework to other gravity models; as discussed, we have already started to do this for the nDGP model.

Applying the pipeline to real survey data will involve additional considerations. For example, our mock cluster samples were generated assuming a fixed cut in the observable and a constant fractional measurement uncertainty. Using real survey data will require more accurate modellings, including a selection function which describes the probability of a cluster with a given redshift and flux being selected. Therefore, tests should be carried out using mocks with more realistic selection criteria to confirm that the pipeline can infer stable and reliable constraints. It will be interesting to apply our pipeline to previous data (for example, the Planck SZ cluster sample), to check whether our detailed modellings of the dynamical mass, concentration and thermal gas properties significantly impact the final constraints. If so, this would indicate that previous constraints inferred using similar samples may have been subject to biases. Another interesting check will be to infer constraints using both X-ray and SZ selected samples, to check if one can give more powerful constraints than the other (for example, SZ samples are known to have nearly redshift-independent selection functions). Our pipeline can then be applied to much larger catalogues from ongoing surveys, such as the SZ AdvACT cluster survey and the X-ray eROSITA survey, and as mentioned above it will be helpful to make use of weak lensing data from additional surveys like DES and Euclid.

Our models for the enhancements of the dynamical mass and halo concentration in $f(R)$ gravity are important components of the constraint pipeline. However, these have only been calibrated using DMO simulations. It would therefore be useful to carefully verify these results using cluster data from full-physics hydrodynamical simulations run for $f(R)$ gravity. We have already carried out preliminary checks of the dynamical mass model for masses $M_{500}\lesssim10^{14.5}M_{\odot}$ using the full-physics \textsc{shybone} simulations (Fig.~\ref{fig:mdyn_mtrue}). With the baryonic model presented in Chapter \ref{chapter:baryonic_fine_tuning}, which can be used to run much larger full-physics simulations of screened modified gravity, it will now be possible to carry out these checks with many more haloes spanning masses $10^{13}M_{\odot}\lesssim M_{500}\lesssim10^{15}M_{\odot}$. 

On a similar note, our baryonic model can also be used to run full-physics simulations of nDGP for boxes that are much larger than the current $62h^{-1}{\rm Mpc}$ runs. These could be used for a robust study of the observable-mass scaling relations in nDGP up to $M_{500}\sim10^{15}M_{\odot}$, and they could also be used to check the accuracy of our general models for the enhancements of the concentration and the HMF (which were calibrated using DMO simulations) in the presence of full baryonic physics. 

Finally, in our study of the tSZ and kSZ angular power spectra in $f(R)$ gravity and nDGP, we were unable to study larger angular scales ($l\lesssim500$), again due to the relatively small box size of the \textsc{shybone} simulations. This can potentially be rectified by using the larger simulations discussed above. Our results in Chapter \ref{chapter:sz_power_spectrum} indicate that the gravity model differences are less sensitive to baryonic physics at larger scales. The precise scales at which the SZ power spectra can be used to reliably probe gravity should be confirmed by using a combination of full-physics and non-radiative simulations, as well as a range of baryonic physics parameters. This will also help to better-assess the potential constraining power from current and upcoming CMB experiments.
\graphicspath{{./gfx/}}

\appendix

\chapter{\boldmath Modelling the dynamical mass of haloes in \texorpdfstring{$f(R)$}{f(R)} gravity}
\label{appendix:mdyn}

\section{Weighted fitting of tanh curve}
\label{appendix:mdyn:weighted_fitting}

\begin{figure*}
\centering
\includegraphics[width=0.86\textwidth]{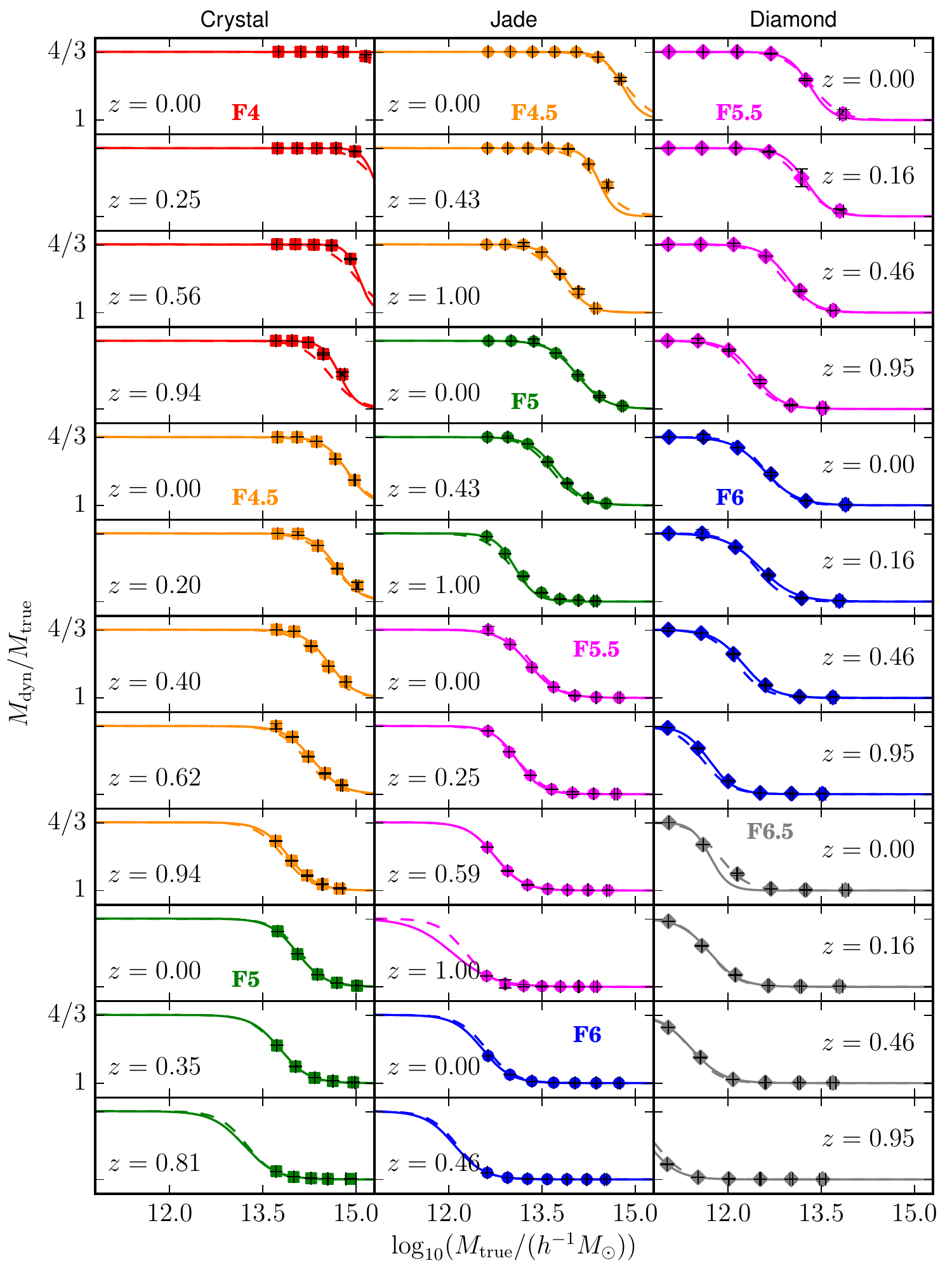}
\caption[Dynamical mass enhancement for haloes in HS $f(R)$ gravity with weighted least squares fits of Eq.~(\ref{eq:mdyn_enhancement}).]{Dynamical mass to lensing mass ratio as a function of the lensing mass for F4 (\textit{red}), F4.5 (\textit{orange}), F5 (\textit{green}), F5.5 (\textit{magenta}), F6 (\textit{blue}) and F6.5 (\textit{grey}) at various redshifts as annotated. The data has been generated using the Crystal (\textit{left column}), Jade (\textit{middle column}) and Diamond (\textit{right column}) modified \textsc{ecosmog} simulations (see Table \ref{table:simulations}). The data points, corresponding to mass bins represented by their median ratio and mean mass, and their one standard deviation error bars are produced using jackknife resampling. Jackknife errors less than $10^{-4}$ are replaced with half of the range between the 16th and 84th percentiles. \textit{Solid line}: Eq.~(\ref{eq:mdyn_enhancement}) with $p_1$ and $p_2$ determined by weighted least squares fitting for the given snapshot; \textit{Dashed line}: Eq.~(\ref{eq:mdyn_enhancement}) with best-fit constant $p_1$ result ($p_1=2.23$) and linear $p_2$ result (Eq.~(\ref{eq:weighted_p2_fit})) from Figs.~\ref{fig:p_1} and \ref{fig:p_2}, respectively.}
\label{fig:matrix}
\end{figure*}


In Sec.~\ref{measure_m_dyn}, we discussed and compared, for a few selected cases, two schemes to fit the $M_{\rm dyn}/M_{\rm true}$ mass ratio data using a $\tanh$ curve. We found that, although the weighted and unweighted fitting schemes give broadly consistent results, the latter scheme, by assuming that all data points have the same error, leads to fitted $\tanh$ curves that have better visual agreement with the data points. This is because in some snapshots the data for the median $M_{\rm dyn}/M_{\rm true}$ ratio has big disparities in the uncertainties because there are few high-mass haloes due to box size constraints, or because the ratio data in screened and unscreened regimes shows too little variation. The estimated median ratio values therein are not biased because of this, and so we presented our main results (see Sec.~\ref{results}) using the unweighted scheme. This gives all bins equal weight regardless of the large disparities in the uncertainty, allowing the fitted curve to more easily go through the data points. However, one could still argue that the strong variation of median $M_{\rm dyn}/M_{\rm true}$ ratio uncertainties in the different mass bins is at least partly physical (e.g., in the completely unscreened regime there is intrinsically little uncertainty in the ratio). Therefore here we present our results from using the weighted approach, which show that the choice of method does not have a significant effect on the final results, namely on the constant and linear fits of $p_1$ and $p_2$ respectively.

To check the reliability of the weighted fit across all redshifts, field strengths and simulations, Fig.~\ref{fig:matrix} has been produced, which is analogous to Fig.~\ref{fig:unweighted_matrix} and covers the same snapshots. The solid line {trends} are the weighted fits of the simulation data at the given snapshots, and in general these show very good agreement with the simulation data. However the disparities in the sizes of the error bars now have a stronger impact on the fit and significant deviation from the simulation data is observed for several snapshots, including the Crystal F4 $z=0.00$, the Jade F4.5 snapshots and Diamond F6.5 $z=0.00$. 

\begin{figure}
\centering
\includegraphics[width=0.8\textwidth]{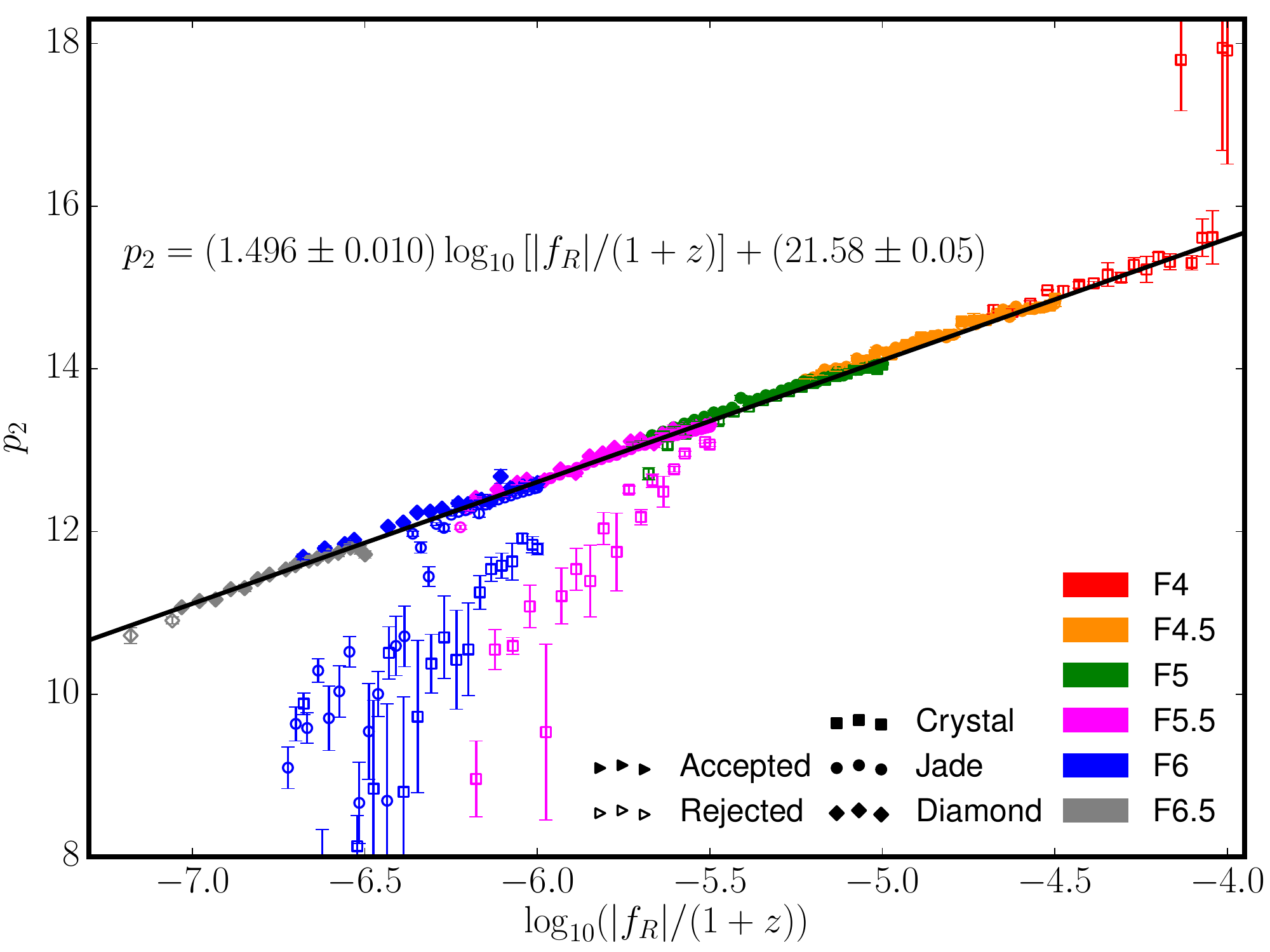}
\caption[Parameter $p_2$ of Eq.~(\ref{eq:mdyn_enhancement}), measured using weighted least squares fitting, as a function of $\bar{f}_R(z)/(1+z)$.]{Parameter $p_2$ in Eq.~(\ref{eq:mdyn_enhancement}) plotted as a function of the background scalar field at redshift $z$, $\bar{f}_R(z)$, divided by $(1+z)$, for several present day field strengths $f_{R0}$ (see legends) of Hu-Sawicki $f(R)$ gravity with $n=1$. $p_2$ is measured via a weighted least squares optimization of Eq.~(\ref{eq:mdyn_enhancement}) to data from modified \textsc{ecosmog} simulations, described by Table \ref{table:simulations}, at simulation snapshots with redshift $z<1$. $\bar{f}_R(z)$ is calculated for each snapshot using Eq.~(\ref{eq:fR_background}). The trend line has been produced via a weighted least squares linear fit, using the one standard deviation error bars, of the solid data points, which correspond to snapshots for which the mass bins contain at least half of the median mass ratio range 1 to 4/3. The hollow data does not meet this criterion, so is deemed unreliable and neglected from the fit, which is given by Eq.~(\ref{eq:weighted_p2_fit}).}
\label{fig:p_2}
\end{figure}

The results for $p_2$, produced through the weighted approach, are shown in Fig.~\ref{fig:p_2}. The lowest redshift snapshots of F4 are now observed to peel off from the linear trend due to the large disparities in the uncertainties of the mass bin data, as discussed in Sec.~\ref{measure_m_dyn} (see Fig.~\ref{fig:raw_fits}). The disparity in uncertainty in part results from the limited number of high-mass haloes which could be screened in F4; such massive haloes are very rare and the only way to resolve this issue is to have a simulation with a much larger box size. However, as is shown in Fig.~\ref{fig:unweighted_p_2} in Sec.~\ref{results}, using unweighted least squares to measure $p_2$ has the effect of smoothing out the F4 data for $p_2$, although this does not reduce the general scatter in F4. In general the data is more scattered across all models in Fig.~\ref{fig:p_2} than in Fig.~\ref{fig:unweighted_p_2}, although for F4 there is now a more even scatter, with the data showing better alignment with the trend line than for the unweighted case.

The criterion for the rejection of the measured $p_2$ values is the same as for the unweighted approach, and so the outliers for low-redshift F4 in Fig.~\ref{fig:p_2} do not affect the linear fit of this data. As can be seen from Fig.~\ref{fig:p_2}, all of the solid data points, which meet this criterion, lie along a clear linear trend, while the hollow data points of F5.5 and F6 are all observed to peel off from this trend in a similar manner to the data in Fig.~\ref{fig:unweighted_p_2}. The result of the linear fit, using the one standard deviation error bars, is:
\begin{equation}
    p_2=(1.496\pm0.010)\log_{10}\left(\frac{|\bar{f}_R|}{1+z}\right)+(21.58\pm0.05).
    \label{eq:weighted_p2_fit}
\end{equation}
Agreement of the slope with the theoretical prediction of $1.5$ from Eq.~(\ref{p_2}) is excellent. The best-fit linear parameters of $1.496\pm0.010$ and $21.58\pm0.05$ also show strong agreement with the linear fit of the unweighted results (see Fig.~\ref{fig:unweighted_p_2}), implying that the choice of whether to use weighted or unweighted least squares fitting of Eq.~(\ref{eq:mdyn_enhancement}) is not of particular importance as far as $p_2$ is concerned.

\begin{figure}
\centering
\includegraphics[width = 0.8\textwidth]{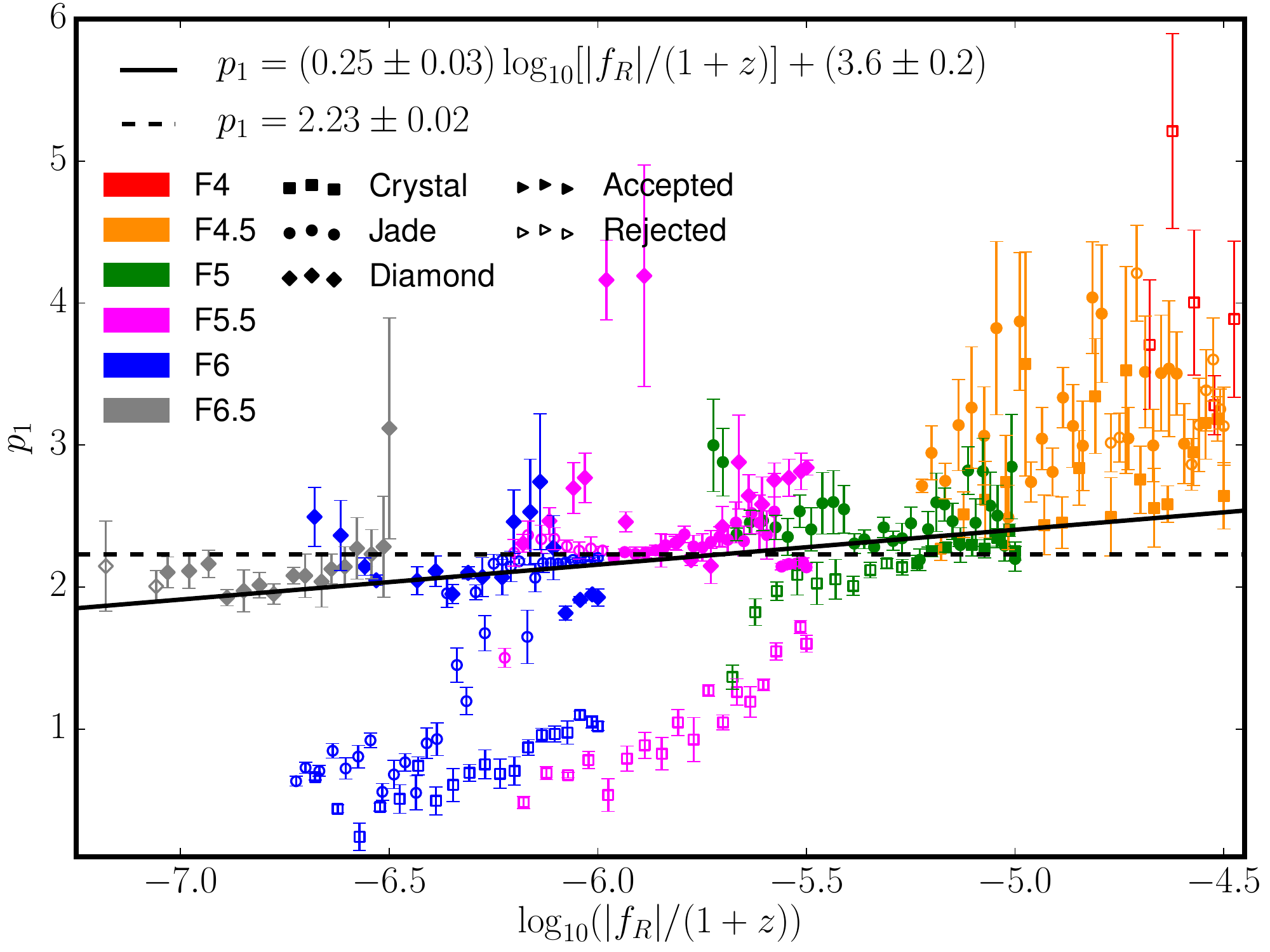}
\caption[Parameter $p_1$ of Eq.~(\ref{eq:mdyn_enhancement}), measured using weighted least squares fitting, as a function of $\bar{f}_R(z)/(1+z)$.]{Parameter $p_1$ in Eq.~(\ref{eq:mdyn_enhancement}) plotted as a function of the background scalar field at redshift $z$, $\bar{f}_R(z)$, divided by $(1+z)$, for several present day field strengths $f_{R0}$ (see legends) of Hu-Sawicki $f(R)$ gravity with $n=1$. $p_1$ is measured via a weighted least squares optimization of Eq.~(\ref{eq:mdyn_enhancement}) to data from modified \textsc{ecosmog} simulations, described by Table \ref{table:simulations}, at simulation snapshots with redshift $z<1$. $\bar{f}_R(z)$ is calculated for each snapshot using Eq.~(\ref{eq:fR_background}). Weighted least squares linear (\textit{solid line}) and constant (\textit{dashed line}) fits, using the one standard deviation error bars, of the solid data points, which correspond to snapshots for which the mass bins contain at least half of the median mass ratio range 1 to 4/3, are shown. The hollow data does not meet this criteria, so is deemed unreliable and neglected from the fits, which are given by Eq.~(\ref{eq:weighted_p1_fit}) and $p_1=(2.23\pm0.02)$ respectively.}
\label{fig:p_1}
\end{figure}

The results for $p_1$, measured via weighted least squares, are given in Fig.~\ref{fig:p_1}, which is plotted on the same axes range as Fig.~\ref{fig:unweighted_p_1}. Once again, the same selection criteria is used as for the unweighted least squares approach, and the hollow data points are left out of any fitting. The points are now significantly more scattered, and all models now contain notable outliers which include several of the solid data points. 

Taking F6.5 $z=0.00$ as an example, we can clearly see from Fig.~\ref{fig:matrix} that the width of the mass transition has been under-estimated by the weighted least squares approach, probably because of the large error bar on one of the data points lying within the transition. A similar effect applies to the other strongly over-estimated data points in Fig.~\ref{fig:p_1}, and as discussed above this comes down to limitations in using a weighted least squares fit.  

The result of the constant fit, which is motivated by the theoretical prediction of Eq.~(\ref{p_1}), using the solid data points only, is $p_1=(2.23\pm0.02)$, which is shown by the dashed line. This shows excellent agreement with the constant fit to the unweighted data of Fig.~\ref{fig:unweighted_p_1}. Again, a linear model was also fitted, shown by the solid line, and is given by:
\begin{equation}
    p_1=(0.25\pm0.03)\log_{10}\left(\frac{|\bar{f}_R|}{1+z}\right)+(3.6\pm0.2).
    \label{eq:weighted_p1_fit}
\end{equation}
The gradient is still not in agreement with the prediction of zero. Accounting for higher order effects, e.g., environmental screening and the non-sphericity of haloes, may bring these results into better agreement with the theoretical predictions; however, since we are interested in an empirical fitting function that can be of practical use, we prefer a simple over a sophisticated theoretical model.

As with the unweighted least squares fitting, the validity of these fits of $p_1$ and $p_2$ can be checked through an examination of Fig.~\ref{fig:matrix}. This time the dashed lines are produced using Eq.~(\ref{eq:mdyn_enhancement}) along with the linear fit of $p_2$ from Fig.~\ref{fig:p_2} (solid line) and the constant fit of $p_1$ from Fig.~\ref{fig:p_1} (dashed line), which are motivated by theory. Agreement is now not quite as strong between the dashed and solid lines as in Fig.~\ref{fig:unweighted_matrix}, though still very good for most snapshots shown. Disagreement with the simulation data still exists for F4, which partly results from the lack of high-mass haloes and the flatness of the data in the unscreened regime, as for the unweighted approach. However, in Fig.~\ref{fig:matrix}, disparities in F4 also result from the limitations in the weighted least squares fit in finding agreement with mass bins of large error, and this affects other models as well. Examples include the Jade F4.5 snapshots, Jade F5.5 $z=1.00$ and Diamond F6.5 $z=0.00$. In these panels the theoretical dashed line fits actually show better agreement with the simulation data than the solid lines, as they depend on fits from all snapshots and are therefore effectively not error bar dependent. On the whole, the dashed lines show excellent agreement with the simulation data, providing further validation of the analytical model given by Eqs.~(\ref{eq:mdyn_enhancement}), (\ref{p_1}) and (\ref{p_2}), even if agreement is not quite as strong as for the unweighted approach.

\begin{figure*}
\centering
\includegraphics[width=1.0\textwidth]{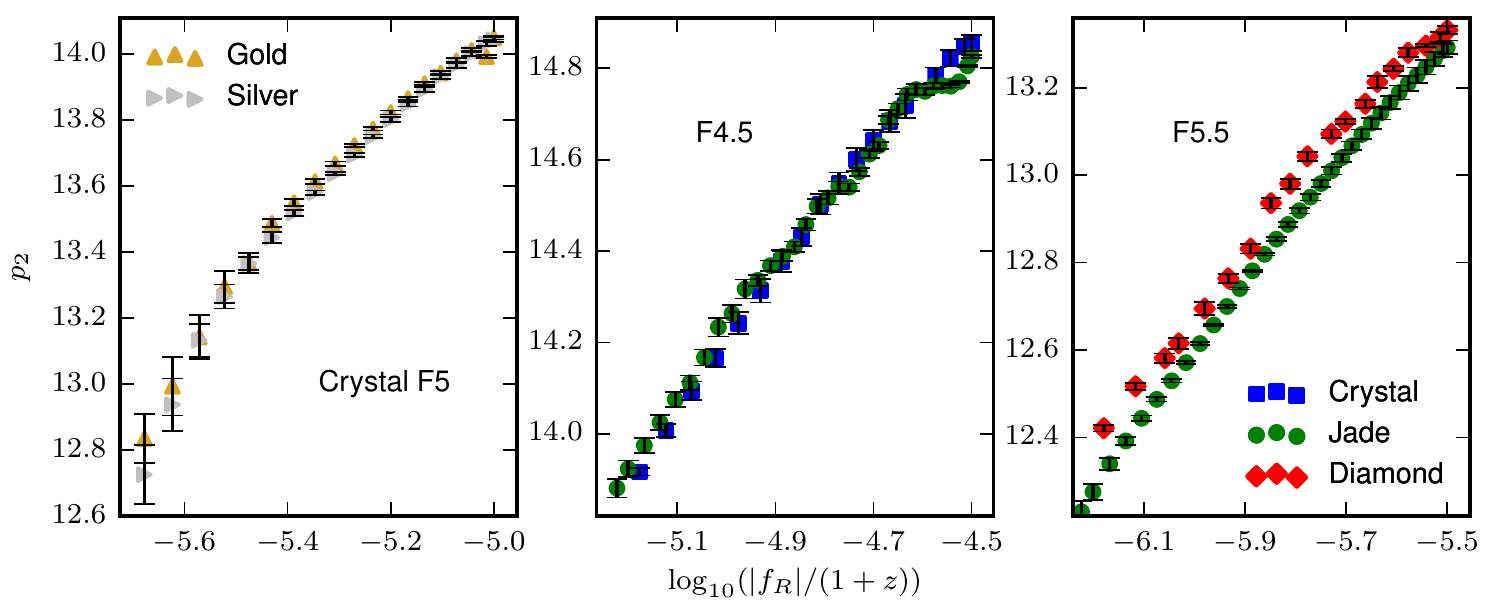} 
\caption[Comparison of measurements of the parameter $p_2$ of Eq.~(\ref{eq:mdyn_enhancement}), using different simulations and datasets.]{Parameter $p_2$ in Eq.~(\ref{eq:mdyn_enhancement}) plotted as a function of the background scalar field at redshift $z$, $\bar{f}_R(z)$, divided by $(1+z)$, for Hu-Sawicki $f(R)$ gravity with $n=1$. $p_2$ is measured via an unweighted least squares optimization of Eq.~(\ref{eq:mdyn_enhancement}) to data from modified \textsc{ecosmog} simulations, described by Table \ref{table:simulations}, at simulation snapshots with redshift $z<1$. The one standard deviation error bars are included. \textit{Left to right}: comparison of gold and silver data from the Crystal simulation with present day scalar field value $|f_{R0}|=10^{-5}$; comparison of the Crystal and Jade data with $|f_{R0}|=10^{-4.5}$;  comparison of the Jade and Diamond data with $10^{-5.5}$. The legend in the right plot applies to both the middle and right plots.}
\label{fig:consistency_tests}
\end{figure*}

\section{Consistency tests}
\label{appendix:mdyn:consistency}

As was explained in the main text, the issue of an insufficient mass range is resolved through the use of three simulations with varying resolutions, whereas the use of silver data ensures an extended set of present-day scalar field values from $|f_{R0}|=10^{-4}$ right down to $|f_{R0}|=10^{-6.5}$. This allows the theoretical model to be rigorously tested for all present-day field strengths in this range, not just for F4, F5 and F6, for which full simulation data are available. 

The validity of using silver data was tested by generating F5 silver data from the Crystal simulation $\Lambda$CDM data, to be directly compared with the F5 gold data from the same simulation. A comparison of the values of the Eq.~(\ref{eq:mdyn_enhancement}) parameter $p_2$ is shown in the left panel of Fig.~\ref{fig:consistency_tests}, where the percentage error is measured at around $0.1\%$ for the unweighted approach. This is clearly low enough so that the use of silver data is justified. Physically, this makes sense, because major differences between a full $f(R)$ simulation (used to generate gold data) and its $\Lambda$CDM counterpart (used to generate silver data) include the halo density profile and halo mass, but the difference is generally small enough to not have a strong impact on the scalar field profile. The averaging of the halo mass distribution in the top-hat approximation is shown to be a very good approximation, and further makes the differences in the halo density profiles irrelevant from the point of view of thin-shell modelling.

When combining simulations of different resolutions, the dispersion between these simulations can also lead to a significant systematic source of uncertainty. This can be tested by looking at a few model parameters $f_{R0}$ for which the mass range necessary to fit $p_1$ and $p_2$ as a function of $\bar{f}_R(z)/(1+z)$ for $0\leq z\leq1$ is offered by simulations of different resolutions. In the middle panel of Fig.~\ref{fig:consistency_tests} the Crystal and Jade simulations are compared for F4.5, and found to agree to within an accuracy of $0.3\%$. A similar test on the Jade and Diamond simulations for F5.5 yielded an error of $0.4$-$0.8\%$ (right panel of Fig.~\ref{fig:consistency_tests}). These agreements are good 
enough that the disparity between the results of the simulations is negligible and combination of different simulations is justified. Note that these two checks are also done using the unweighted least squares approach. 

A limitation of the current study is that we do not have simulations that allow us to test the fitting functions of $p_1$ and $p_2$ for other cosmological parameters, such as $\Omega_{\rm M}$ and
$\sigma_8$, as these are fixed in the original simulations and cannot be changed for producing the silver data. While this is something that would be good to explicitly check in future work, we believe that the excellent agreement between the physically motivated thin-shell modelling and the simulation data, in spite of the approximations employed, has indicated that the theoretical model has successfully captured the essential physics. Therefore we expect the fitting functions we found in this paper to apply to other values of $\Omega_{\rm M}$ and $\sigma_8$ as well. For example, in the paragraph below Eqs.~(\ref{p_1}, \ref{p_2}) we have discussed that, according to the thin-shell model, $p_1$ and $p_2$ should depend only on Newton's constant $G$ and the halo mass definition $\Delta$ (with $H_0=100h$~kms$^{-1}$Mpc$^{-1}$), and in particular they do not depend on cosmological parameters such as $\Omega_m$ and $\sigma_8$. Note that varying $\Omega_{\rm M}$ and $\sigma_8$ will modify the halo abundances and density profiles, and in the check of silver vs.~gold data above we have already confirmed that slight changes to these quantities do not affect our fitting functions noticeably.

Another check that is not included in this study is whether the fitting functions work for forms of $f(R)$ other than Hu-Sawicki as well. While a detailed investigation of this is of interest, we do not find a compelling justification to make substantial effort to include it here, for two reasons. First, as for the case of varying $\Omega_{\rm M}$ and $\sigma_8$, the effects on the modelling of $M_{\rm dyn}/M_{\rm true}$ through a modified halo abundance and density profile are expected to be small/negligible. Second, the different $f(R)$ models generally have a different transition from screened to unscreened regimes, though the details of this transition depends on the model itself and its parameters. This indicates that, even though the slope of $p_2$, which is $1.5$, is expected to remain for general $f(R)$ models, the intercept of $p_2$ could be model dependent. For $p_1$, which denotes how the transition from screened to unscreened regimes takes place, the discussion after Eqs.~(\ref{p_1}, \ref{p_2}) implies it does not depend on the details of $f(R)$, though more explicit checks using simulations are necessary to confirm this or to calibrate its (probably constant) value for general $f(R)$ models. As mentioned above, it is not feasible to do simulations for all possible models. And nor is this necessary, given that any $f(R)$ model studied in a cosmological context is phenomenological and not fundamental, and the focus should really be how to get precise stringent constraints on a representative example, which can then be interpreted in the context of general cases.

\chapter{\boldmath A self-consistent pipeline for unbiased constraints of \texorpdfstring{$f(R)$}{f(R)} gravity}
\label{appendix:pipeline}

\section{Modelling the dynamical mass scatter}
\label{sec:appendix:pipeline:mdyn_scatter}

In Fig.~\ref{fig:rms_scatter}, the data points show the binned mass ratio scatter as a function of the rescaled logarithmic mass,  $\log_{10}(M_{500}M_{\odot}^{-1}h)-p_2=\log_{10}(M_{500}/10^{p_2})$. To generate this, we have evaluated the difference between the actual dynamical mass enhancement and the value predicted by Eq.~(\ref{eq:pipeline:mdyn_enhancement}) for each halo, and measured the root-mean-square difference within the same mass bins as used to fit Eq.~(\ref{eq:pipeline:mdyn_enhancement}) in Chapter \ref{chapter:mdyn}. We have modelled this data using a 6-parameter fitting formula which is made up of two parts. A skewed normal distribution is used to capture the shape of the peak: this includes parameters for the normalisation $\lambda_{\rm s}$, the position $\mu_{\rm s}$ and width $\sigma_{\rm s}$ with respect to the $x=\log_{10}(M_{500}/10^{p_2})$ axis, and a parameter $\alpha$ quantifying the skewness. On its own, this distribution would fall to zero at both low and high $x$; however, we see from Fig.~\ref{fig:rms_scatter} that the scatter is slightly greater on average at high $x$ than at low $x$. To account for this, we add on a $\tanh$ function with two parameters: an amplitude $\lambda_{\rm t}$ and a shift $y_{\rm t}$ along the vertical axis. Our full model is then given by:
\begin{equation}
\sigma_{\mathcal{R}} = \frac{\lambda_{\rm s}}{\sigma_{\rm s}}\phi(x')\left[1+\rm{erf}\left(\frac{\alpha x'}{\sqrt[]{2}}\right)\right] + \left(\lambda_{\rm t}\tanh(x)+y_{\rm t}\right),
\label{eq:scatter_model}
\end{equation}
where $x'=(x-\mu_{\rm s})/\sigma_{\rm s}$. $\phi(x')$ represents the normal distribution:
\begin{equation}
\phi(x') = \frac{1}{\sqrt[]{2\pi}}\exp\left(-\frac{x'^2}{2}\right),
\label{eq:appendix:normal_dist}
\end{equation}
and ${\rm erf}(x')$ is the error function:
\begin{equation}
{\rm erf}(x') = \frac{2}{\sqrt[]{\pi}}\int_0^{x'}e^{-t^2}{\rm d}t.
\label{eq:err_func}
\end{equation}
Since we have many more data points at higher masses than at lower masses in Fig.~\ref{fig:rms_scatter}, we have used a weighted least squares approach which ensures that different parts of the $\log_{10}(M_{500}/10^{p_2})$ range have an equal contribution to the fitting of Eq.~(\ref{eq:scatter_model}). To do this, we have split the rescaled mass range into 10 equal-width bins and counted the number, $N_i$, of data points within each bin $i$. In the least squares fitting, each data point is then weighted by $1/N_i$. This means that points found at lower masses, where there are fewer data points, are each given a greater weight than points found at higher masses. The resulting best-fit parameter values are: $\lambda_{\rm s}=0.0532\pm0.0008$, $\sigma_{\rm s}=0.58\pm0.03$, $\mu_{\rm s}=-0.35\pm0.03$, $\alpha=1.09\pm0.18$, $\lambda_{\rm t}=0.0012\pm0.0003$ and $y_{\rm t}=0.0019\pm0.0002$.

\section{Mass conversions}
\label{sec:appendix:pipeline:mass_conversions}

The following formula can be used to convert the HMF from mass definition $M_{\Delta}$ to a new definition $M_{\Delta'}$:
\begin{equation}
    n'(M_{\Delta'}) = n(M_{\Delta}(M_{\Delta'}))\left(\frac{{\rm d}\ln M_{\Delta'}}{{\rm d} \ln M_{\Delta}}\right)^{-1},
    \label{eq:hmf_conversion}
\end{equation}
where $n'$ is the HMF in the new mass definition and $n$ is the HMF in the old definition. This requires a relation between the mass definitions. For this, we use the following \citep{Hu:2002we}:
\begin{equation}
    \frac{M_{\Delta}}{M_{200}} = \frac{\Delta}{200}\left(\frac{c_{\Delta}}{c_{200}}\right)^3,
    \label{eq:mass_conversion}
\end{equation}
where $c_{\Delta}$ is the concentration with respect to generic overdensity $\Delta$. The latter can be computed from $c_{200}$ using:
\begin{equation}
    \frac{1}{c_{\Delta}} = x\left[f_{\Delta}=\frac{\Delta}{200}f\left(\frac{1}{c_{200}}\right)\right],
    \label{eq:c_delta}
\end{equation}
where the function $f(x)$ is given by:
\begin{equation}
    f(x) = x^3\left[\ln(1+x^{-1}) - (1+x)^{-1}\right].
    \label{eq:f_of_x}
\end{equation}
Eq.~(\ref{eq:c_delta}) is computed using the inverse of this function. \citet{Hu:2002we} provide an analytical formula which can accurately solve this:
\begin{equation}
    x(f) = \left[a_1f^{2p} + \left(\frac{3}{4}\right)^2\right]^{-\frac{1}{2}} + 2f,
    \label{eq:x_of_f}
\end{equation}
where $p = a_2 + a_3\ln f + a_4(\ln f)^2$ and the parameters have values $a_1=0.5116$, $a_2=-0.4283$, $a_3=-3.13\times10^{-3}$ and $a_4=-3.52\times10^{-5}$. The authors state that this formula has $\sim0.3\%$ accuracy for galaxy and cluster scales.

\section{Test of the constraint pipeline on a stronger \texorpdfstring{$\lowercase{f}(R)$}{f(R)} model}
\label{sec:appendix:pipeline:F4.5}

\begin{figure*}
\centering
\includegraphics[width=\textwidth]{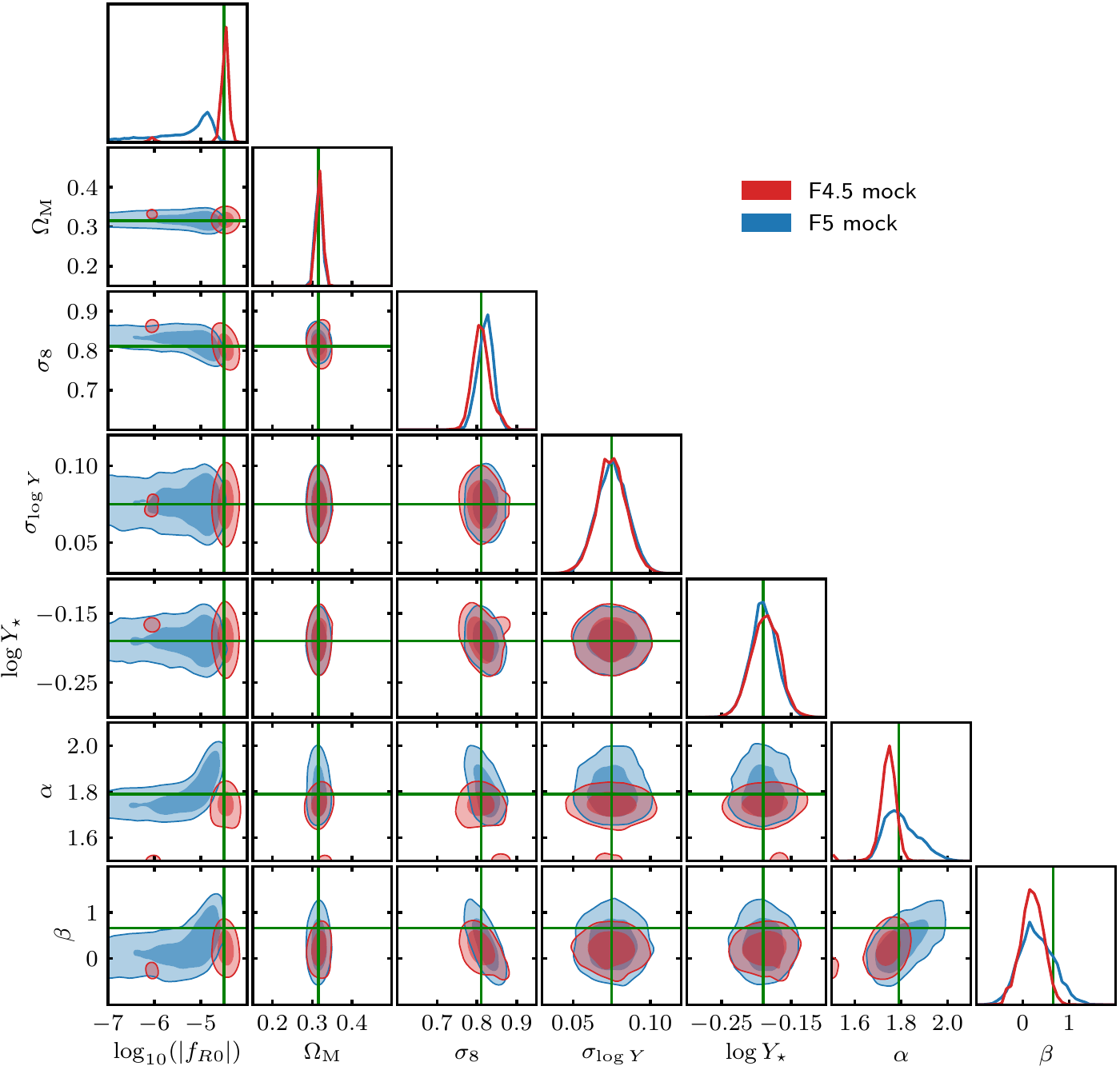}
\caption[Parameter constraints using our full $f(R)$ pipeline for both F4.5 and F5 fiducial cosmologies.]{Parameter constraints obtained by applying our full $f(R)$ pipeline, as detailed in Sec.~\ref{sec:methods_pipeline}, to F4.5 (\textit{red}) and F5 (\textit{blue}) mocks, with observational flux threshold $Y_{\rm cut}=1.5\times10^{-5}{\rm Mpc}^2$. The dark and light regions of the contours represent 68\% and 95\% confidences, respectively. The marginalised distributions of the sampled parameter values are shown in the top panels of each column. The fiducial cosmological parameter values of the F4.5 mock are indicated by the green lines.}
\label{fig:F4.5_constraints}
\end{figure*}

For the main results of Chapter \ref{chapter:constraint_pipeline}, we have tested our constraint pipeline using GR and F5 mocks. For the F5 mock (cf.~Fig.~\ref{fig:full_fr_pipeline}), our pipeline produces a marginalised distribution of $\log_{10}|f_{R0}|$ which peaks close to $-5$, but features a long tail extending to $-7$, which is the lowest value of $\log_{10}|f_{R0}|$ considered in this work. As discussed in Sec.~\ref{sec:results_pipeline:fr_pipeline}, this can be explained by parameter degeneracies, which can make it more difficult to fully distinguish this model from GR.

To check whether our pipeline can successfully distinguish stronger $f(R)$ models than F5, and whether such models suffer from the same degeneracies, we show, in Fig.~\ref{fig:F4.5_constraints}, constraints obtained using an F4.5 ($\log_{10}|f_{R0}|=-4.5$) mock along with the F5 results from Fig.~\ref{fig:full_fr_pipeline}. The F4.5 constraint features smaller contours and a tight peak at $\log_{10}|f_{R0}|\approx-4.5$ which does not feature long tails towards lower or higher values of $\log_{10}|f_{R0}|$. The median and 68\% range is given by $-4.47^{+0.06}_{-0.07}$, which is in excellent agreement with the fiducial value of $-4.50$. This indicates that our pipeline can clearly distinguish different values of $|f_{R0}|$ and it provides further evidence that it can distinguish $f(R)$ models from GR in an unbiased manner.

\chapter{Cluster and halo properties in DGP gravity}
\label{appendix:dgp_clusters}

\section{Simulation consistency}
\label{sec:appendix:dgp_clusters:consistency}

\begin{figure*}
\centering
\includegraphics[width=1.0\textwidth]{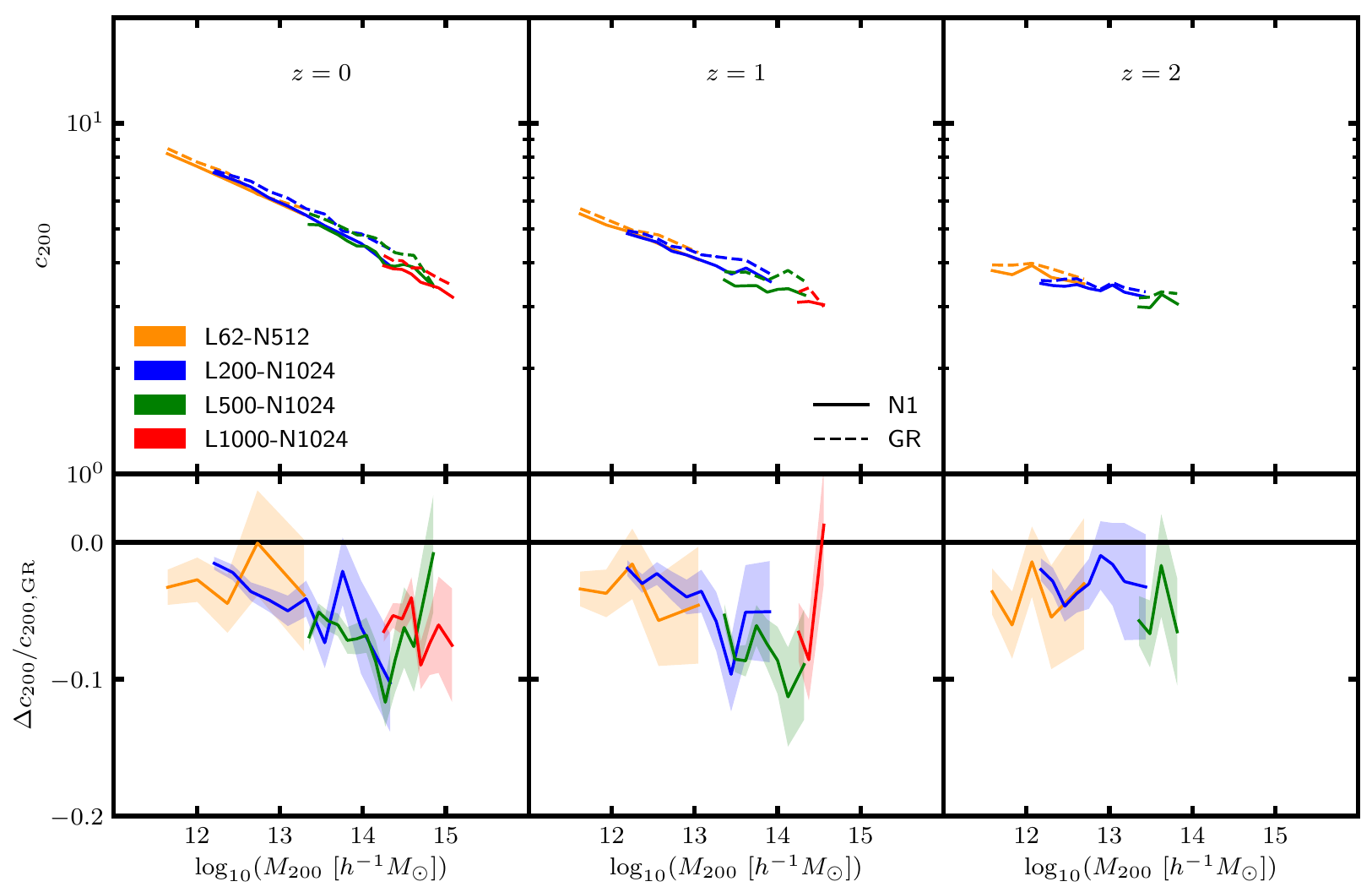}
\caption[Comparison of measurements of the median halo concentration in GR and N1 using simulations with different resolutions.]{Median halo concentration (\textit{top row}) and relative difference with respect to GR (\textit{bottom row}) as a function of the mean logarithm of the halo mass at redshifts $0$, $1$ and $2$. The data is generated using the dark-matter-only simulations L62 (\textit{orange}), L200 (\textit{blue}), L500 (\textit{green}) and L1000 (\textit{red}), the specifications of which are given in Table \ref{table:simulations}. Data is shown for GR (\textit{dashed lines}) and the nDGP model N1 (\textit{solid lines}). The shaded regions in the lower panels show the $1\sigma$ uncertainty in the relative difference.}
\label{fig:c_consistency}
\end{figure*}

\begin{figure*}
\centering
\includegraphics[width=1.0\textwidth]{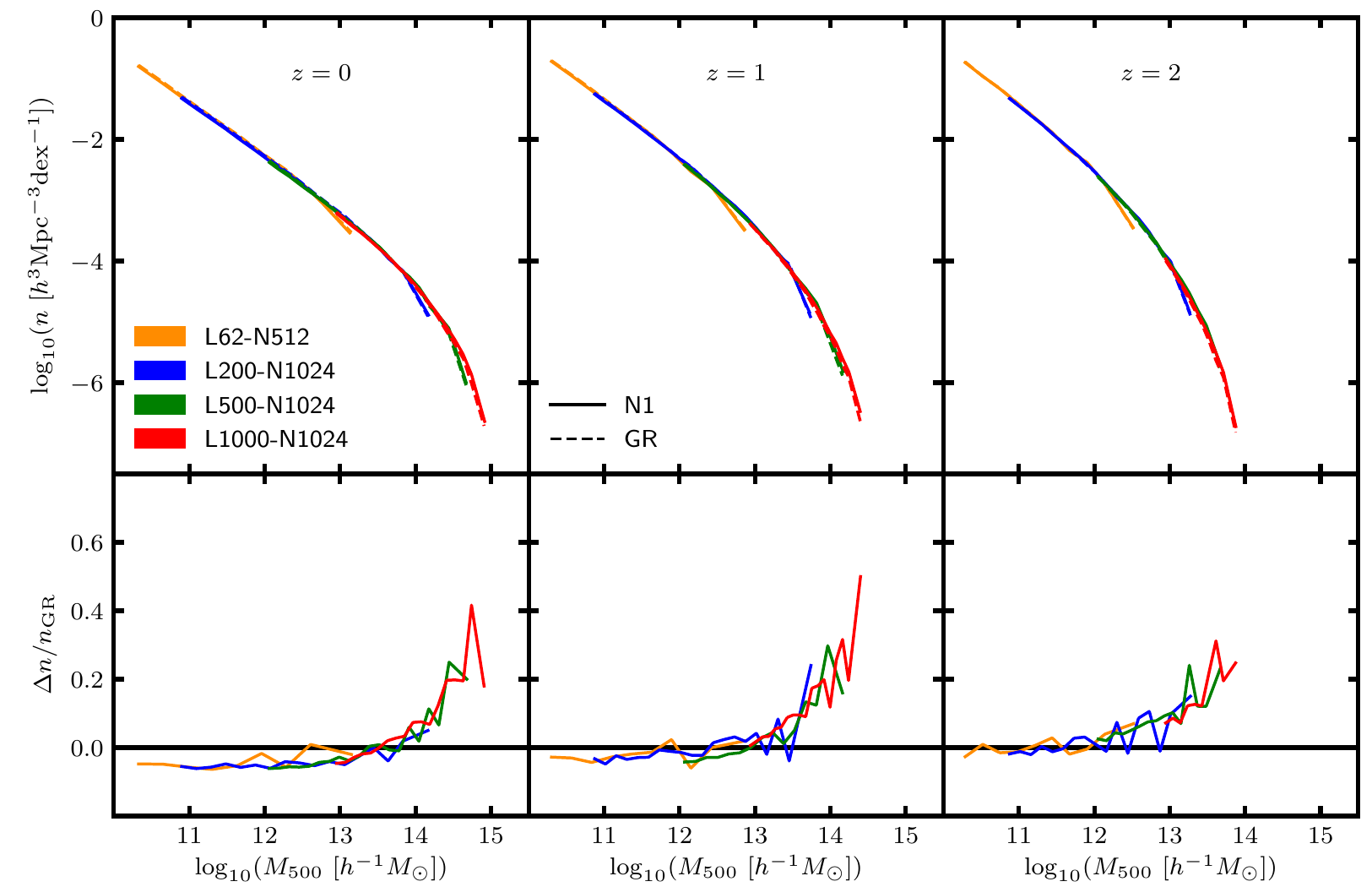}
\caption[Comparison of measurements of the halo mass function in GR and N1 using simulations with different resolutions.]{Halo mass function (\textit{top row}) and its relative difference in nDGP with respect to GR (\textit{bottom row}), as a function of the mean logarithm of the halo mass at redshifts $0$, $1$ and $2$. The data is generated using the dark-matter-only simulations L62 (\textit{orange}), L200 (\textit{blue}), L500 (\textit{green}) and L1000 (\textit{red}), the specifications of which are given in Table \ref{table:simulations}. Data is shown for GR (\textit{dashed lines}) and the nDGP model N1 (\textit{solid lines}).}
\label{fig:hmf_consistency}
\end{figure*}

In Secs.~\ref{sec:results:dgp:concentration} and \ref{sec:results:dgp:hmf}, we combined the halo data from our DMO simulations in order to study the effects of nDGP on the halo concentration and the HMF over a wide mass range. In doing this, it is important to verify that the data from the simulations, which have different resolutions, are consistent. We therefore show, in Figs.~\ref{fig:c_consistency} and \ref{fig:hmf_consistency}, the concentration and HMF data, respectively, from each of our DMO simulations for GR and N1.

In Fig.~\ref{fig:c_consistency}, we show the binned concentration from all four of our DMO simulations, including L1000 which was excluded from our results in Sec.~\ref{sec:results:dgp:concentration}. At redshift 0, where the simulations all have sufficient resolution, the concentration follows a smooth power-law relation as a function of the mass, with the simulations showing excellent agreement at overlapping masses for both GR and N1. The agreement is not as strong at redshifts 1 and 2, where we see, for example, gaps between the L200 (blue) and L500 (green) concentrations. The concentration is slightly underestimated for haloes that are not well-resolved, affecting the data at the low-mass end (close to the lower mass cut of 2000 particles) of the L500 and L1000 data at $z=1$ and the L200 and L500 data at $z=2$. 

These resolution issues are potentially problematic for studies of the absolute concentration; however, in this work, we are more interested in the relative difference between the nDGP and GR concentration. From the lower panels of Fig.~\ref{fig:c_consistency}, it appears that the L62, L200 and L500 simulations give consistent predictions of the relative difference at overlapping masses for each redshift shown. This justifies using a halo mass cut of 2000 particles to study and model the relative difference in Sec.~\ref{sec:results:dgp:concentration}. This cut ensures that there are plenty of haloes at overlapping masses, which is important for the combined binning of the halo data, while it does not give rise to inconsistencies in the relative difference for these three simulations. We decided to exclude the L1000 simulation for a couple of reasons: the concentration suppression does not appear to be fully consistent with the data from the higher-resolution simulations -- for example, at $z=0$, the suppression in L1000 appears to be lower than the predictions from L500 at low masses and greater at high masses -- and at higher redshifts it does not have many resolved haloes.

In Fig.~\ref{fig:hmf_consistency}, we show the binned HMF from DMO simulations. The predictions of the absolute HMF, shown in the top row, agree very well. The HMF is slightly underestimated at the high-mass end of each simulation: this is a natural consequence of the limited box sizes, which causes the high-mass HMF to be incomplete. We note that combining the halo data of the four simulations and summing the total volume in the way that we have described in Sec.~\ref{sec:results:dgp:hmf} means that incompleteness is only really present for the highest-mass bins shown in Fig.~\ref{fig:hmf_combined}. The lower panels of Fig.~\ref{fig:hmf_consistency} show the relative differences between GR and N1. The predictions from the four simulations show excellent agreement, again indicating that these simulations can be safely combined.

\section{\texorpdfstring{$M_{200}$}{M200} mass function}
\label{sec:appendix:dgp_clusters:hmf}

\begin{figure*}
\centering
\includegraphics[width=1.0\textwidth]{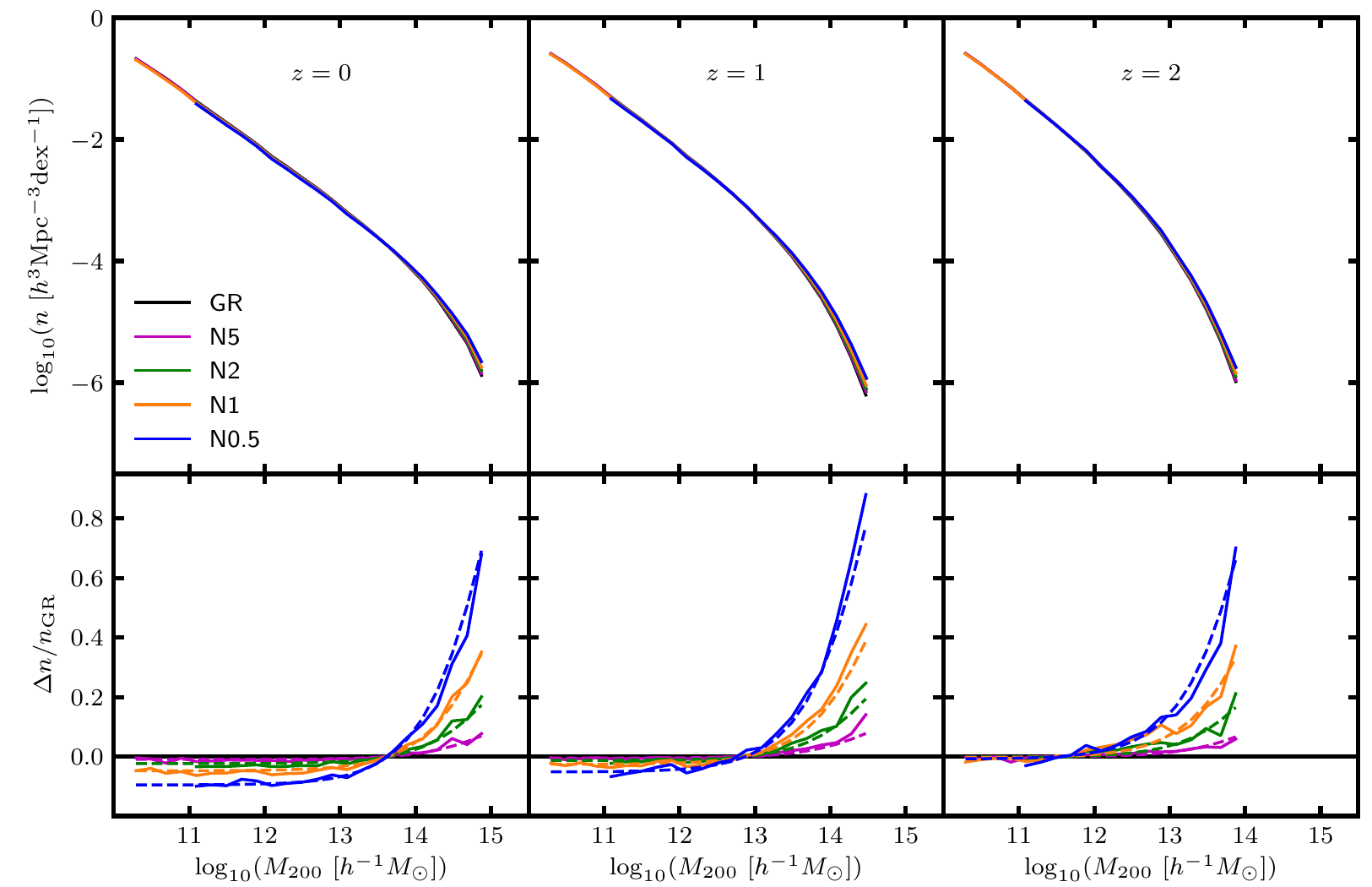}
\caption[Halo mass function in nDGP and GR, measured and fitted using mass definition $M_{200}$.]{Halo mass function (\textit{top row}) and its relative difference in nDGP with respect to GR (\textit{bottom row}), as a function of the mean logarithm of the halo mass at redshifts $0$, $1$ and $2$. The results shown are similar to Fig.~\ref{fig:hmf_combined}; however, here we use mass definition $M_{200}$ instead of $M_{500}$, and the dashed lines show the predictions from the model given by Eqs.~(\ref{eq:hmf}, \ref{eq:hmf_params_200}).}
\label{fig:hmf_combined_200c}
\end{figure*}

In Sec.~\ref{sec:results:dgp:hmf}, we presented our results and model for the nDGP HMF in terms of the $M_{500}$ mass definition. For completeness, we also show, in Fig.~\ref{fig:hmf_combined_200c}, the HMF in terms of the $M_{200}$ mass definition. This has again been calculated by combining the haloes from all four DMO simulations, although here we impose a lower mass threshold of 100 particles within the radius $R_{200}$ rather than $R_{500}$. We use the same set of logarithmic mass bins (with fixed width 0.2) and again show all bins that contain at least 100 haloes. 

The results in Fig.~\ref{fig:hmf_combined_200c} are very similar to Fig.~\ref{fig:hmf_combined}, with the nDGP fifth force suppressing the HMF at lower masses and enhancing the HMF at higher masses. Therefore, we are able to use the same fitting formula to model the relative difference. Replacing $M_{500}$ with $M_{200}$ in Eq.~(\ref{eq:hmf}), the best-fit parameter 
are now:
\begin{equation}
    \begin{split}
    &A(H_0r_{\rm c}) = (0.59\pm0.03){\left(H_0r_{\rm c}\right)^{-1}},\\
    &B(z) = (15.22\pm0.03) - (0.441\pm0.006)z,\\
    &C(z) = (0.919\pm0.005) + (0.037\pm0.003)z.
    \end{split}
    \label{eq:hmf_params_200}
\end{equation}
The predictions of this model are also in very good agreement with the simulation measurement. As in the case of $M_{500}$, we note that this model should only be used to predict the HMF within the mass range $10^{11}h^{-1}M_{\odot}\lesssim M_{200}\lesssim M_{\rm max}(z)$, where $M_{\rm max}(z)$ is the maximum mass used for the calibration. For definition $M_{200}$, this can be estimated using:
\begin{equation}
    \log_{10}\left(M_{\rm max}M_{\odot}^{-1}h\right) = 14.93 - 0.52z.
\end{equation}

\chapter{Realistic galaxy formation simulations to study clusters in modified gravity} 
\label{appendix:baryonic_fine_tuning}

In Chapter \ref{chapter:baryonic_fine_tuning}, we presented our new baryonic model for low-resolution, full-physics cosmological simulations. In particular, we focused on the changes that we made to the IllustrisTNG model and described only a small subset of the $\sim200$ calibration runs. In this appendix, we will provide a more detailed description of the calibration procedure, including an outline of our simulations and details of the parameter search.

Table \ref{table:simulations:fine_tuning} shows the specifications of the simulations used to tune the baryonic model. The primary goal of the tuning was to find a model that can produce sufficient star formation in low-resolution simulations to match galaxy observations. We studied in detail simulations with three different mass resolutions before settling on the resolution of the L68-N256 simulations, which have already been mentioned in Chapter \ref{chapter:baryonic_fine_tuning}.

\begin{table*}
\centering

\small
\begin{tabular}{ c@{\hskip 0.5in}cccc } 
 \toprule
 
 Specifications & \multicolumn{4}{c}{Simulations} \\
  & L100-N256 & L86-N256 & L68-N256 & L136-N512 \\

 \midrule

 box size / $h^{-1}$Mpc & 100 & 86 & 68 & 136 \\ 
 DM particle number & $256^3$ & $256^3$ & $256^3$ & $512^3$ \\
 $m_{\rm DM}$ / $h^{-1}M_{\odot}$ & $4.29\times10^9$ & $2.73\times10^9$ & $1.35\times10^9$ & $1.35\times10^9$ \\
 $m_{\rm gas}$ / $h^{-1}M_{\odot}$ & $8.3\times10^8$ & $5.3\times10^8$ & $2.6\times10^8$ & $2.6\times10^8$ \\
 number of runs & $\sim100$ & $\sim60$ & $\sim50$ & $3$ \\
 
 \bottomrule
 
\end{tabular}

\caption[Specifications of the simulations used to tune our baryonic model.]{Specifications of the \textsc{arepo} simulations that have been used to tune our baryonic model. These are labelled L100-N256, L86-N256, L68-N256 and L136-N512, according to their box size and dark matter particle number (we note that there are initially the same number of gas cells as dark matter particles). The simulations have all been run with standard gravity (GR).}
\label{table:simulations:fine_tuning}

\end{table*}

\section{L100-N256 simulations}

To start with, we used simulations with a box size of $100h^{-1}{\rm Mpc}$, containing $256^3$ dark matter particles and (initially) the same number of gas cells (L100-N256). With an average gas cell mass of $\sim8.3\times10^8h^{-1}M_{\odot}$, these have 512 times lower mass resolution than the simulations used to calibrate the fiducial TNG model and the same resolution as the BAHAMAS simulations \citep{McCarthy:2016mry}, which were run using \textsc{gadget-3} \citep{Springel:2005mi} rather than \textsc{arepo}. We ran $\sim100$ simulations at this resolution, varying the following baryonic parameters: the threshold gas density for star formation $\rho_{\star}$ (see Sec.~\ref{sec:fine_tuning:star_formation_model}) was varied in the range $[0.00,0.13]~{\rm cm^{-3}}$; the parameter $\bar{e}_{\rm w}$ controlling the stellar wind energy (see Eq.~(\ref{eq:wind_energy})) was varied in the range $[0.0,3.6]$; the parameters $\kappa_{\rm w}$ and $v_{\rm w,min}$ controlling the stellar wind speed (see Eq.~(\ref{eq:wind_speed})) were varied over ranges $[0.0,29.6]$ and $[0,500]~{\rm kms^{-1}}$, respectively; and the black hole radiative efficiency $\epsilon_{\rm r}$ (see Sec.~\ref{sec:fine_tuning:bh_feedback}) was varied in the range [0.02,0.20]. We also tested gravitational softening lengths in the range 1/40 to 1/10 times the mean inter-particle separation.

These runs provided a very useful insight into the effects of changing each parameter, however all of the tested parameter combinations at the L100-N256 resolution resulted in insufficient star formation within haloes of mass $M_{200}\lesssim10^{13}M_{\odot}$, and at higher halo masses it was difficult to simultaneously match observations for different galaxy properties. For example, parameter combinations which yielded a sufficiently high stellar mass function typically resulted in the stellar mass fraction being overestimated, and in order to match the SFRD observations it was necessary to set either $\rho_{\star}$ or the stellar wind energy close to zero. We therefore decided to look at higher resolutions.

\section{L86-N256 simulations}


Keeping the dark matter particle number (and initial gas cell number) unchanged, we initially reduced the box size to $86h^{-1}{\rm Mpc}$ (L86-N256), and executed $\sim60$ runs at this higher resolution. Our best model used a gravitational softening length of 1/20 times the mean inter-particle separation and the following parameter combination: $\rho_{\star}=0.05{\rm cm^{-3}}$, $\bar{e}_{\rm w}=0.5$, $\kappa_{\rm w}=2$, $v_{\rm w,min}=200~{\rm kms^{-1}}$ and $\epsilon_{\rm r}=0.15$. This gave a reasonable match with high-mass observations of the stellar mass fraction ($M_{200}\gtrsim2\times10^{12}h^{-1}M_{\odot}$) and stellar mass function ($M_{\star}\gtrsim10^{11}h^{-1}M_{\odot}$), however the agreement was still poor at lower masses and the SFRD was significantly underestimated for redshifts $z\gtrsim2$. We made some efforts to rectify this. For example, we switched off feedback entirely by setting the wind energy to zero and preventing the formation of black hole particles, and we tried using much lower values of $\rho_{\star}$. While these efforts resulted in more star formation at lower masses, it was still not enough to match observational data, while at higher masses and lower redshifts there was far too much star formation.

\section{L68-N256 and L136-N512 simulations}

We finally settled on the $68h^{-1}{\rm Mpc}$ (L68-N256) box size, where we ran a further $\sim50$ runs to calibrate the final model presented in Chapter \ref{chapter:baryonic_fine_tuning}. Our final model, with $\rho_{\star}=0.08{\rm cm^{-3}}$, $\bar{e}_{\rm w}=0.5$, $\epsilon_{\rm r}=0.22$ and a softening length of 1/20 times the mean inter-particle separation, is able to produce sufficient star formation at lower masses and higher redshifts than the above L86-N256 model, and only requires changes to three of the TNG model parameters (we take the TNG values for $\kappa_{\rm w}$ and $v_{\rm w,min}$). The predictions from five of the L68-N256 runs are shown in Fig.~\ref{fig:L68_calibration} to illustrate the effects of each of the changes to the fiducial TNG parameter values.

In order to assess the effects of sample variance, we also ran three simulations with an increased box size of $136h^{-1}{\rm Mpc}$ (L136-N512) and the same mass resolution as the L68-N256 runs. The results from these simulations, which were run using our three most promising baryonic models, indicated that the stellar mass fraction and stellar mass function are slightly reduced in the larger box. This is why we selected the above model, even though it slightly overestimates the stellar mass function in Fig.~\ref{fig:L68_calibration}.

\cleardoublepage

\bibliography{thesis}

%

\cleardoublepage


\end{document}